\documentclass[12pt]{article}
\usepackage{epsfig,amsfonts,amssymb}
\usepackage{hyperref}
\usepackage{cite}
\input epsf.sty
\usepackage{cite}

\newcommand{\m}{m}
\newcommand{\n}{n}

\def\ZZZ{{\hbox{ Z\kern-1.6mm Z}}}
\def\RRR{{\hbox{ R\kern-2.4mm R}}}
\def\CCC{{\hbox{ C\kern-2.0mm C}}}
\def\zzz{{\hbox{z\kern-1mm z}}}

\newcommand{\vt}{\vartheta}

\newcommand{\qeq}{{\hbox{=\kern-2.3mm ? \kern.5mm }}}
\renewcommand{\qeq}{=}

\newcommand{\eps}{\epsilon}

\newcommand{\ra}{\rangle}
\newcommand{\la}{\langle}

\newcommand{\vp}{\varphi}

\newcommand{\VV}{{\cal V}}
\newcommand{\BB}{{\cal B}}

\newcommand{\AAA}{{\cal A}}
\newcommand{\GG}{{\cal G}}

\newcommand{\HH}{{\cal H}}
\newcommand{\MM}{{\cal M}}

\newcommand{\OO}{{\cal O}}

\newcommand{\PP}{{\cal P}}
\newcommand{\EE}{{\cal E}}
\newcommand{\LL}{{\cal L}}

\newcommand{\XX}{{\cal X}}

\newcommand{\wt}{\widetilde}
\newcommand{\wh}{\widehat}

\newcommand{\RR}{{\cal R}}

\newcommand{\UU}{{\cal U}}

\newcommand{\SSS}{{\cal S}}

\newcommand{\be}{\begin{equation}}
\newcommand{\ee}{\end{equation}}
\newcommand{\ben}{\begin{eqnarray}\displaystyle}
\newcommand{\een}{\end{eqnarray}}

\newcommand{\refb}[1]{(\ref{#1})}
\newcommand{\p}{\partial}
\newcommand{\sectiono}[1]{\section{#1}\setcounter{equation}{0}}

\def\one{{\hbox{ 1\kern-.8mm l}}}
\def\zero{{\hbox{ 0\kern-1.5mm 0}}}

\newcommand{\bea}[1]{\begin{eqnarray}\label{#1} }
\newcommand{\eea}{\end{eqnarray}}

\newcommand{\eqref}{\refb}

\newcommand{\PPP}{{\cal P}}

\newcommand{\Chi}{\Pi}

\usepackage{bm}
\usepackage[table]{xcolor}

\newcommand{\cL} {\{\hskip -4pt\{}
\newcommand{\cR} {\}\hskip -4pt\}}

\newcommand{\oR}{{\overline{\RR}}}

\def\figpgmn{

\def\JPicScale{0.8}
\ifx\JPicScale\undefined\def\JPicScale{1}\fi
\unitlength \JPicScale mm
\begin{picture}(150,80)(0,0)
\linethickness{0.3mm}
\put(10,40){\line(1,0){140}}
\linethickness{0.3mm}
\put(80,40){\line(0,1){40}}
\linethickness{0.3mm}
\multiput(145,45)(0.12,-0.12){42}{\line(1,0){0.12}}
\linethickness{0.3mm}
\multiput(145,35)(0.12,0.12){42}{\line(1,0){0.12}}
\linethickness{0.3mm}
\multiput(75,75)(0.12,0.12){42}{\line(1,0){0.12}}
\linethickness{0.3mm}
\multiput(80,80)(0.12,-0.12){42}{\line(1,0){0.12}}
\put(145,32){\makebox(0,0)[cc]{${\cal M}_{g,m,n}$}}

\put(80,85){\makebox(0,0)[cc]{Local coordinates and PCO locations}}

\put(125,70){\makebox(0,0)[cc]{$\wt {\cal P}_{g,m,n}$}}

\end{picture}

}

\def\figC{

\def\JPicScale{0.5}
\ifx\JPicScale\undefined\def\JPicScale{1}\fi
\unitlength \JPicScale mm

}

\def\figBB{

\def\JPicScale{0.6}
\ifx\JPicScale\undefined\def\JPicScale{1}\fi
\unitlength \JPicScale mm
\begin{picture}(74.24,80)(0,0)
\linethickness{0.3mm}
{\put(74.24,44.75){\line(0,1){0.5}}
\multiput(74.22,45.76)(0.01,-0.5){1}{\line(0,-1){0.5}}
\multiput(74.2,46.26)(0.03,-0.5){1}{\line(0,-1){0.5}}
\multiput(74.16,46.76)(0.04,-0.5){1}{\line(0,-1){0.5}}
\multiput(74.11,47.26)(0.05,-0.5){1}{\line(0,-1){0.5}}
\multiput(74.04,47.76)(0.07,-0.5){1}{\line(0,-1){0.5}}
\multiput(73.96,48.26)(0.08,-0.5){1}{\line(0,-1){0.5}}
\multiput(73.87,48.75)(0.09,-0.5){1}{\line(0,-1){0.5}}
\multiput(73.77,49.25)(0.1,-0.49){1}{\line(0,-1){0.49}}
\multiput(73.65,49.74)(0.12,-0.49){1}{\line(0,-1){0.49}}
\multiput(73.52,50.22)(0.13,-0.49){1}{\line(0,-1){0.49}}
\multiput(73.37,50.71)(0.14,-0.48){1}{\line(0,-1){0.48}}
\multiput(73.22,51.18)(0.16,-0.48){1}{\line(0,-1){0.48}}
\multiput(73.05,51.66)(0.17,-0.47){1}{\line(0,-1){0.47}}
\multiput(72.87,52.13)(0.09,-0.24){2}{\line(0,-1){0.24}}
\multiput(72.68,52.59)(0.1,-0.23){2}{\line(0,-1){0.23}}
\multiput(72.47,53.05)(0.1,-0.23){2}{\line(0,-1){0.23}}
\multiput(72.26,53.51)(0.11,-0.23){2}{\line(0,-1){0.23}}
\multiput(72.03,53.96)(0.11,-0.22){2}{\line(0,-1){0.22}}
\multiput(71.79,54.4)(0.12,-0.22){2}{\line(0,-1){0.22}}
\multiput(71.53,54.84)(0.13,-0.22){2}{\line(0,-1){0.22}}
\multiput(71.27,55.27)(0.13,-0.21){2}{\line(0,-1){0.21}}
\multiput(71,55.69)(0.14,-0.21){2}{\line(0,-1){0.21}}
\multiput(70.71,56.1)(0.14,-0.21){2}{\line(0,-1){0.21}}
\multiput(70.42,56.51)(0.15,-0.2){2}{\line(0,-1){0.2}}
\multiput(70.11,56.91)(0.1,-0.13){3}{\line(0,-1){0.13}}
\multiput(69.79,57.3)(0.11,-0.13){3}{\line(0,-1){0.13}}
\multiput(69.47,57.69)(0.11,-0.13){3}{\line(0,-1){0.13}}
\multiput(69.13,58.06)(0.11,-0.12){3}{\line(0,-1){0.12}}
\multiput(68.78,58.43)(0.12,-0.12){3}{\line(0,-1){0.12}}
\multiput(68.43,58.78)(0.12,-0.12){3}{\line(1,0){0.12}}
\multiput(68.06,59.13)(0.12,-0.12){3}{\line(1,0){0.12}}
\multiput(67.69,59.47)(0.12,-0.11){3}{\line(1,0){0.12}}
\multiput(67.3,59.79)(0.13,-0.11){3}{\line(1,0){0.13}}
\multiput(66.91,60.11)(0.13,-0.11){3}{\line(1,0){0.13}}
\multiput(66.51,60.42)(0.13,-0.1){3}{\line(1,0){0.13}}
\multiput(66.1,60.71)(0.2,-0.15){2}{\line(1,0){0.2}}
\multiput(65.69,61)(0.21,-0.14){2}{\line(1,0){0.21}}
\multiput(65.27,61.27)(0.21,-0.14){2}{\line(1,0){0.21}}
\multiput(64.84,61.53)(0.21,-0.13){2}{\line(1,0){0.21}}
\multiput(64.4,61.79)(0.22,-0.13){2}{\line(1,0){0.22}}
\multiput(63.96,62.03)(0.22,-0.12){2}{\line(1,0){0.22}}
\multiput(63.51,62.26)(0.22,-0.11){2}{\line(1,0){0.22}}
\multiput(63.05,62.47)(0.23,-0.11){2}{\line(1,0){0.23}}
\multiput(62.59,62.68)(0.23,-0.1){2}{\line(1,0){0.23}}
\multiput(62.13,62.87)(0.23,-0.1){2}{\line(1,0){0.23}}
\multiput(61.66,63.05)(0.24,-0.09){2}{\line(1,0){0.24}}
\multiput(61.18,63.22)(0.47,-0.17){1}{\line(1,0){0.47}}
\multiput(60.71,63.37)(0.48,-0.16){1}{\line(1,0){0.48}}
\multiput(60.22,63.52)(0.48,-0.14){1}{\line(1,0){0.48}}
\multiput(59.74,63.65)(0.49,-0.13){1}{\line(1,0){0.49}}
\multiput(59.25,63.77)(0.49,-0.12){1}{\line(1,0){0.49}}
\multiput(58.75,63.87)(0.49,-0.1){1}{\line(1,0){0.49}}
\multiput(58.26,63.96)(0.5,-0.09){1}{\line(1,0){0.5}}
\multiput(57.76,64.04)(0.5,-0.08){1}{\line(1,0){0.5}}
\multiput(57.26,64.11)(0.5,-0.07){1}{\line(1,0){0.5}}
\multiput(56.76,64.16)(0.5,-0.05){1}{\line(1,0){0.5}}
\multiput(56.26,64.2)(0.5,-0.04){1}{\line(1,0){0.5}}
\multiput(55.76,64.22)(0.5,-0.03){1}{\line(1,0){0.5}}
\multiput(55.25,64.24)(0.5,-0.01){1}{\line(1,0){0.5}}
\put(54.75,64.24){\line(1,0){0.5}}
\multiput(54.24,64.22)(0.5,0.01){1}{\line(1,0){0.5}}
\multiput(53.74,64.2)(0.5,0.03){1}{\line(1,0){0.5}}
\multiput(53.24,64.16)(0.5,0.04){1}{\line(1,0){0.5}}
\multiput(52.74,64.11)(0.5,0.05){1}{\line(1,0){0.5}}
\multiput(52.24,64.04)(0.5,0.07){1}{\line(1,0){0.5}}
\multiput(51.74,63.96)(0.5,0.08){1}{\line(1,0){0.5}}
\multiput(51.25,63.87)(0.5,0.09){1}{\line(1,0){0.5}}
\multiput(50.75,63.77)(0.49,0.1){1}{\line(1,0){0.49}}
\multiput(50.26,63.65)(0.49,0.12){1}{\line(1,0){0.49}}
\multiput(49.78,63.52)(0.49,0.13){1}{\line(1,0){0.49}}
\multiput(49.29,63.37)(0.48,0.14){1}{\line(1,0){0.48}}
\multiput(48.82,63.22)(0.48,0.16){1}{\line(1,0){0.48}}
\multiput(48.34,63.05)(0.47,0.17){1}{\line(1,0){0.47}}
\multiput(47.87,62.87)(0.24,0.09){2}{\line(1,0){0.24}}
\multiput(47.41,62.68)(0.23,0.1){2}{\line(1,0){0.23}}
\multiput(46.95,62.47)(0.23,0.1){2}{\line(1,0){0.23}}
\multiput(46.49,62.26)(0.23,0.11){2}{\line(1,0){0.23}}
\multiput(46.04,62.03)(0.22,0.11){2}{\line(1,0){0.22}}
\multiput(45.6,61.79)(0.22,0.12){2}{\line(1,0){0.22}}
\multiput(45.16,61.53)(0.22,0.13){2}{\line(1,0){0.22}}
\multiput(44.73,61.27)(0.21,0.13){2}{\line(1,0){0.21}}
\multiput(44.31,61)(0.21,0.14){2}{\line(1,0){0.21}}
\multiput(43.9,60.71)(0.21,0.14){2}{\line(1,0){0.21}}
\multiput(43.49,60.42)(0.2,0.15){2}{\line(1,0){0.2}}
\multiput(43.09,60.11)(0.13,0.1){3}{\line(1,0){0.13}}
\multiput(42.7,59.79)(0.13,0.11){3}{\line(1,0){0.13}}
\multiput(42.31,59.47)(0.13,0.11){3}{\line(1,0){0.13}}
\multiput(41.94,59.13)(0.12,0.11){3}{\line(1,0){0.12}}
\multiput(41.57,58.78)(0.12,0.12){3}{\line(1,0){0.12}}
\multiput(41.22,58.43)(0.12,0.12){3}{\line(1,0){0.12}}
\multiput(40.87,58.06)(0.12,0.12){3}{\line(0,1){0.12}}
\multiput(40.53,57.69)(0.11,0.12){3}{\line(0,1){0.12}}
\multiput(40.21,57.3)(0.11,0.13){3}{\line(0,1){0.13}}
\multiput(39.89,56.91)(0.11,0.13){3}{\line(0,1){0.13}}
\multiput(39.58,56.51)(0.1,0.13){3}{\line(0,1){0.13}}
\multiput(39.29,56.1)(0.15,0.2){2}{\line(0,1){0.2}}
\multiput(39,55.69)(0.14,0.21){2}{\line(0,1){0.21}}
\multiput(38.73,55.27)(0.14,0.21){2}{\line(0,1){0.21}}
\multiput(38.47,54.84)(0.13,0.21){2}{\line(0,1){0.21}}
\multiput(38.21,54.4)(0.13,0.22){2}{\line(0,1){0.22}}
\multiput(37.97,53.96)(0.12,0.22){2}{\line(0,1){0.22}}
\multiput(37.74,53.51)(0.11,0.22){2}{\line(0,1){0.22}}
\multiput(37.53,53.05)(0.11,0.23){2}{\line(0,1){0.23}}
\multiput(37.32,52.59)(0.1,0.23){2}{\line(0,1){0.23}}
\multiput(37.13,52.13)(0.1,0.23){2}{\line(0,1){0.23}}
\multiput(36.95,51.66)(0.09,0.24){2}{\line(0,1){0.24}}
\multiput(36.78,51.18)(0.17,0.47){1}{\line(0,1){0.47}}
\multiput(36.63,50.71)(0.16,0.48){1}{\line(0,1){0.48}}
\multiput(36.48,50.22)(0.14,0.48){1}{\line(0,1){0.48}}
\multiput(36.35,49.74)(0.13,0.49){1}{\line(0,1){0.49}}
\multiput(36.23,49.25)(0.12,0.49){1}{\line(0,1){0.49}}
\multiput(36.13,48.75)(0.1,0.49){1}{\line(0,1){0.49}}
\multiput(36.04,48.26)(0.09,0.5){1}{\line(0,1){0.5}}
\multiput(35.96,47.76)(0.08,0.5){1}{\line(0,1){0.5}}
\multiput(35.89,47.26)(0.07,0.5){1}{\line(0,1){0.5}}
\multiput(35.84,46.76)(0.05,0.5){1}{\line(0,1){0.5}}
\multiput(35.8,46.26)(0.04,0.5){1}{\line(0,1){0.5}}
\multiput(35.78,45.76)(0.03,0.5){1}{\line(0,1){0.5}}
\multiput(35.76,45.25)(0.01,0.5){1}{\line(0,1){0.5}}
\put(35.76,44.75){\line(0,1){0.5}}
\multiput(35.76,44.75)(0.01,-0.5){1}{\line(0,-1){0.5}}
\multiput(35.78,44.24)(0.03,-0.5){1}{\line(0,-1){0.5}}
\multiput(35.8,43.74)(0.04,-0.5){1}{\line(0,-1){0.5}}
\multiput(35.84,43.24)(0.05,-0.5){1}{\line(0,-1){0.5}}
\multiput(35.89,42.74)(0.07,-0.5){1}{\line(0,-1){0.5}}
\multiput(35.96,42.24)(0.08,-0.5){1}{\line(0,-1){0.5}}
\multiput(36.04,41.74)(0.09,-0.5){1}{\line(0,-1){0.5}}
\multiput(36.13,41.25)(0.1,-0.49){1}{\line(0,-1){0.49}}
\multiput(36.23,40.75)(0.12,-0.49){1}{\line(0,-1){0.49}}
\multiput(36.35,40.26)(0.13,-0.49){1}{\line(0,-1){0.49}}
\multiput(36.48,39.78)(0.14,-0.48){1}{\line(0,-1){0.48}}
\multiput(36.63,39.29)(0.16,-0.48){1}{\line(0,-1){0.48}}
\multiput(36.78,38.82)(0.17,-0.47){1}{\line(0,-1){0.47}}
\multiput(36.95,38.34)(0.09,-0.24){2}{\line(0,-1){0.24}}
\multiput(37.13,37.87)(0.1,-0.23){2}{\line(0,-1){0.23}}
\multiput(37.32,37.41)(0.1,-0.23){2}{\line(0,-1){0.23}}
\multiput(37.53,36.95)(0.11,-0.23){2}{\line(0,-1){0.23}}
\multiput(37.74,36.49)(0.11,-0.22){2}{\line(0,-1){0.22}}
\multiput(37.97,36.04)(0.12,-0.22){2}{\line(0,-1){0.22}}
\multiput(38.21,35.6)(0.13,-0.22){2}{\line(0,-1){0.22}}
\multiput(38.47,35.16)(0.13,-0.21){2}{\line(0,-1){0.21}}
\multiput(38.73,34.73)(0.14,-0.21){2}{\line(0,-1){0.21}}
\multiput(39,34.31)(0.14,-0.21){2}{\line(0,-1){0.21}}
\multiput(39.29,33.9)(0.15,-0.2){2}{\line(0,-1){0.2}}
\multiput(39.58,33.49)(0.1,-0.13){3}{\line(0,-1){0.13}}
\multiput(39.89,33.09)(0.11,-0.13){3}{\line(0,-1){0.13}}
\multiput(40.21,32.7)(0.11,-0.13){3}{\line(0,-1){0.13}}
\multiput(40.53,32.31)(0.11,-0.12){3}{\line(0,-1){0.12}}
\multiput(40.87,31.94)(0.12,-0.12){3}{\line(0,-1){0.12}}
\multiput(41.22,31.57)(0.12,-0.12){3}{\line(1,0){0.12}}
\multiput(41.57,31.22)(0.12,-0.12){3}{\line(1,0){0.12}}
\multiput(41.94,30.87)(0.12,-0.11){3}{\line(1,0){0.12}}
\multiput(42.31,30.53)(0.13,-0.11){3}{\line(1,0){0.13}}
\multiput(42.7,30.21)(0.13,-0.11){3}{\line(1,0){0.13}}
\multiput(43.09,29.89)(0.13,-0.1){3}{\line(1,0){0.13}}
\multiput(43.49,29.58)(0.2,-0.15){2}{\line(1,0){0.2}}
\multiput(43.9,29.29)(0.21,-0.14){2}{\line(1,0){0.21}}
\multiput(44.31,29)(0.21,-0.14){2}{\line(1,0){0.21}}
\multiput(44.73,28.73)(0.21,-0.13){2}{\line(1,0){0.21}}
\multiput(45.16,28.47)(0.22,-0.13){2}{\line(1,0){0.22}}
\multiput(45.6,28.21)(0.22,-0.12){2}{\line(1,0){0.22}}
\multiput(46.04,27.97)(0.22,-0.11){2}{\line(1,0){0.22}}
\multiput(46.49,27.74)(0.23,-0.11){2}{\line(1,0){0.23}}
\multiput(46.95,27.53)(0.23,-0.1){2}{\line(1,0){0.23}}
\multiput(47.41,27.32)(0.23,-0.1){2}{\line(1,0){0.23}}
\multiput(47.87,27.13)(0.24,-0.09){2}{\line(1,0){0.24}}
\multiput(48.34,26.95)(0.47,-0.17){1}{\line(1,0){0.47}}
\multiput(48.82,26.78)(0.48,-0.16){1}{\line(1,0){0.48}}
\multiput(49.29,26.63)(0.48,-0.14){1}{\line(1,0){0.48}}
\multiput(49.78,26.48)(0.49,-0.13){1}{\line(1,0){0.49}}
\multiput(50.26,26.35)(0.49,-0.12){1}{\line(1,0){0.49}}
\multiput(50.75,26.23)(0.49,-0.1){1}{\line(1,0){0.49}}
\multiput(51.25,26.13)(0.5,-0.09){1}{\line(1,0){0.5}}
\multiput(51.74,26.04)(0.5,-0.08){1}{\line(1,0){0.5}}
\multiput(52.24,25.96)(0.5,-0.07){1}{\line(1,0){0.5}}
\multiput(52.74,25.89)(0.5,-0.05){1}{\line(1,0){0.5}}
\multiput(53.24,25.84)(0.5,-0.04){1}{\line(1,0){0.5}}
\multiput(53.74,25.8)(0.5,-0.03){1}{\line(1,0){0.5}}
\multiput(54.24,25.78)(0.5,-0.01){1}{\line(1,0){0.5}}
\put(54.75,25.76){\line(1,0){0.5}}
\multiput(55.25,25.76)(0.5,0.01){1}{\line(1,0){0.5}}
\multiput(55.76,25.78)(0.5,0.03){1}{\line(1,0){0.5}}
\multiput(56.26,25.8)(0.5,0.04){1}{\line(1,0){0.5}}
\multiput(56.76,25.84)(0.5,0.05){1}{\line(1,0){0.5}}
\multiput(57.26,25.89)(0.5,0.07){1}{\line(1,0){0.5}}
\multiput(57.76,25.96)(0.5,0.08){1}{\line(1,0){0.5}}
\multiput(58.26,26.04)(0.5,0.09){1}{\line(1,0){0.5}}
\multiput(58.75,26.13)(0.49,0.1){1}{\line(1,0){0.49}}
\multiput(59.25,26.23)(0.49,0.12){1}{\line(1,0){0.49}}
\multiput(59.74,26.35)(0.49,0.13){1}{\line(1,0){0.49}}
\multiput(60.22,26.48)(0.48,0.14){1}{\line(1,0){0.48}}
\multiput(60.71,26.63)(0.48,0.16){1}{\line(1,0){0.48}}
\multiput(61.18,26.78)(0.47,0.17){1}{\line(1,0){0.47}}
\multiput(61.66,26.95)(0.24,0.09){2}{\line(1,0){0.24}}
\multiput(62.13,27.13)(0.23,0.1){2}{\line(1,0){0.23}}
\multiput(62.59,27.32)(0.23,0.1){2}{\line(1,0){0.23}}
\multiput(63.05,27.53)(0.23,0.11){2}{\line(1,0){0.23}}
\multiput(63.51,27.74)(0.22,0.11){2}{\line(1,0){0.22}}
\multiput(63.96,27.97)(0.22,0.12){2}{\line(1,0){0.22}}
\multiput(64.4,28.21)(0.22,0.13){2}{\line(1,0){0.22}}
\multiput(64.84,28.47)(0.21,0.13){2}{\line(1,0){0.21}}
\multiput(65.27,28.73)(0.21,0.14){2}{\line(1,0){0.21}}
\multiput(65.69,29)(0.21,0.14){2}{\line(1,0){0.21}}
\multiput(66.1,29.29)(0.2,0.15){2}{\line(1,0){0.2}}
\multiput(66.51,29.58)(0.13,0.1){3}{\line(1,0){0.13}}
\multiput(66.91,29.89)(0.13,0.11){3}{\line(1,0){0.13}}
\multiput(67.3,30.21)(0.13,0.11){3}{\line(1,0){0.13}}
\multiput(67.69,30.53)(0.12,0.11){3}{\line(1,0){0.12}}
\multiput(68.06,30.87)(0.12,0.12){3}{\line(1,0){0.12}}
\multiput(68.43,31.22)(0.12,0.12){3}{\line(0,1){0.12}}
\multiput(68.78,31.57)(0.12,0.12){3}{\line(0,1){0.12}}
\multiput(69.13,31.94)(0.11,0.12){3}{\line(0,1){0.12}}
\multiput(69.47,32.31)(0.11,0.13){3}{\line(0,1){0.13}}
\multiput(69.79,32.7)(0.11,0.13){3}{\line(0,1){0.13}}
\multiput(70.11,33.09)(0.1,0.13){3}{\line(0,1){0.13}}
\multiput(70.42,33.49)(0.15,0.2){2}{\line(0,1){0.2}}
\multiput(70.71,33.9)(0.14,0.21){2}{\line(0,1){0.21}}
\multiput(71,34.31)(0.14,0.21){2}{\line(0,1){0.21}}
\multiput(71.27,34.73)(0.13,0.21){2}{\line(0,1){0.21}}
\multiput(71.53,35.16)(0.13,0.22){2}{\line(0,1){0.22}}
\multiput(71.79,35.6)(0.12,0.22){2}{\line(0,1){0.22}}
\multiput(72.03,36.04)(0.11,0.22){2}{\line(0,1){0.22}}
\multiput(72.26,36.49)(0.11,0.23){2}{\line(0,1){0.23}}
\multiput(72.47,36.95)(0.1,0.23){2}{\line(0,1){0.23}}
\multiput(72.68,37.41)(0.1,0.23){2}{\line(0,1){0.23}}
\multiput(72.87,37.87)(0.09,0.24){2}{\line(0,1){0.24}}
\multiput(73.05,38.34)(0.17,0.47){1}{\line(0,1){0.47}}
\multiput(73.22,38.82)(0.16,0.48){1}{\line(0,1){0.48}}
\multiput(73.37,39.29)(0.14,0.48){1}{\line(0,1){0.48}}
\multiput(73.52,39.78)(0.13,0.49){1}{\line(0,1){0.49}}
\multiput(73.65,40.26)(0.12,0.49){1}{\line(0,1){0.49}}
\multiput(73.77,40.75)(0.1,0.49){1}{\line(0,1){0.49}}
\multiput(73.87,41.25)(0.09,0.5){1}{\line(0,1){0.5}}
\multiput(73.96,41.74)(0.08,0.5){1}{\line(0,1){0.5}}
\multiput(74.04,42.24)(0.07,0.5){1}{\line(0,1){0.5}}
\multiput(74.11,42.74)(0.05,0.5){1}{\line(0,1){0.5}}
\multiput(74.16,43.24)(0.04,0.5){1}{\line(0,1){0.5}}
\multiput(74.2,43.74)(0.03,0.5){1}{\line(0,1){0.5}}
\multiput(74.22,44.24)(0.01,0.5){1}{\line(0,1){0.5}}}

\linethickness{1mm}
\qbezier(36,50)(40.2,65.66)(44.41,72.88)
\qbezier(44.41,72.88)(48.62,80.09)(52.5,80)
\qbezier(52.5,80)(56.4,80.04)(59.41,77.03)
\qbezier(59.41,77.03)(62.41,74.02)(65,67.5)
\qbezier(65,67.5)(67.61,60.98)(68.81,57.97)

\linethickness{0.3mm}

\put(35,43){\line(1,0){40}}

\put(35,47){\line(1,0){40}}

\put(36,51){\line(1,0){38}}

\put(37,55){\line(1,0){35}}

\put(42,60){\line(1,0){25}}

\put(36,39){\line(1,0){38}}

\put(38,35){\line(1,0){34}}

\put(42,31){\line(1,0){26}}

\end{picture}

}

\def\figD{

\def\JPicScale{0.6}
\ifx\JPicScale\undefined\def\JPicScale{1}\fi
\unitlength \JPicScale mm
\begin{picture}(110,70)(0,0)
\linethickness{0.3mm}
\multiput(20,70)(0.18,-0.12){167}{\line(1,0){0.18}}
\linethickness{0.3mm}
\multiput(20,30)(0.18,0.12){167}{\line(1,0){0.18}}
\linethickness{0.3mm}
\put(50,50){\line(1,0){30}}
\linethickness{0.3mm}
\multiput(80,50)(0.18,0.12){167}{\line(1,0){0.18}}
\linethickness{0.3mm}
\multiput(80,50)(0.18,-0.12){167}{\line(1,0){0.18}}
\end{picture}

}

\def\figexampleonesmall{

\def\JPicScale{0.5}
\ifx\JPicScale\undefined\def\JPicScale{1}\fi
\unitlength \JPicScale mm
\begin{picture}(120,88.73)(0,0)
\linethickness{0.3mm}
\multiput(30,70)(0.12,-0.12){167}{\line(1,0){0.12}}
\linethickness{0.3mm}
\multiput(30,30)(0.12,0.12){167}{\line(1,0){0.12}}
\linethickness{0.3mm}
\linethickness{0.3mm}
\put(50,50){\line(1,0){50}}
\linethickness{0.3mm}
\multiput(100,50)(0.12,0.12){167}{\line(1,0){0.12}}
\linethickness{0.3mm}
\multiput(100,50)(0.12,-0.12){167}{\line(1,0){0.12}}
\linethickness{0.3mm}
\put(88.73,74.75){\line(0,1){0.5}}
\multiput(88.71,75.75)(0.02,-0.5){1}{\line(0,-1){0.5}}
\multiput(88.67,76.25)(0.04,-0.5){1}{\line(0,-1){0.5}}
\multiput(88.62,76.75)(0.05,-0.5){1}{\line(0,-1){0.5}}
\multiput(88.54,77.25)(0.07,-0.5){1}{\line(0,-1){0.5}}
\multiput(88.45,77.74)(0.09,-0.49){1}{\line(0,-1){0.49}}
\multiput(88.34,78.23)(0.11,-0.49){1}{\line(0,-1){0.49}}
\multiput(88.22,78.71)(0.13,-0.49){1}{\line(0,-1){0.49}}
\multiput(88.07,79.19)(0.14,-0.48){1}{\line(0,-1){0.48}}
\multiput(87.91,79.67)(0.16,-0.47){1}{\line(0,-1){0.47}}
\multiput(87.73,80.14)(0.18,-0.47){1}{\line(0,-1){0.47}}
\multiput(87.54,80.6)(0.1,-0.23){2}{\line(0,-1){0.23}}
\multiput(87.32,81.05)(0.11,-0.23){2}{\line(0,-1){0.23}}
\multiput(87.09,81.5)(0.11,-0.22){2}{\line(0,-1){0.22}}
\multiput(86.85,81.94)(0.12,-0.22){2}{\line(0,-1){0.22}}
\multiput(86.59,82.37)(0.13,-0.21){2}{\line(0,-1){0.21}}
\multiput(86.31,82.78)(0.14,-0.21){2}{\line(0,-1){0.21}}
\multiput(86.02,83.19)(0.15,-0.2){2}{\line(0,-1){0.2}}
\multiput(85.71,83.59)(0.1,-0.13){3}{\line(0,-1){0.13}}
\multiput(85.39,83.97)(0.11,-0.13){3}{\line(0,-1){0.13}}
\multiput(85.06,84.35)(0.11,-0.12){3}{\line(0,-1){0.12}}
\multiput(84.71,84.71)(0.12,-0.12){3}{\line(0,-1){0.12}}
\multiput(84.35,85.06)(0.12,-0.12){3}{\line(1,0){0.12}}
\multiput(83.97,85.39)(0.12,-0.11){3}{\line(1,0){0.12}}
\multiput(83.59,85.71)(0.13,-0.11){3}{\line(1,0){0.13}}
\multiput(83.19,86.02)(0.13,-0.1){3}{\line(1,0){0.13}}
\multiput(82.78,86.31)(0.2,-0.15){2}{\line(1,0){0.2}}
\multiput(82.37,86.59)(0.21,-0.14){2}{\line(1,0){0.21}}
\multiput(81.94,86.85)(0.21,-0.13){2}{\line(1,0){0.21}}
\multiput(81.5,87.09)(0.22,-0.12){2}{\line(1,0){0.22}}
\multiput(81.05,87.32)(0.22,-0.11){2}{\line(1,0){0.22}}
\multiput(80.6,87.54)(0.23,-0.11){2}{\line(1,0){0.23}}
\multiput(80.14,87.73)(0.23,-0.1){2}{\line(1,0){0.23}}
\multiput(79.67,87.91)(0.47,-0.18){1}{\line(1,0){0.47}}
\multiput(79.19,88.07)(0.47,-0.16){1}{\line(1,0){0.47}}
\multiput(78.71,88.22)(0.48,-0.14){1}{\line(1,0){0.48}}
\multiput(78.23,88.34)(0.49,-0.13){1}{\line(1,0){0.49}}
\multiput(77.74,88.45)(0.49,-0.11){1}{\line(1,0){0.49}}
\multiput(77.25,88.54)(0.49,-0.09){1}{\line(1,0){0.49}}
\multiput(76.75,88.62)(0.5,-0.07){1}{\line(1,0){0.5}}
\multiput(76.25,88.67)(0.5,-0.05){1}{\line(1,0){0.5}}
\multiput(75.75,88.71)(0.5,-0.04){1}{\line(1,0){0.5}}
\multiput(75.25,88.73)(0.5,-0.02){1}{\line(1,0){0.5}}
\put(74.75,88.73){\line(1,0){0.5}}
\multiput(74.25,88.71)(0.5,0.02){1}{\line(1,0){0.5}}
\multiput(73.75,88.67)(0.5,0.04){1}{\line(1,0){0.5}}
\multiput(73.25,88.62)(0.5,0.05){1}{\line(1,0){0.5}}
\multiput(72.75,88.54)(0.5,0.07){1}{\line(1,0){0.5}}
\multiput(72.26,88.45)(0.49,0.09){1}{\line(1,0){0.49}}
\multiput(71.77,88.34)(0.49,0.11){1}{\line(1,0){0.49}}
\multiput(71.29,88.22)(0.49,0.13){1}{\line(1,0){0.49}}
\multiput(70.81,88.07)(0.48,0.14){1}{\line(1,0){0.48}}
\multiput(70.33,87.91)(0.47,0.16){1}{\line(1,0){0.47}}
\multiput(69.86,87.73)(0.47,0.18){1}{\line(1,0){0.47}}
\multiput(69.4,87.54)(0.23,0.1){2}{\line(1,0){0.23}}
\multiput(68.95,87.32)(0.23,0.11){2}{\line(1,0){0.23}}
\multiput(68.5,87.09)(0.22,0.11){2}{\line(1,0){0.22}}
\multiput(68.06,86.85)(0.22,0.12){2}{\line(1,0){0.22}}
\multiput(67.63,86.59)(0.21,0.13){2}{\line(1,0){0.21}}
\multiput(67.22,86.31)(0.21,0.14){2}{\line(1,0){0.21}}
\multiput(66.81,86.02)(0.2,0.15){2}{\line(1,0){0.2}}
\multiput(66.41,85.71)(0.13,0.1){3}{\line(1,0){0.13}}
\multiput(66.03,85.39)(0.13,0.11){3}{\line(1,0){0.13}}
\multiput(65.65,85.06)(0.12,0.11){3}{\line(1,0){0.12}}
\multiput(65.29,84.71)(0.12,0.12){3}{\line(1,0){0.12}}
\multiput(64.94,84.35)(0.12,0.12){3}{\line(0,1){0.12}}
\multiput(64.61,83.97)(0.11,0.12){3}{\line(0,1){0.12}}
\multiput(64.29,83.59)(0.11,0.13){3}{\line(0,1){0.13}}
\multiput(63.98,83.19)(0.1,0.13){3}{\line(0,1){0.13}}
\multiput(63.69,82.78)(0.15,0.2){2}{\line(0,1){0.2}}
\multiput(63.41,82.37)(0.14,0.21){2}{\line(0,1){0.21}}
\multiput(63.15,81.94)(0.13,0.21){2}{\line(0,1){0.21}}
\multiput(62.91,81.5)(0.12,0.22){2}{\line(0,1){0.22}}
\multiput(62.68,81.05)(0.11,0.22){2}{\line(0,1){0.22}}
\multiput(62.46,80.6)(0.11,0.23){2}{\line(0,1){0.23}}
\multiput(62.27,80.14)(0.1,0.23){2}{\line(0,1){0.23}}
\multiput(62.09,79.67)(0.18,0.47){1}{\line(0,1){0.47}}
\multiput(61.93,79.19)(0.16,0.47){1}{\line(0,1){0.47}}
\multiput(61.78,78.71)(0.14,0.48){1}{\line(0,1){0.48}}
\multiput(61.66,78.23)(0.13,0.49){1}{\line(0,1){0.49}}
\multiput(61.55,77.74)(0.11,0.49){1}{\line(0,1){0.49}}
\multiput(61.46,77.25)(0.09,0.49){1}{\line(0,1){0.49}}
\multiput(61.38,76.75)(0.07,0.5){1}{\line(0,1){0.5}}
\multiput(61.33,76.25)(0.05,0.5){1}{\line(0,1){0.5}}
\multiput(61.29,75.75)(0.04,0.5){1}{\line(0,1){0.5}}
\multiput(61.27,75.25)(0.02,0.5){1}{\line(0,1){0.5}}
\put(61.27,74.75){\line(0,1){0.5}}
\multiput(61.27,74.75)(0.02,-0.5){1}{\line(0,-1){0.5}}
\multiput(61.29,74.25)(0.04,-0.5){1}{\line(0,-1){0.5}}
\multiput(61.33,73.75)(0.05,-0.5){1}{\line(0,-1){0.5}}
\multiput(61.38,73.25)(0.07,-0.5){1}{\line(0,-1){0.5}}
\multiput(61.46,72.75)(0.09,-0.49){1}{\line(0,-1){0.49}}
\multiput(61.55,72.26)(0.11,-0.49){1}{\line(0,-1){0.49}}
\multiput(61.66,71.77)(0.13,-0.49){1}{\line(0,-1){0.49}}
\multiput(61.78,71.29)(0.14,-0.48){1}{\line(0,-1){0.48}}
\multiput(61.93,70.81)(0.16,-0.47){1}{\line(0,-1){0.47}}
\multiput(62.09,70.33)(0.18,-0.47){1}{\line(0,-1){0.47}}
\multiput(62.27,69.86)(0.1,-0.23){2}{\line(0,-1){0.23}}
\multiput(62.46,69.4)(0.11,-0.23){2}{\line(0,-1){0.23}}
\multiput(62.68,68.95)(0.11,-0.22){2}{\line(0,-1){0.22}}
\multiput(62.91,68.5)(0.12,-0.22){2}{\line(0,-1){0.22}}
\multiput(63.15,68.06)(0.13,-0.21){2}{\line(0,-1){0.21}}
\multiput(63.41,67.63)(0.14,-0.21){2}{\line(0,-1){0.21}}
\multiput(63.69,67.22)(0.15,-0.2){2}{\line(0,-1){0.2}}
\multiput(63.98,66.81)(0.1,-0.13){3}{\line(0,-1){0.13}}
\multiput(64.29,66.41)(0.11,-0.13){3}{\line(0,-1){0.13}}
\multiput(64.61,66.03)(0.11,-0.12){3}{\line(0,-1){0.12}}
\multiput(64.94,65.65)(0.12,-0.12){3}{\line(0,-1){0.12}}
\multiput(65.29,65.29)(0.12,-0.12){3}{\line(1,0){0.12}}
\multiput(65.65,64.94)(0.12,-0.11){3}{\line(1,0){0.12}}
\multiput(66.03,64.61)(0.13,-0.11){3}{\line(1,0){0.13}}
\multiput(66.41,64.29)(0.13,-0.1){3}{\line(1,0){0.13}}
\multiput(66.81,63.98)(0.2,-0.15){2}{\line(1,0){0.2}}
\multiput(67.22,63.69)(0.21,-0.14){2}{\line(1,0){0.21}}
\multiput(67.63,63.41)(0.21,-0.13){2}{\line(1,0){0.21}}
\multiput(68.06,63.15)(0.22,-0.12){2}{\line(1,0){0.22}}
\multiput(68.5,62.91)(0.22,-0.11){2}{\line(1,0){0.22}}
\multiput(68.95,62.68)(0.23,-0.11){2}{\line(1,0){0.23}}
\multiput(69.4,62.46)(0.23,-0.1){2}{\line(1,0){0.23}}
\multiput(69.86,62.27)(0.47,-0.18){1}{\line(1,0){0.47}}
\multiput(70.33,62.09)(0.47,-0.16){1}{\line(1,0){0.47}}
\multiput(70.81,61.93)(0.48,-0.14){1}{\line(1,0){0.48}}
\multiput(71.29,61.78)(0.49,-0.13){1}{\line(1,0){0.49}}
\multiput(71.77,61.66)(0.49,-0.11){1}{\line(1,0){0.49}}
\multiput(72.26,61.55)(0.49,-0.09){1}{\line(1,0){0.49}}
\multiput(72.75,61.46)(0.5,-0.07){1}{\line(1,0){0.5}}
\multiput(73.25,61.38)(0.5,-0.05){1}{\line(1,0){0.5}}
\multiput(73.75,61.33)(0.5,-0.04){1}{\line(1,0){0.5}}
\multiput(74.25,61.29)(0.5,-0.02){1}{\line(1,0){0.5}}
\put(74.75,61.27){\line(1,0){0.5}}
\multiput(75.25,61.27)(0.5,0.02){1}{\line(1,0){0.5}}
\multiput(75.75,61.29)(0.5,0.04){1}{\line(1,0){0.5}}
\multiput(76.25,61.33)(0.5,0.05){1}{\line(1,0){0.5}}
\multiput(76.75,61.38)(0.5,0.07){1}{\line(1,0){0.5}}
\multiput(77.25,61.46)(0.49,0.09){1}{\line(1,0){0.49}}
\multiput(77.74,61.55)(0.49,0.11){1}{\line(1,0){0.49}}
\multiput(78.23,61.66)(0.49,0.13){1}{\line(1,0){0.49}}
\multiput(78.71,61.78)(0.48,0.14){1}{\line(1,0){0.48}}
\multiput(79.19,61.93)(0.47,0.16){1}{\line(1,0){0.47}}
\multiput(79.67,62.09)(0.47,0.18){1}{\line(1,0){0.47}}
\multiput(80.14,62.27)(0.23,0.1){2}{\line(1,0){0.23}}
\multiput(80.6,62.46)(0.23,0.11){2}{\line(1,0){0.23}}
\multiput(81.05,62.68)(0.22,0.11){2}{\line(1,0){0.22}}
\multiput(81.5,62.91)(0.22,0.12){2}{\line(1,0){0.22}}
\multiput(81.94,63.15)(0.21,0.13){2}{\line(1,0){0.21}}
\multiput(82.37,63.41)(0.21,0.14){2}{\line(1,0){0.21}}
\multiput(82.78,63.69)(0.2,0.15){2}{\line(1,0){0.2}}
\multiput(83.19,63.98)(0.13,0.1){3}{\line(1,0){0.13}}
\multiput(83.59,64.29)(0.13,0.11){3}{\line(1,0){0.13}}
\multiput(83.97,64.61)(0.12,0.11){3}{\line(1,0){0.12}}
\multiput(84.35,64.94)(0.12,0.12){3}{\line(1,0){0.12}}
\multiput(84.71,65.29)(0.12,0.12){3}{\line(0,1){0.12}}
\multiput(85.06,65.65)(0.11,0.12){3}{\line(0,1){0.12}}
\multiput(85.39,66.03)(0.11,0.13){3}{\line(0,1){0.13}}
\multiput(85.71,66.41)(0.1,0.13){3}{\line(0,1){0.13}}
\multiput(86.02,66.81)(0.15,0.2){2}{\line(0,1){0.2}}
\multiput(86.31,67.22)(0.14,0.21){2}{\line(0,1){0.21}}
\multiput(86.59,67.63)(0.13,0.21){2}{\line(0,1){0.21}}
\multiput(86.85,68.06)(0.12,0.22){2}{\line(0,1){0.22}}
\multiput(87.09,68.5)(0.11,0.22){2}{\line(0,1){0.22}}
\multiput(87.32,68.95)(0.11,0.23){2}{\line(0,1){0.23}}
\multiput(87.54,69.4)(0.1,0.23){2}{\line(0,1){0.23}}
\multiput(87.73,69.86)(0.18,0.47){1}{\line(0,1){0.47}}
\multiput(87.91,70.33)(0.16,0.47){1}{\line(0,1){0.47}}
\multiput(88.07,70.81)(0.14,0.48){1}{\line(0,1){0.48}}
\multiput(88.22,71.29)(0.13,0.49){1}{\line(0,1){0.49}}
\multiput(88.34,71.77)(0.11,0.49){1}{\line(0,1){0.49}}
\multiput(88.45,72.26)(0.09,0.49){1}{\line(0,1){0.49}}
\multiput(88.54,72.75)(0.07,0.5){1}{\line(0,1){0.5}}
\multiput(88.62,73.25)(0.05,0.5){1}{\line(0,1){0.5}}
\multiput(88.67,73.75)(0.04,0.5){1}{\line(0,1){0.5}}
\multiput(88.71,74.25)(0.02,0.5){1}{\line(0,1){0.5}}

\linethickness{1mm}
\put(72,50){\line(0,1){11}}
\end{picture}

}

\def\figexampletwosmall{

\def\JPicScale{0.5}
\ifx\JPicScale\undefined\def\JPicScale{1}\fi
\unitlength \JPicScale mm
\begin{picture}(130,85)(0,0)
\linethickness{0.3mm}
\put(54.14,39.75){\line(0,1){0.5}}
\multiput(54.12,40.76)(0.02,-0.5){1}{\line(0,-1){0.5}}
\multiput(54.09,41.26)(0.04,-0.5){1}{\line(0,-1){0.5}}
\multiput(54.03,41.76)(0.05,-0.5){1}{\line(0,-1){0.5}}
\multiput(53.96,42.26)(0.07,-0.5){1}{\line(0,-1){0.5}}
\multiput(53.87,42.76)(0.09,-0.5){1}{\line(0,-1){0.5}}
\multiput(53.76,43.25)(0.11,-0.49){1}{\line(0,-1){0.49}}
\multiput(53.64,43.74)(0.12,-0.49){1}{\line(0,-1){0.49}}
\multiput(53.5,44.23)(0.14,-0.48){1}{\line(0,-1){0.48}}
\multiput(53.34,44.7)(0.16,-0.48){1}{\line(0,-1){0.48}}
\multiput(53.16,45.18)(0.18,-0.47){1}{\line(0,-1){0.47}}
\multiput(52.97,45.64)(0.1,-0.23){2}{\line(0,-1){0.23}}
\multiput(52.76,46.1)(0.1,-0.23){2}{\line(0,-1){0.23}}
\multiput(52.53,46.55)(0.11,-0.23){2}{\line(0,-1){0.23}}
\multiput(52.29,47)(0.12,-0.22){2}{\line(0,-1){0.22}}
\multiput(52.03,47.43)(0.13,-0.22){2}{\line(0,-1){0.22}}
\multiput(51.76,47.86)(0.14,-0.21){2}{\line(0,-1){0.21}}
\multiput(51.47,48.27)(0.14,-0.21){2}{\line(0,-1){0.21}}
\multiput(51.17,48.68)(0.1,-0.13){3}{\line(0,-1){0.13}}
\multiput(50.85,49.07)(0.11,-0.13){3}{\line(0,-1){0.13}}
\multiput(50.52,49.45)(0.11,-0.13){3}{\line(0,-1){0.13}}
\multiput(50.18,49.82)(0.11,-0.12){3}{\line(0,-1){0.12}}
\multiput(49.82,50.18)(0.12,-0.12){3}{\line(0,-1){0.12}}
\multiput(49.45,50.52)(0.12,-0.11){3}{\line(1,0){0.12}}
\multiput(49.07,50.85)(0.13,-0.11){3}{\line(1,0){0.13}}
\multiput(48.68,51.17)(0.13,-0.11){3}{\line(1,0){0.13}}
\multiput(48.27,51.47)(0.13,-0.1){3}{\line(1,0){0.13}}
\multiput(47.86,51.76)(0.21,-0.14){2}{\line(1,0){0.21}}
\multiput(47.43,52.03)(0.21,-0.14){2}{\line(1,0){0.21}}
\multiput(47,52.29)(0.22,-0.13){2}{\line(1,0){0.22}}
\multiput(46.55,52.53)(0.22,-0.12){2}{\line(1,0){0.22}}
\multiput(46.1,52.76)(0.23,-0.11){2}{\line(1,0){0.23}}
\multiput(45.64,52.97)(0.23,-0.1){2}{\line(1,0){0.23}}
\multiput(45.18,53.16)(0.23,-0.1){2}{\line(1,0){0.23}}
\multiput(44.7,53.34)(0.47,-0.18){1}{\line(1,0){0.47}}
\multiput(44.23,53.5)(0.48,-0.16){1}{\line(1,0){0.48}}
\multiput(43.74,53.64)(0.48,-0.14){1}{\line(1,0){0.48}}
\multiput(43.25,53.76)(0.49,-0.12){1}{\line(1,0){0.49}}
\multiput(42.76,53.87)(0.49,-0.11){1}{\line(1,0){0.49}}
\multiput(42.26,53.96)(0.5,-0.09){1}{\line(1,0){0.5}}
\multiput(41.76,54.03)(0.5,-0.07){1}{\line(1,0){0.5}}
\multiput(41.26,54.09)(0.5,-0.05){1}{\line(1,0){0.5}}
\multiput(40.76,54.12)(0.5,-0.04){1}{\line(1,0){0.5}}
\multiput(40.25,54.14)(0.5,-0.02){1}{\line(1,0){0.5}}
\put(39.75,54.14){\line(1,0){0.5}}
\multiput(39.24,54.12)(0.5,0.02){1}{\line(1,0){0.5}}
\multiput(38.74,54.09)(0.5,0.04){1}{\line(1,0){0.5}}
\multiput(38.24,54.03)(0.5,0.05){1}{\line(1,0){0.5}}
\multiput(37.74,53.96)(0.5,0.07){1}{\line(1,0){0.5}}
\multiput(37.24,53.87)(0.5,0.09){1}{\line(1,0){0.5}}
\multiput(36.75,53.76)(0.49,0.11){1}{\line(1,0){0.49}}
\multiput(36.26,53.64)(0.49,0.12){1}{\line(1,0){0.49}}
\multiput(35.77,53.5)(0.48,0.14){1}{\line(1,0){0.48}}
\multiput(35.3,53.34)(0.48,0.16){1}{\line(1,0){0.48}}
\multiput(34.82,53.16)(0.47,0.18){1}{\line(1,0){0.47}}
\multiput(34.36,52.97)(0.23,0.1){2}{\line(1,0){0.23}}
\multiput(33.9,52.76)(0.23,0.1){2}{\line(1,0){0.23}}
\multiput(33.45,52.53)(0.23,0.11){2}{\line(1,0){0.23}}
\multiput(33,52.29)(0.22,0.12){2}{\line(1,0){0.22}}
\multiput(32.57,52.03)(0.22,0.13){2}{\line(1,0){0.22}}
\multiput(32.14,51.76)(0.21,0.14){2}{\line(1,0){0.21}}
\multiput(31.73,51.47)(0.21,0.14){2}{\line(1,0){0.21}}
\multiput(31.32,51.17)(0.13,0.1){3}{\line(1,0){0.13}}
\multiput(30.93,50.85)(0.13,0.11){3}{\line(1,0){0.13}}
\multiput(30.55,50.52)(0.13,0.11){3}{\line(1,0){0.13}}
\multiput(30.18,50.18)(0.12,0.11){3}{\line(1,0){0.12}}
\multiput(29.82,49.82)(0.12,0.12){3}{\line(1,0){0.12}}
\multiput(29.48,49.45)(0.11,0.12){3}{\line(0,1){0.12}}
\multiput(29.15,49.07)(0.11,0.13){3}{\line(0,1){0.13}}
\multiput(28.83,48.68)(0.11,0.13){3}{\line(0,1){0.13}}
\multiput(28.53,48.27)(0.1,0.13){3}{\line(0,1){0.13}}
\multiput(28.24,47.86)(0.14,0.21){2}{\line(0,1){0.21}}
\multiput(27.97,47.43)(0.14,0.21){2}{\line(0,1){0.21}}
\multiput(27.71,47)(0.13,0.22){2}{\line(0,1){0.22}}
\multiput(27.47,46.55)(0.12,0.22){2}{\line(0,1){0.22}}
\multiput(27.24,46.1)(0.11,0.23){2}{\line(0,1){0.23}}
\multiput(27.03,45.64)(0.1,0.23){2}{\line(0,1){0.23}}
\multiput(26.84,45.18)(0.1,0.23){2}{\line(0,1){0.23}}
\multiput(26.66,44.7)(0.18,0.47){1}{\line(0,1){0.47}}
\multiput(26.5,44.23)(0.16,0.48){1}{\line(0,1){0.48}}
\multiput(26.36,43.74)(0.14,0.48){1}{\line(0,1){0.48}}
\multiput(26.24,43.25)(0.12,0.49){1}{\line(0,1){0.49}}
\multiput(26.13,42.76)(0.11,0.49){1}{\line(0,1){0.49}}
\multiput(26.04,42.26)(0.09,0.5){1}{\line(0,1){0.5}}
\multiput(25.97,41.76)(0.07,0.5){1}{\line(0,1){0.5}}
\multiput(25.91,41.26)(0.05,0.5){1}{\line(0,1){0.5}}
\multiput(25.88,40.76)(0.04,0.5){1}{\line(0,1){0.5}}
\multiput(25.86,40.25)(0.02,0.5){1}{\line(0,1){0.5}}
\put(25.86,39.75){\line(0,1){0.5}}
\multiput(25.86,39.75)(0.02,-0.5){1}{\line(0,-1){0.5}}
\multiput(25.88,39.24)(0.04,-0.5){1}{\line(0,-1){0.5}}
\multiput(25.91,38.74)(0.05,-0.5){1}{\line(0,-1){0.5}}
\multiput(25.97,38.24)(0.07,-0.5){1}{\line(0,-1){0.5}}
\multiput(26.04,37.74)(0.09,-0.5){1}{\line(0,-1){0.5}}
\multiput(26.13,37.24)(0.11,-0.49){1}{\line(0,-1){0.49}}
\multiput(26.24,36.75)(0.12,-0.49){1}{\line(0,-1){0.49}}
\multiput(26.36,36.26)(0.14,-0.48){1}{\line(0,-1){0.48}}
\multiput(26.5,35.77)(0.16,-0.48){1}{\line(0,-1){0.48}}
\multiput(26.66,35.3)(0.18,-0.47){1}{\line(0,-1){0.47}}
\multiput(26.84,34.82)(0.1,-0.23){2}{\line(0,-1){0.23}}
\multiput(27.03,34.36)(0.1,-0.23){2}{\line(0,-1){0.23}}
\multiput(27.24,33.9)(0.11,-0.23){2}{\line(0,-1){0.23}}
\multiput(27.47,33.45)(0.12,-0.22){2}{\line(0,-1){0.22}}
\multiput(27.71,33)(0.13,-0.22){2}{\line(0,-1){0.22}}
\multiput(27.97,32.57)(0.14,-0.21){2}{\line(0,-1){0.21}}
\multiput(28.24,32.14)(0.14,-0.21){2}{\line(0,-1){0.21}}
\multiput(28.53,31.73)(0.1,-0.13){3}{\line(0,-1){0.13}}
\multiput(28.83,31.32)(0.11,-0.13){3}{\line(0,-1){0.13}}
\multiput(29.15,30.93)(0.11,-0.13){3}{\line(0,-1){0.13}}
\multiput(29.48,30.55)(0.11,-0.12){3}{\line(0,-1){0.12}}
\multiput(29.82,30.18)(0.12,-0.12){3}{\line(1,0){0.12}}
\multiput(30.18,29.82)(0.12,-0.11){3}{\line(1,0){0.12}}
\multiput(30.55,29.48)(0.13,-0.11){3}{\line(1,0){0.13}}
\multiput(30.93,29.15)(0.13,-0.11){3}{\line(1,0){0.13}}
\multiput(31.32,28.83)(0.13,-0.1){3}{\line(1,0){0.13}}
\multiput(31.73,28.53)(0.21,-0.14){2}{\line(1,0){0.21}}
\multiput(32.14,28.24)(0.21,-0.14){2}{\line(1,0){0.21}}
\multiput(32.57,27.97)(0.22,-0.13){2}{\line(1,0){0.22}}
\multiput(33,27.71)(0.22,-0.12){2}{\line(1,0){0.22}}
\multiput(33.45,27.47)(0.23,-0.11){2}{\line(1,0){0.23}}
\multiput(33.9,27.24)(0.23,-0.1){2}{\line(1,0){0.23}}
\multiput(34.36,27.03)(0.23,-0.1){2}{\line(1,0){0.23}}
\multiput(34.82,26.84)(0.47,-0.18){1}{\line(1,0){0.47}}
\multiput(35.3,26.66)(0.48,-0.16){1}{\line(1,0){0.48}}
\multiput(35.77,26.5)(0.48,-0.14){1}{\line(1,0){0.48}}
\multiput(36.26,26.36)(0.49,-0.12){1}{\line(1,0){0.49}}
\multiput(36.75,26.24)(0.49,-0.11){1}{\line(1,0){0.49}}
\multiput(37.24,26.13)(0.5,-0.09){1}{\line(1,0){0.5}}
\multiput(37.74,26.04)(0.5,-0.07){1}{\line(1,0){0.5}}
\multiput(38.24,25.97)(0.5,-0.05){1}{\line(1,0){0.5}}
\multiput(38.74,25.91)(0.5,-0.04){1}{\line(1,0){0.5}}
\multiput(39.24,25.88)(0.5,-0.02){1}{\line(1,0){0.5}}
\put(39.75,25.86){\line(1,0){0.5}}
\multiput(40.25,25.86)(0.5,0.02){1}{\line(1,0){0.5}}
\multiput(40.76,25.88)(0.5,0.04){1}{\line(1,0){0.5}}
\multiput(41.26,25.91)(0.5,0.05){1}{\line(1,0){0.5}}
\multiput(41.76,25.97)(0.5,0.07){1}{\line(1,0){0.5}}
\multiput(42.26,26.04)(0.5,0.09){1}{\line(1,0){0.5}}
\multiput(42.76,26.13)(0.49,0.11){1}{\line(1,0){0.49}}
\multiput(43.25,26.24)(0.49,0.12){1}{\line(1,0){0.49}}
\multiput(43.74,26.36)(0.48,0.14){1}{\line(1,0){0.48}}
\multiput(44.23,26.5)(0.48,0.16){1}{\line(1,0){0.48}}
\multiput(44.7,26.66)(0.47,0.18){1}{\line(1,0){0.47}}
\multiput(45.18,26.84)(0.23,0.1){2}{\line(1,0){0.23}}
\multiput(45.64,27.03)(0.23,0.1){2}{\line(1,0){0.23}}
\multiput(46.1,27.24)(0.23,0.11){2}{\line(1,0){0.23}}
\multiput(46.55,27.47)(0.22,0.12){2}{\line(1,0){0.22}}
\multiput(47,27.71)(0.22,0.13){2}{\line(1,0){0.22}}
\multiput(47.43,27.97)(0.21,0.14){2}{\line(1,0){0.21}}
\multiput(47.86,28.24)(0.21,0.14){2}{\line(1,0){0.21}}
\multiput(48.27,28.53)(0.13,0.1){3}{\line(1,0){0.13}}
\multiput(48.68,28.83)(0.13,0.11){3}{\line(1,0){0.13}}
\multiput(49.07,29.15)(0.13,0.11){3}{\line(1,0){0.13}}
\multiput(49.45,29.48)(0.12,0.11){3}{\line(1,0){0.12}}
\multiput(49.82,29.82)(0.12,0.12){3}{\line(1,0){0.12}}
\multiput(50.18,30.18)(0.11,0.12){3}{\line(0,1){0.12}}
\multiput(50.52,30.55)(0.11,0.13){3}{\line(0,1){0.13}}
\multiput(50.85,30.93)(0.11,0.13){3}{\line(0,1){0.13}}
\multiput(51.17,31.32)(0.1,0.13){3}{\line(0,1){0.13}}
\multiput(51.47,31.73)(0.14,0.21){2}{\line(0,1){0.21}}
\multiput(51.76,32.14)(0.14,0.21){2}{\line(0,1){0.21}}
\multiput(52.03,32.57)(0.13,0.22){2}{\line(0,1){0.22}}
\multiput(52.29,33)(0.12,0.22){2}{\line(0,1){0.22}}
\multiput(52.53,33.45)(0.11,0.23){2}{\line(0,1){0.23}}
\multiput(52.76,33.9)(0.1,0.23){2}{\line(0,1){0.23}}
\multiput(52.97,34.36)(0.1,0.23){2}{\line(0,1){0.23}}
\multiput(53.16,34.82)(0.18,0.47){1}{\line(0,1){0.47}}
\multiput(53.34,35.3)(0.16,0.48){1}{\line(0,1){0.48}}
\multiput(53.5,35.77)(0.14,0.48){1}{\line(0,1){0.48}}
\multiput(53.64,36.26)(0.12,0.49){1}{\line(0,1){0.49}}
\multiput(53.76,36.75)(0.11,0.49){1}{\line(0,1){0.49}}
\multiput(53.87,37.24)(0.09,0.5){1}{\line(0,1){0.5}}
\multiput(53.96,37.74)(0.07,0.5){1}{\line(0,1){0.5}}
\multiput(54.03,38.24)(0.05,0.5){1}{\line(0,1){0.5}}
\multiput(54.09,38.74)(0.04,0.5){1}{\line(0,1){0.5}}
\multiput(54.12,39.24)(0.02,0.5){1}{\line(0,1){0.5}}

\linethickness{0.3mm}
\multiput(10,20)(0.24,0.12){83}{\line(1,0){0.24}}
\linethickness{1mm}
\multiput(54,44)(0.24,0.12){95}{\line(1,0){0.24}}
\linethickness{0.3mm}
\multiput(30,85)(0.18,-0.12){250}{\line(1,0){0.18}}
\linethickness{0.3mm}
\put(75,55){\line(1,0){30}}
\linethickness{0.3mm}
\multiput(105,55)(0.12,0.12){208}{\line(1,0){0.12}}
\linethickness{0.3mm}
\multiput(105,55)(0.12,-0.12){208}{\line(1,0){0.12}}
\end{picture}

}

\def\figA{

\def\JPicScale{0.6}
\ifx\JPicScale\undefined\def\JPicScale{1}\fi
\unitlength \JPicScale mm
\begin{picture}(140,80)(0,0)
\linethickness{0.3mm}
\qbezier(35,60)(48.02,70.49)(56.44,71.09)
\qbezier(56.44,71.09)(64.86,71.7)(70,62.5)
\qbezier(70,62.5)(75.17,53.33)(81.19,52.13)
\qbezier(81.19,52.13)(87.2,50.92)(95,57.5)
\qbezier(95,57.5)(102.8,64.03)(108.22,66.44)
\qbezier(108.22,66.44)(113.63,68.84)(117.5,67.5)
\qbezier(117.5,67.5)(121.39,66.23)(125,62.63)
\qbezier(125,62.63)(128.61,59.02)(132.5,52.5)
\qbezier(132.5,52.5)(136.45,45.99)(135.84,41.78)
\qbezier(135.84,41.78)(135.24,37.57)(130,35)
\qbezier(130,35)(124.83,32.36)(118.81,33.56)
\qbezier(118.81,33.56)(112.8,34.77)(105,40)
\qbezier(105,40)(97.19,45.22)(92.38,47.62)
\qbezier(92.38,47.62)(87.56,50.03)(85,50)
\qbezier(85,50)(82.43,50.02)(78.22,48.22)
\qbezier(78.22,48.22)(74.01,46.41)(67.5,42.5)
\qbezier(67.5,42.5)(61,38.59)(56.19,36.78)
\qbezier(56.19,36.78)(51.38,34.98)(47.5,35)
\qbezier(47.5,35)(43.62,34.99)(39.41,35.59)
\qbezier(39.41,35.59)(35.2,36.2)(30,37.5)
\qbezier(30,37.5)(24.79,38.8)(21.78,40)
\qbezier(21.78,40)(18.77,41.2)(17.5,42.5)
\qbezier(17.5,42.5)(16.18,43.79)(16.78,45.59)
\qbezier(16.78,45.59)(17.38,47.4)(20,50)
\qbezier(20,50)(22.58,52.59)(26.19,55)
\qbezier(26.19,55)(29.8,57.41)(35,60)
\end{picture}

}

\def\figAA{

\def\JPicScale{0.6}
\ifx\JPicScale\undefined\def\JPicScale{1}\fi
\unitlength \JPicScale mm

}

\def\figoneuni{

\def\JPicScale{0.6}
\ifx\JPicScale\undefined\def\JPicScale{1}\fi
\unitlength \JPicScale mm
\begin{picture}(240,80)(0,0)
\linethickness{0.3mm}
\put(10,50){\line(1,0){100}}
\linethickness{0.3mm}
\multiput(110,50)(0.12,0.12){83}{\line(1,0){0.12}}
\linethickness{0.3mm}
\multiput(110,50)(0.12,-0.12){83}{\line(1,0){0.12}}
\linethickness{0.3mm}
\multiput(0,60)(0.12,-0.12){83}{\line(1,0){0.12}}
\linethickness{0.3mm}
\multiput(0,40)(0.12,0.12){83}{\line(1,0){0.12}}
\linethickness{0.3mm}
\put(140,50){\line(1,0){40}}
\put(170,50){\line(1,0){10}}
\put(200,50){\line(1,0){30}}
\linethickness{0.3mm}
\multiput(230,50)(0.12,0.12){83}{\line(1,0){0.12}}
\linethickness{0.3mm}
\multiput(230,50)(0.12,-0.12){83}{\line(1,0){0.12}}
\linethickness{0.3mm}
\multiput(130,60)(0.12,-0.12){83}{\line(1,0){0.12}}
\linethickness{0.3mm}
\multiput(130,40)(0.12,0.12){83}{\line(1,0){0.12}}
\linethickness{1mm}
\put(60,20){\line(0,1){60}}
\linethickness{1mm}
\put(190,20){\line(0,1){60}}
\linethickness{0.3mm}
\qbezier(20,50)(27.78,55.25)(35,55.25)
\qbezier(35,55.25)(42.22,55.25)(50,50)
\linethickness{0.3mm}
\qbezier(20,50)(27.78,44.75)(35,44.75)
\qbezier(35,44.75)(42.22,44.75)(50,50)
\linethickness{0.3mm}
\qbezier(180,50)(185.19,55.25)(190,55.25)
\qbezier(190,55.25)(194.81,55.25)(200,50)
\linethickness{0.3mm}
\qbezier(180,50)(185.19,44.75)(190,44.75)
\qbezier(190,44.75)(194.81,44.75)(200,50)
\linethickness{0.3mm}
\linethickness{0.3mm}
\put(60,10){\makebox(0,0)[cc]{(a)}}

\put(190,10){\makebox(0,0)[cc]{(b)}}

\end{picture}

}

\def\figtwouni{

\def\JPicScale{0.6}
\ifx\JPicScale\undefined\def\JPicScale{1}\fi
\unitlength \JPicScale mm
\begin{picture}(150,80)(0,0)
\linethickness{0.3mm}
\put(50.68,49.75){\line(0,1){0.49}}
\multiput(50.66,50.74)(0.02,-0.49){1}{\line(0,-1){0.49}}
\multiput(50.61,51.23)(0.05,-0.49){1}{\line(0,-1){0.49}}
\multiput(50.54,51.72)(0.07,-0.49){1}{\line(0,-1){0.49}}
\multiput(50.45,52.21)(0.09,-0.49){1}{\line(0,-1){0.49}}
\multiput(50.34,52.69)(0.11,-0.48){1}{\line(0,-1){0.48}}
\multiput(50.21,53.16)(0.14,-0.47){1}{\line(0,-1){0.47}}
\multiput(50.05,53.63)(0.16,-0.47){1}{\line(0,-1){0.47}}
\multiput(49.87,54.09)(0.18,-0.46){1}{\line(0,-1){0.46}}
\multiput(49.67,54.54)(0.1,-0.23){2}{\line(0,-1){0.23}}
\multiput(49.45,54.98)(0.11,-0.22){2}{\line(0,-1){0.22}}
\multiput(49.21,55.41)(0.12,-0.22){2}{\line(0,-1){0.22}}
\multiput(48.95,55.83)(0.13,-0.21){2}{\line(0,-1){0.21}}
\multiput(48.67,56.24)(0.14,-0.2){2}{\line(0,-1){0.2}}
\multiput(48.38,56.63)(0.15,-0.2){2}{\line(0,-1){0.2}}
\multiput(48.06,57.01)(0.11,-0.13){3}{\line(0,-1){0.13}}
\multiput(47.73,57.38)(0.11,-0.12){3}{\line(0,-1){0.12}}
\multiput(47.38,57.73)(0.12,-0.12){3}{\line(1,0){0.12}}
\multiput(47.01,58.06)(0.12,-0.11){3}{\line(1,0){0.12}}
\multiput(46.63,58.38)(0.13,-0.11){3}{\line(1,0){0.13}}
\multiput(46.24,58.67)(0.2,-0.15){2}{\line(1,0){0.2}}
\multiput(45.83,58.95)(0.2,-0.14){2}{\line(1,0){0.2}}
\multiput(45.41,59.21)(0.21,-0.13){2}{\line(1,0){0.21}}
\multiput(44.98,59.45)(0.22,-0.12){2}{\line(1,0){0.22}}
\multiput(44.54,59.67)(0.22,-0.11){2}{\line(1,0){0.22}}
\multiput(44.09,59.87)(0.23,-0.1){2}{\line(1,0){0.23}}
\multiput(43.63,60.05)(0.46,-0.18){1}{\line(1,0){0.46}}
\multiput(43.16,60.21)(0.47,-0.16){1}{\line(1,0){0.47}}
\multiput(42.69,60.34)(0.47,-0.14){1}{\line(1,0){0.47}}
\multiput(42.21,60.45)(0.48,-0.11){1}{\line(1,0){0.48}}
\multiput(41.72,60.54)(0.49,-0.09){1}{\line(1,0){0.49}}
\multiput(41.23,60.61)(0.49,-0.07){1}{\line(1,0){0.49}}
\multiput(40.74,60.66)(0.49,-0.05){1}{\line(1,0){0.49}}
\multiput(40.25,60.68)(0.49,-0.02){1}{\line(1,0){0.49}}
\put(39.75,60.68){\line(1,0){0.49}}
\multiput(39.26,60.66)(0.49,0.02){1}{\line(1,0){0.49}}
\multiput(38.77,60.61)(0.49,0.05){1}{\line(1,0){0.49}}
\multiput(38.28,60.54)(0.49,0.07){1}{\line(1,0){0.49}}
\multiput(37.79,60.45)(0.49,0.09){1}{\line(1,0){0.49}}
\multiput(37.31,60.34)(0.48,0.11){1}{\line(1,0){0.48}}
\multiput(36.84,60.21)(0.47,0.14){1}{\line(1,0){0.47}}
\multiput(36.37,60.05)(0.47,0.16){1}{\line(1,0){0.47}}
\multiput(35.91,59.87)(0.46,0.18){1}{\line(1,0){0.46}}
\multiput(35.46,59.67)(0.23,0.1){2}{\line(1,0){0.23}}
\multiput(35.02,59.45)(0.22,0.11){2}{\line(1,0){0.22}}
\multiput(34.59,59.21)(0.22,0.12){2}{\line(1,0){0.22}}
\multiput(34.17,58.95)(0.21,0.13){2}{\line(1,0){0.21}}
\multiput(33.76,58.67)(0.2,0.14){2}{\line(1,0){0.2}}
\multiput(33.37,58.38)(0.2,0.15){2}{\line(1,0){0.2}}
\multiput(32.99,58.06)(0.13,0.11){3}{\line(1,0){0.13}}
\multiput(32.62,57.73)(0.12,0.11){3}{\line(1,0){0.12}}
\multiput(32.27,57.38)(0.12,0.12){3}{\line(0,1){0.12}}
\multiput(31.94,57.01)(0.11,0.12){3}{\line(0,1){0.12}}
\multiput(31.62,56.63)(0.11,0.13){3}{\line(0,1){0.13}}
\multiput(31.33,56.24)(0.15,0.2){2}{\line(0,1){0.2}}
\multiput(31.05,55.83)(0.14,0.2){2}{\line(0,1){0.2}}
\multiput(30.79,55.41)(0.13,0.21){2}{\line(0,1){0.21}}
\multiput(30.55,54.98)(0.12,0.22){2}{\line(0,1){0.22}}
\multiput(30.33,54.54)(0.11,0.22){2}{\line(0,1){0.22}}
\multiput(30.13,54.09)(0.1,0.23){2}{\line(0,1){0.23}}
\multiput(29.95,53.63)(0.18,0.46){1}{\line(0,1){0.46}}
\multiput(29.79,53.16)(0.16,0.47){1}{\line(0,1){0.47}}
\multiput(29.66,52.69)(0.14,0.47){1}{\line(0,1){0.47}}
\multiput(29.55,52.21)(0.11,0.48){1}{\line(0,1){0.48}}
\multiput(29.46,51.72)(0.09,0.49){1}{\line(0,1){0.49}}
\multiput(29.39,51.23)(0.07,0.49){1}{\line(0,1){0.49}}
\multiput(29.34,50.74)(0.05,0.49){1}{\line(0,1){0.49}}
\multiput(29.32,50.25)(0.02,0.49){1}{\line(0,1){0.49}}
\put(29.32,49.75){\line(0,1){0.49}}
\multiput(29.32,49.75)(0.02,-0.49){1}{\line(0,-1){0.49}}
\multiput(29.34,49.26)(0.05,-0.49){1}{\line(0,-1){0.49}}
\multiput(29.39,48.77)(0.07,-0.49){1}{\line(0,-1){0.49}}
\multiput(29.46,48.28)(0.09,-0.49){1}{\line(0,-1){0.49}}
\multiput(29.55,47.79)(0.11,-0.48){1}{\line(0,-1){0.48}}
\multiput(29.66,47.31)(0.14,-0.47){1}{\line(0,-1){0.47}}
\multiput(29.79,46.84)(0.16,-0.47){1}{\line(0,-1){0.47}}
\multiput(29.95,46.37)(0.18,-0.46){1}{\line(0,-1){0.46}}
\multiput(30.13,45.91)(0.1,-0.23){2}{\line(0,-1){0.23}}
\multiput(30.33,45.46)(0.11,-0.22){2}{\line(0,-1){0.22}}
\multiput(30.55,45.02)(0.12,-0.22){2}{\line(0,-1){0.22}}
\multiput(30.79,44.59)(0.13,-0.21){2}{\line(0,-1){0.21}}
\multiput(31.05,44.17)(0.14,-0.2){2}{\line(0,-1){0.2}}
\multiput(31.33,43.76)(0.15,-0.2){2}{\line(0,-1){0.2}}
\multiput(31.62,43.37)(0.11,-0.13){3}{\line(0,-1){0.13}}
\multiput(31.94,42.99)(0.11,-0.12){3}{\line(0,-1){0.12}}
\multiput(32.27,42.62)(0.12,-0.12){3}{\line(1,0){0.12}}
\multiput(32.62,42.27)(0.12,-0.11){3}{\line(1,0){0.12}}
\multiput(32.99,41.94)(0.13,-0.11){3}{\line(1,0){0.13}}
\multiput(33.37,41.62)(0.2,-0.15){2}{\line(1,0){0.2}}
\multiput(33.76,41.33)(0.2,-0.14){2}{\line(1,0){0.2}}
\multiput(34.17,41.05)(0.21,-0.13){2}{\line(1,0){0.21}}
\multiput(34.59,40.79)(0.22,-0.12){2}{\line(1,0){0.22}}
\multiput(35.02,40.55)(0.22,-0.11){2}{\line(1,0){0.22}}
\multiput(35.46,40.33)(0.23,-0.1){2}{\line(1,0){0.23}}
\multiput(35.91,40.13)(0.46,-0.18){1}{\line(1,0){0.46}}
\multiput(36.37,39.95)(0.47,-0.16){1}{\line(1,0){0.47}}
\multiput(36.84,39.79)(0.47,-0.14){1}{\line(1,0){0.47}}
\multiput(37.31,39.66)(0.48,-0.11){1}{\line(1,0){0.48}}
\multiput(37.79,39.55)(0.49,-0.09){1}{\line(1,0){0.49}}
\multiput(38.28,39.46)(0.49,-0.07){1}{\line(1,0){0.49}}
\multiput(38.77,39.39)(0.49,-0.05){1}{\line(1,0){0.49}}
\multiput(39.26,39.34)(0.49,-0.02){1}{\line(1,0){0.49}}
\put(39.75,39.32){\line(1,0){0.49}}
\multiput(40.25,39.32)(0.49,0.02){1}{\line(1,0){0.49}}
\multiput(40.74,39.34)(0.49,0.05){1}{\line(1,0){0.49}}
\multiput(41.23,39.39)(0.49,0.07){1}{\line(1,0){0.49}}
\multiput(41.72,39.46)(0.49,0.09){1}{\line(1,0){0.49}}
\multiput(42.21,39.55)(0.48,0.11){1}{\line(1,0){0.48}}
\multiput(42.69,39.66)(0.47,0.14){1}{\line(1,0){0.47}}
\multiput(43.16,39.79)(0.47,0.16){1}{\line(1,0){0.47}}
\multiput(43.63,39.95)(0.46,0.18){1}{\line(1,0){0.46}}
\multiput(44.09,40.13)(0.23,0.1){2}{\line(1,0){0.23}}
\multiput(44.54,40.33)(0.22,0.11){2}{\line(1,0){0.22}}
\multiput(44.98,40.55)(0.22,0.12){2}{\line(1,0){0.22}}
\multiput(45.41,40.79)(0.21,0.13){2}{\line(1,0){0.21}}
\multiput(45.83,41.05)(0.2,0.14){2}{\line(1,0){0.2}}
\multiput(46.24,41.33)(0.2,0.15){2}{\line(1,0){0.2}}
\multiput(46.63,41.62)(0.13,0.11){3}{\line(1,0){0.13}}
\multiput(47.01,41.94)(0.12,0.11){3}{\line(1,0){0.12}}
\multiput(47.38,42.27)(0.12,0.12){3}{\line(0,1){0.12}}
\multiput(47.73,42.62)(0.11,0.12){3}{\line(0,1){0.12}}
\multiput(48.06,42.99)(0.11,0.13){3}{\line(0,1){0.13}}
\multiput(48.38,43.37)(0.15,0.2){2}{\line(0,1){0.2}}
\multiput(48.67,43.76)(0.14,0.2){2}{\line(0,1){0.2}}
\multiput(48.95,44.17)(0.13,0.21){2}{\line(0,1){0.21}}
\multiput(49.21,44.59)(0.12,0.22){2}{\line(0,1){0.22}}
\multiput(49.45,45.02)(0.11,0.22){2}{\line(0,1){0.22}}
\multiput(49.67,45.46)(0.1,0.23){2}{\line(0,1){0.23}}
\multiput(49.87,45.91)(0.18,0.46){1}{\line(0,1){0.46}}
\multiput(50.05,46.37)(0.16,0.47){1}{\line(0,1){0.47}}
\multiput(50.21,46.84)(0.14,0.47){1}{\line(0,1){0.47}}
\multiput(50.34,47.31)(0.11,0.48){1}{\line(0,1){0.48}}
\multiput(50.45,47.79)(0.09,0.49){1}{\line(0,1){0.49}}
\multiput(50.54,48.28)(0.07,0.49){1}{\line(0,1){0.49}}
\multiput(50.61,48.77)(0.05,0.49){1}{\line(0,1){0.49}}
\multiput(50.66,49.26)(0.02,0.49){1}{\line(0,1){0.49}}

\linethickness{0.3mm}
\put(125.68,49.75){\line(0,1){0.49}}
\multiput(125.66,50.74)(0.02,-0.49){1}{\line(0,-1){0.49}}
\multiput(125.61,51.23)(0.05,-0.49){1}{\line(0,-1){0.49}}
\multiput(125.54,51.72)(0.07,-0.49){1}{\line(0,-1){0.49}}
\multiput(125.45,52.21)(0.09,-0.49){1}{\line(0,-1){0.49}}
\multiput(125.34,52.69)(0.11,-0.48){1}{\line(0,-1){0.48}}
\multiput(125.21,53.16)(0.14,-0.47){1}{\line(0,-1){0.47}}
\multiput(125.05,53.63)(0.16,-0.47){1}{\line(0,-1){0.47}}
\multiput(124.87,54.09)(0.18,-0.46){1}{\line(0,-1){0.46}}
\multiput(124.67,54.54)(0.1,-0.23){2}{\line(0,-1){0.23}}
\multiput(124.45,54.98)(0.11,-0.22){2}{\line(0,-1){0.22}}
\multiput(124.21,55.41)(0.12,-0.22){2}{\line(0,-1){0.22}}
\multiput(123.95,55.83)(0.13,-0.21){2}{\line(0,-1){0.21}}
\multiput(123.67,56.24)(0.14,-0.2){2}{\line(0,-1){0.2}}
\multiput(123.38,56.63)(0.15,-0.2){2}{\line(0,-1){0.2}}
\multiput(123.06,57.01)(0.11,-0.13){3}{\line(0,-1){0.13}}
\multiput(122.73,57.38)(0.11,-0.12){3}{\line(0,-1){0.12}}
\multiput(122.38,57.73)(0.12,-0.12){3}{\line(1,0){0.12}}
\multiput(122.01,58.06)(0.12,-0.11){3}{\line(1,0){0.12}}
\multiput(121.63,58.38)(0.13,-0.11){3}{\line(1,0){0.13}}
\multiput(121.24,58.67)(0.2,-0.15){2}{\line(1,0){0.2}}
\multiput(120.83,58.95)(0.2,-0.14){2}{\line(1,0){0.2}}
\multiput(120.41,59.21)(0.21,-0.13){2}{\line(1,0){0.21}}
\multiput(119.98,59.45)(0.22,-0.12){2}{\line(1,0){0.22}}
\multiput(119.54,59.67)(0.22,-0.11){2}{\line(1,0){0.22}}
\multiput(119.09,59.87)(0.23,-0.1){2}{\line(1,0){0.23}}
\multiput(118.63,60.05)(0.46,-0.18){1}{\line(1,0){0.46}}
\multiput(118.16,60.21)(0.47,-0.16){1}{\line(1,0){0.47}}
\multiput(117.69,60.34)(0.47,-0.14){1}{\line(1,0){0.47}}
\multiput(117.21,60.45)(0.48,-0.11){1}{\line(1,0){0.48}}
\multiput(116.72,60.54)(0.49,-0.09){1}{\line(1,0){0.49}}
\multiput(116.23,60.61)(0.49,-0.07){1}{\line(1,0){0.49}}
\multiput(115.74,60.66)(0.49,-0.05){1}{\line(1,0){0.49}}
\multiput(115.25,60.68)(0.49,-0.02){1}{\line(1,0){0.49}}
\put(114.75,60.68){\line(1,0){0.49}}
\multiput(114.26,60.66)(0.49,0.02){1}{\line(1,0){0.49}}
\multiput(113.77,60.61)(0.49,0.05){1}{\line(1,0){0.49}}
\multiput(113.28,60.54)(0.49,0.07){1}{\line(1,0){0.49}}
\multiput(112.79,60.45)(0.49,0.09){1}{\line(1,0){0.49}}
\multiput(112.31,60.34)(0.48,0.11){1}{\line(1,0){0.48}}
\multiput(111.84,60.21)(0.47,0.14){1}{\line(1,0){0.47}}
\multiput(111.37,60.05)(0.47,0.16){1}{\line(1,0){0.47}}
\multiput(110.91,59.87)(0.46,0.18){1}{\line(1,0){0.46}}
\multiput(110.46,59.67)(0.23,0.1){2}{\line(1,0){0.23}}
\multiput(110.02,59.45)(0.22,0.11){2}{\line(1,0){0.22}}
\multiput(109.59,59.21)(0.22,0.12){2}{\line(1,0){0.22}}
\multiput(109.17,58.95)(0.21,0.13){2}{\line(1,0){0.21}}
\multiput(108.76,58.67)(0.2,0.14){2}{\line(1,0){0.2}}
\multiput(108.37,58.38)(0.2,0.15){2}{\line(1,0){0.2}}
\multiput(107.99,58.06)(0.13,0.11){3}{\line(1,0){0.13}}
\multiput(107.62,57.73)(0.12,0.11){3}{\line(1,0){0.12}}
\multiput(107.27,57.38)(0.12,0.12){3}{\line(1,0){0.12}}
\multiput(106.94,57.01)(0.11,0.12){3}{\line(0,1){0.12}}
\multiput(106.62,56.63)(0.11,0.13){3}{\line(0,1){0.13}}
\multiput(106.33,56.24)(0.15,0.2){2}{\line(0,1){0.2}}
\multiput(106.05,55.83)(0.14,0.2){2}{\line(0,1){0.2}}
\multiput(105.79,55.41)(0.13,0.21){2}{\line(0,1){0.21}}
\multiput(105.55,54.98)(0.12,0.22){2}{\line(0,1){0.22}}
\multiput(105.33,54.54)(0.11,0.22){2}{\line(0,1){0.22}}
\multiput(105.13,54.09)(0.1,0.23){2}{\line(0,1){0.23}}
\multiput(104.95,53.63)(0.18,0.46){1}{\line(0,1){0.46}}
\multiput(104.79,53.16)(0.16,0.47){1}{\line(0,1){0.47}}
\multiput(104.66,52.69)(0.14,0.47){1}{\line(0,1){0.47}}
\multiput(104.55,52.21)(0.11,0.48){1}{\line(0,1){0.48}}
\multiput(104.46,51.72)(0.09,0.49){1}{\line(0,1){0.49}}
\multiput(104.39,51.23)(0.07,0.49){1}{\line(0,1){0.49}}
\multiput(104.34,50.74)(0.05,0.49){1}{\line(0,1){0.49}}
\multiput(104.32,50.25)(0.02,0.49){1}{\line(0,1){0.49}}
\put(104.32,49.75){\line(0,1){0.49}}
\multiput(104.32,49.75)(0.02,-0.49){1}{\line(0,-1){0.49}}
\multiput(104.34,49.26)(0.05,-0.49){1}{\line(0,-1){0.49}}
\multiput(104.39,48.77)(0.07,-0.49){1}{\line(0,-1){0.49}}
\multiput(104.46,48.28)(0.09,-0.49){1}{\line(0,-1){0.49}}
\multiput(104.55,47.79)(0.11,-0.48){1}{\line(0,-1){0.48}}
\multiput(104.66,47.31)(0.14,-0.47){1}{\line(0,-1){0.47}}
\multiput(104.79,46.84)(0.16,-0.47){1}{\line(0,-1){0.47}}
\multiput(104.95,46.37)(0.18,-0.46){1}{\line(0,-1){0.46}}
\multiput(105.13,45.91)(0.1,-0.23){2}{\line(0,-1){0.23}}
\multiput(105.33,45.46)(0.11,-0.22){2}{\line(0,-1){0.22}}
\multiput(105.55,45.02)(0.12,-0.22){2}{\line(0,-1){0.22}}
\multiput(105.79,44.59)(0.13,-0.21){2}{\line(0,-1){0.21}}
\multiput(106.05,44.17)(0.14,-0.2){2}{\line(0,-1){0.2}}
\multiput(106.33,43.76)(0.15,-0.2){2}{\line(0,-1){0.2}}
\multiput(106.62,43.37)(0.11,-0.13){3}{\line(0,-1){0.13}}
\multiput(106.94,42.99)(0.11,-0.12){3}{\line(0,-1){0.12}}
\multiput(107.27,42.62)(0.12,-0.12){3}{\line(1,0){0.12}}
\multiput(107.62,42.27)(0.12,-0.11){3}{\line(1,0){0.12}}
\multiput(107.99,41.94)(0.13,-0.11){3}{\line(1,0){0.13}}
\multiput(108.37,41.62)(0.2,-0.15){2}{\line(1,0){0.2}}
\multiput(108.76,41.33)(0.2,-0.14){2}{\line(1,0){0.2}}
\multiput(109.17,41.05)(0.21,-0.13){2}{\line(1,0){0.21}}
\multiput(109.59,40.79)(0.22,-0.12){2}{\line(1,0){0.22}}
\multiput(110.02,40.55)(0.22,-0.11){2}{\line(1,0){0.22}}
\multiput(110.46,40.33)(0.23,-0.1){2}{\line(1,0){0.23}}
\multiput(110.91,40.13)(0.46,-0.18){1}{\line(1,0){0.46}}
\multiput(111.37,39.95)(0.47,-0.16){1}{\line(1,0){0.47}}
\multiput(111.84,39.79)(0.47,-0.14){1}{\line(1,0){0.47}}
\multiput(112.31,39.66)(0.48,-0.11){1}{\line(1,0){0.48}}
\multiput(112.79,39.55)(0.49,-0.09){1}{\line(1,0){0.49}}
\multiput(113.28,39.46)(0.49,-0.07){1}{\line(1,0){0.49}}
\multiput(113.77,39.39)(0.49,-0.05){1}{\line(1,0){0.49}}
\multiput(114.26,39.34)(0.49,-0.02){1}{\line(1,0){0.49}}
\put(114.75,39.32){\line(1,0){0.49}}
\multiput(115.25,39.32)(0.49,0.02){1}{\line(1,0){0.49}}
\multiput(115.74,39.34)(0.49,0.05){1}{\line(1,0){0.49}}
\multiput(116.23,39.39)(0.49,0.07){1}{\line(1,0){0.49}}
\multiput(116.72,39.46)(0.49,0.09){1}{\line(1,0){0.49}}
\multiput(117.21,39.55)(0.48,0.11){1}{\line(1,0){0.48}}
\multiput(117.69,39.66)(0.47,0.14){1}{\line(1,0){0.47}}
\multiput(118.16,39.79)(0.47,0.16){1}{\line(1,0){0.47}}
\multiput(118.63,39.95)(0.46,0.18){1}{\line(1,0){0.46}}
\multiput(119.09,40.13)(0.23,0.1){2}{\line(1,0){0.23}}
\multiput(119.54,40.33)(0.22,0.11){2}{\line(1,0){0.22}}
\multiput(119.98,40.55)(0.22,0.12){2}{\line(1,0){0.22}}
\multiput(120.41,40.79)(0.21,0.13){2}{\line(1,0){0.21}}
\multiput(120.83,41.05)(0.2,0.14){2}{\line(1,0){0.2}}
\multiput(121.24,41.33)(0.2,0.15){2}{\line(1,0){0.2}}
\multiput(121.63,41.62)(0.13,0.11){3}{\line(1,0){0.13}}
\multiput(122.01,41.94)(0.12,0.11){3}{\line(1,0){0.12}}
\multiput(122.38,42.27)(0.12,0.12){3}{\line(0,1){0.12}}
\multiput(122.73,42.62)(0.11,0.12){3}{\line(0,1){0.12}}
\multiput(123.06,42.99)(0.11,0.13){3}{\line(0,1){0.13}}
\multiput(123.38,43.37)(0.15,0.2){2}{\line(0,1){0.2}}
\multiput(123.67,43.76)(0.14,0.2){2}{\line(0,1){0.2}}
\multiput(123.95,44.17)(0.13,0.21){2}{\line(0,1){0.21}}
\multiput(124.21,44.59)(0.12,0.22){2}{\line(0,1){0.22}}
\multiput(124.45,45.02)(0.11,0.22){2}{\line(0,1){0.22}}
\multiput(124.67,45.46)(0.1,0.23){2}{\line(0,1){0.23}}
\multiput(124.87,45.91)(0.18,0.46){1}{\line(0,1){0.46}}
\multiput(125.05,46.37)(0.16,0.47){1}{\line(0,1){0.47}}
\multiput(125.21,46.84)(0.14,0.47){1}{\line(0,1){0.47}}
\multiput(125.34,47.31)(0.11,0.48){1}{\line(0,1){0.48}}
\multiput(125.45,47.79)(0.09,0.49){1}{\line(0,1){0.49}}
\multiput(125.54,48.28)(0.07,0.49){1}{\line(0,1){0.49}}
\multiput(125.61,48.77)(0.05,0.49){1}{\line(0,1){0.49}}
\multiput(125.66,49.26)(0.02,0.49){1}{\line(0,1){0.49}}

\linethickness{0.3mm}
\qbezier(50,55)(65.58,62.88)(78.81,62.88)
\qbezier(78.81,62.88)(92.05,62.88)(105,55)
\linethickness{0.3mm}
\linethickness{0.3mm}
\qbezier(50,45)(65.58,37.12)(78.81,37.12)
\qbezier(78.81,37.12)(92.05,37.12)(105,45)
\linethickness{1mm}
\put(80,20){\line(0,1){60}}
\put(40,50){\makebox(0,0)[cc]{$\Gamma_L$}}

\put(115,50){\makebox(0,0)[cc]{$\Gamma_R$}}

\put(70,58){\makebox(0,0)[cc]{$\bullet$}}

\put(70,50){\makebox(0,0)[cc]{$\bullet$}}

\put(70,42){\makebox(0,0)[cc]{$\bullet$}}

\put(20,54){\makebox(0,0)[cc]{$\bullet$}}

\put(20,48){\makebox(0,0)[cc]{$\bullet$}}

\put(20,40){\makebox(0,0)[cc]{$\bullet$}}

\put(135,58){\makebox(0,0)[cc]{$\bullet$}}

\put(135,50){\makebox(0,0)[cc]{$\bullet$}}

\put(135,42){\makebox(0,0)[cc]{$\bullet$}}

\linethickness{0.3mm}
\multiput(10,65)(0.24,-0.12){83}{\line(1,0){0.24}}
\linethickness{0.3mm}
\linethickness{0.3mm}
\multiput(10,30)(0.3,0.12){84}{\line(1,0){0.3}}
\linethickness{0.3mm}
\multiput(120,60)(0.36,0.12){83}{\line(1,0){0.36}}
\linethickness{0.3mm}
\multiput(120,40)(0.36,-0.12){83}{\line(1,0){0.36}}
\end{picture}

}

\def\figfiber{

\def\JPicScale{0.6}
\ifx\JPicScale\undefined\def\JPicScale{1}\fi
\unitlength \JPicScale mm
\begin{picture}(135,70)(0,0)
\linethickness{0.3mm}
\qbezier(30,50)(55.94,60.5)(80,60.5)
\qbezier(80,60.5)(104.06,60.5)(130,50)
\linethickness{0.3mm}
\qbezier(30,30)(55.94,24.75)(80,24.75)
\qbezier(80,24.75)(104.06,24.75)(130,30)
\linethickness{0.3mm}
\put(30,30){\line(0,1){20}}
\linethickness{0.3mm}
\put(130,30){\line(0,1){20}}
\put(85,40){\makebox(0,0)[cc]{$\UU_{g,m,n}$}}

\put(80,70){\makebox(0,0)[cc]{$\SSS'_{g,m,n}$}}

\put(80,15){\makebox(0,0)[cc]{$\SSS_{g,m,n}$}}

\put(12,40){\makebox(0,0)[cc]{$\VV_{g,m,n}$}}

\put(145,40){\makebox(0,0)[cc]{$\VV_{g,m,n}$}}

\end{picture}

}

\def\figfourpoint{

\def\JPicScale{0.5}
\ifx\JPicScale\undefined\def\JPicScale{1}\fi
\unitlength \JPicScale mm

}

\def\figtorusamp{

\def\JPicScale{0.6}
\ifx\JPicScale\undefined\def\JPicScale{1}\fi
\unitlength \JPicScale mm
\begin{picture}(190,105)(0,0)
\linethickness{0.3mm}
\put(70,100){\line(1,0){120}}
\linethickness{0.3mm}
\multiput(20,50)(0.12,0.12){417}{\line(1,0){0.12}}
\linethickness{0.3mm}
\multiput(140,50)(0.12,0.12){417}{\line(1,0){0.12}}
\linethickness{0.3mm}
\put(20,50){\line(1,0){120}}
\linethickness{0.3mm}
\put(79.01,77.25){\line(0,1){0.51}}
\multiput(78.98,78.26)(0.03,-0.5){1}{\line(0,-1){0.5}}
\multiput(78.93,78.76)(0.06,-0.5){1}{\line(0,-1){0.5}}
\multiput(78.84,79.26)(0.08,-0.5){1}{\line(0,-1){0.5}}
\multiput(78.73,79.75)(0.11,-0.49){1}{\line(0,-1){0.49}}
\multiput(78.59,80.24)(0.14,-0.49){1}{\line(0,-1){0.49}}
\multiput(78.42,80.71)(0.17,-0.48){1}{\line(0,-1){0.48}}
\multiput(78.23,81.18)(0.1,-0.23){2}{\line(0,-1){0.23}}
\multiput(78.01,81.64)(0.11,-0.23){2}{\line(0,-1){0.23}}
\multiput(77.76,82.08)(0.12,-0.22){2}{\line(0,-1){0.22}}
\multiput(77.49,82.51)(0.13,-0.21){2}{\line(0,-1){0.21}}
\multiput(77.2,82.92)(0.15,-0.21){2}{\line(0,-1){0.21}}
\multiput(76.89,83.32)(0.11,-0.13){3}{\line(0,-1){0.13}}
\multiput(76.55,83.69)(0.11,-0.13){3}{\line(0,-1){0.13}}
\multiput(76.19,84.05)(0.12,-0.12){3}{\line(1,0){0.12}}
\multiput(75.82,84.39)(0.13,-0.11){3}{\line(1,0){0.13}}
\multiput(75.42,84.7)(0.13,-0.11){3}{\line(1,0){0.13}}
\multiput(75.01,84.99)(0.21,-0.15){2}{\line(1,0){0.21}}
\multiput(74.58,85.26)(0.21,-0.13){2}{\line(1,0){0.21}}
\multiput(74.14,85.51)(0.22,-0.12){2}{\line(1,0){0.22}}
\multiput(73.68,85.73)(0.23,-0.11){2}{\line(1,0){0.23}}
\multiput(73.21,85.92)(0.23,-0.1){2}{\line(1,0){0.23}}
\multiput(72.74,86.09)(0.48,-0.17){1}{\line(1,0){0.48}}
\multiput(72.25,86.23)(0.49,-0.14){1}{\line(1,0){0.49}}
\multiput(71.76,86.34)(0.49,-0.11){1}{\line(1,0){0.49}}
\multiput(71.26,86.43)(0.5,-0.08){1}{\line(1,0){0.5}}
\multiput(70.76,86.48)(0.5,-0.06){1}{\line(1,0){0.5}}
\multiput(70.25,86.51)(0.5,-0.03){1}{\line(1,0){0.5}}
\put(69.75,86.51){\line(1,0){0.51}}
\multiput(69.24,86.48)(0.5,0.03){1}{\line(1,0){0.5}}
\multiput(68.74,86.43)(0.5,0.06){1}{\line(1,0){0.5}}
\multiput(68.24,86.34)(0.5,0.08){1}{\line(1,0){0.5}}
\multiput(67.75,86.23)(0.49,0.11){1}{\line(1,0){0.49}}
\multiput(67.26,86.09)(0.49,0.14){1}{\line(1,0){0.49}}
\multiput(66.79,85.92)(0.48,0.17){1}{\line(1,0){0.48}}
\multiput(66.32,85.73)(0.23,0.1){2}{\line(1,0){0.23}}
\multiput(65.86,85.51)(0.23,0.11){2}{\line(1,0){0.23}}
\multiput(65.42,85.26)(0.22,0.12){2}{\line(1,0){0.22}}
\multiput(64.99,84.99)(0.21,0.13){2}{\line(1,0){0.21}}
\multiput(64.58,84.7)(0.21,0.15){2}{\line(1,0){0.21}}
\multiput(64.18,84.39)(0.13,0.11){3}{\line(1,0){0.13}}
\multiput(63.81,84.05)(0.13,0.11){3}{\line(1,0){0.13}}
\multiput(63.45,83.69)(0.12,0.12){3}{\line(0,1){0.12}}
\multiput(63.11,83.32)(0.11,0.13){3}{\line(0,1){0.13}}
\multiput(62.8,82.92)(0.11,0.13){3}{\line(0,1){0.13}}
\multiput(62.51,82.51)(0.15,0.21){2}{\line(0,1){0.21}}
\multiput(62.24,82.08)(0.13,0.21){2}{\line(0,1){0.21}}
\multiput(61.99,81.64)(0.12,0.22){2}{\line(0,1){0.22}}
\multiput(61.77,81.18)(0.11,0.23){2}{\line(0,1){0.23}}
\multiput(61.58,80.71)(0.1,0.23){2}{\line(0,1){0.23}}
\multiput(61.41,80.24)(0.17,0.48){1}{\line(0,1){0.48}}
\multiput(61.27,79.75)(0.14,0.49){1}{\line(0,1){0.49}}
\multiput(61.16,79.26)(0.11,0.49){1}{\line(0,1){0.49}}
\multiput(61.07,78.76)(0.08,0.5){1}{\line(0,1){0.5}}
\multiput(61.02,78.26)(0.06,0.5){1}{\line(0,1){0.5}}
\multiput(60.99,77.75)(0.03,0.5){1}{\line(0,1){0.5}}
\put(60.99,77.25){\line(0,1){0.51}}
\multiput(60.99,77.25)(0.03,-0.5){1}{\line(0,-1){0.5}}
\multiput(61.02,76.74)(0.06,-0.5){1}{\line(0,-1){0.5}}
\multiput(61.07,76.24)(0.08,-0.5){1}{\line(0,-1){0.5}}
\multiput(61.16,75.74)(0.11,-0.49){1}{\line(0,-1){0.49}}
\multiput(61.27,75.25)(0.14,-0.49){1}{\line(0,-1){0.49}}
\multiput(61.41,74.76)(0.17,-0.48){1}{\line(0,-1){0.48}}
\multiput(61.58,74.29)(0.1,-0.23){2}{\line(0,-1){0.23}}
\multiput(61.77,73.82)(0.11,-0.23){2}{\line(0,-1){0.23}}
\multiput(61.99,73.36)(0.12,-0.22){2}{\line(0,-1){0.22}}
\multiput(62.24,72.92)(0.13,-0.21){2}{\line(0,-1){0.21}}
\multiput(62.51,72.49)(0.15,-0.21){2}{\line(0,-1){0.21}}
\multiput(62.8,72.08)(0.11,-0.13){3}{\line(0,-1){0.13}}
\multiput(63.11,71.68)(0.11,-0.13){3}{\line(0,-1){0.13}}
\multiput(63.45,71.31)(0.12,-0.12){3}{\line(1,0){0.12}}
\multiput(63.81,70.95)(0.13,-0.11){3}{\line(1,0){0.13}}
\multiput(64.18,70.61)(0.13,-0.11){3}{\line(1,0){0.13}}
\multiput(64.58,70.3)(0.21,-0.15){2}{\line(1,0){0.21}}
\multiput(64.99,70.01)(0.21,-0.13){2}{\line(1,0){0.21}}
\multiput(65.42,69.74)(0.22,-0.12){2}{\line(1,0){0.22}}
\multiput(65.86,69.49)(0.23,-0.11){2}{\line(1,0){0.23}}
\multiput(66.32,69.27)(0.23,-0.1){2}{\line(1,0){0.23}}
\multiput(66.79,69.08)(0.48,-0.17){1}{\line(1,0){0.48}}
\multiput(67.26,68.91)(0.49,-0.14){1}{\line(1,0){0.49}}
\multiput(67.75,68.77)(0.49,-0.11){1}{\line(1,0){0.49}}
\multiput(68.24,68.66)(0.5,-0.08){1}{\line(1,0){0.5}}
\multiput(68.74,68.57)(0.5,-0.06){1}{\line(1,0){0.5}}
\multiput(69.24,68.52)(0.5,-0.03){1}{\line(1,0){0.5}}
\put(69.75,68.49){\line(1,0){0.51}}
\multiput(70.25,68.49)(0.5,0.03){1}{\line(1,0){0.5}}
\multiput(70.76,68.52)(0.5,0.06){1}{\line(1,0){0.5}}
\multiput(71.26,68.57)(0.5,0.08){1}{\line(1,0){0.5}}
\multiput(71.76,68.66)(0.49,0.11){1}{\line(1,0){0.49}}
\multiput(72.25,68.77)(0.49,0.14){1}{\line(1,0){0.49}}
\multiput(72.74,68.91)(0.48,0.17){1}{\line(1,0){0.48}}
\multiput(73.21,69.08)(0.23,0.1){2}{\line(1,0){0.23}}
\multiput(73.68,69.27)(0.23,0.11){2}{\line(1,0){0.23}}
\multiput(74.14,69.49)(0.22,0.12){2}{\line(1,0){0.22}}
\multiput(74.58,69.74)(0.21,0.13){2}{\line(1,0){0.21}}
\multiput(75.01,70.01)(0.21,0.15){2}{\line(1,0){0.21}}
\multiput(75.42,70.3)(0.13,0.11){3}{\line(1,0){0.13}}
\multiput(75.82,70.61)(0.13,0.11){3}{\line(1,0){0.13}}
\multiput(76.19,70.95)(0.12,0.12){3}{\line(1,0){0.12}}
\multiput(76.55,71.31)(0.11,0.13){3}{\line(0,1){0.13}}
\multiput(76.89,71.68)(0.11,0.13){3}{\line(0,1){0.13}}
\multiput(77.2,72.08)(0.15,0.21){2}{\line(0,1){0.21}}
\multiput(77.49,72.49)(0.13,0.21){2}{\line(0,1){0.21}}
\multiput(77.76,72.92)(0.12,0.22){2}{\line(0,1){0.22}}
\multiput(78.01,73.36)(0.11,0.23){2}{\line(0,1){0.23}}
\multiput(78.23,73.82)(0.1,0.23){2}{\line(0,1){0.23}}
\multiput(78.42,74.29)(0.17,0.48){1}{\line(0,1){0.48}}
\multiput(78.59,74.76)(0.14,0.49){1}{\line(0,1){0.49}}
\multiput(78.73,75.25)(0.11,0.49){1}{\line(0,1){0.49}}
\multiput(78.84,75.74)(0.08,0.5){1}{\line(0,1){0.5}}
\multiput(78.93,76.24)(0.06,0.5){1}{\line(0,1){0.5}}
\multiput(78.98,76.74)(0.03,0.5){1}{\line(0,1){0.5}}

\linethickness{0.3mm}
\put(121.18,64.75){\line(0,1){0.5}}
\multiput(121.16,65.75)(0.02,-0.5){1}{\line(0,-1){0.5}}
\multiput(121.11,66.25)(0.04,-0.5){1}{\line(0,-1){0.5}}
\multiput(121.04,66.75)(0.07,-0.5){1}{\line(0,-1){0.5}}
\multiput(120.95,67.24)(0.09,-0.49){1}{\line(0,-1){0.49}}
\multiput(120.84,67.73)(0.11,-0.49){1}{\line(0,-1){0.49}}
\multiput(120.71,68.22)(0.13,-0.48){1}{\line(0,-1){0.48}}
\multiput(120.55,68.69)(0.16,-0.48){1}{\line(0,-1){0.48}}
\multiput(120.38,69.16)(0.18,-0.47){1}{\line(0,-1){0.47}}
\multiput(120.18,69.62)(0.1,-0.23){2}{\line(0,-1){0.23}}
\multiput(119.96,70.08)(0.11,-0.23){2}{\line(0,-1){0.23}}
\multiput(119.72,70.52)(0.12,-0.22){2}{\line(0,-1){0.22}}
\multiput(119.47,70.95)(0.13,-0.22){2}{\line(0,-1){0.22}}
\multiput(119.19,71.37)(0.14,-0.21){2}{\line(0,-1){0.21}}
\multiput(118.9,71.77)(0.15,-0.2){2}{\line(0,-1){0.2}}
\multiput(118.58,72.17)(0.1,-0.13){3}{\line(0,-1){0.13}}
\multiput(118.25,72.54)(0.11,-0.13){3}{\line(0,-1){0.13}}
\multiput(117.91,72.91)(0.12,-0.12){3}{\line(0,-1){0.12}}
\multiput(117.54,73.25)(0.12,-0.12){3}{\line(1,0){0.12}}
\multiput(117.17,73.58)(0.13,-0.11){3}{\line(1,0){0.13}}
\multiput(116.77,73.9)(0.13,-0.1){3}{\line(1,0){0.13}}
\multiput(116.37,74.19)(0.2,-0.15){2}{\line(1,0){0.2}}
\multiput(115.95,74.47)(0.21,-0.14){2}{\line(1,0){0.21}}
\multiput(115.52,74.72)(0.22,-0.13){2}{\line(1,0){0.22}}
\multiput(115.08,74.96)(0.22,-0.12){2}{\line(1,0){0.22}}
\multiput(114.62,75.18)(0.23,-0.11){2}{\line(1,0){0.23}}
\multiput(114.16,75.38)(0.23,-0.1){2}{\line(1,0){0.23}}
\multiput(113.69,75.55)(0.47,-0.18){1}{\line(1,0){0.47}}
\multiput(113.22,75.71)(0.48,-0.16){1}{\line(1,0){0.48}}
\multiput(112.73,75.84)(0.48,-0.13){1}{\line(1,0){0.48}}
\multiput(112.24,75.95)(0.49,-0.11){1}{\line(1,0){0.49}}
\multiput(111.75,76.04)(0.49,-0.09){1}{\line(1,0){0.49}}
\multiput(111.25,76.11)(0.5,-0.07){1}{\line(1,0){0.5}}
\multiput(110.75,76.16)(0.5,-0.04){1}{\line(1,0){0.5}}
\multiput(110.25,76.18)(0.5,-0.02){1}{\line(1,0){0.5}}
\put(109.75,76.18){\line(1,0){0.5}}
\multiput(109.25,76.16)(0.5,0.02){1}{\line(1,0){0.5}}
\multiput(108.75,76.11)(0.5,0.04){1}{\line(1,0){0.5}}
\multiput(108.25,76.04)(0.5,0.07){1}{\line(1,0){0.5}}
\multiput(107.76,75.95)(0.49,0.09){1}{\line(1,0){0.49}}
\multiput(107.27,75.84)(0.49,0.11){1}{\line(1,0){0.49}}
\multiput(106.78,75.71)(0.48,0.13){1}{\line(1,0){0.48}}
\multiput(106.31,75.55)(0.48,0.16){1}{\line(1,0){0.48}}
\multiput(105.84,75.38)(0.47,0.18){1}{\line(1,0){0.47}}
\multiput(105.38,75.18)(0.23,0.1){2}{\line(1,0){0.23}}
\multiput(104.92,74.96)(0.23,0.11){2}{\line(1,0){0.23}}
\multiput(104.48,74.72)(0.22,0.12){2}{\line(1,0){0.22}}
\multiput(104.05,74.47)(0.22,0.13){2}{\line(1,0){0.22}}
\multiput(103.63,74.19)(0.21,0.14){2}{\line(1,0){0.21}}
\multiput(103.23,73.9)(0.2,0.15){2}{\line(1,0){0.2}}
\multiput(102.83,73.58)(0.13,0.1){3}{\line(1,0){0.13}}
\multiput(102.46,73.25)(0.13,0.11){3}{\line(1,0){0.13}}
\multiput(102.09,72.91)(0.12,0.12){3}{\line(1,0){0.12}}
\multiput(101.75,72.54)(0.12,0.12){3}{\line(0,1){0.12}}
\multiput(101.42,72.17)(0.11,0.13){3}{\line(0,1){0.13}}
\multiput(101.1,71.77)(0.1,0.13){3}{\line(0,1){0.13}}
\multiput(100.81,71.37)(0.15,0.2){2}{\line(0,1){0.2}}
\multiput(100.53,70.95)(0.14,0.21){2}{\line(0,1){0.21}}
\multiput(100.28,70.52)(0.13,0.22){2}{\line(0,1){0.22}}
\multiput(100.04,70.08)(0.12,0.22){2}{\line(0,1){0.22}}
\multiput(99.82,69.62)(0.11,0.23){2}{\line(0,1){0.23}}
\multiput(99.62,69.16)(0.1,0.23){2}{\line(0,1){0.23}}
\multiput(99.45,68.69)(0.18,0.47){1}{\line(0,1){0.47}}
\multiput(99.29,68.22)(0.16,0.48){1}{\line(0,1){0.48}}
\multiput(99.16,67.73)(0.13,0.48){1}{\line(0,1){0.48}}
\multiput(99.05,67.24)(0.11,0.49){1}{\line(0,1){0.49}}
\multiput(98.96,66.75)(0.09,0.49){1}{\line(0,1){0.49}}
\multiput(98.89,66.25)(0.07,0.5){1}{\line(0,1){0.5}}
\multiput(98.84,65.75)(0.04,0.5){1}{\line(0,1){0.5}}
\multiput(98.82,65.25)(0.02,0.5){1}{\line(0,1){0.5}}
\put(98.82,64.75){\line(0,1){0.5}}
\multiput(98.82,64.75)(0.02,-0.5){1}{\line(0,-1){0.5}}
\multiput(98.84,64.25)(0.04,-0.5){1}{\line(0,-1){0.5}}
\multiput(98.89,63.75)(0.07,-0.5){1}{\line(0,-1){0.5}}
\multiput(98.96,63.25)(0.09,-0.49){1}{\line(0,-1){0.49}}
\multiput(99.05,62.76)(0.11,-0.49){1}{\line(0,-1){0.49}}
\multiput(99.16,62.27)(0.13,-0.48){1}{\line(0,-1){0.48}}
\multiput(99.29,61.78)(0.16,-0.48){1}{\line(0,-1){0.48}}
\multiput(99.45,61.31)(0.18,-0.47){1}{\line(0,-1){0.47}}
\multiput(99.62,60.84)(0.1,-0.23){2}{\line(0,-1){0.23}}
\multiput(99.82,60.38)(0.11,-0.23){2}{\line(0,-1){0.23}}
\multiput(100.04,59.92)(0.12,-0.22){2}{\line(0,-1){0.22}}
\multiput(100.28,59.48)(0.13,-0.22){2}{\line(0,-1){0.22}}
\multiput(100.53,59.05)(0.14,-0.21){2}{\line(0,-1){0.21}}
\multiput(100.81,58.63)(0.15,-0.2){2}{\line(0,-1){0.2}}
\multiput(101.1,58.23)(0.1,-0.13){3}{\line(0,-1){0.13}}
\multiput(101.42,57.83)(0.11,-0.13){3}{\line(0,-1){0.13}}
\multiput(101.75,57.46)(0.12,-0.12){3}{\line(0,-1){0.12}}
\multiput(102.09,57.09)(0.12,-0.12){3}{\line(1,0){0.12}}
\multiput(102.46,56.75)(0.13,-0.11){3}{\line(1,0){0.13}}
\multiput(102.83,56.42)(0.13,-0.1){3}{\line(1,0){0.13}}
\multiput(103.23,56.1)(0.2,-0.15){2}{\line(1,0){0.2}}
\multiput(103.63,55.81)(0.21,-0.14){2}{\line(1,0){0.21}}
\multiput(104.05,55.53)(0.22,-0.13){2}{\line(1,0){0.22}}
\multiput(104.48,55.28)(0.22,-0.12){2}{\line(1,0){0.22}}
\multiput(104.92,55.04)(0.23,-0.11){2}{\line(1,0){0.23}}
\multiput(105.38,54.82)(0.23,-0.1){2}{\line(1,0){0.23}}
\multiput(105.84,54.62)(0.47,-0.18){1}{\line(1,0){0.47}}
\multiput(106.31,54.45)(0.48,-0.16){1}{\line(1,0){0.48}}
\multiput(106.78,54.29)(0.48,-0.13){1}{\line(1,0){0.48}}
\multiput(107.27,54.16)(0.49,-0.11){1}{\line(1,0){0.49}}
\multiput(107.76,54.05)(0.49,-0.09){1}{\line(1,0){0.49}}
\multiput(108.25,53.96)(0.5,-0.07){1}{\line(1,0){0.5}}
\multiput(108.75,53.89)(0.5,-0.04){1}{\line(1,0){0.5}}
\multiput(109.25,53.84)(0.5,-0.02){1}{\line(1,0){0.5}}
\put(109.75,53.82){\line(1,0){0.5}}
\multiput(110.25,53.82)(0.5,0.02){1}{\line(1,0){0.5}}
\multiput(110.75,53.84)(0.5,0.04){1}{\line(1,0){0.5}}
\multiput(111.25,53.89)(0.5,0.07){1}{\line(1,0){0.5}}
\multiput(111.75,53.96)(0.49,0.09){1}{\line(1,0){0.49}}
\multiput(112.24,54.05)(0.49,0.11){1}{\line(1,0){0.49}}
\multiput(112.73,54.16)(0.48,0.13){1}{\line(1,0){0.48}}
\multiput(113.22,54.29)(0.48,0.16){1}{\line(1,0){0.48}}
\multiput(113.69,54.45)(0.47,0.18){1}{\line(1,0){0.47}}
\multiput(114.16,54.62)(0.23,0.1){2}{\line(1,0){0.23}}
\multiput(114.62,54.82)(0.23,0.11){2}{\line(1,0){0.23}}
\multiput(115.08,55.04)(0.22,0.12){2}{\line(1,0){0.22}}
\multiput(115.52,55.28)(0.22,0.13){2}{\line(1,0){0.22}}
\multiput(115.95,55.53)(0.21,0.14){2}{\line(1,0){0.21}}
\multiput(116.37,55.81)(0.2,0.15){2}{\line(1,0){0.2}}
\multiput(116.77,56.1)(0.13,0.1){3}{\line(1,0){0.13}}
\multiput(117.17,56.42)(0.13,0.11){3}{\line(1,0){0.13}}
\multiput(117.54,56.75)(0.12,0.12){3}{\line(1,0){0.12}}
\multiput(117.91,57.09)(0.12,0.12){3}{\line(0,1){0.12}}
\multiput(118.25,57.46)(0.11,0.13){3}{\line(0,1){0.13}}
\multiput(118.58,57.83)(0.1,0.13){3}{\line(0,1){0.13}}
\multiput(118.9,58.23)(0.15,0.2){2}{\line(0,1){0.2}}
\multiput(119.19,58.63)(0.14,0.21){2}{\line(0,1){0.21}}
\multiput(119.47,59.05)(0.13,0.22){2}{\line(0,1){0.22}}
\multiput(119.72,59.48)(0.12,0.22){2}{\line(0,1){0.22}}
\multiput(119.96,59.92)(0.11,0.23){2}{\line(0,1){0.23}}
\multiput(120.18,60.38)(0.1,0.23){2}{\line(0,1){0.23}}
\multiput(120.38,60.84)(0.18,0.47){1}{\line(0,1){0.47}}
\multiput(120.55,61.31)(0.16,0.48){1}{\line(0,1){0.48}}
\multiput(120.71,61.78)(0.13,0.48){1}{\line(0,1){0.48}}
\multiput(120.84,62.27)(0.11,0.49){1}{\line(0,1){0.49}}
\multiput(120.95,62.76)(0.09,0.49){1}{\line(0,1){0.49}}
\multiput(121.04,63.25)(0.07,0.5){1}{\line(0,1){0.5}}
\multiput(121.11,63.75)(0.04,0.5){1}{\line(0,1){0.5}}
\multiput(121.16,64.25)(0.02,0.5){1}{\line(0,1){0.5}}

\linethickness{0.3mm}
\put(60,90){\line(1,0){120}}
\put(70,76){\makebox(0,0)[cc]{$D_1$}}

\put(109,65){\makebox(0,0)[cc]{$D_2$}}

\put(95,80){\makebox(0,0)[cc]{$S_1$}}

\put(105,95){\makebox(0,0)[cc]{$S_2$}}

\put(68,64){\makebox(0,0)[cc]{$C_1$}}

\put(127,65){\makebox(0,0)[cc]{$C_2$}}

\put(140,85){\makebox(0,0)[cc]{$C_3$}}

\put(125,105){\makebox(0,0)[cc]{$C_4$}}

\put(88,45){\makebox(0,0)[cc]{$C_4$}}

\end{picture}

}

\def\figtwocut{

\def\JPicScale{0.45}
\ifx\JPicScale\undefined\def\JPicScale{1}\fi
\unitlength \JPicScale mm
\begin{picture}(155,90)(0,0)
\linethickness{0.3mm}
\put(80,0){\line(0,1){40}}
\linethickness{0.3mm}
\put(80,40){\line(1,0){40}}
\linethickness{0.3mm}
\put(120,40){\line(0,1){20}}
\linethickness{0.3mm}
\put(80,60){\line(1,0){40}}
\linethickness{0.3mm}
\put(80,60){\line(0,1){30}}
\put(130,50){\makebox(0,0)[cc]{x}}

\put(155,50){\makebox(0,0)[cc]{x}}

\put(40,50){\makebox(0,0)[cc]{x}}

\put(100,50){\makebox(0,0)[cc]{x}}

\put(35,45){\makebox(0,0)[cc]{$Q_2$}}

\put(130,45){\makebox(0,0)[cc]{$Q_1$}}

\put(100,45){\makebox(0,0)[cc]{$Q_4$}}

\put(155,45){\makebox(0,0)[cc]{$Q_3$}}

\put(100,-10){\makebox(0,0)[cc]{(a)}}

\end{picture}

}

\def \figthreecut{

\def\JPicScale{0.45}
\ifx\JPicScale\undefined\def\JPicScale{1}\fi
\unitlength \JPicScale mm
\begin{picture}(150,105)(0,0)
\put(30,50){\makebox(0,0)[cc]{x}}

\put(90,50){\makebox(0,0)[cc]{x}}

\put(110,50){\makebox(0,0)[cc]{x}}

\put(150,50){\makebox(0,0)[cc]{x}}

\put(25,45){\makebox(0,0)[cc]{$Q_2$}}

\put(85,45){\makebox(0,0)[cc]{$Q_1$}}

\put(110,45){\makebox(0,0)[cc]{$Q_4$}}

\put(150,45){\makebox(0,0)[cc]{$Q_3$}}

\linethickness{0.3mm}
\qbezier(60,105)(60,102.41)(60,100)
\qbezier(60,100)(60,97.59)(60,95)
\qbezier(60,95)(60,92.44)(60,87.62)
\qbezier(60,87.62)(60,82.81)(60,75)
\qbezier(60,75)(59.95,67.22)(64.16,60)
\qbezier(64.16,60)(68.37,52.78)(77.5,45)
\qbezier(77.5,45)(86.59,37.1)(93.81,38.91)
\qbezier(93.81,38.91)(101.03,40.71)(107.5,52.5)
\qbezier(107.5,52.5)(114.02,64.31)(117.03,64.31)
\qbezier(117.03,64.31)(120.04,64.31)(120,52.5)
\qbezier(120,52.5)(120.02,40.78)(118.22,33.56)
\qbezier(118.22,33.56)(116.41,26.34)(112.5,22.5)
\qbezier(112.5,22.5)(108.62,18.58)(104.41,17.38)
\qbezier(104.41,17.38)(100.2,16.17)(95,17.5)
\qbezier(95,17.5)(89.8,18.77)(85.59,22.38)
\qbezier(85.59,22.38)(81.38,25.98)(77.5,32.5)
\qbezier(77.5,32.5)(73.6,39.03)(70.59,41.44)
\qbezier(70.59,41.44)(67.59,43.84)(65,42.5)
\qbezier(65,42.5)(62.39,41.21)(61.19,39.41)
\qbezier(61.19,39.41)(59.98,37.6)(60,35)
\qbezier(60,35)(60,32.41)(60,30)
\qbezier(60,30)(60,27.59)(60,25)
\qbezier(60,25)(60,22.42)(60,18.81)
\qbezier(60,18.81)(60,15.2)(60,10)

\put(100,-10){\makebox(0,0)[cc]{(b)}}

\end{picture}

}

\def\figfieldredef{
\def\JPicScale{0.8}
\ifx\JPicScale\undefined\def\JPicScale{1}\fi
\unitlength \JPicScale mm
\begin{picture}(155,65)(0,0)
\linethickness{0.3mm}
\put(20,60){\line(1,0){120}}
\put(20,60){\vector(-1,0){0.12}}
\linethickness{0.3mm}
\put(20,40){\line(0,1){20}}
\put(20,40){\vector(0,-1){0.12}}
\linethickness{0.3mm}
\put(20,40){\line(1,0){120}}
\put(140,40){\vector(1,0){0.12}}
\linethickness{0.3mm}
\put(140,40){\line(0,1){20}}
\put(140,60){\vector(0,1){0.12}}
\put(80,65){\makebox(0,0)[cc]{$\RR'_{g,m,n}$}}

\put(80,50){\makebox(0,0)[cc]{$\wt\RR_{g,m,n}$}}

\put(80,35){\makebox(0,0)[cc]{$-\RR_{g,m,n}$}}

\linethickness{0.3mm}
\put(155,40){\line(0,1){20}}
\put(155,60){\vector(0,1){0.12}}
\put(156,35){\makebox(0,0)[cc]{$\wh U_{g,m,n}$}}

\end{picture}

}

\def\figfourcut{

\def\JPicScale{0.45}
\ifx\JPicScale\undefined\def\JPicScale{1}\fi
\unitlength \JPicScale mm
\begin{picture}(150,90)(0,0)
\put(30,50){\makebox(0,0)[cc]{x}}

\put(90,50){\makebox(0,0)[cc]{x}}

\put(110,50){\makebox(0,0)[cc]{x}}

\put(150,50){\makebox(0,0)[cc]{x}}

\put(35,45){\makebox(0,0)[cc]{$Q_2$}}

\put(85,45){\makebox(0,0)[cc]{$Q_1$}}

\put(115,45){\makebox(0,0)[cc]{$Q_4$}}

\put(150,45){\makebox(0,0)[cc]{$Q_3$}}

\linethickness{0.3mm}
\qbezier(60,90)(60,84.8)(60,81.19)
\qbezier(60,81.19)(60,77.58)(60,75)
\qbezier(60,75)(60,72.41)(60,70)
\qbezier(60,70)(60,67.59)(60,65)
\qbezier(60,65)(59.98,62.4)(61.19,60.59)
\qbezier(61.19,60.59)(62.39,58.79)(65,57.5)
\qbezier(65,57.5)(67.59,56.16)(70.59,58.56)
\qbezier(70.59,58.56)(73.6,60.97)(77.5,67.5)
\qbezier(77.5,67.5)(81.38,74.02)(85.59,77.62)
\qbezier(85.59,77.62)(89.8,81.23)(95,82.5)
\qbezier(95,82.5)(100.2,83.83)(104.41,82.62)
\qbezier(104.41,82.62)(108.62,81.42)(112.5,77.5)
\qbezier(112.5,77.5)(116.41,73.66)(118.22,66.44)
\qbezier(118.22,66.44)(120.02,59.22)(120,47.5)
\qbezier(120,47.5)(120.04,35.69)(117.03,35.69)
\qbezier(117.03,35.69)(114.02,35.69)(107.5,47.5)
\qbezier(107.5,47.5)(101.03,59.29)(93.81,61.09)

\qbezier(93.81,61.09)(86.59,62.9)(77.5,55)
\qbezier(77.5,55)(68.37,47.22)(64.16,40)
\qbezier(64.16,40)(59.95,32.78)(60,25)
\qbezier(60,25)(60,17.19)(60,12.38)
\qbezier(60,12.38)(60,7.56)(60,5)
\qbezier(60,5)(60,2.41)(60,0)
\qbezier(60,0)(60,-2.41)(60,-5)

\put(100,-10){\makebox(0,0)[cc]{(b)}}

\end{picture}

}

\def\figqft{

\def\JPicScale{0.4}
\ifx\JPicScale\undefined\def\JPicScale{1}\fi
\unitlength \JPicScale mm
\begin{picture}(135,80)(0,0)
\linethickness{0.3mm}
\put(100.35,48.61){\line(0,1){0.5}}
\multiput(100.34,49.62)(0.01,-0.5){1}{\line(0,-1){0.5}}
\multiput(100.32,50.12)(0.02,-0.5){1}{\line(0,-1){0.5}}
\multiput(100.28,50.62)(0.04,-0.5){1}{\line(0,-1){0.5}}
\multiput(100.23,51.12)(0.05,-0.5){1}{\line(0,-1){0.5}}
\multiput(100.17,51.62)(0.06,-0.5){1}{\line(0,-1){0.5}}
\multiput(100.1,52.12)(0.07,-0.5){1}{\line(0,-1){0.5}}
\multiput(100.02,52.62)(0.08,-0.5){1}{\line(0,-1){0.5}}
\multiput(99.92,53.11)(0.1,-0.49){1}{\line(0,-1){0.49}}
\multiput(99.82,53.61)(0.11,-0.49){1}{\line(0,-1){0.49}}
\multiput(99.7,54.09)(0.12,-0.49){1}{\line(0,-1){0.49}}
\multiput(99.57,54.58)(0.13,-0.49){1}{\line(0,-1){0.49}}
\multiput(99.43,55.06)(0.14,-0.48){1}{\line(0,-1){0.48}}
\multiput(99.27,55.54)(0.15,-0.48){1}{\line(0,-1){0.48}}
\multiput(99.11,56.02)(0.16,-0.48){1}{\line(0,-1){0.48}}
\multiput(98.93,56.49)(0.18,-0.47){1}{\line(0,-1){0.47}}
\multiput(98.75,56.96)(0.09,-0.23){2}{\line(0,-1){0.23}}
\multiput(98.55,57.42)(0.1,-0.23){2}{\line(0,-1){0.23}}
\multiput(98.34,57.88)(0.1,-0.23){2}{\line(0,-1){0.23}}
\multiput(98.12,58.33)(0.11,-0.23){2}{\line(0,-1){0.23}}
\multiput(97.89,58.78)(0.12,-0.22){2}{\line(0,-1){0.22}}
\multiput(97.65,59.22)(0.12,-0.22){2}{\line(0,-1){0.22}}
\multiput(97.39,59.66)(0.13,-0.22){2}{\line(0,-1){0.22}}
\multiput(97.13,60.09)(0.13,-0.21){2}{\line(0,-1){0.21}}
\multiput(96.86,60.51)(0.14,-0.21){2}{\line(0,-1){0.21}}
\multiput(96.58,60.93)(0.14,-0.21){2}{\line(0,-1){0.21}}
\multiput(96.29,61.34)(0.15,-0.21){2}{\line(0,-1){0.21}}
\multiput(95.99,61.74)(0.1,-0.13){3}{\line(0,-1){0.13}}
\multiput(95.67,62.14)(0.1,-0.13){3}{\line(0,-1){0.13}}
\multiput(95.35,62.52)(0.11,-0.13){3}{\line(0,-1){0.13}}
\multiput(95.02,62.9)(0.11,-0.13){3}{\line(0,-1){0.13}}
\multiput(94.69,63.28)(0.11,-0.12){3}{\line(0,-1){0.12}}
\multiput(94.34,63.64)(0.12,-0.12){3}{\line(0,-1){0.12}}
\multiput(93.98,64)(0.12,-0.12){3}{\line(0,-1){0.12}}
\multiput(93.62,64.34)(0.12,-0.12){3}{\line(1,0){0.12}}
\multiput(93.24,64.68)(0.12,-0.11){3}{\line(1,0){0.12}}
\multiput(92.86,65.01)(0.13,-0.11){3}{\line(1,0){0.13}}
\multiput(92.48,65.33)(0.13,-0.11){3}{\line(1,0){0.13}}
\multiput(92.08,65.64)(0.13,-0.1){3}{\line(1,0){0.13}}
\multiput(91.68,65.95)(0.13,-0.1){3}{\line(1,0){0.13}}
\multiput(91.27,66.24)(0.21,-0.15){2}{\line(1,0){0.21}}
\multiput(90.85,66.52)(0.21,-0.14){2}{\line(1,0){0.21}}
\multiput(90.43,66.79)(0.21,-0.14){2}{\line(1,0){0.21}}
\multiput(90,67.05)(0.21,-0.13){2}{\line(1,0){0.21}}
\multiput(89.56,67.31)(0.22,-0.13){2}{\line(1,0){0.22}}
\multiput(89.12,67.55)(0.22,-0.12){2}{\line(1,0){0.22}}
\multiput(88.67,67.78)(0.22,-0.12){2}{\line(1,0){0.22}}
\multiput(88.22,68)(0.23,-0.11){2}{\line(1,0){0.23}}
\multiput(87.76,68.21)(0.23,-0.1){2}{\line(1,0){0.23}}
\multiput(87.3,68.4)(0.23,-0.1){2}{\line(1,0){0.23}}
\multiput(86.83,68.59)(0.23,-0.09){2}{\line(1,0){0.23}}
\multiput(86.36,68.77)(0.47,-0.18){1}{\line(1,0){0.47}}
\multiput(85.88,68.93)(0.48,-0.16){1}{\line(1,0){0.48}}
\multiput(85.4,69.09)(0.48,-0.15){1}{\line(1,0){0.48}}
\multiput(84.92,69.23)(0.48,-0.14){1}{\line(1,0){0.48}}
\multiput(84.44,69.36)(0.49,-0.13){1}{\line(1,0){0.49}}
\multiput(83.95,69.48)(0.49,-0.12){1}{\line(1,0){0.49}}
\multiput(83.45,69.58)(0.49,-0.11){1}{\line(1,0){0.49}}
\multiput(82.96,69.68)(0.49,-0.1){1}{\line(1,0){0.49}}
\multiput(82.46,69.76)(0.5,-0.08){1}{\line(1,0){0.5}}
\multiput(81.97,69.83)(0.5,-0.07){1}{\line(1,0){0.5}}
\multiput(81.47,69.89)(0.5,-0.06){1}{\line(1,0){0.5}}
\multiput(80.96,69.94)(0.5,-0.05){1}{\line(1,0){0.5}}
\multiput(80.46,69.98)(0.5,-0.04){1}{\line(1,0){0.5}}
\multiput(79.96,70)(0.5,-0.02){1}{\line(1,0){0.5}}
\multiput(79.46,70.01)(0.5,-0.01){1}{\line(1,0){0.5}}
\put(78.95,70.01){\line(1,0){0.5}}
\multiput(78.45,70)(0.5,0.01){1}{\line(1,0){0.5}}
\multiput(77.95,69.98)(0.5,0.02){1}{\line(1,0){0.5}}
\multiput(77.44,69.94)(0.5,0.04){1}{\line(1,0){0.5}}
\multiput(76.94,69.89)(0.5,0.05){1}{\line(1,0){0.5}}
\multiput(76.44,69.83)(0.5,0.06){1}{\line(1,0){0.5}}
\multiput(75.95,69.76)(0.5,0.07){1}{\line(1,0){0.5}}
\multiput(75.45,69.68)(0.5,0.08){1}{\line(1,0){0.5}}
\multiput(74.95,69.58)(0.49,0.1){1}{\line(1,0){0.49}}
\multiput(74.46,69.48)(0.49,0.11){1}{\line(1,0){0.49}}
\multiput(73.97,69.36)(0.49,0.12){1}{\line(1,0){0.49}}
\multiput(73.49,69.23)(0.49,0.13){1}{\line(1,0){0.49}}
\multiput(73,69.09)(0.48,0.14){1}{\line(1,0){0.48}}
\multiput(72.52,68.93)(0.48,0.15){1}{\line(1,0){0.48}}
\multiput(72.05,68.77)(0.48,0.16){1}{\line(1,0){0.48}}
\multiput(71.58,68.59)(0.47,0.18){1}{\line(1,0){0.47}}
\multiput(71.11,68.4)(0.23,0.09){2}{\line(1,0){0.23}}
\multiput(70.65,68.21)(0.23,0.1){2}{\line(1,0){0.23}}
\multiput(70.19,68)(0.23,0.1){2}{\line(1,0){0.23}}
\multiput(69.74,67.78)(0.23,0.11){2}{\line(1,0){0.23}}
\multiput(69.29,67.55)(0.22,0.12){2}{\line(1,0){0.22}}
\multiput(68.85,67.31)(0.22,0.12){2}{\line(1,0){0.22}}
\multiput(68.41,67.05)(0.22,0.13){2}{\line(1,0){0.22}}
\multiput(67.98,66.79)(0.21,0.13){2}{\line(1,0){0.21}}
\multiput(67.56,66.52)(0.21,0.14){2}{\line(1,0){0.21}}
\multiput(67.14,66.24)(0.21,0.14){2}{\line(1,0){0.21}}
\multiput(66.73,65.95)(0.21,0.15){2}{\line(1,0){0.21}}
\multiput(66.33,65.64)(0.13,0.1){3}{\line(1,0){0.13}}
\multiput(65.93,65.33)(0.13,0.1){3}{\line(1,0){0.13}}
\multiput(65.54,65.01)(0.13,0.11){3}{\line(1,0){0.13}}
\multiput(65.16,64.68)(0.13,0.11){3}{\line(1,0){0.13}}
\multiput(64.79,64.34)(0.12,0.11){3}{\line(1,0){0.12}}
\multiput(64.43,64)(0.12,0.12){3}{\line(1,0){0.12}}
\multiput(64.07,63.64)(0.12,0.12){3}{\line(1,0){0.12}}
\multiput(63.72,63.28)(0.12,0.12){3}{\line(0,1){0.12}}
\multiput(63.39,62.9)(0.11,0.12){3}{\line(0,1){0.12}}
\multiput(63.06,62.52)(0.11,0.13){3}{\line(0,1){0.13}}
\multiput(62.74,62.14)(0.11,0.13){3}{\line(0,1){0.13}}
\multiput(62.42,61.74)(0.1,0.13){3}{\line(0,1){0.13}}
\multiput(62.12,61.34)(0.1,0.13){3}{\line(0,1){0.13}}
\multiput(61.83,60.93)(0.15,0.21){2}{\line(0,1){0.21}}
\multiput(61.55,60.51)(0.14,0.21){2}{\line(0,1){0.21}}
\multiput(61.28,60.09)(0.14,0.21){2}{\line(0,1){0.21}}
\multiput(61.01,59.66)(0.13,0.21){2}{\line(0,1){0.21}}
\multiput(60.76,59.22)(0.13,0.22){2}{\line(0,1){0.22}}
\multiput(60.52,58.78)(0.12,0.22){2}{\line(0,1){0.22}}
\multiput(60.29,58.33)(0.12,0.22){2}{\line(0,1){0.22}}
\multiput(60.07,57.88)(0.11,0.23){2}{\line(0,1){0.23}}
\multiput(59.86,57.42)(0.1,0.23){2}{\line(0,1){0.23}}
\multiput(59.66,56.96)(0.1,0.23){2}{\line(0,1){0.23}}
\multiput(59.48,56.49)(0.09,0.23){2}{\line(0,1){0.23}}
\multiput(59.3,56.02)(0.18,0.47){1}{\line(0,1){0.47}}
\multiput(59.14,55.54)(0.16,0.48){1}{\line(0,1){0.48}}
\multiput(58.98,55.06)(0.15,0.48){1}{\line(0,1){0.48}}
\multiput(58.84,54.58)(0.14,0.48){1}{\line(0,1){0.48}}
\multiput(58.71,54.09)(0.13,0.49){1}{\line(0,1){0.49}}
\multiput(58.59,53.61)(0.12,0.49){1}{\line(0,1){0.49}}
\multiput(58.48,53.11)(0.11,0.49){1}{\line(0,1){0.49}}
\multiput(58.39,52.62)(0.1,0.49){1}{\line(0,1){0.49}}
\multiput(58.31,52.12)(0.08,0.5){1}{\line(0,1){0.5}}
\multiput(58.23,51.62)(0.07,0.5){1}{\line(0,1){0.5}}
\multiput(58.17,51.12)(0.06,0.5){1}{\line(0,1){0.5}}
\multiput(58.13,50.62)(0.05,0.5){1}{\line(0,1){0.5}}
\multiput(58.09,50.12)(0.04,0.5){1}{\line(0,1){0.5}}
\multiput(58.07,49.62)(0.02,0.5){1}{\line(0,1){0.5}}
\multiput(58.05,49.12)(0.01,0.5){1}{\line(0,1){0.5}}
\put(58.05,48.61){\line(0,1){0.5}}
\multiput(58.05,48.61)(0.01,-0.5){1}{\line(0,-1){0.5}}
\multiput(58.07,48.11)(0.02,-0.5){1}{\line(0,-1){0.5}}
\multiput(58.09,47.61)(0.04,-0.5){1}{\line(0,-1){0.5}}
\multiput(58.13,47.1)(0.05,-0.5){1}{\line(0,-1){0.5}}
\multiput(58.17,46.6)(0.06,-0.5){1}{\line(0,-1){0.5}}
\multiput(58.23,46.1)(0.07,-0.5){1}{\line(0,-1){0.5}}
\multiput(58.31,45.6)(0.08,-0.5){1}{\line(0,-1){0.5}}
\multiput(58.39,45.11)(0.1,-0.49){1}{\line(0,-1){0.49}}
\multiput(58.48,44.61)(0.11,-0.49){1}{\line(0,-1){0.49}}
\multiput(58.59,44.12)(0.12,-0.49){1}{\line(0,-1){0.49}}
\multiput(58.71,43.63)(0.13,-0.49){1}{\line(0,-1){0.49}}
\multiput(58.84,43.15)(0.14,-0.48){1}{\line(0,-1){0.48}}
\multiput(58.98,42.66)(0.15,-0.48){1}{\line(0,-1){0.48}}
\multiput(59.14,42.18)(0.16,-0.48){1}{\line(0,-1){0.48}}
\multiput(59.3,41.71)(0.18,-0.47){1}{\line(0,-1){0.47}}
\multiput(59.48,41.24)(0.09,-0.23){2}{\line(0,-1){0.23}}
\multiput(59.66,40.77)(0.1,-0.23){2}{\line(0,-1){0.23}}
\multiput(59.86,40.31)(0.1,-0.23){2}{\line(0,-1){0.23}}
\multiput(60.07,39.85)(0.11,-0.23){2}{\line(0,-1){0.23}}
\multiput(60.29,39.4)(0.12,-0.22){2}{\line(0,-1){0.22}}
\multiput(60.52,38.95)(0.12,-0.22){2}{\line(0,-1){0.22}}
\multiput(60.76,38.51)(0.13,-0.22){2}{\line(0,-1){0.22}}
\multiput(61.01,38.07)(0.13,-0.21){2}{\line(0,-1){0.21}}
\multiput(61.28,37.64)(0.14,-0.21){2}{\line(0,-1){0.21}}
\multiput(61.55,37.22)(0.14,-0.21){2}{\line(0,-1){0.21}}
\multiput(61.83,36.8)(0.15,-0.21){2}{\line(0,-1){0.21}}
\multiput(62.12,36.39)(0.1,-0.13){3}{\line(0,-1){0.13}}
\multiput(62.42,35.99)(0.1,-0.13){3}{\line(0,-1){0.13}}
\multiput(62.74,35.59)(0.11,-0.13){3}{\line(0,-1){0.13}}
\multiput(63.06,35.2)(0.11,-0.13){3}{\line(0,-1){0.13}}
\multiput(63.39,34.82)(0.11,-0.12){3}{\line(0,-1){0.12}}
\multiput(63.72,34.45)(0.12,-0.12){3}{\line(0,-1){0.12}}
\multiput(64.07,34.09)(0.12,-0.12){3}{\line(0,-1){0.12}}
\multiput(64.43,33.73)(0.12,-0.12){3}{\line(1,0){0.12}}
\multiput(64.79,33.38)(0.12,-0.11){3}{\line(1,0){0.12}}
\multiput(65.16,33.04)(0.13,-0.11){3}{\line(1,0){0.13}}
\multiput(65.54,32.71)(0.13,-0.11){3}{\line(1,0){0.13}}
\multiput(65.93,32.39)(0.13,-0.1){3}{\line(1,0){0.13}}
\multiput(66.33,32.08)(0.13,-0.1){3}{\line(1,0){0.13}}
\multiput(66.73,31.78)(0.21,-0.15){2}{\line(1,0){0.21}}
\multiput(67.14,31.49)(0.21,-0.14){2}{\line(1,0){0.21}}
\multiput(67.56,31.21)(0.21,-0.14){2}{\line(1,0){0.21}}
\multiput(67.98,30.94)(0.21,-0.13){2}{\line(1,0){0.21}}
\multiput(68.41,30.67)(0.22,-0.13){2}{\line(1,0){0.22}}
\multiput(68.85,30.42)(0.22,-0.12){2}{\line(1,0){0.22}}
\multiput(69.29,30.18)(0.22,-0.12){2}{\line(1,0){0.22}}
\multiput(69.74,29.95)(0.23,-0.11){2}{\line(1,0){0.23}}
\multiput(70.19,29.73)(0.23,-0.1){2}{\line(1,0){0.23}}
\multiput(70.65,29.52)(0.23,-0.1){2}{\line(1,0){0.23}}
\multiput(71.11,29.32)(0.23,-0.09){2}{\line(1,0){0.23}}
\multiput(71.58,29.14)(0.47,-0.18){1}{\line(1,0){0.47}}
\multiput(72.05,28.96)(0.48,-0.16){1}{\line(1,0){0.48}}
\multiput(72.52,28.79)(0.48,-0.15){1}{\line(1,0){0.48}}
\multiput(73,28.64)(0.48,-0.14){1}{\line(1,0){0.48}}
\multiput(73.49,28.5)(0.49,-0.13){1}{\line(1,0){0.49}}
\multiput(73.97,28.37)(0.49,-0.12){1}{\line(1,0){0.49}}
\multiput(74.46,28.25)(0.49,-0.11){1}{\line(1,0){0.49}}
\multiput(74.95,28.14)(0.49,-0.1){1}{\line(1,0){0.49}}
\multiput(75.45,28.05)(0.5,-0.08){1}{\line(1,0){0.5}}
\multiput(75.95,27.96)(0.5,-0.07){1}{\line(1,0){0.5}}
\multiput(76.44,27.89)(0.5,-0.06){1}{\line(1,0){0.5}}
\multiput(76.94,27.83)(0.5,-0.05){1}{\line(1,0){0.5}}
\multiput(77.44,27.79)(0.5,-0.04){1}{\line(1,0){0.5}}
\multiput(77.95,27.75)(0.5,-0.02){1}{\line(1,0){0.5}}
\multiput(78.45,27.73)(0.5,-0.01){1}{\line(1,0){0.5}}
\put(78.95,27.71){\line(1,0){0.5}}
\multiput(79.46,27.71)(0.5,0.01){1}{\line(1,0){0.5}}
\multiput(79.96,27.73)(0.5,0.02){1}{\line(1,0){0.5}}
\multiput(80.46,27.75)(0.5,0.04){1}{\line(1,0){0.5}}
\multiput(80.96,27.79)(0.5,0.05){1}{\line(1,0){0.5}}
\multiput(81.47,27.83)(0.5,0.06){1}{\line(1,0){0.5}}
\multiput(81.97,27.89)(0.5,0.07){1}{\line(1,0){0.5}}
\multiput(82.46,27.96)(0.5,0.08){1}{\line(1,0){0.5}}
\multiput(82.96,28.05)(0.49,0.1){1}{\line(1,0){0.49}}
\multiput(83.45,28.14)(0.49,0.11){1}{\line(1,0){0.49}}
\multiput(83.95,28.25)(0.49,0.12){1}{\line(1,0){0.49}}
\multiput(84.44,28.37)(0.49,0.13){1}{\line(1,0){0.49}}
\multiput(84.92,28.5)(0.48,0.14){1}{\line(1,0){0.48}}
\multiput(85.4,28.64)(0.48,0.15){1}{\line(1,0){0.48}}
\multiput(85.88,28.79)(0.48,0.16){1}{\line(1,0){0.48}}
\multiput(86.36,28.96)(0.47,0.18){1}{\line(1,0){0.47}}
\multiput(86.83,29.14)(0.23,0.09){2}{\line(1,0){0.23}}
\multiput(87.3,29.32)(0.23,0.1){2}{\line(1,0){0.23}}
\multiput(87.76,29.52)(0.23,0.1){2}{\line(1,0){0.23}}
\multiput(88.22,29.73)(0.23,0.11){2}{\line(1,0){0.23}}
\multiput(88.67,29.95)(0.22,0.12){2}{\line(1,0){0.22}}
\multiput(89.12,30.18)(0.22,0.12){2}{\line(1,0){0.22}}
\multiput(89.56,30.42)(0.22,0.13){2}{\line(1,0){0.22}}
\multiput(90,30.67)(0.21,0.13){2}{\line(1,0){0.21}}
\multiput(90.43,30.94)(0.21,0.14){2}{\line(1,0){0.21}}
\multiput(90.85,31.21)(0.21,0.14){2}{\line(1,0){0.21}}
\multiput(91.27,31.49)(0.21,0.15){2}{\line(1,0){0.21}}
\multiput(91.68,31.78)(0.13,0.1){3}{\line(1,0){0.13}}
\multiput(92.08,32.08)(0.13,0.1){3}{\line(1,0){0.13}}
\multiput(92.48,32.39)(0.13,0.11){3}{\line(1,0){0.13}}
\multiput(92.86,32.71)(0.13,0.11){3}{\line(1,0){0.13}}
\multiput(93.24,33.04)(0.12,0.11){3}{\line(1,0){0.12}}
\multiput(93.62,33.38)(0.12,0.12){3}{\line(1,0){0.12}}
\multiput(93.98,33.73)(0.12,0.12){3}{\line(1,0){0.12}}
\multiput(94.34,34.09)(0.12,0.12){3}{\line(0,1){0.12}}
\multiput(94.69,34.45)(0.11,0.12){3}{\line(0,1){0.12}}
\multiput(95.02,34.82)(0.11,0.13){3}{\line(0,1){0.13}}
\multiput(95.35,35.2)(0.11,0.13){3}{\line(0,1){0.13}}
\multiput(95.67,35.59)(0.1,0.13){3}{\line(0,1){0.13}}
\multiput(95.99,35.99)(0.1,0.13){3}{\line(0,1){0.13}}
\multiput(96.29,36.39)(0.15,0.21){2}{\line(0,1){0.21}}
\multiput(96.58,36.8)(0.14,0.21){2}{\line(0,1){0.21}}
\multiput(96.86,37.22)(0.14,0.21){2}{\line(0,1){0.21}}
\multiput(97.13,37.64)(0.13,0.21){2}{\line(0,1){0.21}}
\multiput(97.39,38.07)(0.13,0.22){2}{\line(0,1){0.22}}
\multiput(97.65,38.51)(0.12,0.22){2}{\line(0,1){0.22}}
\multiput(97.89,38.95)(0.12,0.22){2}{\line(0,1){0.22}}
\multiput(98.12,39.4)(0.11,0.23){2}{\line(0,1){0.23}}
\multiput(98.34,39.85)(0.1,0.23){2}{\line(0,1){0.23}}
\multiput(98.55,40.31)(0.1,0.23){2}{\line(0,1){0.23}}
\multiput(98.75,40.77)(0.09,0.23){2}{\line(0,1){0.23}}
\multiput(98.93,41.24)(0.18,0.47){1}{\line(0,1){0.47}}
\multiput(99.11,41.71)(0.16,0.48){1}{\line(0,1){0.48}}
\multiput(99.27,42.18)(0.15,0.48){1}{\line(0,1){0.48}}
\multiput(99.43,42.66)(0.14,0.48){1}{\line(0,1){0.48}}
\multiput(99.57,43.15)(0.13,0.49){1}{\line(0,1){0.49}}
\multiput(99.7,43.63)(0.12,0.49){1}{\line(0,1){0.49}}
\multiput(99.82,44.12)(0.11,0.49){1}{\line(0,1){0.49}}
\multiput(99.92,44.61)(0.1,0.49){1}{\line(0,1){0.49}}
\multiput(100.02,45.11)(0.08,0.5){1}{\line(0,1){0.5}}
\multiput(100.1,45.6)(0.07,0.5){1}{\line(0,1){0.5}}
\multiput(100.17,46.1)(0.06,0.5){1}{\line(0,1){0.5}}
\multiput(100.23,46.6)(0.05,0.5){1}{\line(0,1){0.5}}
\multiput(100.28,47.1)(0.04,0.5){1}{\line(0,1){0.5}}
\multiput(100.32,47.61)(0.02,0.5){1}{\line(0,1){0.5}}
\multiput(100.34,48.11)(0.01,0.5){1}{\line(0,1){0.5}}

\linethickness{0.9mm}
\put(30,50){\line(1,0){28}}
\linethickness{0.9mm}
\put(100,50){\line(1,0){28}}
\put(40,55){\makebox(0,0)[cc]{p}}

\put(115,55){\makebox(0,0)[cc]{p}}

\put(78,77){\makebox(0,0)[cc]{k}}

\put(79,21){\makebox(0,0)[cc]{p-k}}

\end{picture}

}

\def\figblind{

\def\JPicScale{0.6}
\ifx\JPicScale\undefined\def\JPicScale{1}\fi
\unitlength \JPicScale mm

}

\def\figcutPI{

\def\JPicScale{0.8}
\ifx\JPicScale\undefined\def\JPicScale{1}\fi
\unitlength \JPicScale mm

}

\def\figspdeg{

\def\JPicScale{0.6}
\ifx\JPicScale\undefined\def\JPicScale{1}\fi
\unitlength \JPicScale mm
\begin{picture}(100.09,86.01)(0,0)
\linethickness{0.3mm}
\put(100.09,62.66){\line(0,1){0.5}}
\multiput(100.08,63.65)(0.01,-0.5){1}{\line(0,-1){0.5}}
\multiput(100.06,64.15)(0.02,-0.5){1}{\line(0,-1){0.5}}
\multiput(100.03,64.65)(0.03,-0.5){1}{\line(0,-1){0.5}}
\multiput(99.99,65.14)(0.04,-0.5){1}{\line(0,-1){0.5}}
\multiput(99.93,65.64)(0.05,-0.49){1}{\line(0,-1){0.49}}
\multiput(99.87,66.13)(0.06,-0.49){1}{\line(0,-1){0.49}}
\multiput(99.79,66.62)(0.07,-0.49){1}{\line(0,-1){0.49}}
\multiput(99.71,67.11)(0.09,-0.49){1}{\line(0,-1){0.49}}
\multiput(99.61,67.6)(0.1,-0.49){1}{\line(0,-1){0.49}}
\multiput(99.51,68.08)(0.11,-0.49){1}{\line(0,-1){0.49}}
\multiput(99.39,68.57)(0.12,-0.48){1}{\line(0,-1){0.48}}
\multiput(99.26,69.05)(0.13,-0.48){1}{\line(0,-1){0.48}}
\multiput(99.13,69.53)(0.14,-0.48){1}{\line(0,-1){0.48}}
\multiput(98.98,70)(0.15,-0.47){1}{\line(0,-1){0.47}}
\multiput(98.82,70.47)(0.16,-0.47){1}{\line(0,-1){0.47}}
\multiput(98.65,70.94)(0.17,-0.47){1}{\line(0,-1){0.47}}
\multiput(98.48,71.4)(0.18,-0.46){1}{\line(0,-1){0.46}}
\multiput(98.29,71.86)(0.09,-0.23){2}{\line(0,-1){0.23}}
\multiput(98.09,72.32)(0.1,-0.23){2}{\line(0,-1){0.23}}
\multiput(97.88,72.77)(0.1,-0.23){2}{\line(0,-1){0.23}}
\multiput(97.67,73.22)(0.11,-0.22){2}{\line(0,-1){0.22}}
\multiput(97.44,73.66)(0.11,-0.22){2}{\line(0,-1){0.22}}
\multiput(97.2,74.1)(0.12,-0.22){2}{\line(0,-1){0.22}}
\multiput(96.96,74.53)(0.12,-0.22){2}{\line(0,-1){0.22}}
\multiput(96.7,74.96)(0.13,-0.21){2}{\line(0,-1){0.21}}
\multiput(96.44,75.38)(0.13,-0.21){2}{\line(0,-1){0.21}}
\multiput(96.17,75.8)(0.14,-0.21){2}{\line(0,-1){0.21}}
\multiput(95.88,76.21)(0.14,-0.2){2}{\line(0,-1){0.2}}
\multiput(95.59,76.61)(0.15,-0.2){2}{\line(0,-1){0.2}}
\multiput(95.3,77.01)(0.15,-0.2){2}{\line(0,-1){0.2}}
\multiput(94.99,77.4)(0.1,-0.13){3}{\line(0,-1){0.13}}
\multiput(94.67,77.78)(0.11,-0.13){3}{\line(0,-1){0.13}}
\multiput(94.35,78.16)(0.11,-0.13){3}{\line(0,-1){0.13}}
\multiput(94.02,78.53)(0.11,-0.12){3}{\line(0,-1){0.12}}
\multiput(93.68,78.89)(0.11,-0.12){3}{\line(0,-1){0.12}}
\multiput(93.33,79.25)(0.12,-0.12){3}{\line(0,-1){0.12}}
\multiput(92.97,79.59)(0.12,-0.12){3}{\line(1,0){0.12}}
\multiput(92.61,79.93)(0.12,-0.11){3}{\line(1,0){0.12}}
\multiput(92.24,80.27)(0.12,-0.11){3}{\line(1,0){0.12}}
\multiput(91.86,80.59)(0.13,-0.11){3}{\line(1,0){0.13}}
\multiput(91.48,80.91)(0.13,-0.11){3}{\line(1,0){0.13}}
\multiput(91.09,81.21)(0.13,-0.1){3}{\line(1,0){0.13}}
\multiput(90.69,81.51)(0.2,-0.15){2}{\line(1,0){0.2}}
\multiput(90.29,81.8)(0.2,-0.15){2}{\line(1,0){0.2}}
\multiput(89.88,82.08)(0.2,-0.14){2}{\line(1,0){0.2}}
\multiput(89.46,82.36)(0.21,-0.14){2}{\line(1,0){0.21}}
\multiput(89.04,82.62)(0.21,-0.13){2}{\line(1,0){0.21}}
\multiput(88.61,82.88)(0.21,-0.13){2}{\line(1,0){0.21}}
\multiput(88.18,83.12)(0.22,-0.12){2}{\line(1,0){0.22}}
\multiput(87.74,83.36)(0.22,-0.12){2}{\line(1,0){0.22}}
\multiput(87.3,83.58)(0.22,-0.11){2}{\line(1,0){0.22}}
\multiput(86.85,83.8)(0.22,-0.11){2}{\line(1,0){0.22}}
\multiput(86.4,84.01)(0.23,-0.1){2}{\line(1,0){0.23}}
\multiput(85.95,84.21)(0.23,-0.1){2}{\line(1,0){0.23}}
\multiput(85.49,84.39)(0.23,-0.09){2}{\line(1,0){0.23}}
\multiput(85.02,84.57)(0.46,-0.18){1}{\line(1,0){0.46}}
\multiput(84.55,84.74)(0.47,-0.17){1}{\line(1,0){0.47}}
\multiput(84.08,84.9)(0.47,-0.16){1}{\line(1,0){0.47}}
\multiput(83.61,85.05)(0.47,-0.15){1}{\line(1,0){0.47}}
\multiput(83.13,85.18)(0.48,-0.14){1}{\line(1,0){0.48}}
\multiput(82.65,85.31)(0.48,-0.13){1}{\line(1,0){0.48}}
\multiput(82.17,85.43)(0.48,-0.12){1}{\line(1,0){0.48}}
\multiput(81.68,85.53)(0.49,-0.11){1}{\line(1,0){0.49}}
\multiput(81.19,85.63)(0.49,-0.1){1}{\line(1,0){0.49}}
\multiput(80.7,85.71)(0.49,-0.09){1}{\line(1,0){0.49}}
\multiput(80.21,85.79)(0.49,-0.07){1}{\line(1,0){0.49}}
\multiput(79.72,85.85)(0.49,-0.06){1}{\line(1,0){0.49}}
\multiput(79.22,85.91)(0.49,-0.05){1}{\line(1,0){0.49}}
\multiput(78.73,85.95)(0.5,-0.04){1}{\line(1,0){0.5}}
\multiput(78.23,85.98)(0.5,-0.03){1}{\line(1,0){0.5}}
\multiput(77.74,86)(0.5,-0.02){1}{\line(1,0){0.5}}
\multiput(77.24,86.01)(0.5,-0.01){1}{\line(1,0){0.5}}
\put(76.74,86.01){\line(1,0){0.5}}
\multiput(76.24,86)(0.5,0.01){1}{\line(1,0){0.5}}
\multiput(75.75,85.98)(0.5,0.02){1}{\line(1,0){0.5}}
\multiput(75.25,85.95)(0.5,0.03){1}{\line(1,0){0.5}}
\multiput(74.76,85.91)(0.5,0.04){1}{\line(1,0){0.5}}
\multiput(74.26,85.85)(0.49,0.05){1}{\line(1,0){0.49}}
\multiput(73.77,85.79)(0.49,0.06){1}{\line(1,0){0.49}}
\multiput(73.28,85.71)(0.49,0.07){1}{\line(1,0){0.49}}
\multiput(72.79,85.63)(0.49,0.09){1}{\line(1,0){0.49}}
\multiput(72.3,85.53)(0.49,0.1){1}{\line(1,0){0.49}}
\multiput(71.81,85.43)(0.49,0.11){1}{\line(1,0){0.49}}
\multiput(71.33,85.31)(0.48,0.12){1}{\line(1,0){0.48}}
\multiput(70.85,85.18)(0.48,0.13){1}{\line(1,0){0.48}}
\multiput(70.37,85.05)(0.48,0.14){1}{\line(1,0){0.48}}
\multiput(69.9,84.9)(0.47,0.15){1}{\line(1,0){0.47}}
\multiput(69.43,84.74)(0.47,0.16){1}{\line(1,0){0.47}}
\multiput(68.96,84.57)(0.47,0.17){1}{\line(1,0){0.47}}
\multiput(68.49,84.39)(0.46,0.18){1}{\line(1,0){0.46}}
\multiput(68.03,84.21)(0.23,0.09){2}{\line(1,0){0.23}}
\multiput(67.58,84.01)(0.23,0.1){2}{\line(1,0){0.23}}
\multiput(67.13,83.8)(0.23,0.1){2}{\line(1,0){0.23}}
\multiput(66.68,83.58)(0.22,0.11){2}{\line(1,0){0.22}}
\multiput(66.24,83.36)(0.22,0.11){2}{\line(1,0){0.22}}
\multiput(65.8,83.12)(0.22,0.12){2}{\line(1,0){0.22}}
\multiput(65.37,82.88)(0.22,0.12){2}{\line(1,0){0.22}}
\multiput(64.94,82.62)(0.21,0.13){2}{\line(1,0){0.21}}
\multiput(64.52,82.36)(0.21,0.13){2}{\line(1,0){0.21}}
\multiput(64.1,82.08)(0.21,0.14){2}{\line(1,0){0.21}}
\multiput(63.69,81.8)(0.2,0.14){2}{\line(1,0){0.2}}
\multiput(63.29,81.51)(0.2,0.15){2}{\line(1,0){0.2}}
\multiput(62.89,81.21)(0.2,0.15){2}{\line(1,0){0.2}}
\multiput(62.5,80.91)(0.13,0.1){3}{\line(1,0){0.13}}
\multiput(62.12,80.59)(0.13,0.11){3}{\line(1,0){0.13}}
\multiput(61.74,80.27)(0.13,0.11){3}{\line(1,0){0.13}}
\multiput(61.37,79.93)(0.12,0.11){3}{\line(1,0){0.12}}
\multiput(61.01,79.59)(0.12,0.11){3}{\line(1,0){0.12}}
\multiput(60.65,79.25)(0.12,0.12){3}{\line(1,0){0.12}}
\multiput(60.3,78.89)(0.12,0.12){3}{\line(0,1){0.12}}
\multiput(59.96,78.53)(0.11,0.12){3}{\line(0,1){0.12}}
\multiput(59.63,78.16)(0.11,0.12){3}{\line(0,1){0.12}}
\multiput(59.31,77.78)(0.11,0.13){3}{\line(0,1){0.13}}
\multiput(58.99,77.4)(0.11,0.13){3}{\line(0,1){0.13}}
\multiput(58.68,77.01)(0.1,0.13){3}{\line(0,1){0.13}}
\multiput(58.39,76.61)(0.15,0.2){2}{\line(0,1){0.2}}
\multiput(58.09,76.21)(0.15,0.2){2}{\line(0,1){0.2}}
\multiput(57.81,75.8)(0.14,0.2){2}{\line(0,1){0.2}}
\multiput(57.54,75.38)(0.14,0.21){2}{\line(0,1){0.21}}
\multiput(57.28,74.96)(0.13,0.21){2}{\line(0,1){0.21}}
\multiput(57.02,74.53)(0.13,0.21){2}{\line(0,1){0.21}}
\multiput(56.78,74.1)(0.12,0.22){2}{\line(0,1){0.22}}
\multiput(56.54,73.66)(0.12,0.22){2}{\line(0,1){0.22}}
\multiput(56.31,73.22)(0.11,0.22){2}{\line(0,1){0.22}}
\multiput(56.1,72.77)(0.11,0.22){2}{\line(0,1){0.22}}
\multiput(55.89,72.32)(0.1,0.23){2}{\line(0,1){0.23}}
\multiput(55.69,71.86)(0.1,0.23){2}{\line(0,1){0.23}}
\multiput(55.5,71.4)(0.09,0.23){2}{\line(0,1){0.23}}
\multiput(55.33,70.94)(0.18,0.46){1}{\line(0,1){0.46}}
\multiput(55.16,70.47)(0.17,0.47){1}{\line(0,1){0.47}}
\multiput(55,70)(0.16,0.47){1}{\line(0,1){0.47}}
\multiput(54.85,69.53)(0.15,0.47){1}{\line(0,1){0.47}}
\multiput(54.72,69.05)(0.14,0.48){1}{\line(0,1){0.48}}
\multiput(54.59,68.57)(0.13,0.48){1}{\line(0,1){0.48}}
\multiput(54.47,68.08)(0.12,0.48){1}{\line(0,1){0.48}}
\multiput(54.37,67.6)(0.11,0.49){1}{\line(0,1){0.49}}
\multiput(54.27,67.11)(0.1,0.49){1}{\line(0,1){0.49}}
\multiput(54.18,66.62)(0.09,0.49){1}{\line(0,1){0.49}}
\multiput(54.11,66.13)(0.07,0.49){1}{\line(0,1){0.49}}
\multiput(54.05,65.64)(0.06,0.49){1}{\line(0,1){0.49}}
\multiput(53.99,65.14)(0.05,0.49){1}{\line(0,1){0.49}}
\multiput(53.95,64.65)(0.04,0.5){1}{\line(0,1){0.5}}
\multiput(53.92,64.15)(0.03,0.5){1}{\line(0,1){0.5}}
\multiput(53.9,63.65)(0.02,0.5){1}{\line(0,1){0.5}}
\multiput(53.89,63.16)(0.01,0.5){1}{\line(0,1){0.5}}
\put(53.89,62.66){\line(0,1){0.5}}
\multiput(53.89,62.66)(0.01,-0.5){1}{\line(0,-1){0.5}}
\multiput(53.9,62.16)(0.02,-0.5){1}{\line(0,-1){0.5}}
\multiput(53.92,61.67)(0.03,-0.5){1}{\line(0,-1){0.5}}
\multiput(53.95,61.17)(0.04,-0.5){1}{\line(0,-1){0.5}}
\multiput(53.99,60.67)(0.05,-0.49){1}{\line(0,-1){0.49}}
\multiput(54.05,60.18)(0.06,-0.49){1}{\line(0,-1){0.49}}
\multiput(54.11,59.69)(0.07,-0.49){1}{\line(0,-1){0.49}}
\multiput(54.18,59.2)(0.09,-0.49){1}{\line(0,-1){0.49}}
\multiput(54.27,58.71)(0.1,-0.49){1}{\line(0,-1){0.49}}
\multiput(54.37,58.22)(0.11,-0.49){1}{\line(0,-1){0.49}}
\multiput(54.47,57.73)(0.12,-0.48){1}{\line(0,-1){0.48}}
\multiput(54.59,57.25)(0.13,-0.48){1}{\line(0,-1){0.48}}
\multiput(54.72,56.77)(0.14,-0.48){1}{\line(0,-1){0.48}}
\multiput(54.85,56.29)(0.15,-0.47){1}{\line(0,-1){0.47}}
\multiput(55,55.82)(0.16,-0.47){1}{\line(0,-1){0.47}}
\multiput(55.16,55.34)(0.17,-0.47){1}{\line(0,-1){0.47}}
\multiput(55.33,54.88)(0.18,-0.46){1}{\line(0,-1){0.46}}
\multiput(55.5,54.41)(0.09,-0.23){2}{\line(0,-1){0.23}}
\multiput(55.69,53.95)(0.1,-0.23){2}{\line(0,-1){0.23}}
\multiput(55.89,53.5)(0.1,-0.23){2}{\line(0,-1){0.23}}
\multiput(56.1,53.04)(0.11,-0.22){2}{\line(0,-1){0.22}}
\multiput(56.31,52.6)(0.11,-0.22){2}{\line(0,-1){0.22}}
\multiput(56.54,52.15)(0.12,-0.22){2}{\line(0,-1){0.22}}
\multiput(56.78,51.72)(0.12,-0.22){2}{\line(0,-1){0.22}}
\multiput(57.02,51.28)(0.13,-0.21){2}{\line(0,-1){0.21}}
\multiput(57.28,50.86)(0.13,-0.21){2}{\line(0,-1){0.21}}
\multiput(57.54,50.44)(0.14,-0.21){2}{\line(0,-1){0.21}}
\multiput(57.81,50.02)(0.14,-0.2){2}{\line(0,-1){0.2}}
\multiput(58.09,49.61)(0.15,-0.2){2}{\line(0,-1){0.2}}
\multiput(58.39,49.21)(0.15,-0.2){2}{\line(0,-1){0.2}}
\multiput(58.68,48.81)(0.1,-0.13){3}{\line(0,-1){0.13}}
\multiput(58.99,48.42)(0.11,-0.13){3}{\line(0,-1){0.13}}
\multiput(59.31,48.04)(0.11,-0.13){3}{\line(0,-1){0.13}}
\multiput(59.63,47.66)(0.11,-0.12){3}{\line(0,-1){0.12}}
\multiput(59.96,47.29)(0.11,-0.12){3}{\line(0,-1){0.12}}
\multiput(60.3,46.93)(0.12,-0.12){3}{\line(0,-1){0.12}}
\multiput(60.65,46.57)(0.12,-0.12){3}{\line(1,0){0.12}}
\multiput(61.01,46.22)(0.12,-0.11){3}{\line(1,0){0.12}}
\multiput(61.37,45.88)(0.12,-0.11){3}{\line(1,0){0.12}}
\multiput(61.74,45.55)(0.13,-0.11){3}{\line(1,0){0.13}}
\multiput(62.12,45.23)(0.13,-0.11){3}{\line(1,0){0.13}}
\multiput(62.5,44.91)(0.13,-0.1){3}{\line(1,0){0.13}}
\multiput(62.89,44.6)(0.2,-0.15){2}{\line(1,0){0.2}}
\multiput(63.29,44.3)(0.2,-0.15){2}{\line(1,0){0.2}}
\multiput(63.69,44.01)(0.2,-0.14){2}{\line(1,0){0.2}}
\multiput(64.1,43.73)(0.21,-0.14){2}{\line(1,0){0.21}}
\multiput(64.52,43.46)(0.21,-0.13){2}{\line(1,0){0.21}}
\multiput(64.94,43.19)(0.21,-0.13){2}{\line(1,0){0.21}}
\multiput(65.37,42.94)(0.22,-0.12){2}{\line(1,0){0.22}}
\multiput(65.8,42.69)(0.22,-0.12){2}{\line(1,0){0.22}}
\multiput(66.24,42.46)(0.22,-0.11){2}{\line(1,0){0.22}}
\multiput(66.68,42.23)(0.22,-0.11){2}{\line(1,0){0.22}}
\multiput(67.13,42.01)(0.23,-0.1){2}{\line(1,0){0.23}}
\multiput(67.58,41.81)(0.23,-0.1){2}{\line(1,0){0.23}}
\multiput(68.03,41.61)(0.23,-0.09){2}{\line(1,0){0.23}}
\multiput(68.49,41.42)(0.46,-0.18){1}{\line(1,0){0.46}}
\multiput(68.96,41.24)(0.47,-0.17){1}{\line(1,0){0.47}}
\multiput(69.43,41.08)(0.47,-0.16){1}{\line(1,0){0.47}}
\multiput(69.9,40.92)(0.47,-0.15){1}{\line(1,0){0.47}}
\multiput(70.37,40.77)(0.48,-0.14){1}{\line(1,0){0.48}}
\multiput(70.85,40.63)(0.48,-0.13){1}{\line(1,0){0.48}}
\multiput(71.33,40.51)(0.48,-0.12){1}{\line(1,0){0.48}}
\multiput(71.81,40.39)(0.49,-0.11){1}{\line(1,0){0.49}}
\multiput(72.3,40.28)(0.49,-0.1){1}{\line(1,0){0.49}}
\multiput(72.79,40.19)(0.49,-0.09){1}{\line(1,0){0.49}}
\multiput(73.28,40.1)(0.49,-0.07){1}{\line(1,0){0.49}}
\multiput(73.77,40.03)(0.49,-0.06){1}{\line(1,0){0.49}}
\multiput(74.26,39.96)(0.49,-0.05){1}{\line(1,0){0.49}}
\multiput(74.76,39.91)(0.5,-0.04){1}{\line(1,0){0.5}}
\multiput(75.25,39.87)(0.5,-0.03){1}{\line(1,0){0.5}}
\multiput(75.75,39.84)(0.5,-0.02){1}{\line(1,0){0.5}}
\multiput(76.24,39.82)(0.5,-0.01){1}{\line(1,0){0.5}}
\put(76.74,39.8){\line(1,0){0.5}}
\multiput(77.24,39.8)(0.5,0.01){1}{\line(1,0){0.5}}
\multiput(77.74,39.82)(0.5,0.02){1}{\line(1,0){0.5}}
\multiput(78.23,39.84)(0.5,0.03){1}{\line(1,0){0.5}}
\multiput(78.73,39.87)(0.5,0.04){1}{\line(1,0){0.5}}
\multiput(79.22,39.91)(0.49,0.05){1}{\line(1,0){0.49}}
\multiput(79.72,39.96)(0.49,0.06){1}{\line(1,0){0.49}}
\multiput(80.21,40.03)(0.49,0.07){1}{\line(1,0){0.49}}
\multiput(80.7,40.1)(0.49,0.09){1}{\line(1,0){0.49}}
\multiput(81.19,40.19)(0.49,0.1){1}{\line(1,0){0.49}}
\multiput(81.68,40.28)(0.49,0.11){1}{\line(1,0){0.49}}
\multiput(82.17,40.39)(0.48,0.12){1}{\line(1,0){0.48}}
\multiput(82.65,40.51)(0.48,0.13){1}{\line(1,0){0.48}}
\multiput(83.13,40.63)(0.48,0.14){1}{\line(1,0){0.48}}
\multiput(83.61,40.77)(0.47,0.15){1}{\line(1,0){0.47}}
\multiput(84.08,40.92)(0.47,0.16){1}{\line(1,0){0.47}}
\multiput(84.55,41.08)(0.47,0.17){1}{\line(1,0){0.47}}
\multiput(85.02,41.24)(0.46,0.18){1}{\line(1,0){0.46}}
\multiput(85.49,41.42)(0.23,0.09){2}{\line(1,0){0.23}}
\multiput(85.95,41.61)(0.23,0.1){2}{\line(1,0){0.23}}
\multiput(86.4,41.81)(0.23,0.1){2}{\line(1,0){0.23}}
\multiput(86.85,42.01)(0.22,0.11){2}{\line(1,0){0.22}}
\multiput(87.3,42.23)(0.22,0.11){2}{\line(1,0){0.22}}
\multiput(87.74,42.46)(0.22,0.12){2}{\line(1,0){0.22}}
\multiput(88.18,42.69)(0.22,0.12){2}{\line(1,0){0.22}}
\multiput(88.61,42.94)(0.21,0.13){2}{\line(1,0){0.21}}
\multiput(89.04,43.19)(0.21,0.13){2}{\line(1,0){0.21}}
\multiput(89.46,43.46)(0.21,0.14){2}{\line(1,0){0.21}}
\multiput(89.88,43.73)(0.2,0.14){2}{\line(1,0){0.2}}
\multiput(90.29,44.01)(0.2,0.15){2}{\line(1,0){0.2}}
\multiput(90.69,44.3)(0.2,0.15){2}{\line(1,0){0.2}}
\multiput(91.09,44.6)(0.13,0.1){3}{\line(1,0){0.13}}
\multiput(91.48,44.91)(0.13,0.11){3}{\line(1,0){0.13}}
\multiput(91.86,45.23)(0.13,0.11){3}{\line(1,0){0.13}}
\multiput(92.24,45.55)(0.12,0.11){3}{\line(1,0){0.12}}
\multiput(92.61,45.88)(0.12,0.11){3}{\line(1,0){0.12}}
\multiput(92.97,46.22)(0.12,0.12){3}{\line(1,0){0.12}}
\multiput(93.33,46.57)(0.12,0.12){3}{\line(0,1){0.12}}
\multiput(93.68,46.93)(0.11,0.12){3}{\line(0,1){0.12}}
\multiput(94.02,47.29)(0.11,0.12){3}{\line(0,1){0.12}}
\multiput(94.35,47.66)(0.11,0.13){3}{\line(0,1){0.13}}
\multiput(94.67,48.04)(0.11,0.13){3}{\line(0,1){0.13}}
\multiput(94.99,48.42)(0.1,0.13){3}{\line(0,1){0.13}}
\multiput(95.3,48.81)(0.15,0.2){2}{\line(0,1){0.2}}
\multiput(95.59,49.21)(0.15,0.2){2}{\line(0,1){0.2}}
\multiput(95.88,49.61)(0.14,0.2){2}{\line(0,1){0.2}}
\multiput(96.17,50.02)(0.14,0.21){2}{\line(0,1){0.21}}
\multiput(96.44,50.44)(0.13,0.21){2}{\line(0,1){0.21}}
\multiput(96.7,50.86)(0.13,0.21){2}{\line(0,1){0.21}}
\multiput(96.96,51.28)(0.12,0.22){2}{\line(0,1){0.22}}
\multiput(97.2,51.72)(0.12,0.22){2}{\line(0,1){0.22}}
\multiput(97.44,52.15)(0.11,0.22){2}{\line(0,1){0.22}}
\multiput(97.67,52.6)(0.11,0.22){2}{\line(0,1){0.22}}
\multiput(97.88,53.04)(0.1,0.23){2}{\line(0,1){0.23}}
\multiput(98.09,53.5)(0.1,0.23){2}{\line(0,1){0.23}}
\multiput(98.29,53.95)(0.09,0.23){2}{\line(0,1){0.23}}
\multiput(98.48,54.41)(0.18,0.46){1}{\line(0,1){0.46}}
\multiput(98.65,54.88)(0.17,0.47){1}{\line(0,1){0.47}}
\multiput(98.82,55.34)(0.16,0.47){1}{\line(0,1){0.47}}
\multiput(98.98,55.82)(0.15,0.47){1}{\line(0,1){0.47}}
\multiput(99.13,56.29)(0.14,0.48){1}{\line(0,1){0.48}}
\multiput(99.26,56.77)(0.13,0.48){1}{\line(0,1){0.48}}
\multiput(99.39,57.25)(0.12,0.48){1}{\line(0,1){0.48}}
\multiput(99.51,57.73)(0.11,0.49){1}{\line(0,1){0.49}}
\multiput(99.61,58.22)(0.1,0.49){1}{\line(0,1){0.49}}
\multiput(99.71,58.71)(0.09,0.49){1}{\line(0,1){0.49}}
\multiput(99.79,59.2)(0.07,0.49){1}{\line(0,1){0.49}}
\multiput(99.87,59.69)(0.06,0.49){1}{\line(0,1){0.49}}
\multiput(99.93,60.18)(0.05,0.49){1}{\line(0,1){0.49}}
\multiput(99.99,60.67)(0.04,0.5){1}{\line(0,1){0.5}}
\multiput(100.03,61.17)(0.03,0.5){1}{\line(0,1){0.5}}
\multiput(100.06,61.67)(0.02,0.5){1}{\line(0,1){0.5}}
\multiput(100.08,62.16)(0.01,0.5){1}{\line(0,1){0.5}}

\linethickness{0.3mm}
\put(80,10){\line(0,1){30}}
\put(85,35){\makebox(0,0)[cc]{$V$}}

\put(77,90){\makebox(0,0)[cc]{$P$}}

\end{picture}

}

\def\figinsert{

\def\JPicScale{0.8}
\ifx\JPicScale\undefined\def\JPicScale{1}\fi
\unitlength \JPicScale mm
\begin{picture}(130,70)(0,0)
\linethickness{0.3mm}
\qbezier(30,70)(55.94,59.5)(80,59.5)
\qbezier(80,59.5)(104.06,59.5)(130,70)
\linethickness{0.3mm}
\qbezier(30,20)(55.94,30.5)(80,30.5)
\qbezier(80,30.5)(104.06,30.5)(130,20)
\linethickness{0.3mm}
\qbezier(45,65)(39.75,54.62)(39.75,45)
\qbezier(39.75,45)(39.75,35.38)(45,25)
\linethickness{0.3mm}
\qbezier(45,65)(50.25,54.62)(50.25,45)
\qbezier(50.25,45)(50.25,35.38)(45,25)
\linethickness{0.3mm}
\qbezier(115,65)(109.75,54.62)(109.75,45)
\qbezier(109.75,45)(109.75,35.38)(115,25)
\linethickness{0.3mm}
\qbezier(115,65)(120.25,54.62)(120.25,45)
\qbezier(120.25,45)(120.25,35.38)(115,25)

\put(45,19){\makebox(0,0)[cc]{$|\chi_s\rangle \langle\chi_s'|$}}

\put(115,19){\makebox(0,0)[cc]{$|\chi_r\rangle\langle \chi_r'|$}}

\end{picture}

}

\def\figvert{

\def\JPicScale{0.6}
\ifx\JPicScale\undefined\def\JPicScale{1}\fi
\unitlength \JPicScale mm
\begin{picture}(130,80)(0,0)
\linethickness{0.3mm}
\put(20,20){\line(0,1){60}}
\linethickness{0.3mm}
\put(20,20){\line(1,0){110}}
\linethickness{0.3mm}
\qbezier(20,40)(20.53,39.43)(27.14,35.2)
\qbezier(27.14,35.2)(33.74,30.98)(40,30)
\qbezier(40,30)(48.96,30.12)(59.05,34.71)
\qbezier(59.05,34.71)(69.14,39.3)(70,40)
\linethickness{0.3mm}
\qbezier(70,60)(70.86,60.62)(80.93,65)
\qbezier(80.93,65)(91,69.38)(100,70)
\qbezier(100,70)(109,69.38)(119.07,65)
\qbezier(119.07,65)(129.14,60.62)(130,60)
\linethickness{0.3mm}
\put(70,40){\line(0,1){20}}

\linethickness{0.1mm}
\put(70,20){\line(0,1){5}}
\put(70,30){\line(0,1){5}}

\put(20,85){\makebox(0,0)[cc]{$y_1$}}

\put(145,20){\makebox(0,0)[cc]{$\MM_{g,m,n}$}}

\put(40,35){\makebox(0,0)[cc]{$S_i$}}

\put(75,50){\makebox(0,0)[cc]{$V$}}

\put(95,75){\makebox(0,0)[cc]{$S_j$}}

\put(40,15){\makebox(0,0)[cc]{$R_i$}}

\put(90,15){\makebox(0,0)[cc]{$R_j$}}

\end{picture}

}

\def\figscsc{

\def\JPicScale{0.6}
\ifx\JPicScale\undefined\def\JPicScale{1}\fi
\unitlength \JPicScale mm
\begin{picture}(120,70)(0,0)
\linethickness{0.3mm}
\put(101.46,41.62){\line(0,1){0.5}}
\multiput(101.45,42.63)(0.01,-0.5){1}{\line(0,-1){0.5}}
\multiput(101.43,43.13)(0.02,-0.5){1}{\line(0,-1){0.5}}
\multiput(101.39,43.63)(0.03,-0.5){1}{\line(0,-1){0.5}}
\multiput(101.35,44.13)(0.04,-0.5){1}{\line(0,-1){0.5}}
\multiput(101.3,44.63)(0.05,-0.5){1}{\line(0,-1){0.5}}
\multiput(101.23,45.13)(0.06,-0.5){1}{\line(0,-1){0.5}}
\multiput(101.16,45.62)(0.08,-0.5){1}{\line(0,-1){0.5}}
\multiput(101.07,46.12)(0.09,-0.49){1}{\line(0,-1){0.49}}
\multiput(100.97,46.61)(0.1,-0.49){1}{\line(0,-1){0.49}}
\multiput(100.87,47.1)(0.11,-0.49){1}{\line(0,-1){0.49}}
\multiput(100.75,47.59)(0.12,-0.49){1}{\line(0,-1){0.49}}
\multiput(100.62,48.08)(0.13,-0.49){1}{\line(0,-1){0.49}}
\multiput(100.48,48.56)(0.14,-0.48){1}{\line(0,-1){0.48}}
\multiput(100.33,49.04)(0.15,-0.48){1}{\line(0,-1){0.48}}
\multiput(100.17,49.51)(0.16,-0.48){1}{\line(0,-1){0.48}}
\multiput(100,49.99)(0.17,-0.47){1}{\line(0,-1){0.47}}
\multiput(99.83,50.46)(0.18,-0.47){1}{\line(0,-1){0.47}}
\multiput(99.64,50.92)(0.09,-0.23){2}{\line(0,-1){0.23}}
\multiput(99.44,51.38)(0.1,-0.23){2}{\line(0,-1){0.23}}
\multiput(99.23,51.84)(0.1,-0.23){2}{\line(0,-1){0.23}}
\multiput(99.01,52.29)(0.11,-0.23){2}{\line(0,-1){0.23}}
\multiput(98.78,52.74)(0.11,-0.22){2}{\line(0,-1){0.22}}
\multiput(98.54,53.18)(0.12,-0.22){2}{\line(0,-1){0.22}}
\multiput(98.29,53.61)(0.12,-0.22){2}{\line(0,-1){0.22}}
\multiput(98.03,54.05)(0.13,-0.22){2}{\line(0,-1){0.22}}
\multiput(97.77,54.47)(0.13,-0.21){2}{\line(0,-1){0.21}}
\multiput(97.49,54.89)(0.14,-0.21){2}{\line(0,-1){0.21}}
\multiput(97.21,55.31)(0.14,-0.21){2}{\line(0,-1){0.21}}
\multiput(96.91,55.71)(0.15,-0.2){2}{\line(0,-1){0.2}}
\multiput(96.61,56.11)(0.1,-0.13){3}{\line(0,-1){0.13}}
\multiput(96.3,56.51)(0.1,-0.13){3}{\line(0,-1){0.13}}
\multiput(95.98,56.9)(0.11,-0.13){3}{\line(0,-1){0.13}}
\multiput(95.66,57.28)(0.11,-0.13){3}{\line(0,-1){0.13}}
\multiput(95.32,57.65)(0.11,-0.12){3}{\line(0,-1){0.12}}
\multiput(94.98,58.02)(0.11,-0.12){3}{\line(0,-1){0.12}}
\multiput(94.63,58.38)(0.12,-0.12){3}{\line(0,-1){0.12}}
\multiput(94.27,58.73)(0.12,-0.12){3}{\line(1,0){0.12}}
\multiput(93.9,59.07)(0.12,-0.11){3}{\line(1,0){0.12}}
\multiput(93.53,59.41)(0.12,-0.11){3}{\line(1,0){0.12}}
\multiput(93.15,59.73)(0.13,-0.11){3}{\line(1,0){0.13}}
\multiput(92.76,60.05)(0.13,-0.11){3}{\line(1,0){0.13}}
\multiput(92.36,60.36)(0.13,-0.1){3}{\line(1,0){0.13}}
\multiput(91.96,60.66)(0.13,-0.1){3}{\line(1,0){0.13}}
\multiput(91.56,60.96)(0.2,-0.15){2}{\line(1,0){0.2}}
\multiput(91.14,61.24)(0.21,-0.14){2}{\line(1,0){0.21}}
\multiput(90.72,61.52)(0.21,-0.14){2}{\line(1,0){0.21}}
\multiput(90.3,61.78)(0.21,-0.13){2}{\line(1,0){0.21}}
\multiput(89.86,62.04)(0.22,-0.13){2}{\line(1,0){0.22}}
\multiput(89.43,62.29)(0.22,-0.12){2}{\line(1,0){0.22}}
\multiput(88.99,62.53)(0.22,-0.12){2}{\line(1,0){0.22}}
\multiput(88.54,62.76)(0.22,-0.11){2}{\line(1,0){0.22}}
\multiput(88.09,62.98)(0.23,-0.11){2}{\line(1,0){0.23}}
\multiput(87.63,63.19)(0.23,-0.1){2}{\line(1,0){0.23}}
\multiput(87.17,63.39)(0.23,-0.1){2}{\line(1,0){0.23}}
\multiput(86.71,63.58)(0.23,-0.09){2}{\line(1,0){0.23}}
\multiput(86.24,63.75)(0.47,-0.18){1}{\line(1,0){0.47}}
\multiput(85.76,63.92)(0.47,-0.17){1}{\line(1,0){0.47}}
\multiput(85.29,64.08)(0.48,-0.16){1}{\line(1,0){0.48}}
\multiput(84.81,64.23)(0.48,-0.15){1}{\line(1,0){0.48}}
\multiput(84.33,64.37)(0.48,-0.14){1}{\line(1,0){0.48}}
\multiput(83.84,64.5)(0.49,-0.13){1}{\line(1,0){0.49}}
\multiput(83.35,64.62)(0.49,-0.12){1}{\line(1,0){0.49}}
\multiput(82.86,64.72)(0.49,-0.11){1}{\line(1,0){0.49}}
\multiput(82.37,64.82)(0.49,-0.1){1}{\line(1,0){0.49}}
\multiput(81.87,64.91)(0.49,-0.09){1}{\line(1,0){0.49}}
\multiput(81.38,64.98)(0.5,-0.08){1}{\line(1,0){0.5}}
\multiput(80.88,65.05)(0.5,-0.06){1}{\line(1,0){0.5}}
\multiput(80.38,65.1)(0.5,-0.05){1}{\line(1,0){0.5}}
\multiput(79.88,65.14)(0.5,-0.04){1}{\line(1,0){0.5}}
\multiput(79.38,65.18)(0.5,-0.03){1}{\line(1,0){0.5}}
\multiput(78.88,65.2)(0.5,-0.02){1}{\line(1,0){0.5}}
\multiput(78.38,65.21)(0.5,-0.01){1}{\line(1,0){0.5}}
\put(77.87,65.21){\line(1,0){0.5}}
\multiput(77.37,65.2)(0.5,0.01){1}{\line(1,0){0.5}}
\multiput(76.87,65.18)(0.5,0.02){1}{\line(1,0){0.5}}
\multiput(76.37,65.14)(0.5,0.03){1}{\line(1,0){0.5}}
\multiput(75.87,65.1)(0.5,0.04){1}{\line(1,0){0.5}}
\multiput(75.37,65.05)(0.5,0.05){1}{\line(1,0){0.5}}
\multiput(74.87,64.98)(0.5,0.06){1}{\line(1,0){0.5}}
\multiput(74.38,64.91)(0.5,0.08){1}{\line(1,0){0.5}}
\multiput(73.88,64.82)(0.49,0.09){1}{\line(1,0){0.49}}
\multiput(73.39,64.72)(0.49,0.1){1}{\line(1,0){0.49}}
\multiput(72.9,64.62)(0.49,0.11){1}{\line(1,0){0.49}}
\multiput(72.41,64.5)(0.49,0.12){1}{\line(1,0){0.49}}
\multiput(71.92,64.37)(0.49,0.13){1}{\line(1,0){0.49}}
\multiput(71.44,64.23)(0.48,0.14){1}{\line(1,0){0.48}}
\multiput(70.96,64.08)(0.48,0.15){1}{\line(1,0){0.48}}
\multiput(70.49,63.92)(0.48,0.16){1}{\line(1,0){0.48}}
\multiput(70.01,63.75)(0.47,0.17){1}{\line(1,0){0.47}}
\multiput(69.54,63.58)(0.47,0.18){1}{\line(1,0){0.47}}
\multiput(69.08,63.39)(0.23,0.09){2}{\line(1,0){0.23}}
\multiput(68.62,63.19)(0.23,0.1){2}{\line(1,0){0.23}}
\multiput(68.16,62.98)(0.23,0.1){2}{\line(1,0){0.23}}
\multiput(67.71,62.76)(0.23,0.11){2}{\line(1,0){0.23}}
\multiput(67.26,62.53)(0.22,0.11){2}{\line(1,0){0.22}}
\multiput(66.82,62.29)(0.22,0.12){2}{\line(1,0){0.22}}
\multiput(66.39,62.04)(0.22,0.12){2}{\line(1,0){0.22}}
\multiput(65.95,61.78)(0.22,0.13){2}{\line(1,0){0.22}}
\multiput(65.53,61.52)(0.21,0.13){2}{\line(1,0){0.21}}
\multiput(65.11,61.24)(0.21,0.14){2}{\line(1,0){0.21}}
\multiput(64.69,60.96)(0.21,0.14){2}{\line(1,0){0.21}}
\multiput(64.29,60.66)(0.2,0.15){2}{\line(1,0){0.2}}
\multiput(63.89,60.36)(0.13,0.1){3}{\line(1,0){0.13}}
\multiput(63.49,60.05)(0.13,0.1){3}{\line(1,0){0.13}}
\multiput(63.1,59.73)(0.13,0.11){3}{\line(1,0){0.13}}
\multiput(62.72,59.41)(0.13,0.11){3}{\line(1,0){0.13}}
\multiput(62.35,59.07)(0.12,0.11){3}{\line(1,0){0.12}}
\multiput(61.98,58.73)(0.12,0.11){3}{\line(1,0){0.12}}
\multiput(61.62,58.38)(0.12,0.12){3}{\line(1,0){0.12}}
\multiput(61.27,58.02)(0.12,0.12){3}{\line(0,1){0.12}}
\multiput(60.93,57.65)(0.11,0.12){3}{\line(0,1){0.12}}
\multiput(60.59,57.28)(0.11,0.12){3}{\line(0,1){0.12}}
\multiput(60.27,56.9)(0.11,0.13){3}{\line(0,1){0.13}}
\multiput(59.95,56.51)(0.11,0.13){3}{\line(0,1){0.13}}
\multiput(59.64,56.11)(0.1,0.13){3}{\line(0,1){0.13}}
\multiput(59.34,55.71)(0.1,0.13){3}{\line(0,1){0.13}}
\multiput(59.04,55.31)(0.15,0.2){2}{\line(0,1){0.2}}
\multiput(58.76,54.89)(0.14,0.21){2}{\line(0,1){0.21}}
\multiput(58.48,54.47)(0.14,0.21){2}{\line(0,1){0.21}}
\multiput(58.22,54.05)(0.13,0.21){2}{\line(0,1){0.21}}
\multiput(57.96,53.61)(0.13,0.22){2}{\line(0,1){0.22}}
\multiput(57.71,53.18)(0.12,0.22){2}{\line(0,1){0.22}}
\multiput(57.47,52.74)(0.12,0.22){2}{\line(0,1){0.22}}
\multiput(57.24,52.29)(0.11,0.22){2}{\line(0,1){0.22}}
\multiput(57.02,51.84)(0.11,0.23){2}{\line(0,1){0.23}}
\multiput(56.81,51.38)(0.1,0.23){2}{\line(0,1){0.23}}
\multiput(56.61,50.92)(0.1,0.23){2}{\line(0,1){0.23}}
\multiput(56.42,50.46)(0.09,0.23){2}{\line(0,1){0.23}}
\multiput(56.25,49.99)(0.18,0.47){1}{\line(0,1){0.47}}
\multiput(56.08,49.51)(0.17,0.47){1}{\line(0,1){0.47}}
\multiput(55.92,49.04)(0.16,0.48){1}{\line(0,1){0.48}}
\multiput(55.77,48.56)(0.15,0.48){1}{\line(0,1){0.48}}
\multiput(55.63,48.08)(0.14,0.48){1}{\line(0,1){0.48}}
\multiput(55.5,47.59)(0.13,0.49){1}{\line(0,1){0.49}}
\multiput(55.38,47.1)(0.12,0.49){1}{\line(0,1){0.49}}
\multiput(55.28,46.61)(0.11,0.49){1}{\line(0,1){0.49}}
\multiput(55.18,46.12)(0.1,0.49){1}{\line(0,1){0.49}}
\multiput(55.09,45.62)(0.09,0.49){1}{\line(0,1){0.49}}
\multiput(55.02,45.13)(0.08,0.5){1}{\line(0,1){0.5}}
\multiput(54.95,44.63)(0.06,0.5){1}{\line(0,1){0.5}}
\multiput(54.9,44.13)(0.05,0.5){1}{\line(0,1){0.5}}
\multiput(54.86,43.63)(0.04,0.5){1}{\line(0,1){0.5}}
\multiput(54.82,43.13)(0.03,0.5){1}{\line(0,1){0.5}}
\multiput(54.8,42.63)(0.02,0.5){1}{\line(0,1){0.5}}
\multiput(54.79,42.13)(0.01,0.5){1}{\line(0,1){0.5}}
\put(54.79,41.62){\line(0,1){0.5}}
\multiput(54.79,41.62)(0.01,-0.5){1}{\line(0,-1){0.5}}
\multiput(54.8,41.12)(0.02,-0.5){1}{\line(0,-1){0.5}}
\multiput(54.82,40.62)(0.03,-0.5){1}{\line(0,-1){0.5}}
\multiput(54.86,40.12)(0.04,-0.5){1}{\line(0,-1){0.5}}
\multiput(54.9,39.62)(0.05,-0.5){1}{\line(0,-1){0.5}}
\multiput(54.95,39.12)(0.06,-0.5){1}{\line(0,-1){0.5}}
\multiput(55.02,38.62)(0.08,-0.5){1}{\line(0,-1){0.5}}
\multiput(55.09,38.13)(0.09,-0.49){1}{\line(0,-1){0.49}}
\multiput(55.18,37.63)(0.1,-0.49){1}{\line(0,-1){0.49}}
\multiput(55.28,37.14)(0.11,-0.49){1}{\line(0,-1){0.49}}
\multiput(55.38,36.65)(0.12,-0.49){1}{\line(0,-1){0.49}}
\multiput(55.5,36.16)(0.13,-0.49){1}{\line(0,-1){0.49}}
\multiput(55.63,35.67)(0.14,-0.48){1}{\line(0,-1){0.48}}
\multiput(55.77,35.19)(0.15,-0.48){1}{\line(0,-1){0.48}}
\multiput(55.92,34.71)(0.16,-0.48){1}{\line(0,-1){0.48}}
\multiput(56.08,34.24)(0.17,-0.47){1}{\line(0,-1){0.47}}
\multiput(56.25,33.76)(0.18,-0.47){1}{\line(0,-1){0.47}}
\multiput(56.42,33.29)(0.09,-0.23){2}{\line(0,-1){0.23}}
\multiput(56.61,32.83)(0.1,-0.23){2}{\line(0,-1){0.23}}
\multiput(56.81,32.37)(0.1,-0.23){2}{\line(0,-1){0.23}}
\multiput(57.02,31.91)(0.11,-0.23){2}{\line(0,-1){0.23}}
\multiput(57.24,31.46)(0.11,-0.22){2}{\line(0,-1){0.22}}
\multiput(57.47,31.01)(0.12,-0.22){2}{\line(0,-1){0.22}}
\multiput(57.71,30.57)(0.12,-0.22){2}{\line(0,-1){0.22}}
\multiput(57.96,30.14)(0.13,-0.22){2}{\line(0,-1){0.22}}
\multiput(58.22,29.7)(0.13,-0.21){2}{\line(0,-1){0.21}}
\multiput(58.48,29.28)(0.14,-0.21){2}{\line(0,-1){0.21}}
\multiput(58.76,28.86)(0.14,-0.21){2}{\line(0,-1){0.21}}
\multiput(59.04,28.44)(0.15,-0.2){2}{\line(0,-1){0.2}}
\multiput(59.34,28.04)(0.1,-0.13){3}{\line(0,-1){0.13}}
\multiput(59.64,27.64)(0.1,-0.13){3}{\line(0,-1){0.13}}
\multiput(59.95,27.24)(0.11,-0.13){3}{\line(0,-1){0.13}}
\multiput(60.27,26.85)(0.11,-0.13){3}{\line(0,-1){0.13}}
\multiput(60.59,26.47)(0.11,-0.12){3}{\line(0,-1){0.12}}
\multiput(60.93,26.1)(0.11,-0.12){3}{\line(0,-1){0.12}}
\multiput(61.27,25.73)(0.12,-0.12){3}{\line(0,-1){0.12}}
\multiput(61.62,25.37)(0.12,-0.12){3}{\line(1,0){0.12}}
\multiput(61.98,25.02)(0.12,-0.11){3}{\line(1,0){0.12}}
\multiput(62.35,24.68)(0.12,-0.11){3}{\line(1,0){0.12}}
\multiput(62.72,24.34)(0.13,-0.11){3}{\line(1,0){0.13}}
\multiput(63.1,24.02)(0.13,-0.11){3}{\line(1,0){0.13}}
\multiput(63.49,23.7)(0.13,-0.1){3}{\line(1,0){0.13}}
\multiput(63.89,23.39)(0.13,-0.1){3}{\line(1,0){0.13}}
\multiput(64.29,23.09)(0.2,-0.15){2}{\line(1,0){0.2}}
\multiput(64.69,22.79)(0.21,-0.14){2}{\line(1,0){0.21}}
\multiput(65.11,22.51)(0.21,-0.14){2}{\line(1,0){0.21}}
\multiput(65.53,22.23)(0.21,-0.13){2}{\line(1,0){0.21}}
\multiput(65.95,21.97)(0.22,-0.13){2}{\line(1,0){0.22}}
\multiput(66.39,21.71)(0.22,-0.12){2}{\line(1,0){0.22}}
\multiput(66.82,21.46)(0.22,-0.12){2}{\line(1,0){0.22}}
\multiput(67.26,21.22)(0.22,-0.11){2}{\line(1,0){0.22}}
\multiput(67.71,20.99)(0.23,-0.11){2}{\line(1,0){0.23}}
\multiput(68.16,20.77)(0.23,-0.1){2}{\line(1,0){0.23}}
\multiput(68.62,20.56)(0.23,-0.1){2}{\line(1,0){0.23}}
\multiput(69.08,20.36)(0.23,-0.09){2}{\line(1,0){0.23}}
\multiput(69.54,20.17)(0.47,-0.18){1}{\line(1,0){0.47}}
\multiput(70.01,20)(0.47,-0.17){1}{\line(1,0){0.47}}
\multiput(70.49,19.83)(0.48,-0.16){1}{\line(1,0){0.48}}
\multiput(70.96,19.67)(0.48,-0.15){1}{\line(1,0){0.48}}
\multiput(71.44,19.52)(0.48,-0.14){1}{\line(1,0){0.48}}
\multiput(71.92,19.38)(0.49,-0.13){1}{\line(1,0){0.49}}
\multiput(72.41,19.25)(0.49,-0.12){1}{\line(1,0){0.49}}
\multiput(72.9,19.13)(0.49,-0.11){1}{\line(1,0){0.49}}
\multiput(73.39,19.03)(0.49,-0.1){1}{\line(1,0){0.49}}
\multiput(73.88,18.93)(0.49,-0.09){1}{\line(1,0){0.49}}
\multiput(74.38,18.84)(0.5,-0.08){1}{\line(1,0){0.5}}
\multiput(74.87,18.77)(0.5,-0.06){1}{\line(1,0){0.5}}
\multiput(75.37,18.7)(0.5,-0.05){1}{\line(1,0){0.5}}
\multiput(75.87,18.65)(0.5,-0.04){1}{\line(1,0){0.5}}
\multiput(76.37,18.61)(0.5,-0.03){1}{\line(1,0){0.5}}
\multiput(76.87,18.57)(0.5,-0.02){1}{\line(1,0){0.5}}
\multiput(77.37,18.55)(0.5,-0.01){1}{\line(1,0){0.5}}
\put(77.87,18.54){\line(1,0){0.5}}
\multiput(78.38,18.54)(0.5,0.01){1}{\line(1,0){0.5}}
\multiput(78.88,18.55)(0.5,0.02){1}{\line(1,0){0.5}}
\multiput(79.38,18.57)(0.5,0.03){1}{\line(1,0){0.5}}
\multiput(79.88,18.61)(0.5,0.04){1}{\line(1,0){0.5}}
\multiput(80.38,18.65)(0.5,0.05){1}{\line(1,0){0.5}}
\multiput(80.88,18.7)(0.5,0.06){1}{\line(1,0){0.5}}
\multiput(81.38,18.77)(0.5,0.08){1}{\line(1,0){0.5}}
\multiput(81.87,18.84)(0.49,0.09){1}{\line(1,0){0.49}}
\multiput(82.37,18.93)(0.49,0.1){1}{\line(1,0){0.49}}
\multiput(82.86,19.03)(0.49,0.11){1}{\line(1,0){0.49}}
\multiput(83.35,19.13)(0.49,0.12){1}{\line(1,0){0.49}}
\multiput(83.84,19.25)(0.49,0.13){1}{\line(1,0){0.49}}
\multiput(84.33,19.38)(0.48,0.14){1}{\line(1,0){0.48}}
\multiput(84.81,19.52)(0.48,0.15){1}{\line(1,0){0.48}}
\multiput(85.29,19.67)(0.48,0.16){1}{\line(1,0){0.48}}
\multiput(85.76,19.83)(0.47,0.17){1}{\line(1,0){0.47}}
\multiput(86.24,20)(0.47,0.18){1}{\line(1,0){0.47}}
\multiput(86.71,20.17)(0.23,0.09){2}{\line(1,0){0.23}}
\multiput(87.17,20.36)(0.23,0.1){2}{\line(1,0){0.23}}
\multiput(87.63,20.56)(0.23,0.1){2}{\line(1,0){0.23}}
\multiput(88.09,20.77)(0.23,0.11){2}{\line(1,0){0.23}}
\multiput(88.54,20.99)(0.22,0.11){2}{\line(1,0){0.22}}
\multiput(88.99,21.22)(0.22,0.12){2}{\line(1,0){0.22}}
\multiput(89.43,21.46)(0.22,0.12){2}{\line(1,0){0.22}}
\multiput(89.86,21.71)(0.22,0.13){2}{\line(1,0){0.22}}
\multiput(90.3,21.97)(0.21,0.13){2}{\line(1,0){0.21}}
\multiput(90.72,22.23)(0.21,0.14){2}{\line(1,0){0.21}}
\multiput(91.14,22.51)(0.21,0.14){2}{\line(1,0){0.21}}
\multiput(91.56,22.79)(0.2,0.15){2}{\line(1,0){0.2}}
\multiput(91.96,23.09)(0.13,0.1){3}{\line(1,0){0.13}}
\multiput(92.36,23.39)(0.13,0.1){3}{\line(1,0){0.13}}
\multiput(92.76,23.7)(0.13,0.11){3}{\line(1,0){0.13}}
\multiput(93.15,24.02)(0.13,0.11){3}{\line(1,0){0.13}}
\multiput(93.53,24.34)(0.12,0.11){3}{\line(1,0){0.12}}
\multiput(93.9,24.68)(0.12,0.11){3}{\line(1,0){0.12}}
\multiput(94.27,25.02)(0.12,0.12){3}{\line(1,0){0.12}}
\multiput(94.63,25.37)(0.12,0.12){3}{\line(0,1){0.12}}
\multiput(94.98,25.73)(0.11,0.12){3}{\line(0,1){0.12}}
\multiput(95.32,26.1)(0.11,0.12){3}{\line(0,1){0.12}}
\multiput(95.66,26.47)(0.11,0.13){3}{\line(0,1){0.13}}
\multiput(95.98,26.85)(0.11,0.13){3}{\line(0,1){0.13}}
\multiput(96.3,27.24)(0.1,0.13){3}{\line(0,1){0.13}}
\multiput(96.61,27.64)(0.1,0.13){3}{\line(0,1){0.13}}
\multiput(96.91,28.04)(0.15,0.2){2}{\line(0,1){0.2}}
\multiput(97.21,28.44)(0.14,0.21){2}{\line(0,1){0.21}}
\multiput(97.49,28.86)(0.14,0.21){2}{\line(0,1){0.21}}
\multiput(97.77,29.28)(0.13,0.21){2}{\line(0,1){0.21}}
\multiput(98.03,29.7)(0.13,0.22){2}{\line(0,1){0.22}}
\multiput(98.29,30.14)(0.12,0.22){2}{\line(0,1){0.22}}
\multiput(98.54,30.57)(0.12,0.22){2}{\line(0,1){0.22}}
\multiput(98.78,31.01)(0.11,0.22){2}{\line(0,1){0.22}}
\multiput(99.01,31.46)(0.11,0.23){2}{\line(0,1){0.23}}
\multiput(99.23,31.91)(0.1,0.23){2}{\line(0,1){0.23}}
\multiput(99.44,32.37)(0.1,0.23){2}{\line(0,1){0.23}}
\multiput(99.64,32.83)(0.09,0.23){2}{\line(0,1){0.23}}
\multiput(99.83,33.29)(0.18,0.47){1}{\line(0,1){0.47}}
\multiput(100,33.76)(0.17,0.47){1}{\line(0,1){0.47}}
\multiput(100.17,34.24)(0.16,0.48){1}{\line(0,1){0.48}}
\multiput(100.33,34.71)(0.15,0.48){1}{\line(0,1){0.48}}
\multiput(100.48,35.19)(0.14,0.48){1}{\line(0,1){0.48}}
\multiput(100.62,35.67)(0.13,0.49){1}{\line(0,1){0.49}}
\multiput(100.75,36.16)(0.12,0.49){1}{\line(0,1){0.49}}
\multiput(100.87,36.65)(0.11,0.49){1}{\line(0,1){0.49}}
\multiput(100.97,37.14)(0.1,0.49){1}{\line(0,1){0.49}}
\multiput(101.07,37.63)(0.09,0.49){1}{\line(0,1){0.49}}
\multiput(101.16,38.13)(0.08,0.5){1}{\line(0,1){0.5}}
\multiput(101.23,38.62)(0.06,0.5){1}{\line(0,1){0.5}}
\multiput(101.3,39.12)(0.05,0.5){1}{\line(0,1){0.5}}
\multiput(101.35,39.62)(0.04,0.5){1}{\line(0,1){0.5}}
\multiput(101.39,40.12)(0.03,0.5){1}{\line(0,1){0.5}}
\multiput(101.43,40.62)(0.02,0.5){1}{\line(0,1){0.5}}
\multiput(101.45,41.12)(0.01,0.5){1}{\line(0,1){0.5}}

\linethickness{0.3mm}
\multiput(40,70)(0.16,-0.12){125}{\line(1,0){0.16}}
\linethickness{0.3mm}
\multiput(97,55)(0.16,0.12){125}{\line(1,0){0.16}}
\linethickness{0.3mm}
\multiput(38,5)(0.12,0.15){167}{\line(0,1){0.15}}
\linethickness{0.3mm}
\multiput(98,30)(0.12,-0.12){167}{\line(1,0){0.12}}
\put(58,68){\makebox(0,0)[cc]{$\wh Q_B C_s$}}

\put(100,67){\makebox(0,0)[cc]{$B_s$}}

\put(47,10){\makebox(0,0)[cc]{$\AAA_1$}}

\put(108,10){\makebox(0,0)[cc]{$\AAA_N$}}

\put(75,10){\makebox(0,0)[cc]{$\cdots$}}

\put(78,42){\makebox(0,0)[cc]{Full}}

\put(20,45){\makebox(0,0)[cc]{$\delta \alpha \, \, \times {\displaystyle 
\sum}_s$}}

\put(43,60){\makebox(0,0)[cc]{$\searrow$}}

\put(49,53){\makebox(0,0)[cc]{$k$}}

\put(105,57){\makebox(0,0)[cc]{$\nearrow$}}

\put(110,57){\makebox(0,0)[cc]{$k$}}

\end{picture}

}

\def\figscscb{

\def\JPicScale{0.6}
\ifx\JPicScale\undefined\def\JPicScale{1}\fi
\unitlength \JPicScale mm
\begin{picture}(120,70)(0,0)
\linethickness{0.3mm}
\put(101.46,41.62){\line(0,1){0.5}}
\multiput(101.45,42.63)(0.01,-0.5){1}{\line(0,-1){0.5}}
\multiput(101.43,43.13)(0.02,-0.5){1}{\line(0,-1){0.5}}
\multiput(101.39,43.63)(0.03,-0.5){1}{\line(0,-1){0.5}}
\multiput(101.35,44.13)(0.04,-0.5){1}{\line(0,-1){0.5}}
\multiput(101.3,44.63)(0.05,-0.5){1}{\line(0,-1){0.5}}
\multiput(101.23,45.13)(0.06,-0.5){1}{\line(0,-1){0.5}}
\multiput(101.16,45.62)(0.08,-0.5){1}{\line(0,-1){0.5}}
\multiput(101.07,46.12)(0.09,-0.49){1}{\line(0,-1){0.49}}
\multiput(100.97,46.61)(0.1,-0.49){1}{\line(0,-1){0.49}}
\multiput(100.87,47.1)(0.11,-0.49){1}{\line(0,-1){0.49}}
\multiput(100.75,47.59)(0.12,-0.49){1}{\line(0,-1){0.49}}
\multiput(100.62,48.08)(0.13,-0.49){1}{\line(0,-1){0.49}}
\multiput(100.48,48.56)(0.14,-0.48){1}{\line(0,-1){0.48}}
\multiput(100.33,49.04)(0.15,-0.48){1}{\line(0,-1){0.48}}
\multiput(100.17,49.51)(0.16,-0.48){1}{\line(0,-1){0.48}}
\multiput(100,49.99)(0.17,-0.47){1}{\line(0,-1){0.47}}
\multiput(99.83,50.46)(0.18,-0.47){1}{\line(0,-1){0.47}}
\multiput(99.64,50.92)(0.09,-0.23){2}{\line(0,-1){0.23}}
\multiput(99.44,51.38)(0.1,-0.23){2}{\line(0,-1){0.23}}
\multiput(99.23,51.84)(0.1,-0.23){2}{\line(0,-1){0.23}}
\multiput(99.01,52.29)(0.11,-0.23){2}{\line(0,-1){0.23}}
\multiput(98.78,52.74)(0.11,-0.22){2}{\line(0,-1){0.22}}
\multiput(98.54,53.18)(0.12,-0.22){2}{\line(0,-1){0.22}}
\multiput(98.29,53.61)(0.12,-0.22){2}{\line(0,-1){0.22}}
\multiput(98.03,54.05)(0.13,-0.22){2}{\line(0,-1){0.22}}
\multiput(97.77,54.47)(0.13,-0.21){2}{\line(0,-1){0.21}}
\multiput(97.49,54.89)(0.14,-0.21){2}{\line(0,-1){0.21}}
\multiput(97.21,55.31)(0.14,-0.21){2}{\line(0,-1){0.21}}
\multiput(96.91,55.71)(0.15,-0.2){2}{\line(0,-1){0.2}}
\multiput(96.61,56.11)(0.1,-0.13){3}{\line(0,-1){0.13}}
\multiput(96.3,56.51)(0.1,-0.13){3}{\line(0,-1){0.13}}
\multiput(95.98,56.9)(0.11,-0.13){3}{\line(0,-1){0.13}}
\multiput(95.66,57.28)(0.11,-0.13){3}{\line(0,-1){0.13}}
\multiput(95.32,57.65)(0.11,-0.12){3}{\line(0,-1){0.12}}
\multiput(94.98,58.02)(0.11,-0.12){3}{\line(0,-1){0.12}}
\multiput(94.63,58.38)(0.12,-0.12){3}{\line(0,-1){0.12}}
\multiput(94.27,58.73)(0.12,-0.12){3}{\line(1,0){0.12}}
\multiput(93.9,59.07)(0.12,-0.11){3}{\line(1,0){0.12}}
\multiput(93.53,59.41)(0.12,-0.11){3}{\line(1,0){0.12}}
\multiput(93.15,59.73)(0.13,-0.11){3}{\line(1,0){0.13}}
\multiput(92.76,60.05)(0.13,-0.11){3}{\line(1,0){0.13}}
\multiput(92.36,60.36)(0.13,-0.1){3}{\line(1,0){0.13}}
\multiput(91.96,60.66)(0.13,-0.1){3}{\line(1,0){0.13}}
\multiput(91.56,60.96)(0.2,-0.15){2}{\line(1,0){0.2}}
\multiput(91.14,61.24)(0.21,-0.14){2}{\line(1,0){0.21}}
\multiput(90.72,61.52)(0.21,-0.14){2}{\line(1,0){0.21}}
\multiput(90.3,61.78)(0.21,-0.13){2}{\line(1,0){0.21}}
\multiput(89.86,62.04)(0.22,-0.13){2}{\line(1,0){0.22}}
\multiput(89.43,62.29)(0.22,-0.12){2}{\line(1,0){0.22}}
\multiput(88.99,62.53)(0.22,-0.12){2}{\line(1,0){0.22}}
\multiput(88.54,62.76)(0.22,-0.11){2}{\line(1,0){0.22}}
\multiput(88.09,62.98)(0.23,-0.11){2}{\line(1,0){0.23}}
\multiput(87.63,63.19)(0.23,-0.1){2}{\line(1,0){0.23}}
\multiput(87.17,63.39)(0.23,-0.1){2}{\line(1,0){0.23}}
\multiput(86.71,63.58)(0.23,-0.09){2}{\line(1,0){0.23}}
\multiput(86.24,63.75)(0.47,-0.18){1}{\line(1,0){0.47}}
\multiput(85.76,63.92)(0.47,-0.17){1}{\line(1,0){0.47}}
\multiput(85.29,64.08)(0.48,-0.16){1}{\line(1,0){0.48}}
\multiput(84.81,64.23)(0.48,-0.15){1}{\line(1,0){0.48}}
\multiput(84.33,64.37)(0.48,-0.14){1}{\line(1,0){0.48}}
\multiput(83.84,64.5)(0.49,-0.13){1}{\line(1,0){0.49}}
\multiput(83.35,64.62)(0.49,-0.12){1}{\line(1,0){0.49}}
\multiput(82.86,64.72)(0.49,-0.11){1}{\line(1,0){0.49}}
\multiput(82.37,64.82)(0.49,-0.1){1}{\line(1,0){0.49}}
\multiput(81.87,64.91)(0.49,-0.09){1}{\line(1,0){0.49}}
\multiput(81.38,64.98)(0.5,-0.08){1}{\line(1,0){0.5}}
\multiput(80.88,65.05)(0.5,-0.06){1}{\line(1,0){0.5}}
\multiput(80.38,65.1)(0.5,-0.05){1}{\line(1,0){0.5}}
\multiput(79.88,65.14)(0.5,-0.04){1}{\line(1,0){0.5}}
\multiput(79.38,65.18)(0.5,-0.03){1}{\line(1,0){0.5}}
\multiput(78.88,65.2)(0.5,-0.02){1}{\line(1,0){0.5}}
\multiput(78.38,65.21)(0.5,-0.01){1}{\line(1,0){0.5}}
\put(77.87,65.21){\line(1,0){0.5}}
\multiput(77.37,65.2)(0.5,0.01){1}{\line(1,0){0.5}}
\multiput(76.87,65.18)(0.5,0.02){1}{\line(1,0){0.5}}
\multiput(76.37,65.14)(0.5,0.03){1}{\line(1,0){0.5}}
\multiput(75.87,65.1)(0.5,0.04){1}{\line(1,0){0.5}}
\multiput(75.37,65.05)(0.5,0.05){1}{\line(1,0){0.5}}
\multiput(74.87,64.98)(0.5,0.06){1}{\line(1,0){0.5}}
\multiput(74.38,64.91)(0.5,0.08){1}{\line(1,0){0.5}}
\multiput(73.88,64.82)(0.49,0.09){1}{\line(1,0){0.49}}
\multiput(73.39,64.72)(0.49,0.1){1}{\line(1,0){0.49}}
\multiput(72.9,64.62)(0.49,0.11){1}{\line(1,0){0.49}}
\multiput(72.41,64.5)(0.49,0.12){1}{\line(1,0){0.49}}
\multiput(71.92,64.37)(0.49,0.13){1}{\line(1,0){0.49}}
\multiput(71.44,64.23)(0.48,0.14){1}{\line(1,0){0.48}}
\multiput(70.96,64.08)(0.48,0.15){1}{\line(1,0){0.48}}
\multiput(70.49,63.92)(0.48,0.16){1}{\line(1,0){0.48}}
\multiput(70.01,63.75)(0.47,0.17){1}{\line(1,0){0.47}}
\multiput(69.54,63.58)(0.47,0.18){1}{\line(1,0){0.47}}
\multiput(69.08,63.39)(0.23,0.09){2}{\line(1,0){0.23}}
\multiput(68.62,63.19)(0.23,0.1){2}{\line(1,0){0.23}}
\multiput(68.16,62.98)(0.23,0.1){2}{\line(1,0){0.23}}
\multiput(67.71,62.76)(0.23,0.11){2}{\line(1,0){0.23}}
\multiput(67.26,62.53)(0.22,0.11){2}{\line(1,0){0.22}}
\multiput(66.82,62.29)(0.22,0.12){2}{\line(1,0){0.22}}
\multiput(66.39,62.04)(0.22,0.12){2}{\line(1,0){0.22}}
\multiput(65.95,61.78)(0.22,0.13){2}{\line(1,0){0.22}}
\multiput(65.53,61.52)(0.21,0.13){2}{\line(1,0){0.21}}
\multiput(65.11,61.24)(0.21,0.14){2}{\line(1,0){0.21}}
\multiput(64.69,60.96)(0.21,0.14){2}{\line(1,0){0.21}}
\multiput(64.29,60.66)(0.2,0.15){2}{\line(1,0){0.2}}
\multiput(63.89,60.36)(0.13,0.1){3}{\line(1,0){0.13}}
\multiput(63.49,60.05)(0.13,0.1){3}{\line(1,0){0.13}}
\multiput(63.1,59.73)(0.13,0.11){3}{\line(1,0){0.13}}
\multiput(62.72,59.41)(0.13,0.11){3}{\line(1,0){0.13}}
\multiput(62.35,59.07)(0.12,0.11){3}{\line(1,0){0.12}}
\multiput(61.98,58.73)(0.12,0.11){3}{\line(1,0){0.12}}
\multiput(61.62,58.38)(0.12,0.12){3}{\line(1,0){0.12}}
\multiput(61.27,58.02)(0.12,0.12){3}{\line(0,1){0.12}}
\multiput(60.93,57.65)(0.11,0.12){3}{\line(0,1){0.12}}
\multiput(60.59,57.28)(0.11,0.12){3}{\line(0,1){0.12}}
\multiput(60.27,56.9)(0.11,0.13){3}{\line(0,1){0.13}}
\multiput(59.95,56.51)(0.11,0.13){3}{\line(0,1){0.13}}
\multiput(59.64,56.11)(0.1,0.13){3}{\line(0,1){0.13}}
\multiput(59.34,55.71)(0.1,0.13){3}{\line(0,1){0.13}}
\multiput(59.04,55.31)(0.15,0.2){2}{\line(0,1){0.2}}
\multiput(58.76,54.89)(0.14,0.21){2}{\line(0,1){0.21}}
\multiput(58.48,54.47)(0.14,0.21){2}{\line(0,1){0.21}}
\multiput(58.22,54.05)(0.13,0.21){2}{\line(0,1){0.21}}
\multiput(57.96,53.61)(0.13,0.22){2}{\line(0,1){0.22}}
\multiput(57.71,53.18)(0.12,0.22){2}{\line(0,1){0.22}}
\multiput(57.47,52.74)(0.12,0.22){2}{\line(0,1){0.22}}
\multiput(57.24,52.29)(0.11,0.22){2}{\line(0,1){0.22}}
\multiput(57.02,51.84)(0.11,0.23){2}{\line(0,1){0.23}}
\multiput(56.81,51.38)(0.1,0.23){2}{\line(0,1){0.23}}
\multiput(56.61,50.92)(0.1,0.23){2}{\line(0,1){0.23}}
\multiput(56.42,50.46)(0.09,0.23){2}{\line(0,1){0.23}}
\multiput(56.25,49.99)(0.18,0.47){1}{\line(0,1){0.47}}
\multiput(56.08,49.51)(0.17,0.47){1}{\line(0,1){0.47}}
\multiput(55.92,49.04)(0.16,0.48){1}{\line(0,1){0.48}}
\multiput(55.77,48.56)(0.15,0.48){1}{\line(0,1){0.48}}
\multiput(55.63,48.08)(0.14,0.48){1}{\line(0,1){0.48}}
\multiput(55.5,47.59)(0.13,0.49){1}{\line(0,1){0.49}}
\multiput(55.38,47.1)(0.12,0.49){1}{\line(0,1){0.49}}
\multiput(55.28,46.61)(0.11,0.49){1}{\line(0,1){0.49}}
\multiput(55.18,46.12)(0.1,0.49){1}{\line(0,1){0.49}}
\multiput(55.09,45.62)(0.09,0.49){1}{\line(0,1){0.49}}
\multiput(55.02,45.13)(0.08,0.5){1}{\line(0,1){0.5}}
\multiput(54.95,44.63)(0.06,0.5){1}{\line(0,1){0.5}}
\multiput(54.9,44.13)(0.05,0.5){1}{\line(0,1){0.5}}
\multiput(54.86,43.63)(0.04,0.5){1}{\line(0,1){0.5}}
\multiput(54.82,43.13)(0.03,0.5){1}{\line(0,1){0.5}}
\multiput(54.8,42.63)(0.02,0.5){1}{\line(0,1){0.5}}
\multiput(54.79,42.13)(0.01,0.5){1}{\line(0,1){0.5}}
\put(54.79,41.62){\line(0,1){0.5}}
\multiput(54.79,41.62)(0.01,-0.5){1}{\line(0,-1){0.5}}
\multiput(54.8,41.12)(0.02,-0.5){1}{\line(0,-1){0.5}}
\multiput(54.82,40.62)(0.03,-0.5){1}{\line(0,-1){0.5}}
\multiput(54.86,40.12)(0.04,-0.5){1}{\line(0,-1){0.5}}
\multiput(54.9,39.62)(0.05,-0.5){1}{\line(0,-1){0.5}}
\multiput(54.95,39.12)(0.06,-0.5){1}{\line(0,-1){0.5}}
\multiput(55.02,38.62)(0.08,-0.5){1}{\line(0,-1){0.5}}
\multiput(55.09,38.13)(0.09,-0.49){1}{\line(0,-1){0.49}}
\multiput(55.18,37.63)(0.1,-0.49){1}{\line(0,-1){0.49}}
\multiput(55.28,37.14)(0.11,-0.49){1}{\line(0,-1){0.49}}
\multiput(55.38,36.65)(0.12,-0.49){1}{\line(0,-1){0.49}}
\multiput(55.5,36.16)(0.13,-0.49){1}{\line(0,-1){0.49}}
\multiput(55.63,35.67)(0.14,-0.48){1}{\line(0,-1){0.48}}
\multiput(55.77,35.19)(0.15,-0.48){1}{\line(0,-1){0.48}}
\multiput(55.92,34.71)(0.16,-0.48){1}{\line(0,-1){0.48}}
\multiput(56.08,34.24)(0.17,-0.47){1}{\line(0,-1){0.47}}
\multiput(56.25,33.76)(0.18,-0.47){1}{\line(0,-1){0.47}}
\multiput(56.42,33.29)(0.09,-0.23){2}{\line(0,-1){0.23}}
\multiput(56.61,32.83)(0.1,-0.23){2}{\line(0,-1){0.23}}
\multiput(56.81,32.37)(0.1,-0.23){2}{\line(0,-1){0.23}}
\multiput(57.02,31.91)(0.11,-0.23){2}{\line(0,-1){0.23}}
\multiput(57.24,31.46)(0.11,-0.22){2}{\line(0,-1){0.22}}
\multiput(57.47,31.01)(0.12,-0.22){2}{\line(0,-1){0.22}}
\multiput(57.71,30.57)(0.12,-0.22){2}{\line(0,-1){0.22}}
\multiput(57.96,30.14)(0.13,-0.22){2}{\line(0,-1){0.22}}
\multiput(58.22,29.7)(0.13,-0.21){2}{\line(0,-1){0.21}}
\multiput(58.48,29.28)(0.14,-0.21){2}{\line(0,-1){0.21}}
\multiput(58.76,28.86)(0.14,-0.21){2}{\line(0,-1){0.21}}
\multiput(59.04,28.44)(0.15,-0.2){2}{\line(0,-1){0.2}}
\multiput(59.34,28.04)(0.1,-0.13){3}{\line(0,-1){0.13}}
\multiput(59.64,27.64)(0.1,-0.13){3}{\line(0,-1){0.13}}
\multiput(59.95,27.24)(0.11,-0.13){3}{\line(0,-1){0.13}}
\multiput(60.27,26.85)(0.11,-0.13){3}{\line(0,-1){0.13}}
\multiput(60.59,26.47)(0.11,-0.12){3}{\line(0,-1){0.12}}
\multiput(60.93,26.1)(0.11,-0.12){3}{\line(0,-1){0.12}}
\multiput(61.27,25.73)(0.12,-0.12){3}{\line(0,-1){0.12}}
\multiput(61.62,25.37)(0.12,-0.12){3}{\line(1,0){0.12}}
\multiput(61.98,25.02)(0.12,-0.11){3}{\line(1,0){0.12}}
\multiput(62.35,24.68)(0.12,-0.11){3}{\line(1,0){0.12}}
\multiput(62.72,24.34)(0.13,-0.11){3}{\line(1,0){0.13}}
\multiput(63.1,24.02)(0.13,-0.11){3}{\line(1,0){0.13}}
\multiput(63.49,23.7)(0.13,-0.1){3}{\line(1,0){0.13}}
\multiput(63.89,23.39)(0.13,-0.1){3}{\line(1,0){0.13}}
\multiput(64.29,23.09)(0.2,-0.15){2}{\line(1,0){0.2}}
\multiput(64.69,22.79)(0.21,-0.14){2}{\line(1,0){0.21}}
\multiput(65.11,22.51)(0.21,-0.14){2}{\line(1,0){0.21}}
\multiput(65.53,22.23)(0.21,-0.13){2}{\line(1,0){0.21}}
\multiput(65.95,21.97)(0.22,-0.13){2}{\line(1,0){0.22}}
\multiput(66.39,21.71)(0.22,-0.12){2}{\line(1,0){0.22}}
\multiput(66.82,21.46)(0.22,-0.12){2}{\line(1,0){0.22}}
\multiput(67.26,21.22)(0.22,-0.11){2}{\line(1,0){0.22}}
\multiput(67.71,20.99)(0.23,-0.11){2}{\line(1,0){0.23}}
\multiput(68.16,20.77)(0.23,-0.1){2}{\line(1,0){0.23}}
\multiput(68.62,20.56)(0.23,-0.1){2}{\line(1,0){0.23}}
\multiput(69.08,20.36)(0.23,-0.09){2}{\line(1,0){0.23}}
\multiput(69.54,20.17)(0.47,-0.18){1}{\line(1,0){0.47}}
\multiput(70.01,20)(0.47,-0.17){1}{\line(1,0){0.47}}
\multiput(70.49,19.83)(0.48,-0.16){1}{\line(1,0){0.48}}
\multiput(70.96,19.67)(0.48,-0.15){1}{\line(1,0){0.48}}
\multiput(71.44,19.52)(0.48,-0.14){1}{\line(1,0){0.48}}
\multiput(71.92,19.38)(0.49,-0.13){1}{\line(1,0){0.49}}
\multiput(72.41,19.25)(0.49,-0.12){1}{\line(1,0){0.49}}
\multiput(72.9,19.13)(0.49,-0.11){1}{\line(1,0){0.49}}
\multiput(73.39,19.03)(0.49,-0.1){1}{\line(1,0){0.49}}
\multiput(73.88,18.93)(0.49,-0.09){1}{\line(1,0){0.49}}
\multiput(74.38,18.84)(0.5,-0.08){1}{\line(1,0){0.5}}
\multiput(74.87,18.77)(0.5,-0.06){1}{\line(1,0){0.5}}
\multiput(75.37,18.7)(0.5,-0.05){1}{\line(1,0){0.5}}
\multiput(75.87,18.65)(0.5,-0.04){1}{\line(1,0){0.5}}
\multiput(76.37,18.61)(0.5,-0.03){1}{\line(1,0){0.5}}
\multiput(76.87,18.57)(0.5,-0.02){1}{\line(1,0){0.5}}
\multiput(77.37,18.55)(0.5,-0.01){1}{\line(1,0){0.5}}
\put(77.87,18.54){\line(1,0){0.5}}
\multiput(78.38,18.54)(0.5,0.01){1}{\line(1,0){0.5}}
\multiput(78.88,18.55)(0.5,0.02){1}{\line(1,0){0.5}}
\multiput(79.38,18.57)(0.5,0.03){1}{\line(1,0){0.5}}
\multiput(79.88,18.61)(0.5,0.04){1}{\line(1,0){0.5}}
\multiput(80.38,18.65)(0.5,0.05){1}{\line(1,0){0.5}}
\multiput(80.88,18.7)(0.5,0.06){1}{\line(1,0){0.5}}
\multiput(81.38,18.77)(0.5,0.08){1}{\line(1,0){0.5}}
\multiput(81.87,18.84)(0.49,0.09){1}{\line(1,0){0.49}}
\multiput(82.37,18.93)(0.49,0.1){1}{\line(1,0){0.49}}
\multiput(82.86,19.03)(0.49,0.11){1}{\line(1,0){0.49}}
\multiput(83.35,19.13)(0.49,0.12){1}{\line(1,0){0.49}}
\multiput(83.84,19.25)(0.49,0.13){1}{\line(1,0){0.49}}
\multiput(84.33,19.38)(0.48,0.14){1}{\line(1,0){0.48}}
\multiput(84.81,19.52)(0.48,0.15){1}{\line(1,0){0.48}}
\multiput(85.29,19.67)(0.48,0.16){1}{\line(1,0){0.48}}
\multiput(85.76,19.83)(0.47,0.17){1}{\line(1,0){0.47}}
\multiput(86.24,20)(0.47,0.18){1}{\line(1,0){0.47}}
\multiput(86.71,20.17)(0.23,0.09){2}{\line(1,0){0.23}}
\multiput(87.17,20.36)(0.23,0.1){2}{\line(1,0){0.23}}
\multiput(87.63,20.56)(0.23,0.1){2}{\line(1,0){0.23}}
\multiput(88.09,20.77)(0.23,0.11){2}{\line(1,0){0.23}}
\multiput(88.54,20.99)(0.22,0.11){2}{\line(1,0){0.22}}
\multiput(88.99,21.22)(0.22,0.12){2}{\line(1,0){0.22}}
\multiput(89.43,21.46)(0.22,0.12){2}{\line(1,0){0.22}}
\multiput(89.86,21.71)(0.22,0.13){2}{\line(1,0){0.22}}
\multiput(90.3,21.97)(0.21,0.13){2}{\line(1,0){0.21}}
\multiput(90.72,22.23)(0.21,0.14){2}{\line(1,0){0.21}}
\multiput(91.14,22.51)(0.21,0.14){2}{\line(1,0){0.21}}
\multiput(91.56,22.79)(0.2,0.15){2}{\line(1,0){0.2}}
\multiput(91.96,23.09)(0.13,0.1){3}{\line(1,0){0.13}}
\multiput(92.36,23.39)(0.13,0.1){3}{\line(1,0){0.13}}
\multiput(92.76,23.7)(0.13,0.11){3}{\line(1,0){0.13}}
\multiput(93.15,24.02)(0.13,0.11){3}{\line(1,0){0.13}}
\multiput(93.53,24.34)(0.12,0.11){3}{\line(1,0){0.12}}
\multiput(93.9,24.68)(0.12,0.11){3}{\line(1,0){0.12}}
\multiput(94.27,25.02)(0.12,0.12){3}{\line(1,0){0.12}}
\multiput(94.63,25.37)(0.12,0.12){3}{\line(0,1){0.12}}
\multiput(94.98,25.73)(0.11,0.12){3}{\line(0,1){0.12}}
\multiput(95.32,26.1)(0.11,0.12){3}{\line(0,1){0.12}}
\multiput(95.66,26.47)(0.11,0.13){3}{\line(0,1){0.13}}
\multiput(95.98,26.85)(0.11,0.13){3}{\line(0,1){0.13}}
\multiput(96.3,27.24)(0.1,0.13){3}{\line(0,1){0.13}}
\multiput(96.61,27.64)(0.1,0.13){3}{\line(0,1){0.13}}
\multiput(96.91,28.04)(0.15,0.2){2}{\line(0,1){0.2}}
\multiput(97.21,28.44)(0.14,0.21){2}{\line(0,1){0.21}}
\multiput(97.49,28.86)(0.14,0.21){2}{\line(0,1){0.21}}
\multiput(97.77,29.28)(0.13,0.21){2}{\line(0,1){0.21}}
\multiput(98.03,29.7)(0.13,0.22){2}{\line(0,1){0.22}}
\multiput(98.29,30.14)(0.12,0.22){2}{\line(0,1){0.22}}
\multiput(98.54,30.57)(0.12,0.22){2}{\line(0,1){0.22}}
\multiput(98.78,31.01)(0.11,0.22){2}{\line(0,1){0.22}}
\multiput(99.01,31.46)(0.11,0.23){2}{\line(0,1){0.23}}
\multiput(99.23,31.91)(0.1,0.23){2}{\line(0,1){0.23}}
\multiput(99.44,32.37)(0.1,0.23){2}{\line(0,1){0.23}}
\multiput(99.64,32.83)(0.09,0.23){2}{\line(0,1){0.23}}
\multiput(99.83,33.29)(0.18,0.47){1}{\line(0,1){0.47}}
\multiput(100,33.76)(0.17,0.47){1}{\line(0,1){0.47}}
\multiput(100.17,34.24)(0.16,0.48){1}{\line(0,1){0.48}}
\multiput(100.33,34.71)(0.15,0.48){1}{\line(0,1){0.48}}
\multiput(100.48,35.19)(0.14,0.48){1}{\line(0,1){0.48}}
\multiput(100.62,35.67)(0.13,0.49){1}{\line(0,1){0.49}}
\multiput(100.75,36.16)(0.12,0.49){1}{\line(0,1){0.49}}
\multiput(100.87,36.65)(0.11,0.49){1}{\line(0,1){0.49}}
\multiput(100.97,37.14)(0.1,0.49){1}{\line(0,1){0.49}}
\multiput(101.07,37.63)(0.09,0.49){1}{\line(0,1){0.49}}
\multiput(101.16,38.13)(0.08,0.5){1}{\line(0,1){0.5}}
\multiput(101.23,38.62)(0.06,0.5){1}{\line(0,1){0.5}}
\multiput(101.3,39.12)(0.05,0.5){1}{\line(0,1){0.5}}
\multiput(101.35,39.62)(0.04,0.5){1}{\line(0,1){0.5}}
\multiput(101.39,40.12)(0.03,0.5){1}{\line(0,1){0.5}}
\multiput(101.43,40.62)(0.02,0.5){1}{\line(0,1){0.5}}
\multiput(101.45,41.12)(0.01,0.5){1}{\line(0,1){0.5}}

\linethickness{0.3mm}
\multiput(40,70)(0.16,-0.12){125}{\line(1,0){0.16}}
\linethickness{0.3mm}
\multiput(97,55)(0.16,0.12){125}{\line(1,0){0.16}}
\linethickness{0.3mm}
\multiput(38,5)(0.12,0.15){167}{\line(0,1){0.15}}
\linethickness{0.3mm}
\multiput(98,30)(0.12,-0.12){167}{\line(1,0){0.12}}
\put(56,64){\makebox(0,0)[cc]{$C_s$}}

\put(98,67){\makebox(0,0)[cc]{$\wh Q_B B_s$}}

\put(47,10){\makebox(0,0)[cc]{$\AAA_1$}}

\put(108,10){\makebox(0,0)[cc]{$\AAA_N$}}

\put(75,10){\makebox(0,0)[cc]{$\cdots$}}

\put(78,42){\makebox(0,0)[cc]{Full}}

\put(15,45){\makebox(0,0)[cc]{$+ \quad  \delta \alpha \, \, \times {\displaystyle 
\sum}_s
(-1)^{C_s}$}}

\put(43,60){\makebox(0,0)[cc]{$\searrow$}}

\put(49,53){\makebox(0,0)[cc]{$k$}}

\put(105,57){\makebox(0,0)[cc]{$\nearrow$}}

\put(110,57){\makebox(0,0)[cc]{$k$}}

\end{picture}

}

\newcommand{\plu}{'s}

\begin{document}

\begin{flushright}
HRI/ST/1701 \\
LPTENS/17/04 \\
\end{flushright}

\baselineskip 24pt

\begin{center}
{\Large \bf  Closed Superstring  Field Theory and its  Applications}

\end{center}

\vskip .6cm
\medskip

\vspace*{4.0ex}

\baselineskip=18pt

\begin{center}

{\large 
\rm Corinne de Lacroix$^a$, Harold Erbin$^a$, Sitender Pratap Kashyap$^{b}$, 

Ashoke Sen$^{b}$, Mritunjay Verma$^{b,c}$ }

\end{center}

\vspace*{4.0ex}

\centerline{\it \small $^a$Ecole normale superieure,
24 rue Lhomond,
F-75230 Paris cedex 05, France}
\centerline{\it \small $^b$Harish-Chandra Research Institute, HBNI,
Chhatnag Road, Jhusi,
Allahabad 211019, India}
\centerline{ \it \small $^c$International Centre for Theoretical Sciences,
 Hesarghatta,
Bengaluru 560089, India.}

\vspace*{1.0ex}
\centerline{\small E-mail:  corinne.delacroix,harold.erbin@lpt.ens.fr,
sitenderpratap,sen,mritunjayverma@hri.res.in}

\vspace*{5.0ex}

\centerline{\bf Abstract} \bigskip

We review recent developments in the construction of heterotic and type II string field
theories and their various applications. These include systematic procedures for 
determining the shifts in the vacuum expectation values of fields under quantum
corrections, 
computing renormalized masses and S-matrix of the theory around the shifted vacuum
and a proof of unitarity of the S-matrix. The S-matrix computed this way is free from all
divergences when there are more than 4 non-compact space-time dimensions, but
suffers from the usual infrared divergences when the number of non-compact space-time
dimensions is 4 or less. 

\vfill


\vfill \eject

\baselineskip 18pt

\tableofcontents

\sectiono{Introduction and Motivation} \label{sintro}

In string theory the observables are S-matrix elements -- also called 
amplitudes.\footnote{Throughout this review
string theory will mean superstring theory, which in turn will include the two heterotic 
string theories and the two type II string theories, possibly compactified on some
manifold with NS (NSNS) background fields. We shall assume that there are some
non-compact dimensions with flat Minkowski metric that can be used to define the
S-matrix. \label{foint}}  These are the observables in field theories as well.
However, the prescription for computing the S-matrix 
in string theory is apparently different from that in
 quantum field theories.
A $g$-loop, $N$-point amplitude is given by an expression of the form:
\be \label{estring}
\int dm_1\ldots dm_{6g-6+2N} \, F(m_1,m_2,\ldots  , m_{6g-6+2N})
\ee
where $m_i$ are the parameters labelling the moduli space of two dimensional 
Riemann surfaces of genus $g$ and $n$ marked points -- also known as
punctures. $F(\{m_i\})$ denotes
a correlation function of a two dimensional conformal
field theory on the Riemann surface, with vertex operators for external states
inserted at the punctures and additional insertions of ghost fields and picture
changing operators (PCO)\cite{FMS} that do not depend on the external 
states. In particular, at any given loop order there is
only one term, while in a quantum field theory for a similar amplitude, there will be many
terms representing contributions from different Feynman diagrams.  
Given these differences, we may wonder if there is any similarity 
between string theory amplitudes and ordinary quantum field theory amplitudes.

The closest comparison between string theory amplitudes and the amplitudes in an
ordinary quantum
field theory can be made in Schwinger parameter representation
of the latter, in which we replace the denominator factors of each propagator by
an integral:
\be \label{esch}
(k^2 + m^2)^{-1} = \int_0^\infty ds \, e^{-s(k^2+m^2)}\, .
\ee
With this replacement, the integration over loop momenta takes the form of gaussian integrals,
possibly multiplied by a polynomial in momenta arising from vertices and
propagators, and the integrals can be
easily performed. The result takes the form
\be \label{efield}
\int ds_1\ldots ds_n \, f(s_1,\ldots, s_n)
\ee
where $s_1,\ldots ,s_n$ are the Schwinger parameters for the $n$ propagators
and $f(\{s_i\})$ is some function of these parameters that we obtain after integration over
momenta. At a very crude level, in  string theory the
parameters $\{m_i\}$  labelling the moduli space of Riemann surfaces play the role
of the Schwinger parameters $\{s_i\}$,  and the integrand $F$ appearing in \refb{estring}
plays the role of the
function $f$ appearing in \refb{efield}. 

\begin{figure}
\begin{center}
\includegraphics[width=6.0 in]{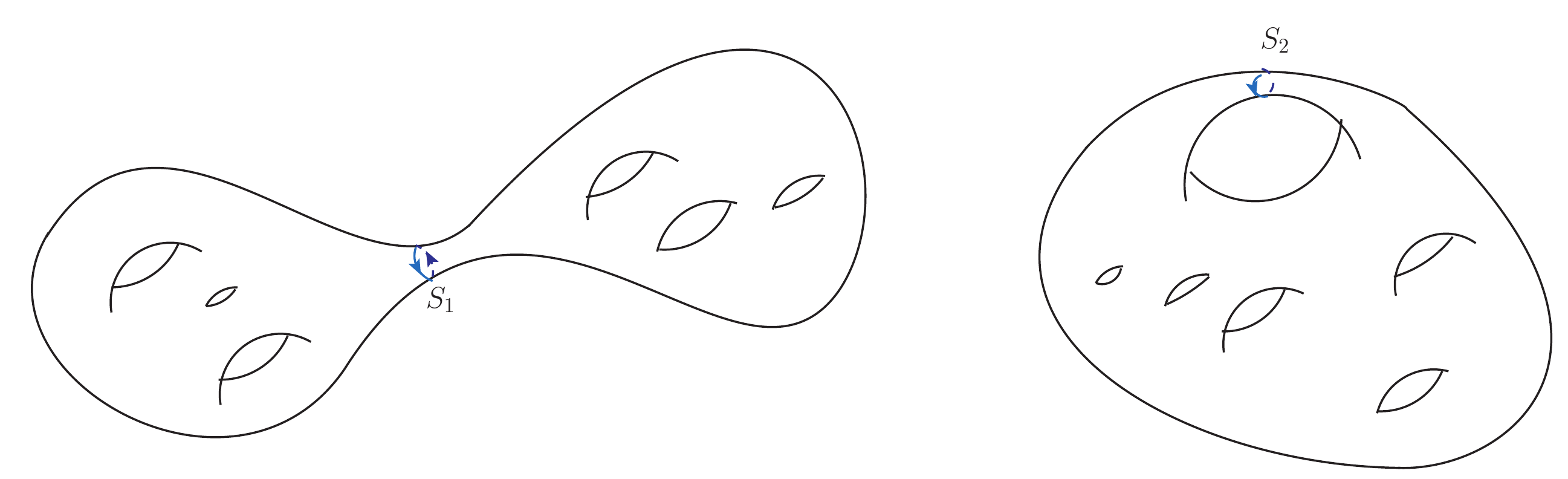}
\end{center}

\vskip -.2in

\caption{The left figure shows a Riemann surface near a separating type 
degeneration and the right figure shows a Riemann surface near a non-separating type 
degeneration. The degeneration 
happens when $S_1$ and $S_2$ shrink to a point. \label{f1}}
\end{figure}

In quantum field theories we typically have both ultraviolet (UV) and infrared (IR)
divergences. The UV divergences arise from regions of integration where one or
more loop momenta become large, while IR divergences arise from 
regions of integration where one or
more propagators have vanishing denominator. In the Schwinger parameter
representation where all loop momenta integrations 
have already been performed, the UV
divergences arise from the region where one or more Schwinger parameter
vanishes, and the IR divergences arise from the region where one or more 
Schwinger parameter becomes infinite.

\begin{figure}
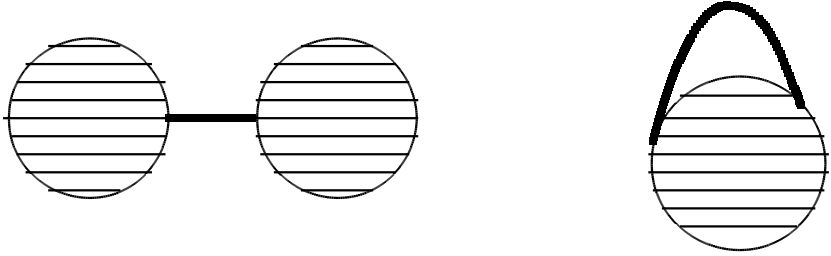

\begin{center}
\hbox{\hspace{.75in}\figAA \hskip 20pt \figBB}
\end{center}

\vskip -.6in

\caption{The left figure shows the Feynman diagrams in field theory analogous to a
separating type degeneration and the right figure shows the Feynman diagrams in
field theory analogous to a non-separating type degeneration. The blobs represent
arbitrary Feynman diagrams, and the thick lines represent propagators whose Schwinger
parameters go to infinity in the degeneration limit.
\label{f2}}
\end{figure}

The string theory amplitudes \refb{estring} also suffer from divergences. These
divergences come from near the boundary of the moduli space where the 
Riemann surface degenerates. As shown in Fig.~\ref{f1}, 
the degeneration can be of two types -- 
separating type degeneration in which the Riemann surface
breaks apart into two parts and non-separating type degeneration
in which
the Riemann surface breaks into a lower genus surface with two extra
punctures. Examination of the integrand $F$ in \refb{estring} in this limit shows that
the integrand behaves in a way similar to the integrand $f$ in the field theory
expression in the limit where the Schwinger parameter of a propagator
approaches infinity. The corresponding field theory Feynman diagrams have been
shown in Fig.~\ref{f2}, where the thick lines represent the propagators with large
Schwinger parameters. 

\begin{figure}
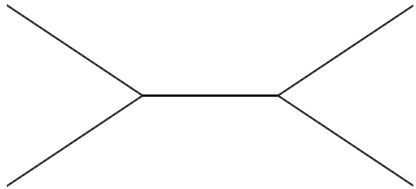

\begin{center}
\hbox{\hspace{1.5in}\figD}
\end{center}
\caption{A tree level diagram that encounters type 1 divergence
when the total energy flowing along the
horizontal line exceeds the threshold for producing an on-shell single particle
intermediate state.
\label{f3}}
\end{figure}

\begin{figure}
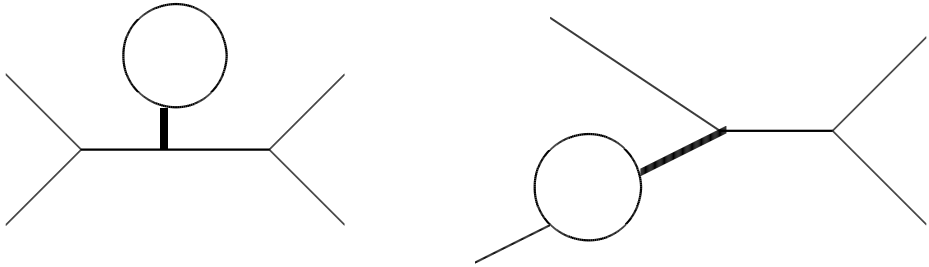

\begin{center}
\hbox{\figexampleonesmall \qquad \figexampletwosmall}
\end{center}
\caption{The figure on the left shows a
divergence associated with massless tadpoles. The thick line is forced to carry zero
momentum due to momentum conservation. Therefore if it represents a massless particle,
the propagator diverges. The diagram on the right shows
divergences associated with mass renormalization.
Requiring the external line to be on-shell also puts the internal line marked by the thick
line on-shell, causing a divergence.  
\label{f4}}
\end{figure}

Since, in field theory, divergences for large Schwinger parameters represent IR 
divergences, we conclude that the divergences in string theory, arising from 
degenerate Riemann surfaces, are IR divergences. Therefore, in order to deal with
IR divergences in string theory, it will be instructive to see what kind of divergences
arise in quantum field theories in the large Schwinger parameter regime and how
they are resolved. They can be classified into two categories:
\begin{enumerate}
\item For $k^2+m^2<0$, the left hand side of \refb{esch} is finite but the
right hand side diverges.  As
shown in Fig.~\ref{f3},
such a divergence can arise even at the tree level. 
In a quantum field theory this is easily dealt with by working
directly with the left hand side, i.e.\ in the momentum space representation of the
Feynman amplitudes. This option does not exist in the conventional formulation of
superstring perturbation theory. The second option, which can be generalized to
string theory\cite{berera,1307.5124}, is to treat the Schwinger parameters as
complex variable and treat the integration over these variables as contour integrals
with the upper limit taken to be $i\infty$ instead of $\infty$. A closely related third
approach is to write the amplitude with the external
momenta in the region where such divergences are absent and then define the 
amplitude in other regions via analytic continuation. Examples of such divergences
in string theory include those arising from two or more 
vertex operators in the world-sheet
coming close, {\it e.g.} the apparent divergences in the integral representation of
Virasoro-Shapiro amplitude in certain kinematic regime.
\item For $(k^2+m^2)=0$, both the left hand side and the right hand side of
\refb{esch} diverge. These are genuine divergences in quantum field theories
in which some internal propagator is forced to be on-shell. Examples of such
diagrams are mass renormalization diagrams and massless tadpole diagrams as
shown in Fig.~\ref{f4}. In quantum field theory, these divergences have standard
remedies. For example, the presence of tadpole diagrams in a quantum field theory
indicates that the tree level vacuum is modified by quantum corrections. 
We deal with these divergences by first constructing the 
one particle irreducible (1PI) effective action, finding its extremum and then expanding
the action around the new extremum to reorganize the perturbation expansion.
Similarly, the 
divergences associated with the mass renormalization diagrams
are removed by first finding the solution to the linearized equations of motion
of 1PI effective action around the extremum to determine the renormalized mass
of the particle, and then computing 
the S-matrix using the LSZ prescription. 
In this approach, diagrams of the type shown in Fig.~\ref{f4} never
appear, but we have to compensate for it in other ways that involve correcting
the interaction terms and/or masses. 
However, in conventional superstring perturbation theory, there is no well defined
procedure for removing these divergences, essentially due to the fact that at each loop
there is a single term, and there is no fully systematic procedure 
for removing some parts of this
contribution and compensating for this in other ways\cite{earlyref,FS,Weinberg,Seiberg,OoguriSakai,
Yamamoto,AS,Das,Rey,Lee,berera2,0410101,0812.3129,0903.3979}. 
Even when these divergences
are absent, the final results for S-matrix computed using standard rules have
apparent
ambiguities\cite{ARS,global,1209.5461} 
which need to be absorbed into redefinitions of moduli fields and / or wave-function
renormalization factors\cite{catoptric,1209.5461}.
\end{enumerate}
One of our goals in this review will be to describe how superstring field theory can be used
to remove these divergences. Along the way, we shall also see various other applications
of superstring field theory. We shall mostly follow the approach described in
\cite{1411.7478,1501.00988,1508.02481,1508.05387,1604.01783,1606.03455,1607.08244,
1609.00459}.

What is superstring field theory?
By requirement, superstring field theory is a quantum field theory whose amplitudes, 
computed with Feynman diagrams, have the following properties:
\begin{enumerate}
\item They agree with standard superstring amplitudes when the latter are finite.
\item They agree with \underline{analytic continuation} of 
 standard superstring amplitudes when the latter are finite.
\item They formally agree with  standard superstring amplitudes when the latter have
\underline{genuine} \underline{divergences}. However, in superstring field theory we should be
able to deal with these divergences using
standard field theory techniques like mass renormalization and shift of vacuum.
\end{enumerate}
The question is: Does such a theory exist? For open and 
closed bosonic string theory such a theory
has been known to exist for a long time\cite{wittensft,saadi,kugo,sonoda,9206084,9705241}.
There have been various approaches
to constructing {\it tree level} open superstring and closed heterotic string field 
theories\cite{wittenssft,9202087,9503099,0109100,0406212,0409018,0412215,
0911.2962,1201.1761,1303.2323,1305.3893,1312.1677,
1312.2948,1312.7197,1403.0940,1407.8485,1412.5281,1505.01659,1505.02069,
1506.05774,1506.06657,1508.00366,1510.00364,1510.06023,1512.03379,
1602.02582,1602.02583,1606.07194,1612.00777,1612.08827}. 
Some of these have been discussed briefly in section \ref{sother}.
However, there is an apparent no go theorem ruling
out the existence of such theories for type IIB superstrings.
It goes as follows. If we can construct an action for type IIB 
superstring theory then by taking its low
energy limit we should get an action for type IIB supergravity.
However, it is known that it is impossible  to construct such an action due to the existence
of the four form gauge field with self-dual field strength in this theory. Therefore, it
follows that we should not be able to construct an action for type IIB superstring
field theory. While this does not rule out the possibility of having type IIA or heterotic
string field theories, it shows that there cannot be a generic formalism covering all
superstring theories.

It turns out that there is a way to circumvent this no-go theorem as 
follows\cite{1501.00988,1508.05387}. 
It is possible to construct actions for heterotic and type II string field theories, 
but each of these theories contains an additional set of 
``ghost''-like  particles which are free.
These additional particles are unobservable since they do not scatter. Therefore,
their existence can be ignored for all practical purposes except that the fields 
corresponding to these particles are necessary to construct the kinetic term of
the action. Using this formalism one can now construct heterotic and type II
superstring field theories -- collectively called superstring field theory -- not only at the
tree level but also at the full quantum level\cite{1411.7478,1501.00988,1508.05387}.
In the following sections, we shall
describe the structure of these theories in detail. This
construction closely follows the structure of  the closed bosonic 
string field theory\cite{9206084}, with few additional 
twists.

Once a superstring field theory is formulated, the divergences associated with
massless tadpoles and mass renormalizations, illustrated in Fig.~\ref{f4}, can be
dealt with using standard techniques of quantum field 
theory\cite{1411.7478,1501.00988,1508.02481}. 
This leads to an unambiguous, divergence free definition of S-matrix 
elements when the number of non-compact space-time dimensions is $\ge 5$.
Furthermore, this S-matrix can be shown to be 
unitary\cite{1604.01783,1606.03455,1607.08244}.
When the number of
non-compact space-time dimensions is four or less, there is another kind of infrared
divergence that comes from loops involving massless particles. These are reflections 
of the fact that we cannot distinguish between a final state with no massless particles
from a final state with massless particles if the energy carried by the massless particles
is sufficiently low\footnote{Such particles are called \textit{soft} particles.}
or if the opening angle between two or more 
massless particles in the
final state is sufficiently low. 
In quantum field theory one can show that these infrared divergences
go away if in the cross section we sum over final state soft particles 
and collinear massless
particles-- i.e.\ not
calculate the cross section for a fixed final state but a fixed final state accompanied by
arbitrary number of soft particles carrying total energy below some fixed value and/or 
almost collinear massless particles with opening angle below some fixed value
-- 
and average over initial state soft and collinear 
particles\cite{kinoshita,lee,bloch,sterman}.   
The analogue of this result for superstring field
theory has not yet been established, but we do not expect any unsurmountable difficulty
in establishing this.

The rest of this review is organized as follows. In section \ref{soff} we describe the
construction of off-shell amplitudes of superstring field theory, without worrying 
whether
they come from an underlying superstring field theory. In section \ref{ssft} we describe the
condition under which the off-shell amplitudes arise from the Feynman diagram of a
superstring field theory, and explicitly construct the
action of the gauge fixed superstring field
theory. In section \ref{ssftinv} we describe the quantum master action, whose 
Batalin-Vilkovisky (BV) quantization gives the gauge fixed action of
section \ref{ssft}. In section \ref{sward} we derive the Ward identities for the off-shell
amputated Green's functions of this superstring field theory.
At this stage this still remains a formal derivation, since this Green's function is 
divergent in the presence of massless tadpoles. We also describe the
construction of the effective action
obtained by integrating out a subset of the fields
of the theory and also construct the gauge invariant 1PI effective action. 
These are free from
all divergences. In section \ref{svac} we describe how using the 1PI action constructed
in section \ref{sward},
we can find the vacuum solution and expand the action around it to find the renormalized masses
and the unbroken (super-)symmetries. We also construct the Siegel gauge propagator
and the interaction terms of the action expanded around the vacuum solution so that
the Feynman diagrams computed using these vertices and propagators are free from
tadpole and mass renormalization divergences. In section \ref{swardshift}
we derive the 
Ward
identities of the {\it divergence free} amplitudes computed from this new action. 
In section \ref{smom} we 
formulate the Feynman rules of string field theory in momentum space as in
conventional quantum field theories and show that the rules for integration over the
loop energies need to be modified in order to get UV finite results. In section \ref{sunitary}
we make use of the momentum space Feynman rules of section \ref{smom} to prove unitarity
of the S-matrix of superstring field theory.  
In section \ref{sother} we briefly discuss
some of the other approaches used in 
the construction of superstring field theories.
 Appendix \ref{saconv} 
contains a summary of
notations and conventions while the rest of the appendices
provide 
various supplementary material containing some details that were left out in the 
main text and also some simple examples illustrating some of the 
points discussed in the text.

We end this section by describing some of the notations and conventions we shall use,
as well as the scope and limitations of this review.
As already mentioned in footnote 
\ref{foint}, superstring will refer to either of the heterotic
string theories or either of the type II string theories, possibly compactified on a manifold
with NS (NSNS for type II) background. The latter restriction is due to
the fact that conformal invariance of the world-sheet theory will play an important role
in this construction and at present the world-sheet description of string theory in an RR
background has not been fully understood. If the pure spinor 
approach\cite{0001035,0609012} can be made into a
fully workable formalism that works to all orders in superstring perturbation theory, then
the present approach may be extendable to RR background as well. We shall also keep
away from type I string theory, but we expect that this formalism can be generalized to
type I theories with minor changes.\footnote{At loop level one needs to 
construct a field
theory containing both open and closed string fields along the lines described in
\cite{9705241}.}
We shall use the formalism of picture changing operators (PCO)
to define amplitudes in superstring theory.  For on-shell 
amplitudes
there is an alternative formalism based on
integration over supermoduli 
space\cite{9609220,9703183,9706033,dp,1209.2199,1209.2459,1209.5461,1304.2832,
1304.7798,1306.3621,1307.1749,1403.5494,1404.6257,
1501.02499,1501.02675,1502.03673}.
So far, off-shell generalization
of these amplitudes have not been written down except for partial construction of
tree level open string field theory\cite{okawa}, but in future it may be possible to
reformulate the whole analysis described here by expressing the off-shell amplitudes as
integrals over supermoduli spaces.

While our approach will be based on a manifestly Lorentz covariant formulation of
superstring field theory, there is an alternative approach, known as light-cone string
field theory, that only manifestly preserves the $SO(d-1)$ subgroup of the $SO(d,1)$
Lorentz group. This approach has been successful for bosonic string 
theory\cite{mandelstam1,mandelstam2}, but there
are various contact term ambiguities when we consider superstring field 
theory which have not been completely 
resolved\cite{gr1,gr2,gr3,greenseiberg,dhoker,1605.04666,1611.06340}.

We shall set $\alpha'=1$ and define the mass$^2$ level of a 
state carrying momentum $k$ to be the eigenvalue of the operator
$2 (L_0+\bar L_0)-k^2$ where $L_0$ and 
$\bar L_0$ denote zero modes of the total Virasoro generators. 
Physically this gives the squared mass 
of the state at tree level if it corresponds to a physical state of string theory. 

Finally we would like to remark that in this review our focus will be on the
application of closed superstring field theory in making superstring perturbation theory
well defined. 
For instance, as mentioned earlier, one of the applications of this formalism is in proving the unitarity of the theory in the situations when the perturbation theory can be trusted. Treating the situation beyond perturbation theory, e.g., proving unitarity in the presence of black holes, can't be dealt in this approach. The close cousin of closed superstring theory, namely open string field theory, has been used to construct
non-trivial classical solutions, going beyond what can be achieved in perturbation 
theory\cite{0511286}.
Similar applications of closed string field theory remains beyond reach to this day despite
some tantalizing numerical
results in closed bosonic string field theory\cite{0506077}.

\sectiono{Off-shell amplitudes in superstring theory} \label{soff}

In this section we shall follow \cite{1408.0571,1504.00609} to
describe construction of off-shell amplitudes in superstring 
theory without demanding that they arise from an underlying field theory. 
For definiteness
we shall restrict most of the discussions to heterotic string theory, and later comment on
the additional ingredients necessary for extending the results to type II string theory.

\subsection{World-sheet theory} \label{s2.1}

The world-sheet theory for any heterotic string compactification
at string tree level contains a
matter superconformal field theory with
central charge (26,15) and a ghost system of  total central charge $(-26,-15)$.
For the matter sector, we denote by $T_m$ and $T_F$ the right-moving stress 
tensor and its superpartner
and by $\bar T_m$ the left-moving stress tensor. They satisfy the operator product
expansion
\ben
&& T_m(z) T_m(w) = {15\over 2} {1\over (z-w)^4}+\cdots, \\
&& T_F(z) T_F(w) 
= {5\over 2} {1\over (z-w)^3}+ {1\over 2} {1\over z-w} T_m(w) + \cdots, \nonumber \\ 
&&
T_m(z) T_F(w) = {3\over 2} {1\over (z-w)^2} T_F(w) + {1\over z-w} \p T_F(w)
+\cdots, \nonumber \\
&& \bar T_m(\bar z) \bar T_m(\bar w)
= {26\over 2} {1\over (\bar z - \bar w)^4}+\cdots, \nonumber
\een
where $\cdots$ denote less singular terms.

The ghost system consists of anti-commuting $b$, $c$, $\bar b$, $\bar c$
ghosts and the commuting $\beta, \gamma$ ghosts.
The $(\beta,\gamma)$ system can be bosonized as\cite{FMS}
\be \label{eboserule}
\gamma = \eta\, e^{\phi}, \quad \beta= \p\xi \, e^{-\phi}, \quad \delta(\gamma)
= e^{-\phi}, \quad \delta(\beta) = e^\phi\, ,
\ee
where $\xi, \eta$ are  fermions of conformal weights $(0,0)$ and $(0,1)$ respectively
and $\phi$ is a scalar with background charge.
 The operator products of these fields take the form
\ben \label{eghope}
&& c(z) b(w) =(z-w)^{-1}+\cdots, \\
&& \bar c(\bar z) \bar b(\bar w) = (\bar z-\bar w)^{-1}
+\cdots, \nonumber \\
&& \xi(z)\eta(w) = (z-w)^{-1}+\cdots, \nonumber \\
&& e^{q_1\phi(z)} e^{q_2\phi(w)} = (z-w)^{-q_1q_2} e^{(q_1+q_2)\phi(w)}+ \cdots \, , \nonumber \\
&& \p \phi (z)\,  \p\phi(w) = -{1\over (z-w)^2} +\cdots \nonumber \, ,
\een
where $\cdots$ denote less singular terms. 
The stress tensors of the ghost fields are given by
\be
T_{b,c}=-2 \, b\, \p\, c + c\, \p\, b, \quad \bar T_{\bar b, \bar c} = 
-2 \, \bar b\, \bar\p\, \bar c + \bar c\, \bar\p\, \bar b, 
\ee
\be
T_{\beta,\gamma} (z) = {3\over 2} \beta\p\gamma + {1\over 2} \gamma\p\beta
= T_\phi + T_{\eta,\xi}\, ,
\ee
where
\be
T_{\eta,\xi} = -\eta \p \xi\, ,
\ee
and 
\be \label{e2.4}
T_\phi = -{1\over 2} \p\phi \p\phi - \p^2 \phi\, .
\ee
With this the total $\phi$ charge
needed to get a non-vanishing correlation function on a genus $g$ surface is
$2(g-1)$. We assign (ghost number, picture number, GSO) quantum numbers 
to various fields
as given in table \ref{t1} where we have also given the conformal weights and
Grassmann parities of these fields. 

\begin{table}
\begin{center}
\vspace{0.15in}
\setlength{\arrayrulewidth}{.25mm}
\renewcommand{\arraystretch}{1.5}
\begin{tabular}{ |p{1cm}|p{3cm}|p{2.5cm}|p{2cm}|p{2cm}|p{2cm} | }
\hline
\bf{Field} & \bf{Conformal Weight $(\bar{h},h)$} & \bf{Grassmann Parity} & \bf{Ghost Number} & \bf{Picture Number} & \bf{GSO projection} \\
\hline
$\beta$ & $(0,3/2)$ & even & -1 & 0 & - \\
$\gamma$ & $(0,-{1/2})$ & even & 1 & 0 & - \\
\hline
$b$ & $(0,2)$ & odd & -1 & 0 & + \\
$c$ & $(0,-1)$ & odd & 1 & 0 & + \\
$\bar{b}$ & $(2,0)$ & odd & -1 & 0 & + \\
$\bar{c}$ & $(-1,0)$ & odd & 1 & 0 & + \\
\hline
$\eta$ & (0,1) & odd & 1 & -1 & + \\
$\xi$ & (0,0) & odd & -1 & 1 & +  \\
$\p\phi$ & (0,1) & even & 0 & 0 & + \\
$e^{q\phi}$ & $-q (q+2)/2$ & $(-1)^q$ & 0 & $q$ & $(-1)^q$ \\
\hline
\end{tabular}
\end{center}
\caption{The quantum numbers and conformal weights of various fields.
\label{t1}
} 
\vskip -.1in
\end{table}

The ghost fields have mode expansions
\ben \label{emodeexpan}
&& b(z) =\sum b_n z^{-n-2}, \quad c(z) = \sum_n c_n z^{-n+1}, \quad 
\bar b(\bar z) =\sum \bar b_n \bar z^{-n-2}, \quad \bar c(\bar z) = \sum_n \bar c_n 
\bar z^{-n+1}, \nonumber \\
&& \beta(z) =\sum_n \beta_n z^{-n-{3\over 2}}, \quad \gamma(z) = \sum_n \gamma_n
z^{-n+{1\over 2}},
\quad \eta(z) =\sum_n \eta_n z^{-n-1}, \quad \xi(z) = \sum_n \xi_n z^{-n}\, .
\nonumber \\
\een
Also useful will be the mode expansions of the total stress tensors of the matter + 
ghost
SCFT
\be \label{evira}
T(z) =\sum L_n z^{-n-2},  \quad 
\bar T(\bar z) =\sum \bar L_n \bar z^{-n-2}\, .
\ee

The BRST charge is given by
\be \label{ebrs1}
Q_B = \ointop dz \jmath_B(z) + \ointop d\bar z \bar \jmath_B(\bar z)\, ,
\ee
where
\be\label{ebrs2}
\bar \jmath_B(\bar z) = \bar c(\bar z) \bar T_m(\bar z)
+\bar b(\bar z) \bar c(\bar z) \bar\p \bar c(\bar z)\, ,
\ee
\be \label{ebrstcurrent}
\jmath_B(z) =c(z) (T_{m}(z) + T_{\beta,\gamma}(z) )+ \gamma (z) T_F(z) 
+ b(z) c(z) \p c(z) 
-{1\over 4} \gamma(z)^2 b(z)\, ,
\ee
and $\ointop$ is normalized so that $\ointop dz/z=1$, $\ointop d\bar z/\bar z=1$.
The PCO $\XX$
is defined as\cite{FMS,verlinde}
\be \label{epicture}
\XX(z) = \{Q_B, \xi(z)\} = c \, \partial \xi + 
e^\phi T_F - {1\over 4} \p \eta \, e^{2\phi} \, b
- {1\over 4} \p\left(\eta \, e^{2\phi} \, b\right)\, .
\ee
This is a BRST invariant primary operator of dimension zero
which carries picture number $1$.

We shall be working with 
the so called `small
Hilbert space'\cite{FMS,verlinde} 
where we remove the zero mode of the $\xi$ field from the spectrum.
This means that we only consider states that are 
annihilated by $\eta_0$.
In the vertex operators of such states factors of $\xi$ appear
with at least one derivative acting on them. 
Correlation functions of such vertex operators on any Riemann surface naively vanish
since $\xi$ being a dimension zero field has zero modes on all Riemann surfaces and
there is no factor of $\xi$ to absorb the $\xi$ zero mode. For this reason it will be
understood that in any correlation function of vertex operators in the small Hilbert space
there is an implicit insertion of $\xi(z)$ that absorbs the zero mode. Since only the zero
mode part of $\xi$ is relevant, the result is independent of where we insert $\xi$.
Similarly it will be understood that 
in all inner products we shall insert an implicit factor of $\xi_0$ 
in order to get a non-vanishing result. For definiteness we can take these
insertions to be on the extreme left of the correlation functions. With this 
convention,
a non-vanishing 
correlation function on a genus $g$ Riemann surface must involve equal number
of insertions of $\p\xi$ or its derivatives and $\eta$ and its derivatives\cite{verlinde}.

Finally to get the signs and normalizations of various correlation
functions we need to describe our normalization condition for the $SL(2,C)$
invariant vacuum $|0\rangle$. 
Denoting by $|k\rangle=e^{ik\cdot X}(0)|0\rangle$ the Fock vacuum carrying 
momentum $k$ along the non-compact directions, we choose the normalization
\be \label{evacnorm}
\langle k| c_{-1} \bar c_{-1} c_0 \bar c_0 c_1 \bar c_1
e^{-2\phi(z)}|k'\rangle =(2\pi)^D \delta^{(D)}(k+k')\, .
\ee

For type II string theories the world-sheet theory of matter sector has central charge $(15,15)$.
The ghost system now also includes left-moving $(\bar\beta,\bar\gamma)$ system so that the
total
central charge of the ghost system now is $(-15,-15)$. 
There will now be separate picture numbers and GSO parities associated with the
left- and right-moving sectors.
The left-moving BRST current
$\bar \jmath_B(\bar z)$ now contains extra terms as in \refb{ebrstcurrent} and we have left-handed
PCO $\bar\XX(\bar z)$ given by an expression identical to \refb{epicture} with all right-handed
fields replaced by their left-handed counterpart. 
We work in the small Hilbert space annihilated by $\eta_0$ and $\bar\eta_0$.
The normalization condition
\refb{evacnorm} will be replaced by
\be \label{evacnormA}
\langle k| c_{-1} \bar c_{-1} c_0 \bar c_0 c_1 \bar c_1
e^{-2\phi(z)}e^{-2\bar\phi(\bar w)}|k'\rangle =-(2\pi)^D \delta^{(D)}(k+k')\, . 
\ee
As will be discussed in \S\ref{sreal}, the unusual minus sign on the
right hand side of \refb{evacnormA} allows
us to use a uniform convention for the normalization in the heterotic and type II
string theories.

We denote by $\HH_T$ the Hilbert space of GSO even states 
in the small Hilbert space
of the matter-ghost CFT with arbitrary ghost and
picture numbers, with coefficients taking values in the Grassmann algebra, satisfying the
constraints
\be \label{econd}
|s\rangle \in \HH_T \quad \hbox{iff} \quad b_0^-|s\rangle = 0, 
\quad L_0^{-}|s\rangle =0\, ,
\ee
 where\footnote{The asymmetry due to the factor 
of $\frac{1}{2}$ in the definition of $b_0^\pm$ and $c_0^\pm$ is just a convention which ensures the simple anti-commutation relations $\{b_0^+,c_0^+\}=1=\{b_0^-,c_0^-\}$.}
\be \label{edefbpm}
b_0^\pm \equiv(b_0\pm\bar b_0), \quad L_0^\pm\equiv(L_0\pm\bar L_0), \quad
c_0^\pm \equiv {1\over 2} (c_0\pm\bar c_0)\, .
\ee
The role of the constraints given in \refb{econd} will be 
explained while discussing off-shell amplitudes. 

In the heterotic theory $\HH_T$ decomposes into a direct sum of the 
Neveu-Schwarz (NS) sector 
$\HH_{NS}$ and Ramond (R) sector $\HH_R$. In the type II string theories the corresponding
decomposition is $\HH_T=\HH_{NSNS}\oplus \HH_{NSR}\oplus \HH_{RNS}\oplus \HH_{RR}$.
For our analysis we shall in fact need a finer decomposition.
In the heterotic string theory we shall denote by $\HH_m$ the subspace of states
in $\HH_T$ carrying picture number $m$. $m$ will be integer for NS sector and 
integer + 1/2 for R-sector states. Similarly in type II theory we shall denote by
$\HH_{m,n}$ the subspace of $\HH_T$ carrying left-moving picture number $m$ and
right-moving picture number $n$. 
We also define
\ben 
\hbox{for heterotic} &:& \wh\HH_T \equiv \HH_{-1}\oplus \HH_{-1/2}, \quad
\wt\HH_T\equiv \HH_{-1}\oplus \HH_{-3/2} \, ,
\nonumber \\ 
\hbox{for type II} &:&
\cases{\wh\HH_T \equiv \HH_{-1,-1}\oplus \HH_{-1/2,-1}\oplus \HH_{-1,-1/2}\oplus \HH_{-1/2,-1/2}\cr
\wt \HH_T \equiv \HH_{-1,-1}\oplus \HH_{-3/2,-1}\oplus \HH_{-1,-3/2}\oplus \HH_{-3/2,-3/2}}\, .
\een
The special role of $\wh\HH_T$ and $\wt\HH_T$ can be understood as follows. 
Using the bosonization rules \refb{eboserule},
the operator product expansion \refb{eghope}
and the mode expansion \refb{emodeexpan}, one can see that acting on a
picture number $p$ vacuum $|p\rangle\equiv e^{p\, \phi}(0)|0\rangle$ in the
heterotic string theory, the modes
of $\beta$ and $\gamma$ have the following properties:
\be \label{emodebg}
\beta_n|p\rangle = 0 \quad \hbox{for} \quad n\ge -p-{1\over 2}, 
\qquad \gamma_n|p\rangle = 0 \quad \hbox{for} \quad n\ge p +{3\over 2} \, .
\ee
This shows that in $\HH_{-1}$ all the positive modes of 
$\beta$ and $\gamma$, beginning with $\beta_{1/2}$
and $\gamma_{1/2}$, annihilate the vacuum. For any other integer picture number however
there will be either some positive mode of $\beta$ or positive mode of $\gamma$ that
will not annihilate the vacuum. As a result by acting with these oscillators we can
create states of arbitrary negative dimension.  For on-shell states this does not
cause a severe problem since one can show that the BRST cohomology is the same
in all picture numbers\cite{FMS,9711087}, and therefore we can choose to work in any fixed
picture number sector modulo certain ambiguities related to boundary 
terms\cite{1209.5461}. However,
since in the string field theory all 
off-shell states
will propagate in the loop, presence of states of arbitrary negative weight will make the
theory inconsistent. For this reason, we restrict the off-shell states in the NS sector
to have picture number $-1$. Similar analysis in the R sector shows that only in
picture number $-1/2$ and $-3/2$ sectors we do not have any positive mode of $\beta$
and $\gamma$ that does not annihilate the vacuum.
There is still a milder problem in the R sector since $\gamma_0|-1/2\rangle\ne 0$
and $\beta_0|-3/2\rangle\ne 0$. Therefore we can create infinite number of states
at the same mass$^2$ level by applying these zero mode operators. We shall argue at
the end of \S\ref{scomstub}
that the structure of the propagator in the R sector prevents this from happening.
If we take the
interacting off-shell string states to have picture number $-1/2$, and use
an appropriate
prescription for the propagators of Ramond sector string fields, then
at any mass$^2$ level only a finite number of states can propagate.

The above analysis can be easily generalized to type II string theory to 
illustrate the
special role of $\wh\HH_T$ and $\wt\HH_T$.

For both heterotic and type II string theories we take 
$|\vp_r\rangle\in \wh\HH_T$, $|\vp_r^c\rangle \in
\wt\HH_T$
to be appropriate basis states satisfying
\be \label{ebas}
\langle\vp_r^c |c_0^- | \vp_s\rangle =\delta_{rs}\, ,\quad 
\langle\vp_s |c_0^- | \vp_r^c\rangle =\delta_{rs}\, .
\ee
The second relation follows from the first.
\refb{ebas} implies the completeness relation
\be\label{ecom}
\sum_r |\vp_r\rangle \langle \vp_r^c|c_0^- = {\bf 1}, \quad 
\sum_r |\vp_r^c\rangle \langle \vp_r|c_0^- = {\bf 1}\, ,
\ee
acting on states in $\wh\HH_T$ and $\wt\HH_T$ respectively.
The basis states $\vp_r$ and
$\vp_r^c$ will in general carry non-trivial Grassmann parities which we shall denote
by $(-1)^{\gamma_r}$ and $(-1)^{\gamma_r^c}$ respectively. 
In the NS 
sector of the heterotic theory and the NSNS and RR sector of type II
theory, the Grassmann parity of $\vp_r$ or $\vp_r^c$ is odd (even) if the ghost number
of $\vp_r$  or $\vp_r^c$ is odd (even). In the R sector of the heterotic theory and the
RNS and NSR sector of the type II theory, the Grassmann 
parity of $\vp_r$ or $\vp_r^c$ is odd (even) if the ghost number
of $\vp_r$  or $\vp_r^c$ is even (odd). 
It results from the
ghost number conservation rule following from \refb{evacnorm},
\refb{evacnormA} and \refb{ebas} that
\be \label{egrrule}
(-1)^{\gamma_r+\gamma_r^c}=-1\, .
\ee

We denote by $\XX_0$ and $\bar\XX_0$ the zero modes of the PCOs:
\be \label{edefggr}
\XX_0 \equiv \ointop {dz\over z}\, \XX(z), \quad \bar\XX_0 
\equiv \ointop {d\bar z\over \bar z}\, 
\bar\XX(\bar z).
\ee
In the heterotic string theory we define
\be \label{edefgg}
\GG|s\rangle =\begin{cases} {|s\rangle \quad \hbox{if $|s\rangle\in \HH_{NS}$}\cr
\XX_0\, |s\rangle \quad \hbox{if $|s\rangle\in \HH_R$}}
\end{cases}\, ,
\ee
while in type II string theories we define
\be \label{edefggii}
\GG|s\rangle =\begin{cases} {|s\rangle \quad \hbox{if $|s\rangle\in \HH_{NSNS}$}\cr
\XX_0\, |s\rangle \quad \hbox{if $|s\rangle\in \HH_{NSR}$}\cr 
\bar\XX_0\, |s\rangle \quad \hbox{if $|s\rangle\in \HH_{RNS}$}\cr 
\XX_0\bar\XX_0\, |s\rangle \quad \hbox{if $|s\rangle\in \HH_{RR}$}\cr 
}\, .
\end{cases}
\ee
Note that
\be \label{egrel}
[\GG, L_0^\pm]=0, \qquad [\GG, b_0^\pm]=0\, , \qquad [\GG, Q_B]=0\, ,
\ee
The importance of these operators will become clear from \S\ref{ssft} onwards.

One can define the correlation functions of the local operators of the world-sheet
superconformal field theory on a general Riemann surface
following standard procedure. The ghost and picture number anomalies tell us that 
on a genus $g$ Riemann surface we shall
need total ghost number $6-6g$  and total 
picture number of $2g-2$ 
to get a non-vanishing result for a correlation function. 
In type II theory the required picture number is 
$(2g-2,2g-2)$. This fixes the required
number of PCOs to be inserted on the Riemann surface for a given set of external
states.

Normally correlation function of a set of local 
operators encounters singularities when they
come close to each other. The correlation functions of the $\xi,\eta,\phi$ system have
additional singularities known as spurious poles\cite{verlinde}. 
They occur even when all the vertex
operators are far away from each other and their origin 
can be traced to the appearance of 
$\gamma$ zero modes in the presence of the insertion of the other operators.
There are also more conventional singularities that arise when two PCOs approach
each other or a PCO approaches a vertex operator. 
We shall collectively call these singularities spurious poles
since they are not associated with degenerations of Riemann surfaces with 
punctures. In defining
off-shell amplitudes we have to be careful in avoiding the spurious poles. This is in
contrast to the singularities that arise from collision of vertex operators. These 
correspond to degeneration of Riemann surfaces with punctures, and will appear as
infrared divergences in the underlying superstring field theory that can be dealt with
using standard quantum field theory techniques.

\subsection{Off-shell amplitudes} \label{soffsub}

In this subsection we shall give a precise definition of on-shell and off-shell
amplitudes of superstring theory, but we shall begin our discussion with a qualitative
description of on-shell amplitudes.
A $g$-loop on-shell amplitude in heterotic string theory with $m$ external NS sector 
states and $n$ external R-sector states is expressed as an
integral over the $(6g-6+2m+2n)$ dimensional 
moduli space $\MM_{g,m,n}$ of genus $g$ Riemann surfaces $\Sigma_{g,m,n}$
with $m$ NS and $n$ R punctures.  The integrand is expressed in terms of appropriate 
correlation functions of the vertex
operators of external states inserted at the punctures, 
ghost fields and PCOs inserted at
certain locations on the Riemann surface. The final
result is independent of the locations of the PCOs 
as long as they avoid spurious poles (discussed in 
 appendix \ref{saspurious}) and satisfy certain factorization 
constraints near the boundaries of the moduli space. These factorization conditions
tell us how the PCOs should be distributed among different component Riemann 
surfaces and the neck in the degeneration limit, and will be discussed in 
\S\ref{sconsec}.
For type II string the story is similar except
that there are now four sectors and we have to insert both left and right-moving PCOs.
For simplicity we shall restrict our discussion to the heterotic string theory.

We shall follow a convention in which the sum over spin structures
will be implicit in the integration over $\MM_{g,m,n}$.
If a Ramond puncture is present then the sum over spin structure can be implemented
by extending the range of integration over the location of a Ramond
puncture, since a
translation of the Ramond puncture around a cycle of the Riemann surface 
changes
the boundary condition on the fermions along the dual cycle. Therefore
by doubling the range of integration of the location of a Ramond puncture along
each of the $2\, g$ cycles of the Riemann surface we can get all the $2^{2g}$
spin structures. In order to maintain symmetry under the exchange of all the punctures
we can symmetrize the result with respect to all the punctures following the general
procedure that will be elaborated below (see \refb{egensec}).
If there are no R punctures present then we can implement the sum over
all even (odd) spin structures by starting with a particular even (odd)
spin structure and extending the range of
integration over the moduli. Since 
a modular transformation mixes different even (odd)
spin structures, adding appropriate number of copies of the fundamental
domain is equivalent to summing over different spin structures related by modular
transformation.
But we need to explicitly add the contributions from even and odd spin structures.

\begin{figure}
\begin{center}
\hbox{\figpgmn}
\end{center}

\vskip -1.2in

\caption{The space $\wt\PP_{g,m,n}$ as a fiber bundle. \label{fpgmn}}
\end{figure}

Defining off-shell amplitudes in superstring theory requires extra
data.\footnote{Throughout this paper we shall mean by off-shell amplitude
the analogue of the amputated Green's function in a quantum field theory
where the tree level
propagators for external states are dropped. This is what integral over moduli 
space of Riemann surfaces naturally computes.}
First of all since the vertex operators are not BRST invariant, the result
depends on the choice of PCO locations. Furthermore since the vertex operators 
are not conformally invariant, the result also depends on the choice of world-sheet
metric around the punctures. 
We shall parametrize the metric in terms of the choice of local holomorphic
coordinates around the puncture. If $w$ denotes the local holomorphic coordinate
around a puncture, then we take the metric around the puncture to be $|dw|^2$. But 
the result will now depend on the choice of the local holomorphic coordinate -- if
instead of $w$ we choose the local holomorphic coordinate to be some holomorphic
function $f(w)$ then the metric will be given by $|f'(w)dw|^2$. The only exception is a 
phase rotation of $w$ which does not change the metric. 

The most convenient
way of encoding the dependence on the extra data is to
introduce an infinite dimensional space $\wt\PP_{g,m,n}$ with the structure
of a fiber bundle, whose base is $\MM_{g,m,n}$ and whose (infinite dimensional) 
fiber is parametrized by the possible choices of local
coordinate system around each puncture and the possible choices of
PCO locations on the Riemann surface\cite{nelson,9206084,1408.0571}.  
This has been shown schematically in Fig.~\ref{fpgmn}.
The punctures will be taken to be
distinguishable, i.e. two points in $\MM_{g,m,n}$ related by the exchange of
two punctures will be considered to be distinct points. Since $\MM_{g,m,n}$ has
real dimension $(6g-6+2m+2n)$, a section of $\wt\PP_{g,m,n}$ will have the
same real dimension.
The off-shell amplitude is described as an 
integral of a $(6g-6+2m+2n)$-form over a section of $\wt\PP_{g,m,n}$.\footnote{We
cannot really choose a continuous section -- in order the avoid spurious poles we
have to divide the base $\MM_{g,m,n}$ into small regions, choose different sections
over these different regions and add appropriate correction terms at the boundaries
of these  regions\cite{1408.0571,1504.00609}. This has been discussed briefly in
appendix \ref{saspurious}. 
The net result is that in carrying out various manipulations we
can pretend that we have continuous sections. 
This is how we shall proceed. \label{fo1}}
It will also be convenient to introduce a space $\wh\PP_{g,m,n}$ that is obtained
from $\wt\PP_{g,m,n}$ by forgetting about the PCO locations, i.e., 
$\wh\PP_{g,m,n}$ has a fiber bundle structure whose base is $\MM_{g,m,n}$ 
and whose fiber contains information about the possible choices of local coordinates 
around the punctures.
Then $\wt\PP_{g,m,n}$ can also be regarded as
a fiber bundle with base $\wh\PP_{g,m,n}$ and the fiber parametrized by
possible choices of
PCO locations on the Riemann surface. 

\begin{figure}
	\hspace{0.65in} \includegraphics[width=5in]{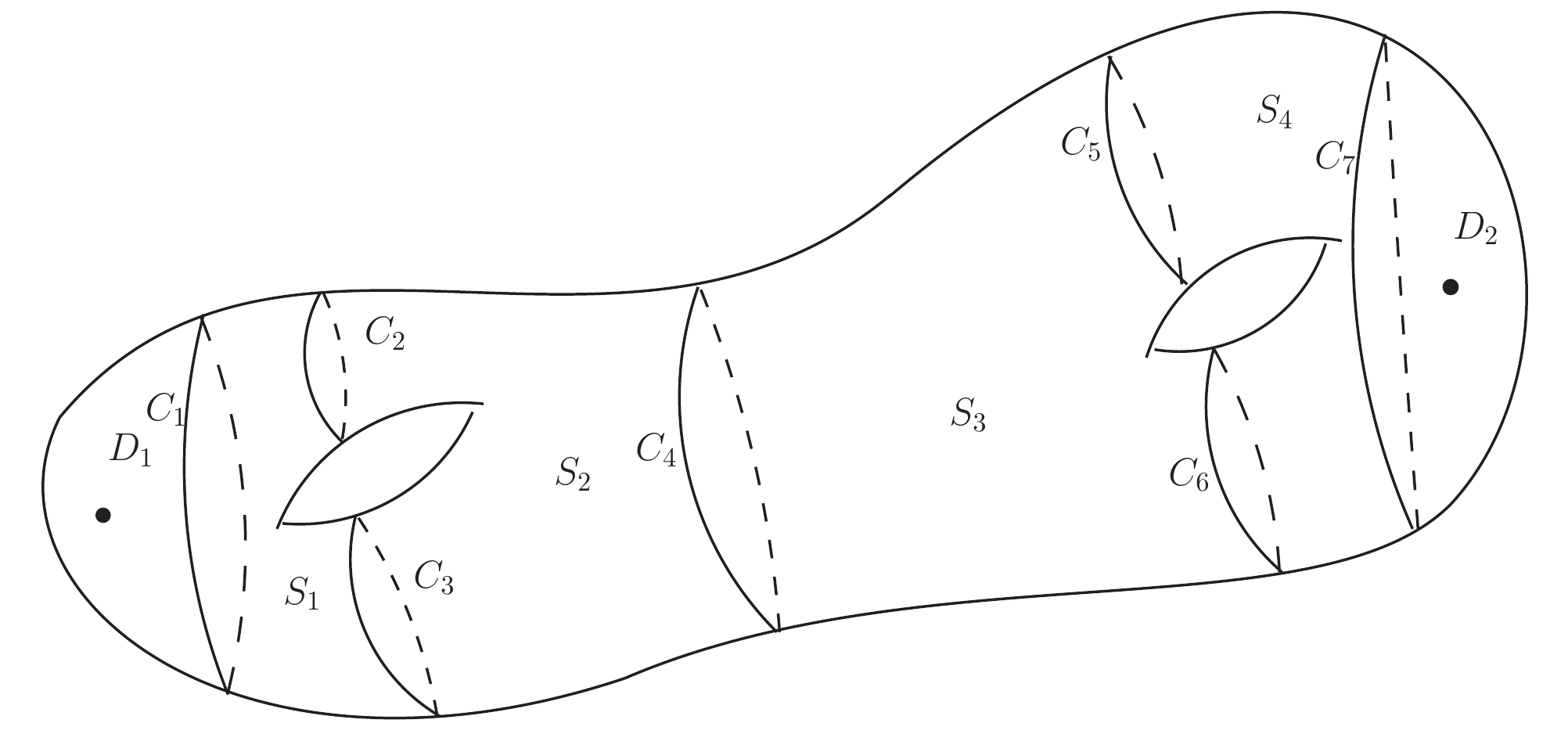}\caption{Two torus with two punctures.}			\label{fig:f2}
\end{figure}

We shall now  turn this qualitative description of on-shell and off-shell
amplitudes into fully quantitative description. Our first task will be to introduce a 
coordinate system on $\wt\PP_{g,m,n}$.
It is easy to see that given a Riemann surface of genus $g$ and $m+n$ punctures, we can
regard this as a union of $m+n$ disks $\{D_a\}$, one around each puncture, and 
$2g-2+m+n$ spheres $\{S_i\}$, each with three holes, joined along $3g-3+2m+2n$ circles
$\{C_s\}$. An example of this for $m+n=2$ and $g=2$ has been shown in Fig.~\ref{fig:f2}.
Let $w_a$ denote the choice of local holomorphic coordinates on $D_a$ such
that the $a$-th puncture is located at $w_a=0$ and
$z_i$ denote the local holomorphic coordinates on $S_i$. Then the Riemann surface
is prescribed completely by specifying the spin structure and the
functional relation between the coordinates
on the two sides of each overlap circle $C_s$. This typically takes the form
\be \label{eintFspre}
z_i = f_{ij}(z_j) \quad \hbox{or} \quad z_i = g_{ia}(w_a)\, .
\ee
In order to use a compact notation, we shall fix some orientation for each 
$C_s$ and call $\sigma_s$ and $\tau_s$ respectively the coordinate systems on the left
and right of $C_s$. Each $\sigma_s$ and $\tau_s$ will correspond to one of the
$z_i$ or one of the $w_a$. Eq.~\refb{eintFspre} may now be reexpressed as
\be \label{eintFs}
\sigma_s = F_s(\tau_s)\, .
\ee
Besides the spin structure,
the functions $\{F_s\}$  contain complete information about the Riemann
surface and the local coordinate system around the punctures (which are taken to
be $w_a$).   Therefore they can be chosen to parametrize $\wh\PP_{g,m,n}$.
 More specifically, 
if $\{u_i\}$ denote the complete set of parameters labelling the functions
$\{F_s\}$, {\it e.g.} coefficients of Laurent series expansion of these functions,
then we can take $\{u_i\}$ to be the coordinates of $\wh\PP_{g,m,n}$. This is
clearly an infinite dimensional space.
Once the coordinate
system on $\wh\PP_{g,m,n}$ is fixed this way, we can introduce coordinate
system on $\wt\PP_{g,m,n}$ by appending to the former
the locations of the PCOs. This 
introduces one complex coordinate for each PCO. If a PCO is located on $S_i$ then
we shall specify its coordinate in the $z_i$ coordinate system while if it is located on
$D_a$ we shall specify its location in the $w_a$ coordinate system. We shall denote
collectively by $\{y_\alpha\}$ the locations of all the PCOs. We shall take the NS
vertex operators to have picture number $-1$ and the R-vertex operators to have
picture number $-1/2$. Then by picture number conservation we need precisely
$2g-2+m+n/2$ PCOs for non zero correlation functions.

The set $\{u_j,y_\alpha\}$ provides a 
highly redundant coordinate system on $\wt\PP_{g,m,n}$, since
a reparametrization of $z_i$ that is non-singular on $S_i$ (with the holes cut out)
changes the function $F_s$ (and hence some of the $u_j$\plu) 
if $C_s$ forms a boundary of $S_i$. This also changes the coordinate
$y_\alpha$ of a PCO if it is situated on $S_i$. On the other hand such a reparametrization
does not
change the Riemann surface or the local coordinates around the punctures or the 
physical location of the PCO. Therefore we must identify points in the $\{u_j,y_\alpha\}$
space related by such reparametrizations. A 
reparametrization of $w_a$ that is non-singular inside $D_a$ and leaves the 
location of the puncture $w_a=0$ unchanged, changes the local coordinate around
the $a$-th puncture but does not change the Riemann surface. Therefore this moves
us along the fiber of $\wt\PP_{g,m,n}$. However, if this transformation is a phase rotation
of $w_a$ then it does not have any action on $\wt\PP_{g,m,n}$ and again we must identify
points in the $\{u_j, y_\alpha\}$ space related by such reparametrizations.

 The  tangent vectors of $\wt \PP_{g,m,n}$ are associated with infinitesimal
motions in $\wt\PP_{g,m,n}$.  One set of tangent
vectors, associated with the changes in the PCO locations keeping moduli
and local coordinates fixed,  are simply 
$\p/\p y_\alpha$. The other tangent vectors $\p/\p u_i$, which are also tangent vectors of
$\wh\PP_{g,m,n}$, are associated with deformation of the
transition functions $F_s$. For later use we define
\be \label{ebdef}
{\bf B} \left[{\p\over \p u_i}\right]\equiv \sum_s \ointop_{C_s} {\p F_s \over \p u_i} d \sigma_s \, b (\sigma_s)
+  \sum_s \ointop_{C_s} {\p \bar F_s \over \p u_i} d \bar \sigma_s \, 
\bar b(\bar \sigma_s)\, ,
\ee
where $\ointop$ includes a factor of $1/2\pi i$ for the first integral and 
$-1/2\pi i$ for the second integral. $b$, $\bar b$ are the usual ghost fields
of the world-sheet theory. By definition, the contour traverses $C_s$
keeping the patch covered by the $\sigma_s$ coordinate system 
to the left. It is easy to verify that this definition is invariant under the reversal of 
the orientation of
$C_s$ that exchanges $\sigma_s$ and $\tau_s$.

 Suppose we want to compute an off-shell amplitude of  
$m$ NS sector vertex operators $K_1,\ldots ,K_m\in\HH_{-1}$
and $n$ R sector
vertex operators
$L_1,\ldots ,L_n\in \HH_{-1/2}$. In order to avoid some cumbersome sign factors we shall
from now on multiply each Grassmann odd vertex operator by a Grassmann odd
c-number so that the vertex operators of external states are always Grassmann 
even. In any equation we can always strip off these Grassmann odd c-numbers 
from both side to recover the necessary sign factors. 
We now describe the construction of a $p$-form $\Omega^{(g,m,n)}_p(\{K_i\}, \{L_j\})$ 
on $\wt\PP_{g,m,n}$ that can be
integrated over a $p$-dimensional subspace of 
$\wt\PP_{g,m,n}$ -- henceforth referred to as an
integration cycle. This is defined by specifying the contraction of this $p$-form
with $p$ arbitrary tangent vectors of $\wt\PP_{g,m,n}$ which could be either of
the type $\p/\p y_\alpha$ or of type $\p/\p u_j$.
Denoting by
$\Omega^{(g,m,n)}_p(\{K_i\}, \{L_j\})\left[\p/\p u_{j_1},\ldots \p/\p u_{j_k}, 
\p/\p y_{\alpha_{k+1}},
\ldots \p/\p y_{\alpha_p}\right]$ the contraction of the $p$-form with such vector
fields, evaluated at some particular point in $\wt\PP_{g,m,n}$, 
we take\footnote{The $(-2\pi i)$ factor in the normalization differs from the 
ones used {\it e.g.} in
\cite{1408.0571} where $2\pi i$ was used.
It was shown in \cite{1508.02481} 
that with the choice given in \cite{1408.0571}, 
for any complex modulus $m=m_R + i m_I$, $dm_I\wedge dm_R$
represented positive integration measure, i.e.\ $\int dm_I\wedge dm_R$ integration
over a region in the complex $m$-plane gives positive result. This is opposite of the
standard convention in which $\int dm_R\wedge dm_I$ over a region gives positive
result. With the normalization convention given in \refb{emeasure} we can use the
more standard normalization for integration measure where 
$\int dm_R\wedge dm_I$ over a region gives positive
result.}
\ben\label{edefomega}
&&\Omega^{(g,m,n)}_p(\{K_i\}, \{L_j\})\left[\p/\p u_{j_1},\ldots ,
\p/\p u_{j_k}, \p/\p y_{\alpha_{k+1}},
\ldots ,\p/\p y_{\alpha_p}\right] \nonumber \\
&=&(-2\pi i)^{-(3g-3+m+n)}\Bigg\langle {\bf B}[\p/\p u_{j_1}]\ldots {\bf B}[\p/\p u_{j_k}] 
(-\p\xi(y_{\alpha_{k+1}})\ldots
(-\p \xi(y_{\alpha_p})) \nonumber \\
&& \hskip 1in \prod_{\alpha=1\atop \alpha\ne \alpha_{k+1},\ldots \alpha_p}^{2g-2
+m + n/2}
\XX(y_\alpha) \, K_1\ldots K_m L_1\ldots L_n\Bigg\rangle_{\Sigma_{g,m,n}}
\, .
\een
We shall now explain the various parts of this formula. First of all this expression 
gives the contraction of the $p$-form with the vector fields at some particular
point in $\wt\PP_{g,m,n}$. Associated with this point there is a specific Riemann
surface $\Sigma_{g,m,n}$
with $m+n$ punctures, local coordinates at each of these punctures and
choice of PCO locations on the Riemann surface. $y_1,\ldots ,y_{2g-2+m+n/2}$
denote these PCO locations.
$\langle\cdots\rangle_{\Sigma_{g,m,n}}$ denotes correlation function on the 
Riemann surface $\Sigma_{g,m,n}$. The actual computation of these 
correlation functions require detailed knowledge of the underlying SCFT. For
simple background, explicit expressions of these correlation functions can be 
found {\it e.g.} in \cite{VerlindeChiral,Verlinde:1988tx}. The vertex operators
$\{K_i\}$ and $\{L_j\}$ are inserted  at the punctures of the Riemann surface using the
chosen local coordinates corresponding to the particular point in $\wt\PP_{g,m,n}$ where
we want to compute the left hand side. 
${\bf B}[\p/\p u_j]$ has been defined in \refb{ebdef}.

The expression \refb{edefomega} clearly depends on the choice of the PCO locations 
$y_\alpha$. It also depends on the choice of local coordinates around the punctures.
 For example, if $K_a$ is a primary operator of dimension $(h,h)$ inserted at the
$a$-th puncture at $w_a=0$, then under a 
change in local coordinates from $w_a$ to $\wt w_a=f(w_a)$, the correlator is multiplied 
by a factor of $|f'(0)|^{-2h}$. For non-primary states the transformation law is more
complicated, involving mixing with  other descendants of the primary.
Since the local coordinates around the punctures are defined only up to a phase
rotation, \refb{edefomega} is well defined only if the external states are
annihilated by $L_0^-$; otherwise
a phase rotation of the local coordinates
will change the correlation function.

 The above definition of $\Omega^{(g,m,n)}_p$ may look somewhat formal since
$\wt\PP_{g,m,n}$ is an infinite dimensional space parametrized by infinite number
of coordinates $\{u_i, y_\alpha\}$.
For any practical computation we shall integrate $\Omega^{(g,m,n)}_p$ 
over a given $p$-dimensional
subspace with fixed set of tangent vectors. In this case \refb{edefomega} can be
used to find a specific $p$-form on this $p$ dimensional subspace of $\wt\PP_{g,m,n}$
as follows. Let us suppose that $t_1,\ldots t_p$ denote the parameters that label
the $p$ dimensional subspace of $\wt\PP_{g,m,n}$. Then in general all the transition
functions $F_s$ and the PCO locations $\{y_\alpha\}$ 
will depend on the parameters $\{t_i\}$. According to \refb{ebdef}, \refb{edefomega}, contraction
of $\Omega^{(g,m,n)}_p$ with a tangent vector $\p/\p t_i$ will correspond to inserting
the operator\footnote{The first two terms are already present in the amplitudes of
the bosonic string theory. The  last factor is needed when the PCO locations vary
with moduli\cite{verlinde}.}
\be \label{edefbbi}
\BB_i = \sum_s \ointop_{C_s} {\p F_s \over \p t_i} d \sigma_s \, b (\sigma_s)
+  \sum_s \ointop_{C_s} {\p \bar F_s \over \p t_i} d \bar \sigma_s \, 
\bar b(\bar \sigma_s) 
- \sum_\alpha {1\over \XX(y_\alpha)} {\p y_\alpha\over \p t_i} \p\xi(y_\alpha)
\ee
into the correlation function. 
Note the formal factor of $1/\XX(y_\alpha)$ -- this simply means that we have to remove
the $\XX(y_\alpha)$ factor from the rest of the operator insertion.
The
net integration measure will be given by
\be \label{emeasure}
(-2\pi i)^{-(3g-3+m+n)}\left\langle \BB_1 dt_1\wedge \BB_2 dt_2 \wedge \ldots \wedge 
\BB_p dt_p\,  \prod_{\alpha=1}^{2g-2
+m + n/2}
\XX(y_\alpha) \, K_1\ldots K_m L_1\ldots L_n\right\rangle_{\Sigma_{g,m,n}}\, .
\ee
This has no $\XX(y_\alpha)$ in the denominator since the same $1/\XX(y_\alpha)$
given in \refb{edefbbi} cannot appear more than once due to the
vanishing of the corresponding wedge product $\p_{t_1} y^\alpha dt_1\wedge 
\p_{t_2} y^\alpha dt_2$. Single factors of $\XX(y_\alpha)$ in the denominator
get cancelled by the $\prod_{\alpha=1}^{2g-2
+m + n/2}
\XX(y_\alpha)$ factor. We emphasize again that the $\XX(y_\alpha)$ in the denominator
of \refb{edefbbi} is only a formal way of writing the final expression.

The $p$-form defined in \refb{edefomega} has several useful properties:
\begin{enumerate}
\item First of all recall that the coordinate system on $\wt\PP_{g,m,n}$ that we
have used is highly redundant. Consequently there are many vectors which 
actually represent zero vectors of $\wt\PP_{g,m,n}$. Examples of such tangent
vectors are those generated by infinitesimal reparametrization of $z_i$  
together with a shift of the PCO locations on $S_i$ that
keeps their physical location unchanged. Such tangent vectors will be represented by
some linear combination of the vectors  $\p/\p u_j$ and $\p/\p y^\alpha$. We need
to ensure that the contraction of $\Omega^{(g,m,n)}_p$ with such vectors must vanish since
they represent zero tangent vector. We shall now show that 
this can be proved using standard properties of 
correlation functions of conformal field theories on Riemann surfaces.

For definiteness, suppose that we make an infinitesimal deformation of the coordinate
system $z_1$ on $S_1$ to $z_1+v(z_1)$.  Let us suppose further that $C_1$, $C_2$ and
$C_3$ form boundaries of $S_1$ keeping $S_1$ on the left,
and that on $S_1$ there are insertions of
PCOs at $y_1,\ldots y_N$.  Then on $C_s$ for $1\le s\le 3$, $F_s$ 
changes by $v(z_1)$, and
for $1\le \alpha\le N$, $y_\alpha$ changes by $v(y_\alpha)$.
The relevant insertion into the correlation function upon contraction
of $\Omega^{(g,m,n)}_p$ with the tangent vector induced by this deformation takes the form
\be
\left[\sum_{i=1}^3 \ointop_{C_i} v(z_1) b(z_1) dz_1 + 
\sum_{i=1}^3 \ointop_{C_i} \bar v(\bar z_1) \bar b(\bar z_1) d\bar z_1\right] 
\prod_{\beta=1}^N \XX(y_\beta) - 
\sum_{\alpha=1}^N (v(y_\alpha)  \p\xi(y_\alpha)) \prod_{\beta=1\atop \beta\ne \alpha}^N\XX(y_\beta) 
\, , 
\ee
with the integration along $C_1,C_2,C_3$ performed by keeping $S_1$ to the left.
We can now deform the integration contours into the interior of $S_1$. The contour
integral over $\bar z_1$ can be shrunk to a point giving vanishing contribution, while
the integral over $z_1$ picks up residues at $y_\alpha$ due to the insertion of 
$\XX(y_\alpha)$. It follows from \refb{epicture} that these residues are given by 
$v(y_\alpha)\p\xi(y_\alpha)$. The sum of all the residues
cancel the last term. Therefore we see that the apparent tangent vector induced by
a change of coordinate on $S_1$ indeed has vanishing contraction with 
$\Omega^{(g,m,n)}_p$.
\item  Similarly contraction of $\Omega^{(g,m,n)}_p$ with tangent vectors associated
with phase rotation $w_a \to w_a + i\eps w_a$ 
of the $w_a$\plu\ can also be shown to vanish.  If there are $M$ insertions
of PCOs at $y_1,\ldots y_M$ on $D_a$ then
the relevant insertion upon contraction with the corresponding tangent vector takes the
form
\ben
&& \left[i \eps \ointop_{C_a} w_a b(w_a) dw_a - i\eps
\ointop_{C_a} \bar w_a \bar b(\bar w_a) d\bar w_a\right] 
\prod_{\beta=1}^M \XX(y_\beta)  V_a(0)  \nonumber \\
&& - i\eps 
\sum_{\alpha=1}^M (y_\alpha  \p\xi(y_\alpha)) \prod_{\beta=1\atop \beta\ne \alpha}^M\XX(y_\beta)  V_a(0)
\, , 
\een
where $C_a$ represents an anti-clockwise contour along
the boundary of the disk $D_a$ around the $a$-th
puncture, and $V_a$ is the vertex operator inserted at the $a$-th puncture at 
$w_a=0$.  We can now deform the contour $C_a$ towards $w_a=0$. Sum of the
residues at
$y_\alpha$ cancel the last term as before, leaving us with the residue at $w_a=0$. This is
proportional to $b_0^- |V_a\rangle$ and vanishes by eq.~\refb{econd} since 
$V_a\in \HH_T$.

\item $\Omega^{(g,m,n)}_p$ satisfies the important identity
\ben \label{emm}
&&\Omega^{(g,m,n)}_p(Q_B K_1, K_2, \ldots ,K_m,L_1,\ldots, L_n)
+\cdots + \Omega^{(g,m,n)}_p(K_1, K_2, \ldots  ,K_m,L_1,\ldots, Q_B L_n) 
\nonumber \\
&=& (-1)^p d \Omega^{(g,m,n)}_{p-1}(K_1, \ldots, K_m,L_1,\ldots, L_n) \, .
\een
The derivation of this formula can be found in \cite{1408.0571}. We shall not repeat it here with
all the details but briefly indicate the general idea behind the proof.
Let us pick some convenient coordinate system $\{u_i\}$ on  $\wh\PP_{g,m,n}$, and
use $\{u_i\}$ and the PCO locations 
$\{y_\alpha\}$ as the coordinates of $\wt\PP_{g,m,n}$. We now
take the contraction of both sides of \refb{emm} with 
$q$ tangent vectors of the form $\p/\p u_i$ and $p-q$ tangent vectors of the form
$\p/\p y_\alpha$, and evaluate both sides using
\refb{edefomega}.
Since on the left hand side we have $Q_B$ acting on all
the states in turn, we can deform the contour of integration defining $Q_B$ into the interior
of the Riemann surface, 
picking up residues from the insertions of $b,\bar b$ ghosts in
the  $B[\p/\p u_i]$ 
factors and also from the $\p\xi$ insertions. $\XX$ insertions of course are
invariant under $Q_B$. One might also worry about possible residues from the
spurious poles mentioned at the end of \S\ref{s2.1}, but as has been reviewed in appendix
\ref{saspurious},  there are no spurious poles in the argument of the BRST 
current\cite{lechtenfeld,morozov}.
Using the relations
\be
\{Q_B, b(z)\} = T(z), \quad \{Q_B, \bar b(\bar z)\} =\bar T(\bar z)\, ,
\ee
where $\bar T$ and $T$ are the left- and right-moving components of the total stress
tensor of the world-sheet theory, 
we can see that the residue at  $B[\p/\p u_i]$ generates a factor similar to that in
\refb{ebdef} with $b,\bar b$ replaced by stress tensors $T,\bar T$.  This generates
derivative of the correlation function 
with respect to $u_i$.
On the other hand the residue at $\p\xi(y_\alpha)$ 
generates a
factor of $\p\XX(y_\alpha)$ and this generates derivative of the correlation function
with respect to the coordinate
$y_\alpha$. Putting all these results
together one finds that the contraction of the left and right hand sides of \refb{emm} with
arbitrary set of tangent vectors agrees. This establishes \refb{emm}.
\item Since on a genus $g$ surface we need 
total ghost number $6-6g$ to get a non-zero
correlator, we see that $\Omega^{(g,m,n)}_p(\{K_i\}, \{L_j\})$ is non-zero only if the total
ghost number carried by $\{K_i\}$ and $\{L_j\}$ is equal to $6-6g+p$. On the other hand
conservation of picture number is automatic due to our choice of picture numbers of 
$\{K_i\}$ and $\{L_j\}$, and the number of PCO insertions in the definition of
$\Omega^{(g,m,n)}_p(\{K_i\}, \{L_j\})$.
\end{enumerate}

We are now ready to define off-shell amplitudes.
The off-shell amplitude of the external states $K_1,\ldots  ,K_m,L_1,\ldots  ,L_n$ is given by
\be \label{eoffshellamp}
\int_{\SSS_{g,m,n}} \Omega^{(g,m,n)}_{6g-6+2m+2n}(K_1, \ldots  ,K_m,L_1,\ldots  ,L_n) \, ,
\ee
where $\SSS_{g,m,n}$ is a section of $\wt\PP_{g,m,n}$. 
For on-shell external states this gives the usual on-shell amplitudes and can be
shown to be formally independent
of the choice of $\SSS_{g,m,n}$.
The proof of this has been reviewed in \S\ref{sformal}.
However for
off-shell external states
the result depends on the choice
of this section since the external states are not BRST invariant. 
We shall describe in \S\ref{scomredef} and appendix \ref{safield}
how physical quantities computed from off-shell
amplitudes become independent of the choice of the section.

As already mentioned in footnote \ref{fo1}, we cannot choose the section to be continuous,
but once we add correct compensating terms we can treat it as continuous in all
manipulations.
We can further generalize the notion of a section by taking weighted averages of sections
-- several sections $\SSS^{(1)}_{g,m,n}, \ldots \SSS^{(k)}_{g,m,n}$ 
with weights $w_1,\ldots w_k$
such that $\sum_{i=1}^k w_i =1$. If we denote by
$\SSS_{g,m,n}$ the formal weighted sum 
$\sum_{i=1}^k w_i \, \SSS^{(i)}_{g,m,n}$ then by definition
\be \label{egensec}
\int_{\SSS_{g,m,n}} \Omega^{(g,m,n)}_{6g-6+2m+2n }(K_1, \ldots  ,K_m,L_1,\ldots  ,L_n) 
=\sum_{i=1}^k w_i \int_{\SSS^{(i)}_{g,m,n}} 
\Omega^{(g,m,n)}_{6g-6+2m+2n } (K_1, \ldots  ,K_m,L_1,\ldots  ,L_n)\, .
\ee
This is also a good definition of off-shell amplitude since for on-shell external states
the result reduces to the usual on-shell amplitude. We shall call sections of this
kind `generalized sections'. In all subsequent analysis whenever we refer to section, we
shall actually mean generalized section.

 Some explicit examples of off-shell amplitudes computed using this
prescription can be found in appendix \ref{saexamples}.

The story in type II string theory is similar. For an amplitude with $m$ NSNS states,
$n$ NSR states, $r$ RNS states and $s$ RR states one has to work with the 
space $\wt\PP_{g,m,n,r,s}$ which has as its fiber the choice of local coordinates
at the punctures and locations of $m+n+(r+s)/2$ left-moving PCOs and 
$m+r+(n+s)/2$ right-moving PCOs. Construction of $\Omega^{(g,m,n,r,s)}_p$
proceeds in a manner identical to that of heterotic string theory, with the contraction
with tangent vectors $\p/\p\tilde y_\alpha$, with $\tilde y_\alpha$ denoting the location
of the left-moving PCO, introducing a factor of 
$-\bar\p \bar \xi(\tilde y_\alpha)$.

\subsection{Formal properties of on-shell amplitudes} \label{sformal}

Using \refb{emm} we can prove some useful formal properties of on-shell 
amplitudes\cite{nelson},
where an on-shell state will refer to a state that is BRST invariant, but not
necessarily a dimension zero primary.
First consider the situation where all the states $K_1,\ldots  ,K_m$, $L_1,\ldots  ,L_n$ are
BRST invariant and one of them, say $K_1$, is BRST exact, i.e.\ $ K_1
= Q_B\Lambda$.
Then using \refb{emm} we get
\be 
\Omega^{(g,m,n)}_{6g-6+2m+2n} (Q_B\Lambda, K_2, \ldots  ,K_m, L_1,\ldots  ,L_n)
= d \Omega^{(g,m,n)}_{6g-6+2m+2n-1} (\Lambda, K_2, \ldots  ,K_m, L_1,\ldots  ,L_n)\, .
\ee
We now integrate both sides over $\SSS_{g,m,n}$. The left hand side gives the
on-shell amplitude in which one state is BRST exact. The right hand side is the integral
of an exact form and hence the integral vanishes {\it provided we can ignore the boundary
terms.} This shows the decoupling of pure gauge states. We emphasize however that
this `derivation' is formal since it ignores possible contribution from the boundary terms.
One of our goals will be to give a complete proof of the decoupling of pure gauge states
with the help of superstring field theory, without having to worry about potential
boundary contributions.

\begin{figure}
\begin{center}
\figfiber
\end{center}

\caption{The region $\UU_{g,m,n}$ interpolating between two sections $\SSS_{g,m,n}$
and $\SSS'_{g,m,n}$.The subspace $\VV_{g,m,n}$ is part of the fiber of $\wt\PP_{g,m,n}$
over the boundary of $\MM_{g,m,n}$. We shall choose the orientation of 
$\SSS'_{g,m,n}$ and $\VV_{g,m,n}$ to be outward from $\UU_{g,m,n}$ and that
of $\SSS_{g,m,n}$ to be inward towards $\UU_{g,m,n}$.
\label{figfiber}}

\end{figure}

Next we shall consider the dependence of the on-shell amplitudes on the choice
of the section $\SSS_{g,m,n}$. For this we again consider a set of BRST invariant
states $K_1,\ldots K_m$, $L_1,\ldots L_n$. The genus $g$ amplitudes of these states
is given by \refb{eoffshellamp}. Now if we choose a different section $\SSS'_{g,m,n}$ then
it is possible to find a dimension $6g-6+2(m+n)+1$ subspace $\UU_{g,m,n}$ 
that interpolates between $\SSS_{g,m,n}$ and $\SSS'_{g,m,n}$. 
Of course $\UU_{g,m,n}$ is not determined uniquely. In this case we have
\be
\p \UU_{g,m,n}=\SSS'_{g,m,n}-\SSS_{g,m,n}+ \VV_{g,m,n}\, ,
\ee
where $\VV_{g,m,n}$ denotes the intersection of $\UU_{g,m,n}$ with the fiber over the
boundary of $\MM_{g,m,n}$. This has been shown pictorially in Fig.~\ref{figfiber}.
We now get
\ben
\int_{\SSS'_{g,m,n}} \Omega^{(g,m,n)}_{6g-6+2(m+n)} -
\int_{\SSS_{g,m,n}} \Omega^{(g,m,n)}_{6g-6+2(m+n)}
&=& \int_{\UU_{g,m,n}} d\Omega^{(g,m,n)}_{6g-6+2(m+n)} -
\int_{\VV_{g,m,n}} \Omega^{(g,m,n)}_{6g-6+2(m+n)}
\nonumber \\
&=& -\int_{\VV_{g,m,n}} \Omega^{(g,m,n)}_{6g-6+2(m+n)}\, .
\een
where in the last step we have used the fact that 
$d\Omega^{(g,m,n)}_{6g-6+2(m+n)}$ with BRST invariant arguments vanishes
due to \refb{emm}. Therefore we see that the difference between the 
on-shell amplitudes computed using the two sections vanishes up to boundary terms.
We shall see in \S\ref{scomredef} and appendix \ref{safield} that 
once on-shell amplitudes are defined using superstring field theory,
the result can be shown to be independent of the choice of sections without having to
make any assumption about the vanishing of the boundary terms.

\sectiono{Superstring field theory: Gauge fixed action} \label{ssft}

A field theory of superstrings will produce off-shell amplitudes but not all off-shell
amplitudes may have field theory interpretation. In order to have a field theory interpretation
the off-shell amplitude must be expressed as the contribution from a sum of Feynman
diagrams. Therefore the section $\SSS_{g,m,n}$ must admit a cell decomposition
with each cell describing a section over a codimension zero
subspace of $\MM_{g,m,n}$, such that 
\begin{itemize}
\item the integral of $\Omega^{(g,m,n)}_{6g-6+2m+2n}$
over a given cell can be interpreted as the contribution from one particular Feynman diagram
of superstring field theory, and 
\item the sum of the contribution from all the cells has the
interpretation of the sum over all the Feynman diagrams. 
\end{itemize}
We shall now describe under
what conditions this holds.  From now on
we shall refer to the segments of $\SSS_{g,m,n}$
corresponding to individual Feynman diagrams as section segments of 
the corresponding Feynman diagrams.

\subsection{Condition on the choice of sections} \label{sconsec}

One of the key properties of Feynman diagrams is that a pair of Feynman diagrams
can be joined by a propagator to make a new Feynman diagram. A
related  property is
that two external legs of a single Feynman diagram can be joined to make a new
Feynman diagram with an additional loop. Now, 
each Feynman diagram of superstring field theory is expected 
to correspond to an
integral of $\Omega$ 
over a section segment.
 This means that there must be an operation that takes the section segments of
two different Feynman diagrams and gives the section segment 
of the new Feynman
diagram obtained by joining the two by a propagator. There must also be
another operation which acts on the section segment of a single
Feynman diagram and generates a section segment of the new 
Feynman diagram
obtained by joining two external legs 
of the original Feynman diagram. Our first task will be
to describe these operations.\footnote{In the usual world-sheet approach to 
string perturbation theory,  these properties
are encoded in the factorization properties of the amplitudes\cite{Polchinski:1988jq}.
}
Again for simplicity of notation, we restrict our analysis to
heterotic string theories; the analysis for type II string theories is more or less 
identical. 

For reasons that will become clear in due course, the operation of joining two legs of two
different Feynman diagrams or two legs of
the same Feynman diagram, is played by an operation
on Riemann surfaces known as
plumbing fixture. First recall that for 
each external leg of a Feynman diagram we have a puncture on the corresponding
Riemann surface, and that for a given choice of section segment, we also have a choice of local
coordinates at the punctures and the PCO locations on the Riemann surface. The operation
of joining a pair of external legs of a Feynman diagram (or of
two Feynman diagrams)
will be represented by the operation of identifying the local coordinates $w_1$ and $w_2$
around the corresponding punctures via the relation
\be \label{eplumb}
w_1 w_2 = e^{-s-i\theta}, \quad 0\le s<\infty, \quad 0\le\theta <2\pi\, .
\ee
This is known as the operation of sewing the regions around the punctures to each
other since as we traverse towards $w_1\to 0$ we emerge in the $w_2$ plane away
from 0. This also induces local coordinates around the punctures of the new
Riemann surface\cite{divecchia,peskin,0708.2591} and 
PCO locations on the new Riemann surface from
those on the original surfaces that are being sewed.
We shall now verify that the new family of Riemann surfaces generated by this
way satisfies the properties that are required to be satisfied by the section segment of
a new Feynman diagram obtained by joining two legs of the same Feynman diagram or
two legs of two different Feynman diagrams.

If the punctures that are being sewed lie on two different Riemann surfaces 
$\Sigma_{g_1,m_1,n_1}$ and 
$\Sigma_{g_2,m_2,n_2}$, then for given $s$, $\theta$ plumbing fixture 
generates a new Riemann surface $\Sigma_{g_1+g_2,
m_1+m_2-2, n_1+n_2}$ if the punctures are NS punctures, i.e.\ they
have NS sector vertex operators inserted. Similarly, the sewing process generates a new Riemann surface
$\Sigma_{g_1+g_2, m_1+m_2, n_1+n_2-2}$ if the punctures that are being sewed
are R punctures.
When we sew the regions around two punctures on the same Riemann
surface $\Sigma_{g,m,n}$, the result is a new Riemann surface 
$\Sigma_{g+1,m-2,n}$ or $\Sigma_{g+1,m,n-2}$.
In all cases this operation also generates
local coordinate systems around the punctures of the new Riemann surface and the
PCO locations on the new Riemann surface from those on the original Riemann
surface(s). There is a slight subtlety 
in the case of sewing R punctures that we shall
illustrate using the case of sewing two different Riemann surfaces.
If we consider two Riemann surfaces $\Sigma_{g_1,m_1,n_1}$ and 
    $\Sigma_{g_2,m_2,n_2}$, then the total number of PCOs is given by
    \ben \label{epcoarrange}
 2g_1 - 2 + m_1 + {n_1\over 2} &+& 2g_2 - 2 + m_2 + {n_2\over 2} \nonumber \\
&=& 2(g_1+g_2)-2 + (m_1 + m_2 - 2) + {n_1+n_2\over 2}  \nonumber \\
&=& 2(g_1+g_2)-2+ (m_1 + m_2) + {n_1+n_2-2\over 2} -1\, .
\een
The equality in the second
line shows that the number on the left hand side is precisely
equal to the required number of PCOs on $\Sigma_{g_1+g_2, m_1+m_2-2,n_1+n_2}$.
However equality in the third line shows that the number on the left is one less 
than the required number of PCOs on 
$\Sigma_{g_1+g_2, m_1+m_2, n_1+n_2-2}$.
In other words when the sewing is done at the R punctures, 
we need to insert an extra PCO on the final surface. We shall use the
symmetric prescription that the new PCO location is taken to be the average over
a circle around one of the punctures, i.e. we insert
\be \label{epcodeg}
\ointop {dw_1\over w_1} \XX(w_1) = \ointop {dw_2\over w_2} \XX(w_2)\, ,
\ee
where $\ointop$ includes a factor of $1/2\pi i$ and the integration is carried out over
an anti-clockwise contour. The equality of the two terms in this equation follows
from \refb{eplumb} and the fact that $\XX$ is a dimension zero primary operator.
A similar analysis involving sewing two punctures on the same Riemann surface
shows that we need a similar insertion when the punctures that are sewed carry
Ramond sector vertex operators.
Note that  for sewing R punctures
this requirement
automatically makes the final section a generalized section in the sense described in
\refb{egensec},
since the extra PCO insertion, instead of being at a single point, is taken to be
an average over
a continuous family as given in \refb{epcodeg}. 

Now a Feynman diagram does not represent a single Riemann surface but a whole
family of Riemann surfaces belonging to the section segment of the Feynman diagram.
Therefore 
given two Feynman diagrams, we can get a subspace of 
$\wt\PP_{g_1+g_2, m_1+m_2-2, n_1+n_2}$ or
$\wt\PP_{g_1+g_2, m_1+m_2, n_1+n_2-2}$ by plumbing fixture of all Riemann 
surfaces corresponding to the first Feynman diagram to all Riemann 
surfaces of the
second Feynman diagram. 
The resulting subspace has dimension
\be\label{eco}
6g_1-6+2m_1+2n_1 + 6g_2-6+2m_2+2n_2 + 2
= 6(g_1+g_2)-6 + 2(m_1+m_2+n_1+n_2-2)\, .
\ee
The last additive factor of 2 on the left hand side comes from the parameters
$(s,\theta)$. 
We shall refer to this operation as plumbing fixture of two section segments.
The right hand side of \refb{eco} is precisely the required dimension of a
section  of $\wt\PP_{g_1+g_2, m_1+m_2-2, n_1+n_2}$ and
$\wt\PP_{g_1+g_2, m_1+m_2, n_1+n_2-2}$.  
Therefore we see that it is consistent to interpret this subspace of
$\wt\PP_{g_1+g_2, m_1+m_2-2, n_1+n_2}$ or
$\wt\PP_{g_1+g_2, m_1+m_2, n_1+n_2-2}$ as the section segment
of the new Feynman diagram obtained by 
joining the original Feynman diagrams by a
propagator.\footnote{It will be understood that we sum over all intermediate states
propagating along the internal propagator.}
A similar analysis shows that when 
we generate a new Feynman diagram by joining two of the legs of a Feynman diagram
by a propagator,
it is consistent to interpret the family of Riemann surfaces,
obtained by the 
plumbing fixture of the section segment of the original diagram at two of its
punctures, as
the section segment of the new Feynman diagram. 
Once we have chosen this interpretation, the section segments of all
Feynman diagrams are obtained from the section segments 
of  Feynman diagrams containing a single interaction vertex and no internal propagators
via repeated operation of plumbing fixture.\footnote{One must distinguish 
between the interaction vertices of
superstring field theory -- a notion in the second quantized theory -- and the vertex
operators acting on the Hilbert space of the first quantized theory. }
We shall denote by $\oR_{g,m,n}$ the section segments of Feynman diagrams
containing a single interaction vertex and no internal propagators, contributing to a genus $g$
amplitude with $m$ external NS sector states and $n$ external R sector states.
To simplify notation, we shall often refer to them as
section segments of the  interaction vertices of 
superstring field theory. 
Of course,
the requirement that the sum of the
section segments of different Feynman diagrams generates a full (generalized) section
$\SSS_{g,m,n}$ puts strong restriction on the choice of  $\oR_{g,m,n}$.

The $s\to\infty$ limit in \refb{eplumb} describes degenerate Riemann surface. We shall
see in \S\ref{sprop} that the
parameter $s$ plays the role of the Schwinger parameter $s$ appearing in
\refb{esch} in quantum field theory. Therefore we see that the type 1 and type
2  divergences
discussed in \S\ref{sintro} are both associated with degenerate Riemann surfaces.

\begin{figure}
	\centering
	\includegraphics[width=5in]{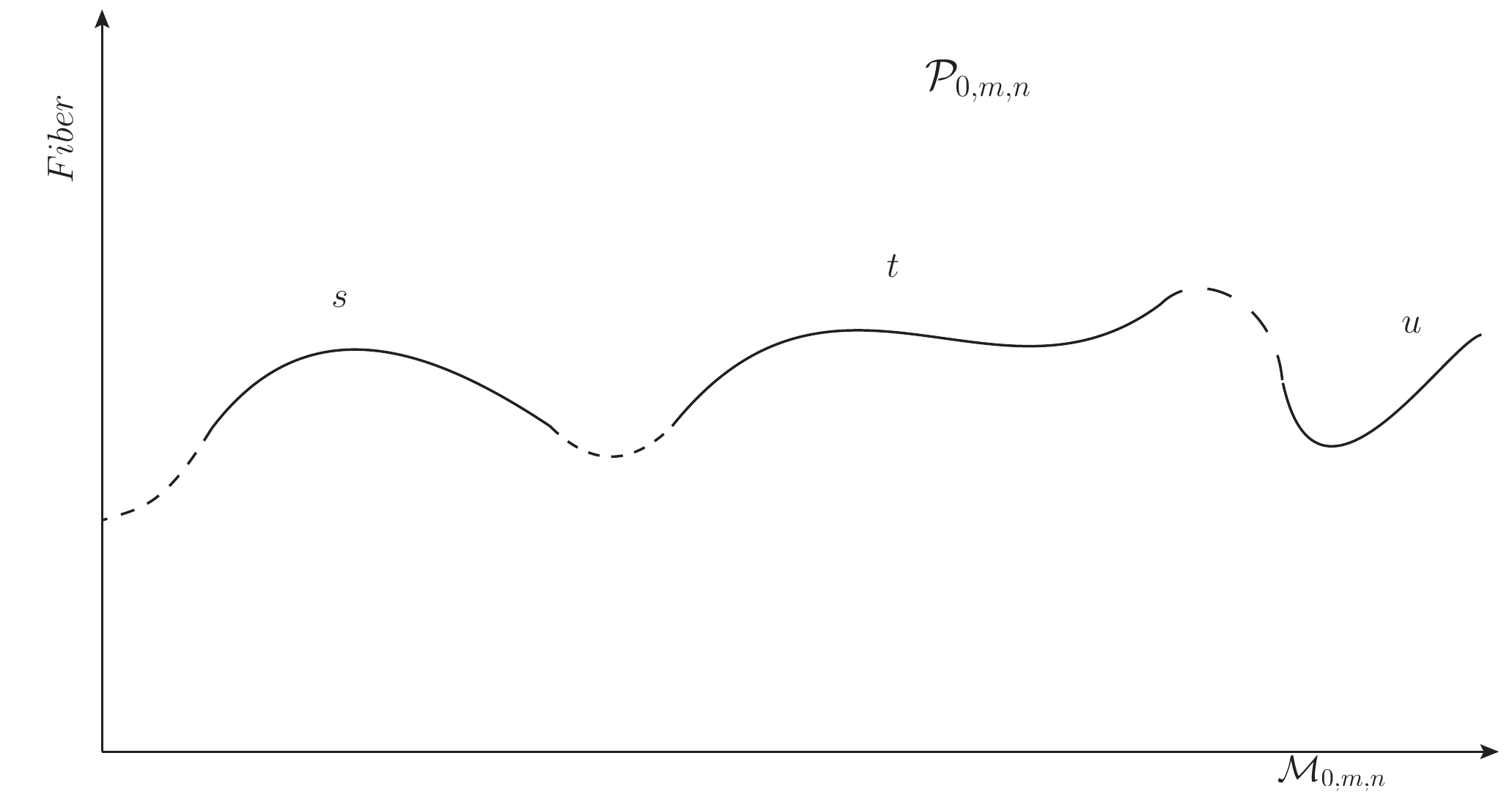}
	\caption{The section of $\wt\PP_{0,m,n}$ for the 4 point amplitude at genus 0.
	The solid lines represent the section segments of the s, t and u-channel diagrams.
	The dashes lines, which are chosen to `fill the gap', represent the section segment 
	of the elementary four point vertex. While in this one dimensional projection
	the dashed lines seem to form disconnected sets, typically they form a 
	connected subspace of $\wt\PP_{0,m,n}$ when we consider their extension
	into other directions. Degenerate Riemann surfaces sit in the interior of the solid
	lines.}
	\label{figusualfill}
\end{figure}

We shall now outline a systematic algorithm for generating section segments
satisfying these requirements, generalizing the corresponding algorithm for bosonic
string field theory\cite{sonoda}. We begin with the section segments of genus zero
three point interaction 
vertex -- they are appropriate subspaces of
$\wt\PP_{0,3,0}$ and $\wt\PP_{0,1,2}$. Since the base is 
zero dimensional, they just represent a point on the fiber. They have 
to be chosen such that they are symmetric under the exchange
of the punctures, {\it e.g.} for NS-NS-NS vertex an SL(2,C) transformation exchanging a pair
of punctures should exchange the corresponding local coordinates and leave the PCO
location unchanged.
This will encode another important property of a Feynman diagram, i.e. a Feynman 
diagram containing a single interaction vertex 
and no internal propagators is symmetric under
the exchange of external legs.
Typically this will require averaging over the choice of PCO locations,
i.e.\ using generalized sections. Furthermore these section segments 
must avoid spurious poles.
For $\wt\PP_{0,1,2}$ this condition is trivial since there are no PCO insertions, while
for $\wt\PP_{0,3,0}$ this condition simply means that the single PCO insertion that is
needed should not coincide with any of the punctures.
We shall call these section segments $\oR_{0,3,0}$ and $\oR_{0,1,2}$ respectively
-- in this case they also represent the full sections $\SSS_{0,3,0}$ and
$\SSS_{0,1,2}$. Now we 
can construct the section segments of tree level four point function
corresponding to s, t and u-channel diagrams by plumbing fixture of two of the 
section segments
of tree level three point functions. These will be appropriate subspaces
of
$\wt\PP_{0,4,0}$,
$\wt\PP_{0,2,2}$ or $\wt\PP_{0,0,4}$.
These together will not, in general, give
full sections of $\wt\PP_{0,4,0}$,
$\wt\PP_{0,2,2}$ and $\wt\PP_{0,0,4}$. We fill the gap by section segments 
$\oR_{0,4,0}$, $\oR_{0,2,2}$ and $\oR_{0,0,4}$ and interpret them as the section segments
of the elementary four point interaction vertices. 
This has been shown schematically in Fig.~\ref{figusualfill}.
Again we have to construct them avoiding 
spurious poles, maintaining
exchange symmetry and the requirement that at the boundary they join smoothly the 
section segments of the s, t and u-channel diagrams so that together they
describe a smooth section of $\wt\PP_{0,m,n}$ 
for $m+n=4 $.\footnote{Of 
course in the interior of
$\oR_{0,4,0}$, $\oR_{0,2,2}$ and $\oR_{0,0,4}$ there may be discontinuities of the kind
mentioned in footnote \ref{fo1} in order to avoid spurious poles.}

\begin{figure}
 \center \includegraphics[width=4in]{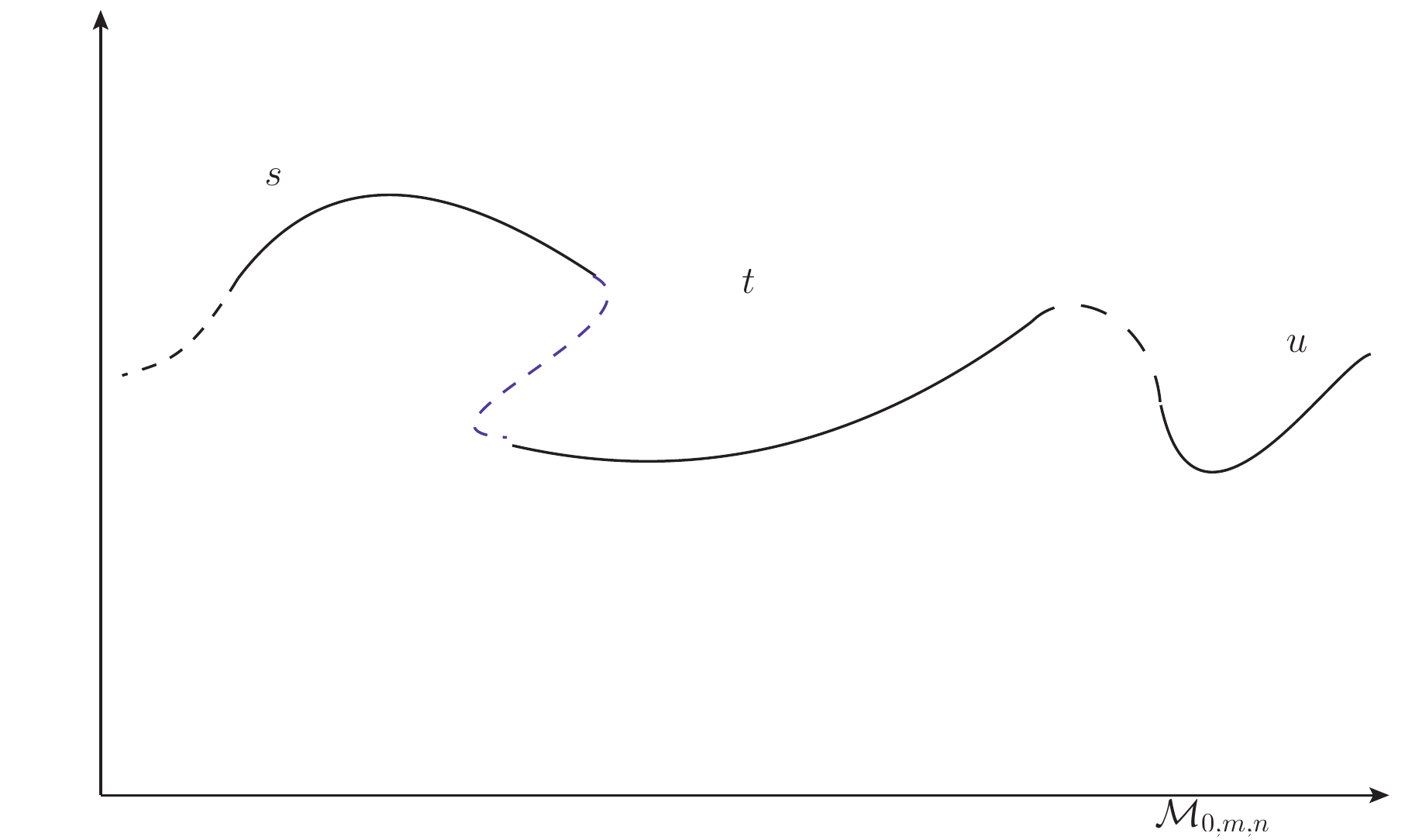} \caption{A case where the
 projection to $\MM_{0,m,n}$
 of section segments of s and t-channel diagrams, shown by the solid
 lines, 
 overlap. We take care of this by joining the end points 
 as shown by dashed line. This
effectively removes the overlap portion by adding negative contribution.}   
\label{figfill}
\end{figure}

Similarly by plumbing fixture of two legs of $\oR_{0,3,0}$ and $\oR_{0,1,2}$ we can generate
section segments of  the one loop tadpole
diagram, given by a subspace of
$\wt\PP_{1,1,0}$.
In general this will not generate a full section of $\wt\PP_{1,1,.0}$ and we
fill the gap by including a section segment $\oR_{1,1,0}$  which
we interpret as the genus 1 contribution to the {\it elementary 1-point vertex} of string 
theory. This procedure can be repeated {\it ad infinitum}, generating,  for all $g,m,n$, 
the section segments $\oR_{g,m,n}$
of the elementary vertices at genus $g$ with $m$ external NS-sector legs and
$n$ external R sector legs. At each stage, we first construct the section
segments of all the Feynman diagrams obtained by joining lower order vertices by
propagators, and then `fill the gap' by appropriate choice of the section segment
 $\oR_{g,m,n}$ of the elementary vertex.

An interesting question is: what happens if the 
section segments of the Feynman diagrams with one or more internal
propagators, 
when projected to $\MM_{g,m,n}$, overlap instead of
leaving gaps. Since we have defined the integration measure on $\wt\PP_{g,m,n}$, there
is in principle no difficulty in filling the gap even in this situation. This has been 
illustrated in
Fig.~\ref{figfill}. However we shall see in \S\ref{scomstub} that by `adding stubs' in the definition
of interaction vertices  it is possible to arrange that the Feynman diagrams containing
one or more internal propagators cover only a small part of $\MM_{g,m,n}$ near the
boundary of the moduli space. In that case there is no overlap of the kind shown
in Fig.~\ref{figfill} in the contribution from
different Feynman diagrams.

There is one more constraint that we need to impose on the section segments 
$\oR_{g,m,n}$.
$\wt\PP_{g,m,n}$ admits a $\ZZZ_2$ action under which all the transition functions
$F_s$ appearing in \refb{eintFs}, the local coordinates around the punctures and the
PCO locations are complex conjugated.  We require $\oR_{g,m,n}$ to be invariant 
under this $\ZZZ_2$ symmetry, i.e. given any point on $\oR_{g,m,n}$, its $\ZZZ_2$ image
must also be in $\oR_{g,m,n}$. Again this may require us to average over section
segments. This condition on $\oR_{g,m,n}$ is necessary for establishing reality of the
superstring field theory action discussed in \S\ref{sreal}.

A key property of the section segments $\oR_{g,m,n}$ is that they do not contain any
degenerate Riemann surface. Indeed all degenerate Riemann surfaces are associated
with Feynman diagrams with at least one propagator, and occur in the limit when
the $s$ parameter of the plumbing fixture 
corresponding to one (or more) of the 
propagators approaches infinity. For example all
degenerate 4-punctured spheres on $\SSS_{0,m,n}$ with
$m+n=4$ come from s, t and u channel Feynman diagrams, and $\oR_{0,m,n}$ is
free from degenerate 4-punctured spheres.

There is one issue that should still worry us. We have mentioned before that in defining
the off-shell amplitudes the sections $\SSS_{g,m,n}$ must be chosen to avoid spurious poles.
Now since the choice of $\oR_{g,m,n}$ is up to us, we can follow the procedure of
\cite{1408.0571,1504.00609} -- reviewed in appendix \ref{saspurious} --
to choose them avoiding spurious poles. But now the section segments 
of Feynman diagrams with one or more propagators are fixed by 
the section segments
$\oR_{g,m,n}$ of the constituent vertices. Therefore we have to 
check if
these section segments also avoid spurious poles. This issue will be addressed
in \S\ref{scomstub} where we shall argue that under certain conditions, 
once we choose $\oR_{g,m,n}$
avoiding spurious poles, the section segments of all other Feynman 
diagrams also avoid spurious poles.

Before concluding this section, we would like to 
emphasize that there are possible choices of sections of $\wt\PP_{g,m,n}$ which
violate these conditions. For example we could choose the 
sections of $\wt\PP_{g,m,n}$
arbitrarily without having any relation to each other. Integrating 
$\Omega^{(g,m,n)}_{6g-6+2m+2n}$ over such sections will define some off-shell amplitudes
but they will not have the interpretation as being given by a sum over Feynman diagrams. 

\subsection{Identities for section segments of interaction vertices } 
\label{selem}

A special role in our analysis will be played by the section segments $\oR_{g,m,n}$
of the elementary vertices. Therefore we shall now analyze some of the
essential  properties of $\oR_{g,\m,\n}$.  As already mentioned, $\oR_{g,m,n}$ is
taken to be
symmetric under the exchange of any pair of NS-punctures and also under the
exchange of any pair of R-punctures. This needs to be achieved, if necessary, by
taking $\oR_{g,m,n}$ to be formal weighted average of subspaces related by these
exchange transformations. 
Plumbing fixture of $\oR_{g_1,\m_1,\n_1}$ and
$\oR_{g_2,\m_2,\n_2}$ at an NS puncture produces a section segment
which we shall denote by $\oR_{g_1,\m_1,\n_1}\circ \oR_{g_2,\m_2,\n_2}$. 
On the other hand,
plumbing fixture of $\oR_{g_1,\m_1,\n_1}$ and
$\oR_{g_2,\m_2,\n_2}$ at an R puncture produces a section segment  which
we shall denote by $\oR_{g_1,\m_1,\n_1}\star \oR_{g_2,\m_2,\n_2}$. 
The information about the insertion of the extra
PCO \refb{epcodeg} will be included in the definition of 
$\oR_{g_1,\m_1,\n_1}\star \oR_{g_2,\m_2,\n_2}$.
Similarly, we denote by $\nabla_{NS}\oR_{g,m,n}$ the section segment  
produced by plumbing fixture of a pair of NS punctures
of $\oR_{g,m,n}$ and by $\nabla_{R}\oR_{g,m,n}$ the section segment
produced by plumbing fixture of a pair of R punctures
of $\oR_{g,m,n}$.
Each of the subspaces $\oR_{g_1,\m_1,\n_1}\circ \oR_{g_2,\m_2,\n_2}$,
$\oR_{g_1,\m_1,\n_1}\star \oR_{g_2,\m_2,\n_2}$, $\nabla_{NS}\oR_{g,m,n}$
and  $\nabla_{R}\oR_{g,m,n}$ has two special boundaries. 
The one corresponding
to $s\to\infty$ corresponds to degenerate Riemann surfaces, and will not be
relevant for our discussion below in this subsection. The other boundary 
contains the Riemann surfaces obtained by setting $s=0$ in the plumbing
fixture relations \refb{eplumb}. 
We shall denote them  by $\{\oR_{g_1,\m_1,\n_1}, \oR_{g_2,\m_2,\n_2}\}$,
$\{\oR_{g_1,\m_1,\n_1}; \oR_{g_2,\m_2,\n_2}\}$,
$\Delta_{NS}\oR_{g,m,n}$
and  $\Delta_{R}\oR_{g,m,n}$
respectively. Therefore
$\{\oR_{g_1,\m_1,\n_1} , \oR_{g_2,\m_2,\n_2}\}$ represents the set of punctured Riemann
surfaces that we obtain by sewing the families of Riemann surfaces  corresponding
to $\oR_{g_1.\m_1,\n_1}$ and $\oR_{g_2,\m_2,\n_2}$ at NS punctures
using plumbing fixture relation \refb{eplumb}
with the parameter
$s$ set to zero. 
$\{\oR_{g_1,\m_1,\n_1} ;\oR_{g_2,\m_2,\n_2}\}$ has a similar interpretation except that
the plumbing fixture is done at Ramond punctures, and we insert an extra PCO
given by \refb{epcodeg}  around the punctures.
Analogous interpretation holds for $\Delta_{NS}\oR_{g,m,n}$
and  $\Delta_{R}\oR_{g,m,n}$.
The orientations of 
$A\circ B$ and $A\star B$ will be defined by taking their volume form to be
$ds\wedge d\theta\wedge dV_A\wedge dV_B$ where $dV_A$ and $dV_B$ 
are volume forms on 
$A$ and $B$ respectively. 
This implies that the volume forms on 
$\{A,B\}$ and $\{A;B\}$
will be given by
$-d\theta\wedge dV_A\wedge dV_B$, the extra minus sign accounting for the fact that
the $s=0$ boundary is a lower bound on the range of $s$. 
Similarly the orientation of $\nabla_{NS}A$
and  $\nabla_{R}A$ will be defined by taking its volume form to be
$ds\wedge d\theta\wedge dV_A$, and consequently the orientation of 
$\Delta_{NS}A$
and  $\Delta_{R}A$ will be given by taking their volume forms to be
$-d\theta\wedge dV_A$.

We have argued before that $\oR_{g,m,n}$ does not contain degenerate Riemann
surfaces, i.e. the base of $\oR_{g,m,n}$ does not extend to the boundaries of
$\MM_{g,m,n}$. However $\oR_{g,m,n}$ does have boundaries, and these are given by
the $s=0$ boundaries of the Feynman diagrams with one propagator, since $\oR_{g,m,n}$
is designed to fill the gap left by Feynman diagrams built by joining 
lower order vertices with propagators.
Therefore
we have
\ben  \label{eboundaryq}
\p \oR_{g,m,n} &=& -{1\over 2} \sum_{g_1,g_2\atop g_1+g_2=g} 
\sum_{m_1,m_2\atop m_1+m_2 = m+2}
\sum_{n_1,n_2\atop n_1+n_2 = n}
{\bf S}[\{\oR_{g_1,m_1,n_1} , \oR_{g_2,m_2,n_2}\}] \nonumber \\ &&
-{1\over 2} \sum_{g_1,g_2\atop g_1+g_2=g} 
\sum_{m_1,m_2\atop m_1+m_2 = m}
\sum_{n_1,n_2\atop n_1+n_2 = n+2}
{\bf S}[\{\oR_{g_1,m_1,n_1} ; \oR_{g_2,m_2,n_2}\}] \nonumber \\ &&
-\Delta_{NS} \oR_{g-1, m+2, n} - \Delta_R \oR_{g-1, m, n+2}
\, ,
\een
where ${\bf S}$ denotes the
operation of summing over inequivalent permutations of external NS-sector punctures and
also external R-sector punctures. Therefore for example 
${\bf S}[\{\oR_{g_1,\m_1,\n_1}, \oR_{g_2,\m_2,\n_2}\}]$ involves sum over
${m_1+m_2-2\choose m_1-1}$ inequivalent permutations of the external NS-sector
punctures and ${\n_1+\n_2\choose \n_1}$ inequivalent permutation of the external R-sector
punctures. These simply reflect sum over inequivalent Feynman diagrams.
The minus sign on the right hand side\footnote{This minus sign will
eventually cancel the minus sign in the integration measure $-d\theta\wedge
dV_A\wedge dV_B$ or $-d\theta\wedge dV_A$. } 
reflects that $\oR_{g,m,n}$,
$\oR_{g_1,\m_1,\n_1}\circ \oR_{g_2,\m_2,\n_2}$,
$\oR_{g_1,\m_1,\n_1}\star \oR_{g_2,\m_2,\n_2}$,
$\nabla_{NS} \oR_{g-1, m+2, n}$ and  $\nabla_R \oR_{g-1, m, n+2}$
 will all have to fit together so they they
form a subspace of the full integration cycle used for defining the off-shell amplitude.
Therefore the boundary of $\oR_{g,m,n}$ will be oppositely oriented to those of 
$\oR_{g_1,\m_1,\n_1}\circ \oR_{g_2,\m_2,\n_2}$,
$\oR_{g_1,\m_1,\n_1}\star \oR_{g_2,\m_2,\n_2}$, 
$\nabla_{NS} \oR_{g-1, m+2, n}$ and  $\nabla_R \oR_{g-1, m, n+2}$.
The factors of $1/2$ in the first two terms on the right hand side
account for the double counting due to the symmetry that exchanges
the two Riemann surfaces corresponding to $\oR_{g_1,m_1,n_1}$ and
$\oR_{g_2,m_2,n_2}$. There are also implicit factors of 1/2 already
included in the definitions of $\Delta_{NS}$ and $\Delta_R$ (and also 
$\nabla_{NS}$ and $\nabla_R$) to account for
the fact that the exchange of two punctures that are being sewed do not
generate new Riemann surface. This will become relevant later, {\it e.g.} in
\refb{evertex}.

\subsection{The multilinear string products and their identities } 
\label{smulti}

Given a set of external NS states $K_1,\ldots, K_m\in \HH_{-1}$
and external R states $L_1,\ldots, L_n\in\HH_{-1/2}$,
all rendered Grassmann even by multiplying the states by 
Grassmann odd c-numbers if
necessary, we now define
\be\label{edefcurl}
\cL K_1\ldots K_m L_1\ldots L_n\cR =\sum_{g=0}^\infty
g_s^{2g} \, \int_{\oR_{g,m,n}}
\Omega^{(g,m,n)}_{6g-6+2m+2n}(K_1,\ldots, K_m,L_1,\ldots, L_n)\, ,
\ee
where $g_s$ is the string coupling.
This has the interpretation as the contribution to the off-shell amplitude with external
states $K_1,\ldots, K_m,L_1,\ldots, L_n$ due to the elementary vertex.
This is by construction symmetric under $K_i\leftrightarrow K_j$ and
$L_i\leftrightarrow L_j$. We shall extend its definition to arbitrary arrangement
of NS and R states inside the product $\cL\cdots\cR$ by declaring the product to be
completely symmetric under the exchange of any states. This means that if we
have an arbitrary arrangement of NS and R vertex operators inside $\cL\cdots\cR$, we
first rearrange them so that all the NS sector vertex operators come to the left of all
the R sector vertex operators, and then use \refb{edefcurl}.

Using \refb{edefomega}, \refb{edefcurl} and 
the ghost number conservation law that says that on
a genus $g$ surface we need total ghost number $6-6g$ to get a non-zero
correlator one can show that $\cL  A_1\ldots A_N \cR$ 
-- where each $A_i$ now
represents either an NS state or an R state -- is non-zero only if
\be \label{eghcon}
\sum_{i=1}^N (n_i-2) = 0\, ,
\ee
where $n_i$ is the ghost number of $A_i$. In the following, a state $A_i$ will denote 
an NS or R sector state, unless mentioned otherwise.

As a consequence of \refb{eboundaryq}, the vertex $\cL\cdots\cR$, for any set of states
$A_1,\ldots, A_N\in \wh\HH_T$, can be shown to satisfy the identity
\ben \label{evertex}
&&\hspace*{-.4in}\sum_{i=1}^N \cL A_1\ldots A_{i-1} (Q_B A_i)
A_{i+1} \ldots A_N\cR  \nonumber \\
&=& -  
{1\over 2} \sum_{\ell,k\ge 0\atop \ell+k=N} \sum_{\{ i_a;a=1,
\ldots \ell\} , \{ j_b;b=1,\ldots k\} \atop
\{ i_a\} \cup \{ j_b\}  = \{ 1,\ldots N\}
}\cL A_{i_1} \ldots A_{i_\ell} \vp_s\cR  \cL \vp_r A_{j_1} \ldots A_{j_k}\cR 
\langle \vp_s^c | c_0^- \GG | \vp_r^c\rangle \nonumber \\
&& -   {1\over 2} \, g_s^2\, \cL A_1 \ldots A_N \vp_s \vp_r \cR \, \langle \vp_s^c | c_0^- \GG 
| \vp_r^c\rangle
\, .
\een

The proof of this goes as 
follows\cite{1408.0571,1411.7478,1501.00988}.\footnote{The 
analysis of \cite{1408.0571,1411.7478,1501.00988} 
was done for the 1PI vertices and hence did not
have the terms involving $\Delta_{NS}$ and $\Delta_R$ in \refb{eboundaryq} and the
last term on the right hand side of \refb{evertex}. But inclusion of these terms is
straightforward.} First using the definition \refb{edefcurl} and the identity 
\refb{emm}  we convert the left hand side of \refb{evertex} into an integral 
of $d\Omega^{(g,m,n)}_{6g-6+2m+2n-1}$ over
$\oR_{g,m,n}$. Using Stokes' theorem, this can now be expressed
as an integral of $\Omega^{(g,m,n)}_{6g-6+2m+2n-1}$
over $\p\oR_{g,m,n}$. We then use \refb{eboundaryq} to express
this as the integration over the $s=0$ boundary of other section segments 
of various Feynman diagrams with a single propagator -- either connecting two 
elementary vertices or connecting two legs of a single elementary vertex. 
 For definiteness let us consider the case where we have a propagator connecting
two elementary vertices -- the analysis in the other case is very similar. We need to
compute correlation
function on the sewed
Riemann surfaces corresponding to this Feynman diagram
and integrate it over the section segments of the constituent
elementary vertices
and $\theta$. There is no integration over $s$ since we integrate
over the $s=0$ boundary. Contraction of $\Omega$ with $\p/\p\theta$ 
inserts an integral 
\be \label{ebfactor}
(-2\pi i)^{-1} (-i) \left[\ointop (w_i b(w_i) dw_i -
\bar w_i \bar b(\bar w_i) d\bar w_i)\right]
\ee
into the correlation function according to
\refb{edefomega}, \refb{ebdef}, \refb{eplumb}.
Here $w_i$ denotes the local coordinate around one of the punctures
that is sewed and the integration contour keeps the region described by $w_i$
coordinate system, $|w_i|\ge e^{-s/2}$, to the left. Therefore the contour
is a clockwise contour around $w_i=0$.
The $(-2\pi i)^{-1}$ factor has its origin in the prefactor in
\refb{edefomega} and the $-i$ factor arises from the factor $-i$
multiplying $\theta$ in the exponent of \refb{eplumb} and the definition of
 ${\bf B}[\p/\p u_i]$ given in \refb{ebdef}.
In R sector we also have to insert the operator
$\int dw_i\, \XX(w_i)/w_i$.

\begin{figure}
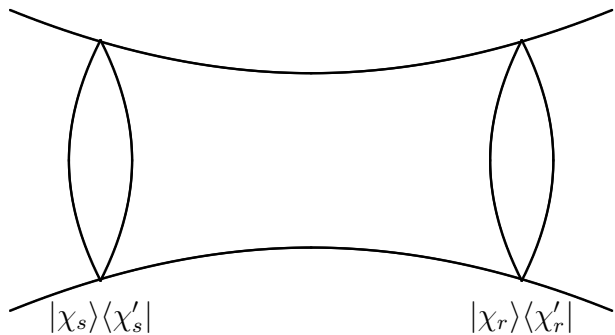


\begin{center}

\figinsert

\end{center}

\vskip -.7in

\caption{Insertion of complete set of states at two ends of a neck.
\label{finsert}}
\end{figure}

We now insert into the correlation function a complete set of states
at $|w_1|=1$ and $|w_2|=1$  using 
\be
\sum_s |\chi_s\rangle \langle \chi'_s| = {\bf 1}
\ee
where $\{|\chi_s\rangle\}$, and independently $\{|\chi_s'\rangle\}$, 
denote a complete set of states in the {\it full Hilbert space
of matter-ghost SCFT} satisfying
\be \label{einnchi}
\langle \chi'_r|\chi_s\rangle=\delta_{rs}\, .
\ee
This has been shown in Fig.~\ref{finsert} for general $s>0$. 
For $s=0$ the circles at $|w_1|=1$ and $|w_2|=1$ coincide with a relative
twist angle $\theta$ and the segment of the Riemann surface between the two
vertical circles in Fig.~\ref{finsert} disappears. Using 
the standard relation involving factorization
of correlation functions on Riemann surfaces, we can now express the 
correlation function on the sewed Riemann surface in terms of product of
correlation functions of $\chi_s$ and $\chi_r'$ inserted
on the original Riemann surfaces and the matrix element
\be \label{emanpre}
(-2\pi i)^{-1} (-i) \left\langle\chi_s'\left| (-b_0^-) \, \int_0^{2\pi} d\theta 
\, e^{-i\theta(L_0-\bar L_0)}
\, \GG \right| |\chi_r\right\rangle = 
\langle\chi_s'|(-b_0^-) \delta_{L_0^-} \GG  |\chi_r\rangle\, .
\ee
The $-b_0^-$ factor comes from terms inside the square bracket in
\refb{ebfactor} after taking into account the fact that the $w_i$ contour runs
clockwise around the origin. $\GG$ simply
encodes the fact that for sewing R punctures we have extra insertion of $\XX_0$. 
Integration
over $\theta$, together with the prefactors on the left hand side of
\refb{emanpre}, produces the factor of 
$(2\pi)^{-1}\int d\theta e^{-i\theta (L_0-\bar L_0)}
= \delta_{L_0,\bar L_0}$. Now the $b_0^-$ factor in \refb{emanpre}
tells us that if we divide the
basis states $\{|\chi_r\rangle\}$, $\{\langle\chi_s'|\}$
into those annihilated by $b_0^-$ and those  annihilated
by $c_0^-$, then both $\langle\chi_s'|$ and $ |\chi_r\rangle$ must belong to the
second set, and the $\delta_{L_0,\bar L_0}$ factor tells us that we can restrict the
basis states to those annihilated by $L_0^-$. Therefore we have 
$|\chi_s'\rangle, |\chi_r\rangle\in c_0^- \HH_T$ and the conjugate states
$|\chi_s\rangle, |\chi_r'\rangle\in  \HH_T$. Finally the picture number conservation,
together with the fact that the number of PCO insertions on each of the component
Riemann surfaces are chosen such that we satisfy picture number conservation when all
external states have picture numbers $-1$ or $-1/2$, tells us that
$|\chi_s\rangle, |\chi_r'\rangle\in  \wh\HH_T$. Therefore we can replace 
$|\chi_s\rangle, |\chi_r'\rangle$ by the basis states $|\vp_s\rangle$ and $|\vp_r\rangle$
of $\wh\HH_T$ satisfying 
\refb{ecom}. Comparing \refb{einnchi} with \refb{ecom}
we see that the corresponding conjugate states $\langle \chi_s'|, |\chi_r\rangle$
are given by $\langle \vp_s^c|c_0^-$ and $c_0^-|\vp_r^c\rangle$ respectively, and we can
replace \refb{emanpre} by
\be \label{emanA}
\langle\vp_s^c|c_0^- (-b_0^-) c_0^- \GG |\vp_r^c\rangle = -\langle\vp_s^c|
c_0^- \GG |\vp_r^c\rangle\, \, .
\ee
A similar analysis can be carried out for the case where the propagator joins two external
lines of a single elementary vertex.

The correlation function(s) on the original Riemann
surface(s),  present before sewing,
are now integrated over the corresponding section segments $\oR_{g',m',n'}$ to 
generate the various factors of $\cL\cdots \cR$ on the right hand side of \refb{evertex}.
The minus signs on the right hand side of \refb{evertex}
arise from the product of three minus signs. The first minus sign originates from
the fact that the integration measure in $\{A,B\}$ and $\{A;B\}$ is $-d\theta\wedge
dV_A\wedge dV_B$ and that in $\Delta_{NS}A$ and $\Delta_RA$ is $-d\theta\wedge
dV_A$. The second minus sign comes from the minus signs on the right hand side of 
\refb{eboundaryq}. The third minus sign arises from the minus sign on the right hand side
of \refb{emanA}.
 
The normalization of the
last term on the right hand side of \refb{evertex} requires additional
explanation. The factor of 1/2 compensates for the  fact that the exchange of the two
punctures that are being sewed gives rise to the same Riemann surface after
sewing. The $g_s^2$ factor
reflects that the operation of sewing two legs of the same vertex increases the
number of loops by one and therefore gives an additional factor of $g_s^2$.
This is correlated with the $g_s^{2g}$ factors in the definition \refb{edefcurl} and could
change in other conventions {\it e.g.} we may replace $g_s^2$ by $-g_s^2$ or
$\pm i g_s^2$ in all formul\ae. Another important difference between the 
second term and the first term on the right hand side of 
\refb{evertex} is that in the first term the
ghost and picture numbers of $\vp_r$ and $\vp_s$ are fixed by the ghost and picture
numbers of the vertex operators $A_i$. In particular we have demonstrated
above that picture number
conservation forces $\vp_r$ and $\vp_s$ to be in $\wh\HH_T$.
However this is not the case for the last term -- we could change the ghost and
picture numbers of $\vp_r$ and $\vp_s$ by opposite amount without violating
ghost or picture number conservation. We need to sum over all ghost number states
consistent with ghost number conservation. However as far as picture number is
concerned, we know from the analysis of \cite{FMS} that every physical state
has a representation in every picture number differing by integers. Therefore 
we need to fix the picture numbers of $\vp_r$ and $\vp_s$ to avoid over counting.
We have taken both $\vp_r$ and $\vp_s$ to be in $\wh\HH_T$, i.e.\ in picture number 
$-1$ for NS sector states and picture number $-1/2$ for R sector states. Consequently
$\vp_s^c$ and $\vp_r^c$ will belong to $\wt\HH_T$. 
As discussed below \refb{emodebg}, 
this choice  avoids states of arbitrarily large negative conformal weight
from propagating in the
loop. This still leaves us with the possibility of 
infinite number of states of the same conformal weight in the R sector 
propagating in the
loop, but we shall argue in \S\ref{scomstub} that this is prevented by the presence
of $\GG$ in  \refb{emanA}.

\subsection{The propagator} \label{sprop}

We shall now 
describe the propagator that represents the plumbing fixture \refb{eplumb} of
section segments as an
algebraic operation on the corresponding Feynman amplitudes obtained by
integrating $\Omega$ on the section segments. 
This analysis is more or less identical to the one that lead to
\refb{emanA}\cite{1408.0571,1411.7478,1501.00988}, so we shall be brief.
The only difference is that instead
of fixing $s$ at zero and integrating over $\theta$ we now also have integration
over $s$.
We insert a complete set of states around $|w_1|=1$ and
$|w_2|=1$ and manipulate the expression as described below \refb{evertex}.
The integration measure requires us to insert an additional factor
of 
\be \label{eb0}
B\left[{\p\over \p s}\right] = - \left[\ointop (w_i b(w_i) dw_i +
\bar w_i \bar b(\bar w_i) d\bar w_i)\right]=b_0^+
\ee 
in \refb{emanA}
due to the contraction of $\Omega$ with $\p/\p s$.
Since the integration measure is $ds\wedge d\theta$, $b_0^+$ will be
inserted to the left of $b_0^-$.
The evolution of the state from $|w_2|=1$ to $|w_2|=e^{-s}$ 
in Fig.~\ref{finsert} is generated by the operator $e^{-s (L_0+\bar L_0)}$. Therefore 
the integration over $s$ produces
a factor of
\be \label{eschsft}
\int_0^\infty ds \, e^{-s (L_0+\bar L_0)} = (L_0+\bar L_0)^{-1}\, .
\ee
After inserting \refb{eb0} and \refb{eschsft} into
\refb{emanA}, we arrive at the following operational expression for the propagator.
Let us suppose that $f(A_1,\ldots  ,A_m, \vp_s)$ denotes the contribution 
to the off-shell amplitude
from a
specific Feynman diagram with external states $A_1,\ldots  ,A_m, \vp_s\in\wh\HH_T$ 
and
$g(B_1,\ldots  ,B_n, \vp_r)$ denotes the contribution from another
Feynman diagram with external states $B_1,\ldots  ,B_n, \vp_r\in\wh\HH_T$. 
Now we can construct a new Feynman diagram with external states 
$A_1,\ldots  ,A_m, B_1,\ldots  ,B_n$ by joining $\vp_s$ and $\vp_r$ by a propagator, and
summing over $s$ and $r$. Its contribution is given by\footnote{The expression for the propagator is somewhat different from the standard one (see e.g.\cite{1508.02481, 1508.05387}) where the propagator contains a $b_0^-$ instead of a $c_0^-$. This difference can be traced to the inclusion of the $c_0^-$ in the normalization \refb{ebas} of the basis states.}
\be \label{efa1}
-f(A_1,\ldots A_m, \vp_s) \, g(B_1,\ldots B_n, \vp_r) \,
\langle \vp_s^c | c_0^- b_0^{+} (L_0^+)^{-1} \GG | \vp_r^c\rangle\, ,
\ee
with the minus sign originating from the minus sign on the right hand side of
\refb{emanA}.
Note that $f$ and/or $g$ may have odd Grassmann parity from the Grassmann odd
numbers
hidden inside the $A_i$\plu, so one should be careful about their relative 
positioning.
Similarly if $f(A_1,\ldots, A_n, \vp_s, \vp_r)$ denotes a Feynman diagram
with external states $A_1,\ldots, A_n,\vp_s,\vp_r$ and if we consider a new
Feynman diagram obtained by joining $\vp_s$ and $\vp_r$ by a propagator and
summing over all choices of $\vp_s$, $\vp_r$, the new Feynman diagram is
given by
\be \label{efa2}
-  {1\over 2}  g_s^2\, f(A_1,\ldots  ,A_m, \vp_s,\vp_r) 
\langle \vp_s^c | c_0^- b_0^{+} (L_0^+)^{-1} \GG | \vp_r^c\rangle\, .
\ee

To summarize, the off-shell amplitudes for given external states can be computed
as the sum of all Feynman diagrams contributing to the amplitude, where the Feynman
diagrams are computed using the elementary $N$-point 
vertices $\cL A_1\ldots A_N\cR$ and the propagator described in \refb{efa1}, 
\refb{efa2}. 

Since the $L_0+\bar L_0$ eigenvalue is given by $(k^2+C)/2$ where $C$
is the mass$^2$ level of a state, by comparing \refb{eschsft} with \refb{esch} we
see that up to a normalization factor of 2, 
the plumbing fixture parameter $s$ corresponds to the Schwinger parameter
of quantum field theories.

\subsection{Action} \label{sAction}

Given the Feynman rules derived above, the next question is: can we write down an
action that gives rise to these Feynman rules? 
In this subsection we shall describe such an action.

The first task will be to introduce the dynamical fields of the theory.
Using the standard identification 
between the wave-function of the first quantized theory and fields
in the second quantized theory, the fields in string field theory are represented as
states in the SCFT.  
Since we have taken the off-shell states to be elements of $\wh\HH_T$, it would
be natural to take the string field $|\Psi\rangle$ to be an element of 
$\wh\HH_T$. We shall impose the further restriction
\be \label{eggg1}
b_0^+|\Psi\rangle = 0\, .
\ee
This can be motivated as follows. We can decompose
$\HH_T$ into a direct sum of two subspaces, one annihilated by $b_0^+$ and the
other annihilated by $c_0^+$, with the BPZ inner product being non-zero only among the
states in different subspaces. Using this we can divide the basis states $\vp_r$ of
$\wh\HH_T$ and $\vp_r^c$ of $\wt\HH_T$ into those annihilated by $b_0^+$ and
those annihilated by $c_0^+$. Now the propagator given in \refb{efa1}, \refb{efa2} is 
non-vanishing
if both $\vp_s^c$ and $\vp_r^c$ are annihilated by $c_0^+$ and therefore if both
$\vp_s$ and $\vp_r$, that are inserted into the amplitudes $f$ and $g$, 
are annihilated by $b_0^+$. This is the reason for restricting
the string field $|\Psi\rangle$, that takes part in the interaction,
to the subspace annihilated by $b_0^+$.

For reasons 
that will be explained below,
we shall introduce another set of string fields
$|\wt\Psi\rangle\in\wt\HH_T$, satisfying 
\be \label{eggg2}
b_0^+|\wt\Psi\rangle = 0\, .
\ee
Both $|\Psi\rangle$ and $|\wt\Psi \rangle$ will be taken to be 
Grassmann even,
in the sense that if we expand these fields as
\be \label{epsexpan}
|\Psi\rangle = \sum_r \psi_r |\vp_r\rangle, \quad |\wt\Psi\rangle 
= \sum_r \wt\psi_r |\vp_r^c\rangle, \qquad b_0^+|\vp_r\rangle=0,
\quad b_0^+|\vp_r^c\rangle=0\, , 
\ee
then $\psi_r$ and $\wt\psi_r$, which are the dynamical variables of the theory, will be
Grassmann even (odd) if the basis state it multiplies is Grassmann even (odd).
Note that the sum over $r$ in \refb{epsexpan} contains integration over momenta and
a sum over infinite number of 
discrete labels. Therefore the set of string field components $\{\psi_r\}$,
$\{\wt\psi_r\}$ actually represent an infinite set of fields in momentum space.
We now take the action\footnote{We shall work in the convention in which the path 
integral
is carried out with the weight factor $e^S$. In the various normalization
and sign conventions that we shall be using, this is the correct sign for the
Euclidean path integral in both the heterotic and the type II 
string theory. This will be discussed in \S\ref{sreal}. The Lorentzian signature case 
requires changing the weight factor to $e^{iS}$. The
effects of this have been discussed in \S\ref{slorentz}.} \label{fo10}
\be \label{esgf}
S_{gf} = {1\over g_s^2} \left[ -{1\over 2} \langle \wt\Psi | c_0^- c_0^+ L_0^+ 
\GG |\wt\Psi\rangle
+ \langle \wt\Psi | c_0^- c_0^+ L_0^+  |\Psi\rangle + \sum_{n=1}^\infty {1\over n!}
\cL \Psi^n\cR \right]
\, .
\ee
The interaction term clearly gives the correct elementary interaction vertex. 
To check that the
propagator comes out correctly we express the kinetic term 
inside the square bracket as
\be \label{e320}
-{1\over 2} \pmatrix{\wt \psi_r & \psi_r} \pmatrix{A_{rs} & B_{rs}\cr B^T_{rs} & 0}
\pmatrix{\wt\psi_s \cr \psi_s}\, ,
\ee
where
\be \label{e321}
A_{rs} =\langle\vp_s^c| c_0^- c_0^+ L_0^+ 
\GG |\vp_r^c\rangle, \quad B_{rs} = -\langle\vp_s| c_0^- c_0^+ L_0^+ 
|\vp_r^c\rangle, \quad B^T_{rs} = -\langle\vp_s^c| c_0^- c_0^+ L_0^+ 
|\vp_r\rangle\, .
\ee
In \refb{e320}, \refb{e321} it is understood that the sum over $r,s$ 
runs over
only those basis states $|\vp_{r}\rangle$, $|\vp_{r}^c\rangle$,
$|\vp_{s}\rangle$, $|\vp_{s}^c\rangle$ that are annihilated by
$b_0^+$. 
It is now easy to see that the propagator, given by the
inverse of the matrix $\pmatrix{A_{rs} & B_{rs}\cr B^T_{rs} & 0}$, takes the form
\be
\pmatrix{0 & P_{st}\cr P^T_{st} & R_{st}}
\ee
where
\be \label{eprop}
P_{st} =  -\langle \vp_t^c | c_0^- b_0^{+} (L_0^+)^{-1} | \vp_s\rangle,
\quad P^T_{st} =  -\langle \vp_t | c_0^- b_0^{+} (L_0^+)^{-1} | \vp_s^c\rangle,
\quad
R_{st} = -\langle \vp_t^c | c_0^- b_0^{+} (L_0^+)^{-1} \GG | \vp_s^c\rangle\, ,
\ee 
where now $s,t$ run over those basis states 
$|\vp_{s}\rangle$, $|\vp_{s}^c\rangle$,
$|\vp_{t}\rangle$, $|\vp_{t}^c\rangle$ that are annihilated by $c_0^+$.
However, using the fact that $b_0^+$ commutes with $L_0^+$ and $\GG$,
we can see that even if we relax this constraint on the basis states, only the
basis states annihilated by $c_0^+$ will give non-vanishing matrix element.
Therefore while computing Feynman diagrams using this propagator, we shall
relax this constraint on the basis states. 

We now 
note that since the interaction term involves
only the $|\Psi\rangle$ field, only the $\psi_r-\psi_s$ component of the propagator is
relevant. Therefore the  relevant component of the propagator is $R_{st}$, 
which agrees with what appears
in \refb{efa1}, \refb{efa2}. On the other hand the $\wt\Psi$ field describes
a set of free field degrees of freedom that completely decouple and have no relevance
for the interacting part of the theory. This will be elaborated further in \S\ref{seom}.

If we had introduced only one set of fields $|\Psi\rangle$ 
instead of two sets of string fields $|\Psi\rangle$ and $|\wt\Psi\rangle$, then
the kinetic term would have been given by the inverse of the propagator
given in \refb{efa1}, \refb{efa2}. However the
operator $\XX_0$ appearing in the Ramond sector 
propagator does not have a well
defined inverse on off-shell states.\footnote{This is related to the difficulty in
writing down a covariant action for type IIB supergravity. With the doubling trick one
can write down such an action\cite{1511.08220}. 
Having two sets of string fields allows us to invert the propagator without having to invert
$\XX_0$.} In the NS sector we do not have any such
difficulty since $\GG$ is the identity operator,  and 
we could set the NS components of $\Psi$ and $\wt\Psi$ to be equal from the beginning.
Here we have kept both components for simplicity of notation.

Finally one remark about the normalization of the propagator.
The presence of $1/g_s^2$
factor in the action \refb{esgf} will give an additional 
multiplicative factor of $g_s^2$ in the propagator.
On the other hand for this action the contribution to the Feynman diagram 
containing a single interaction vertex and no internal
propagators, with external states $A_1,\ldots  ,A_N$, will be $g_s^{-2}
\cL A_1\ldots A_N\cR$. In our analysis in \S\ref{smulti} we have taken this to be
$\cL A_1\ldots A_N\cR$. Therefore the Feynman diagrams of \S\ref{smulti}
correspond to
$g_s^2$ times the Feynman diagrams computed from the action \refb{esgf}. It is easy
to see that this compensates for the absence of the $g_s^2$ factors in the
propagator in \S\ref{sprop} if we express the composition rules 
\refb{efa1} and \refb{efa2} as
\be \label{efa1rule}
-g_s^{-2} \, f(A_1,\ldots  ,A_m, \vp_s) \times g_s^{-2}\, g(B_1,\ldots  ,B_n, \vp_r) \times
g_s^2\, \langle \vp_s^c | c_0^- b_0^{+} (L_0^+)^{-1} \GG | \vp_r^c\rangle \times
g_s^2\, ,
\ee
and
\be \label{efa2rule}
-  {1\over 2}  g_s^{-2}\, f(A_1,\ldots  ,A_m, \vp_s,\vp_r) \times g_s^2\, 
\langle \vp_s^c | c_0^- b_0^{+} (L_0^+)^{-1} \GG | \vp_r^c\rangle \times g_s^2\, .
\ee
Now the first three factors of \refb{efa1rule} and the first two factors of
\refb{efa2rule} have the correct normalization of an amplitude / propagator computed from
the action \refb{esgf}. The multiplication by the final factor of $g_s^2$ converts 
the contribution to the Feynman
diagrams back to the normalization convention of \S\ref{smulti}.

Since the propagators and vertices obtained from this action are exactly what we found
in \S\ref{smulti} and \S\ref{sprop}, 
the off-shell amplitudes computed using this action agree with those obtained by integrating
$\Omega^{(g,m,n)}_{6g-6+2m+2n}$ over $\SSS_{g,m,n}$.

\subsection{Degeneration limit} \label{sdegenerate}

The requirement that the amplitudes arise from the sum over Feynman diagrams of an
underlying field theory puts restriction on the arrangement of the PCOs in the degeneration
limit -- not only for off-shell amplitudes but also for on-shell amplitudes. This restriction comes
from the fact that near separating type degenerations in which a Riemann surface breaks
apart into two Riemann surfaces $\Sigma_{g_1,m_1,n_1}$ and $\Sigma_{g_2,m_2,n_2}$, 
the PCOs must be arranged so that there are $2g_1-2+m_1+n_1/2$ PCOs on the first
Riemann surface and $2g_2-2+m_2+n_2/2$ PCOs on the second Riemann surface. This
rules out some otherwise natural choices. For example for computing the genus one 
two point function of
two NS sector fields, we cannot choose the two PCOs to be on the two NS sector vertex
operators to convert them into zero picture vertex operators. To see this note that in
the degeneration limit in which two NS sector vertex operators come together, 
describing a separating type degeneration in which the Riemann surface splits
into a one punctured torus and a three punctured sphere,  
we shall have both PCOs on the three punctured sphere if we take the two PCOs to be
on the two NS sector vertex operators. This is the wrong
choice since according to the general criterion mentioned above, there should be
only one PCO on the sphere and one PCO on the torus. If we are not careful in making
the right choice we can actually get wrong answer for various physical 
quantities\cite{1508.02481}.
However it is often simpler to first find the wrong answer which is easier to
calculate, and then add to it the difference
between the right answer and the wrong answer. The latter is the integral of a total
derivative in the moduli, and as a result picks up contribution only from the boundary
terms. Analogous situation also arises while computing on-shell amplitudes
as integrals over supermoduli space\cite{1209.5461}.

\subsection{Role of stubs in controlling divergences and
spurious poles}  \label{scomstub}

Although the section segments $\oR_{g,m,n}$ have to satisfy the constraint given in 
\refb{eboundaryq} and the requirement of symmetry under the exchange of punctures,
there is still a lot of ambiguity in choosing these regions. This affects the definition
of the string interaction vertex $\cL\cdots \cR$. As will be discussed in \S\ref{scomredef}
and appendix \ref{safield},
physical quantities do not get affected by this change in the interaction vertex.

There is one particular class of deformations of $\oR_{g,m,n}$, known as the
operation of adding stubs, that is worth 
mentioning\cite{sonoda,9205088,9301097}. 
Adding stubs of length $\ln\lambda$ corresponds to scaling the
local coordinates by some real number  $\lambda>1$, i.e. if $w_i$\plu\ are the 
original local
coordinates at the punctures, we take the new coordinates to be 
$\wt w_i=\lambda \, w_i$.   Comparing the  plumbing fixture relation \refb{eplumb}
with the new relation $\wt w_1 \wt w_2=e^{-\wt s - i\wt\theta}$ we see that 
$\wt s=s-2\ln \lambda$. Therefore when $\wt s $ varies from 0 to $\infty$, the original variable
$s$ varies only in the range $2\ln\lambda\le s<\infty$. For large $\lambda$ this means that
$s$ is large over the whole range $0\le\wt s<\infty$ and hence the Riemann surface is close
to degeneration. Therefore when we add large stubs to the 
interaction vertices, all the Feynman
diagrams with one or more internal propagators will represent Riemann surfaces 
close to degeneration and most of the moduli space away from the degeneration will be
in the Feynman diagrams with a single elementary vertex and no propagators.

If we have a vertex where the external states have conformal weights $h_i$, then 
a change of local coordinates from $w_i$ to $\wt w_i=\lambda\, w_i$ will rescale the
interaction vertex by $\lambda^{-h_i}$. For large $\lambda$, 
this will suppress contribution from states with large $h_i$ propagating
in the internal legs of the Feynman diagram. Due to this suppression factor  
the contribution falls off rapidly for large $h_i$, and for sufficiently large
$\lambda$ the sum
over internal states in a Feynman diagram does not lead to any divergence despite
there being infinite number of fields -- the number of fields grows as 
$\exp[c\sqrt h]$\cite{cardy} for 
some positive number $c$ whereas the suppression factor is of order 
$\exp(-h\ln \lambda)$. In fact this convergence can be made faster by adding
larger stubs.
This suppression factor is also responsible for UV
finiteness of the Feynman amplitudes since it generates an exponential suppression
factor for large Euclidean momenta flowing in the loops.
This will be elaborated further in \S\ref{sloopcon}. 

Having long stubs has a special advantage in superstring field theory where we also have
PCO insertions. Since the choice of $\oR_{g,m,n}$ is up to us subject to the boundary
conditions \refb{eboundaryq}, we can choose the PCO locations in the interior of
$\oR_{g,m,n}$ following the prescription reviewed in appendix \ref{saspurious}
so as to avoid spurious singularities.
However once $\oR_{g,m,n}$\plu\ have been chosen, the section segments 
of all other Feynman diagrams having one or more propagators are completely fixed
by the plumbing fixture rules. The question we need to ask now is: are the associated
PCO locations such that we do not encounter any spurious singularities?

We shall now argue that as long as we attach sufficiently large stubs to the vertices, the
Feynman diagrams with one or more propagators are free from spurious poles as long as
$\oR_{g,m,n}$ avoid spurious poles by a finite margin (i.e.\ do not come too close
to spurious poles). This is simply a consequence of the fact that the contribution from the
Feynman diagrams are obtained by multiplying the propagators and vertices and
summing over internal states and integrating over momenta. We argued above that the
sum over internal states and integral over momenta can be made to converge fast
due to the exponential
suppression factor due to stubs. Therefore as long as the
elementary vertices themselves are finite, the contribution from Feynman diagrams with
propagators will also be finite.

Note that this argument assumes that for fixed momentum, the conformal weight of
the vertex operators is bounded from below. 
As argued below \refb{emodebg},
this is true for states in $\wh \HH_T$ and $\wt \HH_T$, but
may fail in the other picture numbers. 
There is still a possible
subtlety in the Ramond sector, since in the picture number $-1/2$ sector there are infinite
number of states at the same mass$^2$ level created by the action of the $\gamma_0$ 
oscillator. In the conjugate $-3/2$ picture the infinite number of states at a fixed 
mass$^2$
level are created
by the action of the $\beta_0$ oscillator. However we shall now argue that the 
Ramond sector propagator $\langle\vp_s^c|c_0^- b_0^+ (L_0^+)^{-1} \XX_0|\vp_r^c\rangle$
given in \refb{efa1}, \refb{efa2} prevents all but a finite
number of these states from propagating. For this consider the infinite tower of states
created by the action of $\beta_0$ oscillators on a $-3/2$ picture state. Since $\beta_0$
has ghost number $-1$, these states will have arbitrarily low ghost numbers. Now
$\XX_0$ acting on such a state will give a state of picture number $-1/2$ and arbitrarily
low ghost numbers. But such states do not exist in the picture number $-1/2$ sector;
here at a fixed mass$^2$ level we can only have states with arbitrarily large ghost numbers
created by $\gamma_0$ oscillators. This shows that the
$\XX_0$ factor in the propagator
must annihilate all but a finite number of states created by the $\beta_0$ oscillators.
Since now we have a finite sum, our previous argument shows that the Feynman diagrams
computed with this propagator must be free from spurious singularities.

This argument has been somewhat abstract; but we shall illustrate this with a
simple example in appendix \ref{saspdeg}.

\sectiono{Superstring field theory: Master action} \label{ssftinv}

In this section we shall show that the action given in the last section can be regarded
as the gauge fixed version of a more general action satisfying BV
master equation. Our analysis will follow \cite{1508.05387}.

\subsection{Batalin-Vilkovisky quantization} \label{sbvrev}

In the standard Faddeev-Popov quantization of gauge theories, we first fix the gauge,
introduce ghosts and then carry out the path integral. In contrast in the BV
formalism we first introduce ghosts, expand the field space by introducing an anti-field
for every field, construct the master action, then fix the gauge and finally carry out the
path integral. In this subsection
we shall give a lightning review of the BV 
formalism\cite{bv1,9205088,9301097,bv,Henneaux:1992ig}.

For simplicity we shall work with a system with finite number of degrees of freedom 
$\{\phi_a\}$
but this analysis can easily be extended to field theories. $\{\phi_a\}$ could include
both Grassmann even and Grassmann odd variables. Suppose further that the classical
action  has a gauge invariance generated by a set of parameters 
$\{\lambda_\alpha\}$. $\{\lambda_\alpha\}$ may also include Grassmann even and
Grassmann odd variables. Furthermore there may be gauge invariance of the
$\{\lambda_\alpha\}$\plu\ -- deformations of $\{\lambda_\alpha\}$ 
that do not generate any change in $\{\phi_a\}$\plu\ -- generated by a set of parameters
$\{\xi_\ell\}$. This may continue arbitrary number of steps.
The BV prescription tells us to introduce a ghost variable $c^{(1)}_\alpha$
for each $\lambda_\alpha$ carrying Grassmann parity opposite to that of
$\lambda_\alpha$, a ghost variable $c^{(2)}_\ell$ for each $\xi_\ell$ carrying Grassmann
parity equal to that of $\xi_\ell$ and so on. Let us collectively call all these variables
$\{\Phi_r\}$. These will be called `fields'. Next for each field $\Phi_r$ we introduce an
anti-field $\Phi_r^*$ that carries Grassmann parity opposite to that of $\Phi_r$.
Given any pair of functions $F(\Phi, \Phi^*)$ and $G(\Phi,\Phi^*)$ of all the fields and
anti-fields, we now define
the anti-bracket
\be \label{edefanti}
\{ F, G\} = {\p_R F\over \p\Phi_r} {\p_L G \over \p \Phi_r^*} -
{\p_R F\over \p\Phi_r^*} {\p_L G \over \p \Phi_r}\, ,
\ee
and the $\Delta$ operator
\be 
\Delta F = {\p_R\over \p\Phi_r} {\p_L\over \p\Phi_r^*} F\, ,
\ee
where the  subscripts $L$ and $R$ of $\p$ denotes left and right derivatives.

The master action $S(\Phi,\Phi^*)$ is a function of $\Phi$ and
$\Phi^*$ that satisfies
the BV master equation
\be  \label{ebvm}
{1\over 2} \{S,S\} + \Delta S = 0\, ,
\ee
and reduces to the classical action in the  $g_s\to 0$ limit
if we set $\Phi^*_r=0$.
Note that at this stage we have not done any gauge fixing. Since the action has an
overall factor of $g_s^{-2}$, the classical master action $S_{cl}$, obtained from $S$ by
taking the $g_s\to 0$ limit, satisfies the classical master equation
$\{S_{cl},S_{cl}\}=0$.\footnote{The classical master action is to be distinguished
from the classical action. The latter is obtained from the former by setting all the anti-fields
to zero.} It follows from this that
the classical master action  has a
gauge invariance\cite{Henneaux:1992ig}
\be \label{ebvgauge}
\delta\Phi_r = \left\{\Phi_r, {\p_R S_{cl}\over \p\Phi_s}\Lambda_s
+ {\p_R S_{cl}\over \p\Phi_s^*}\Lambda_s^*\right\}, \quad
\delta\Phi_r^* = \left\{\Phi_r^*, {\p_R S_{cl}\over \p\Phi_s}\Lambda_s
+ {\p_R S_{cl}\over \p\Phi_s^*}\Lambda_s^*\right\}
\, ,
\ee
where $\Lambda_s,\Lambda_s^*$ are the infinitesimal gauge transformation parameters
which are independent of the fields and carry Grassmann parities opposite to that
of $\Phi_s,\Phi_s^*$.  
The master action $S$ must be chosen so that \refb{ebvgauge} reproduces the gauge
transformation laws of the classical theory when we set the anti-fields to zero and take the 
classical limit. The quantum generalization of this result
was discussed in \cite{9309027}, but we shall not require it for our analysis.

In the BV formalism
the gauge fixing corresponds to choosing a Lagrangian submanifold defined as
follows. Let us suppose that we find a complete set of new variables 
$\Xi_r(\Phi,\Phi^*)$ and $\Xi_r^*(\Phi,\Phi^*)$ such that
\be \label{ecc1}
\{\Xi_r, \Xi_s\}=0=\{\Xi_r^*, \Xi_s^*\}, \quad \{\Xi_r,\Xi_s^*\}=\delta_{rs}\, ,
\ee
and
\be \label{ecc2}
\prod_r d\Phi_r\wedge d\Phi_r^* = \prod_r d\Xi_r\wedge d\Xi_r^*\, .
\ee
Then $\Xi_r^*=0$ $\forall r$ describes a Lagrangian submanifold. It can be shown that
the physical quantities computed via path integral with integration measure
\be\label{ebvmeas}
\prod_r d\Phi_r\wedge d\Phi_r^* \prod_r \delta(\Xi_r^*) e^S
\ee
is independent of the choice of the Lagrangian submanifold. If we make the trivial
choice $\Xi_r=\Phi_r$, $\Xi_r^*=\Phi_r^*$ then we get back the original integral over
$\Phi_r$ weighted by $e^S$. This has unfixed gauge symmetry and therefore is not
amenable to perturbation theory. On the other hand a judicious choice of 
$\Xi_r$, $\Xi_r^*$ can fix the gauge and give us a 
path integral amenable to perturbation
theory.  The particular choice that will be relevant for our analysis is the exchange of
a certain number of fields with the corresponding anti-fields accompanied by a sign.
This clearly satisfies \refb{ecc1}, \refb{ecc2}. The corresponding gauge fixing
condition involves setting to zero certain set of fields and the anti-fields of the
complementary set. 

We can choose a slightly more general gauge $\Xi_r^*=\bar\Xi_r^*$ where 
$\bar\Xi_r$ are c-number background fields, and construct the 1PI
action for $\Xi_r$\plu\ by first computing the generating function of Green's function 
of $\Xi_r$ 
and then taking its Legendre transform. The resulting action may be 
written as $S_{1PI}(\{\bar\Xi_r\},\{\bar\Xi_r^*\})$ where
$\bar\Xi_r$ are the Legendre transformed variables.
Operationally $S_{1PI}$ can be constructed by summing over 1PI graphs
based on the action $S$, with $\Xi_r^*$ set equal to $\bar\Xi_r^*$
and $\Xi_r$\plu\ regarded as the quantum fields. At the end we set the external
$\Xi_r$\plu\ to $\bar\Xi_r$\plu. 
If we now regard $\bar\Xi_r$ and $\bar\Xi_r^*$ as fields
and conjugate anti-fields respectively, and define a new anti-bracket
between functions of $\bar\Xi_r$, $\bar\Xi_r^*$ by replacing
$\Phi_r$, $\Phi_r^*$ by $\bar\Xi_r$, $\bar\Xi_r^*$ in
\refb{edefanti},
 then $S_{1PI}$ can be
shown to satisfy the classical BV master equation 
$\{S_{1PI},S_{1PI}\}=0$\cite{bv1}. Therefore it will
be invariant under the
gauge transformation \refb{ebvgauge} with $S_{cl}$ replaced by $S_{1PI}$ and
$\Phi_r,\Phi_r^*$ replaced by $\bar\Xi_r,\bar\Xi_r^*$.

\subsection{The master action of superstring field theory} \label{smproof}

In the next three subsections we shall show that the action \refb{esgf} 
arises from gauge fixing of a theory in
the BV formalism\cite{1508.05387},
generalizing the corresponding results in open bosonic string
field theory\cite{thorn,bochicchio} and closed bosonic 
string field theory\cite{sonoda,hata1,hata,9206084}.
For this we have to first specify the full field content of the theory,
identify which of these are fields and which are the conjugate anti-fields and write
down the master action satisfying \refb{ebvm}. It turns out that in the BV formalism 
we have two sets of string fields:
$|\Psi\rangle\in\wh\HH_T$ and $|\wt\Psi\rangle\in\wt\HH_T$ without any further restriction.
The action is given by
\be\label{eactorg}
S={1\over g_s^2} \left[ -{1\over 2} \langle \wt\Psi | c_0^- Q_B \GG |\wt\Psi\rangle
+ \langle \wt\Psi | c_0^- Q_B |\Psi\rangle + \sum_{n=1}^\infty {1\over n!}
\cL \Psi^n\cR \right]
\, .
\ee
The division of the components of $\Psi$ and $\wt\Psi$ into fields
and  anti-fields proceeds as follows\cite{1508.05387}:
\begin{enumerate}
\item 
We divide $\wh \HH_T$ and $\wt\HH_T$ into two subsectors: $\wh\HH_+$
and $\wt\HH_+$ will contain states in $\wh\HH_T$ and $\wt\HH_T$ of
ghost numbers $\ge 3$, while  $\wh\HH_-$
and $\wt\HH_-$ will contain states in $\wh\HH_T$ and $\wt\HH_T$ of
ghost numbers $\le 2$. We introduce basis states $|\wh\vp^-_r\rangle$,
$|\wt\vp^-_r\rangle$, $|\wh\vp_+^r\rangle$ and $|\wt \vp_+^r\rangle$ of
$\wh\HH_-$,
$\wt\HH_-$, $\wh\HH_+$
and $\wt\HH_+$ satisfying orthonormality conditions\footnote{By an abuse
of notation we are using the same indices $r,s$ to label the new basis even though
the label runs over a smaller set for the new basis compared to the old basis $\{|\vp_r\rangle\}$,
$\{|\vp_r^c\rangle\}$. }
\be \label{einner}
\langle \wh\vp^-_r|c_0^- |\wt \vp_+^s \rangle = \delta_r{}^s=
\langle \wt \vp_+^s|c_0^- |\wh\vp^-_r\rangle, \quad 
\langle \wt\vp^-_r|c_0^- |\wh \vp_+^s \rangle = \delta_r{}^s=
\langle \wh \vp_+^s|c_0^- |\wt\vp^-_r\rangle\, ,
\ee
and the completeness relations
\be \label{eortho}
\sum_r |\wh\vp^-_r\rangle \langle \wt \vp_+^r|c_0^-
+ \sum_r |\wh\vp^r_+\rangle\langle \wt \vp_r^-|c_0^- ={\bf 1}, 
\quad 
\sum_r |\wt\vp^-_r\rangle \langle \wh \vp_+^r|c_0^-
+ \sum_r |\wt\vp^r_+\rangle\langle \wh \vp_r^-|c_0^- ={\bf 1}\, ,
\ee
acting on states in $\wh\HH_T$ and $\wt\HH_T$ respectively.
\item We now expand $\Psi$, $\wt\Psi$ as
\ben \label{ephiexpan}
|\wt\Psi\rangle &=& \sum_r |\wt\vp^-_r\rangle \wt\psi^r  
+\sum_r (-1)^{\gamma_r^*+1} |\wt\vp_+^r\rangle \psi_r^*\, , \nonumber \\
|\Psi\rangle -{1\over 2} 
\GG|\wt\Psi\rangle &=& \sum_r |\wh\vp^-_r\rangle \psi^r  
+ \sum_r (-1)^{\wt \gamma_r^*+1} |\wh\vp_+^r\rangle \wt\psi_r^* 
\, .
\een
Here $\gamma^*_r$, $\gamma_r$, $\wt \gamma^*_r$ and $\wt \gamma_r$ 
label the Grassmann
parities of $\psi^*_r$, $\psi^r$, $\wt \psi^*_r$ and $\wt\psi^r$
respectively. They in turn can be determined from the assignment of Grassmann
parities to the basis states as described below \refb{ecom} and the fact that
$|\Psi\rangle$ and $|\wt\Psi\rangle$ are both Grassmann even.  
\item We shall identify the variables 
$\{\psi^r, \wt\psi^r\}$ as `fields' and the
variables $\{\psi^*_r, \wt\psi^*_r\}$ as the conjugate `anti-fields' in the BV
quantization of the theory. It can be easily seen
that $\psi^r$ and $\psi^*_r$
carry opposite Grassmann parities as do $\wt \psi^r$ and $\wt\psi^*_r$. 
This is consistent with their identifications
as fields and conjugate anti-fields.
\item Given two functions $F$ and $G$ of all the fields and anti-fields,
we now define their anti-bracket in the standard way:
\be \label{eantib}
\{F, G\}= {\p_R F\over \p \psi^r} \, {\p_L G\over \p\psi^*_r}
+ 
{\p_R F\over \p \wt\psi^r} \, {\p_L G\over \p\wt\psi^*_r}
- {\p_R F\over \p \psi^*_r} \, {\p_L G\over \p\psi^r}
-
{\p_R F\over \p \wt\psi^*_r} \, {\p_L G\over \delta\wt\psi^r}\, ,
\ee
where the subscripts $R$ and $L$ of $\p$ denote left and right derivatives 
respectively.

The anti-bracket defined above may also be expressed in the following 
way\cite{1508.05387}. If
under an arbitrary variation of $\Psi$ and $\wt\Psi$
\be
\delta F = \langle F_R | c_0^-|\delta\wt\Psi\rangle + \langle \wt 
F_R | c_0^-|\delta\Psi\rangle = 
\langle \delta\wt\Psi | c_0^-|F_L\rangle + \langle
\delta\Psi | c_0^-|\wt F_L\rangle\, ,
\ee
and similarly for $G$, then
\be 
\{ F, G\} = -\left[ \langle \wt F_R | c_0^- \GG |\wt G_L\rangle + 
 \langle \wt F_R | c_0^- | G_L\rangle 
 +  \langle F_R | c_0^-  |\wt G_L\rangle \right]\, .
\ee
\item We also define
\be \label{edefDelta}
\Delta F \equiv {\p_R\over \p\psi^r} {\p_L F\over \p \psi^*_r}
+ {\p_R\over \p\wt\psi^r} {\p_L F\over \p \wt\psi^*_r}\, .
\ee
\end{enumerate}

Using \refb{eactorg} \refb{eortho}, \refb{ephiexpan} and \refb{eantib} one gets, after some algebra,
\be\label{ess}
g_s^4 \{S, S\} = - 2  \sum_{n} {1\over (n-1)!} \cL \Psi^{n-1} Q_B\Psi\cR 
- \sum_{m,n} {1\over m! n!} \cL \vp_ s \Psi^{m}\cR  \cL \vp_ r
\Psi^{n}\cR  \, \langle \vp_ s^c |c_0^- \GG | \vp_ r^c\rangle \, .  
\ee
Here $|\vp_ r\rangle$\plu\ denote the original choice of basis states in $\wh\HH_T$
before splitting it into $\wh\HH_\pm$.
On the other hand using \refb{eactorg} \refb{eortho}, \refb{ephiexpan} and 
\refb{edefDelta} we get
\be \label{edde}
\Delta S  = - {1\over 2\, g_s^2}\sum_n {1\over n!} \cL \Psi^n \vp_s \vp_ r\cR 
\langle \vp_s^c  | c_0^- \GG |\vp^c_ r\rangle \, . 
\ee
Using the identity \refb{evertex}, and eqs.~\refb{ess}, \refb{edde}
one can show that the action $S $ given in  \refb{eactorg}
satisfies the quantum BV master equation
\be \label{eqmaster}
{1\over 2} \{S , S \} + \Delta S  = 0\, .
\ee

With the interpretation of fields and anti-fields described above, we can regard the
ghost number 2 components of $\Psi$, $\wt\Psi$ 
as matter fields, i.e.\ analogue of the fields $\{\phi_a\}$ in
\S\ref{sbvrev}, the ghost number $\le 1$ components as ghosts and ghost number
$\ge 3$ components as anti-fields. If we set all the anti-fields to zero, then it follows from
\refb{eghcon} that the dependence on the ghost fields also drop out and the action
becomes a function of the matter fields only. The $g_s\to 0$ limit of this describes
the classical action. This has the same form as \refb{eactorg} with $\Psi$,
$\wt\Psi$ replaced by $\Psi_{cl}\in \wh \HH_T$, $\wt\Psi_{cl}\in \wt\HH_T$ carrying ghost
number 2, and $\cL\cdots \cR$ replaced by $\cL\cdots \cR_0$ denoting
the genus zero contribution to \refb{edefcurl}:
\be 
S_{cl}={1\over g_s^2} \left[ -{1\over 2} \langle \wt\Psi_{cl} | c_0^- Q_B \GG |\wt\Psi_{cl}\rangle
+ \langle \wt\Psi_{cl} | c_0^- Q_B |\Psi_{cl}\rangle + \sum_{n=3}^\infty {1\over n!}
\cL \Psi_{cl}^n\cR_0 \right]
\, .\ee
The sum over $n$ in the interaction term begins at $n=3$ since one and two point functions
on the sphere vanish in any SCFT. 

\subsection{Perturbation theory in Lorentzian signature space-time} \label{slorentz}

The  Feynman rules 
derived from the action \refb{eactorg}, as described in
\S\ref{ssft}, generate Euclidean Green's functions.
We can get the Lorentzian Green's functions from these by analytic continuation.
If $\{p^E_k\}$ for $k=1,\ldots  ,N$ denote the zero components of the
Euclidean external momenta in a Green's function, 
and if $\{p^0_k\}$ denote the zero components of Lorentzian
momenta, then they are related as $p^0_k=i p^E_k$. This means that up to an
overall normalization, the Lorentzian Green's functions $f(\{p^0_k\})$ are 
related to the Euclidean Green's functions $f_E(\{p^E_k\})$ via the relation
\be 
f(\{p^0_k\}) = f_E(\{p^0_k / i\})\, .
\ee
Therefore for real $\{p_k^0\}$, Euclidean Green's functions compute 
$f(\{u \, p^0_k\}) = f_E(\{p^0_k u /i\})$ with $u$ on the imaginary axis. Given this
we can determine $f(\{p_k^0\})$ via analytic continuation of the function
$f(\{u \, p^0_k) $ from the imaginary $u$ axis to $u=1$ along the first quadrant of the
complex $u$-plane.

However, for some applications it is useful to work directly with the
Feynman diagrams in the Lorentzian theory. For this
we need a weight factor $e^{iS}$ instead of $e^S$ in
the path integral. In this section we shall discuss its effects on the various equations derived earlier.
\begin{enumerate}
\item In the Feynman rules, the effect of replacing $e^S$ by $e^{iS}$ is
to multiply the propagator by $-i$ and the vertices by $i$. Since the amplitudes are
normalized so that the contribution to the amplitude from an elementary vertex is
given by the vertex without any normalization, each amplitude is also multiplied by an
overall factor of $i$. 
Therefore
this would leave \refb{efa1} unchanged, but would
require us to multiply \refb{efa2} by a factor of $-i$.
\item The $S\to iS$ replacement will change the BV master equation \refb{ebvm} to
\be  \label{ebvm1}
{1\over 2} \{S,S\} -i \,  \Delta S = 0\, .
\ee
It will also introduce factors of $\pm i$ into various other equations.
One quick way to determine the various extra factors of $i$ in various expressions is to
note that replacing $S$ by $iS$ in the exponent of the weight factor of the path integral can be
achieved by changing $g_s^2$ to $-i g_s^2$. Therefore in any expression we can recover the 
factors of $i$ by replacing $g_s^2$ by $-i g_s^2$. The  expressions
for the vertices and the propagators are
exceptions to these rules since, 
as has been discussed at the end of \S\ref{sAction}, we have
explicitly stripped off factors of $g_s^2$ from these.
\item 
As an application of the above rules, we see that we need to 
multiply the last term in \refb{evertex} by a factor of $-i$, and 
multiply the 
genus $g$ contribution in \refb{edefcurl} by a factor of
$(-i)^{g}$. One can easily verify that the action \refb{eactorg} satisfies
the modified BV master equation \refb{ebvm1} once we take into account these
changes in \refb{evertex}, and that the modified version of \refb{evertex} holds 
if we use the modified version of \refb{edefcurl}. The factor of
$(-i)^g$ inside the sum in \refb{edefcurl} may appear to be somewhat strange,
but this factor has a straightforward interpretation. In the Lorentzian theory,
while defining the correlation function on a genus $g$ Riemann surface, we have to
trace over states of the SCFT running in the loop. This in particular includes integration
over $g$ loop energies. Each of these integrals can be performed after a 
Euclidean rotation $k^0 \to i k^E$. 
These changes of variables generate a multiplicative factor
of $i^g$ that cancels the $(-i)^g$ factor coming from the effect of changing $S$ to $iS$.
Therefore it is natural to absorb the factor of $(-i)^g$ into the definition of the correlation function
on the genus $g$ Riemann surface and continue to use \refb{edefcurl} without any change. 
\end{enumerate}

\subsection{The reality of the action} \label{sreal}

So far we have not described whether the string field components $\psi_r$ and
$\wt\psi_r$ appearing in the expansion of $\Psi$ and $\wt\Psi$ are real or imaginary,
or whether they have a more complicated transformation under complex conjugation
{\it e.g.} complex conjugate of a particular $\psi_r$ may be related to a linear
combination of the other $\psi_s$\plu. The correct rule is determined by requiring that
the action $S$ is real, i.e.\ it remains invariant under complex conjugation.\footnote{For
real Grassmann variables $\{c_i\}$, the complex conjugate of $c_1\ldots c_n$
is taken to be $c_n\ldots c_1$ as usual.} This rule has been determined in 
\cite{1606.03455} and has been reviewed in appendix \ref{sareal},
but  we do not need the details for the rest of the analysis.
The result that is of importance is that it is possible to assign reality conditions
on the $\psi_r$\plu\ that make the action real.

Once the reality condition is determined one can also check if the action has the
correct sign. For example in the convention described in footnote \ref{fo10},
in which we use $e^S$ as the weight
factor in the Euclidean path integral,
the kinetic term of a physical real boson $\phi$ of mass $m$
in
momentum space must have
the form $-\int d^d k \,\phi(-k)  (k^2+m^2) \phi(k)$. 
It turns out that with the normalization
condition \refb{evacnorm}, \refb{evacnormA} 
the actions for both heterotic and type II string theories have the correct 
sign.\footnote{\cite{1606.03455} used a different sign in \refb{evacnormA} and therefore had 
a different sign of the action for type II string theory.}
This can be checked, for example, by computing the kinetic terms for the graviton field
in both theories using the conventions and reality conditions
described in appendix \ref{sareal}. 

\subsection{Gauge fixing}

In the BV formalism, given the master action we 
compute the quantum amplitudes 
by carrying out the usual path integral over a Lagrangian submanifold
of the full space spanned by $\psi^r$ and $\psi^*_r$. It is most convenient
to work in the Siegel gauge
\be \label{esiegauge}
b_0^+|\Psi\rangle =0, \quad b_0^+|\wt\Psi\rangle=0 \quad \Rightarrow
\quad b_0^+ \left(|\Psi\rangle
-{1\over 2} \GG|\wt\Psi\rangle\right)=0\, .
\ee
To see that this describes a Lagrangian submanifold, we divide the
basis states used in the expansion \refb{ephiexpan} into two
classes: those 
annihilated by $b_0^+$ and those annihilated by $c_0^+$. These two sets
are conjugates of each other under the inner product 
\refb{einner}. Now in the expansion given in \refb{ephiexpan}, Siegel gauge 
condition sets the coefficients of the basis states annihilated by $c_0^+$
to zero. Since in this expansion the fields and their anti-fields multiply
conjugate pairs of basis states, it follows that if the Siegel gauge condition sets
a field to zero then its conjugate anti-field remains unconstrained, and
if it sets an anti-field to zero then its conjugate field remains unconstrained.
Therefore this defines a Lagrangian submanifold.

It is now straightforward to verify that with the  constraint \refb{esiegauge}, 
the action \refb{eactorg}
reduces to the gauge fixed action \refb{esgf}. Therefore it reproduces correctly the
off-shell amplitudes described in \S\ref{ssft}.

\sectiono{Ward identities, 1PI action and effective action} \label{sward}

In this section we shall show how from the superstring field theory described
above, we can derive Ward identities for off-shell Green's functions and
partially integrate
out degrees of freedom to construct 1PI effective action and other types of effective
action.  Our discussion will mainly follow \cite{1609.00459}.

\subsection{Ward identity for off-shell amputated Green's function} \label{sgintro}

Let $G(A_1,\ldots  ,A_N)$ be the full off-shell `semi-amputated'
Green's function with external
states $A_1,\ldots  ,A_N$, 
obtained by summing over all Feynman diagrams with
external states $A_1,\ldots  ,A_N$, but dropping the {\it tree level} propagators of the
external states.\footnote{In the world-sheet description, this is the amplitude computed by the integral of 
$\Omega^{(g,m,n)}_{6g-6+2m+2n}(A_1,\ldots  ,A_N)$ with $N=m+n$
over the full section $\SSS_{g,m,n}$ of
$\wt\PP_{g,m,n}$.}
We impose Siegel gauge condition on the internal states,
but take the external states $A_1,\ldots  ,A_N$ to be arbitrary elements of 
$\wh \HH_T$. These Green's functions can suffer from divergences 
of type 1 mentioned in 
\S\ref{sintro}, but since we have a field theory they can be handled using representation
of the Feynman diagrams as integrals over momenta and sum over fields instead
of the Schwinger parameter representation. This will be discussed in more 
details in \S\ref{sloopcon}. $G$ does not suffer from divergences 
associated with mass renormalization since the external states are off-shell. However it
may suffer from tadpole divergences if  massless tadpoles are present.
In such cases the manipulations described below are formal. Nevertheless they
are useful since later we shall carry out similar manipulation with quantities which
do not suffer from such divergences.

Let us now
consider the combination $\sum_{i=1}^N G(A_1,\ldots  ,A_{i-1}, Q_B A_i,  A_{i+1},
\ldots  ,A_N)$. Since in a given Feynman diagram each $A_i$ must come from some vertex
$\cL\cdots \cR$,
the sum over $i$ can be organized into subsets, where in a given subset $Q_B$ acts
on different external states of the same vertex. This can then be simplified
using \refb{evertex}. This gives
\ben \label{egidpre} 
&&\hskip .2in \sum_{i=1}^N  G(A_1,\ldots  ,A_{i-1}, Q_B A_i,
A_{i+1} \ldots  ,A_N)  \nonumber \\
&& = -  
{1\over 2} \sum_{\ell,k\ge 0\atop \ell+k=N} \sum_{\{ i_a;a=1,
\ldots \ell\} , \{ j_b;b=1,\ldots k\} \atop
\{ i_a\} \cup \{ j_b\}  = \{ 1,\ldots N\}
}G( A_{i_1}, \ldots  ,A_{i_\ell} ,\vp_s)  G( \vp_r, A_{j_1} \ldots  ,A_{j_k})
\langle \vp_s^c | c_0^- \GG | \vp_r^c\rangle \nonumber \\
&& -   {1\over 2} \, g_s^2\, G(A_1, \ldots  ,A_N ,\vp_s, \vp_r) \, \langle \vp_s^c | c_0^- \GG 
| \vp_r^c\rangle \nonumber \\
&& -  
{1\over 2} \sum_{\ell,k\ge 0\atop \ell+k=N} \sum_{\{ i_a;a=1,
\ldots \ell\} , \{ j_b;b=1,\ldots k\} \atop
\{ i_a\} \cup \{ j_b\}  = \{ 1,\ldots N\}
}\bigg[ - G( A_{i_1}, \ldots  ,A_{i_\ell} ,Q_B\vp_s)  G( \vp_r, A_{j_1} \ldots  ,A_{j_k})
\nonumber \\
&& \hskip 1in - (-1)^{\gamma_s}
G( A_{i_1}, \ldots  ,A_{i_\ell} ,\vp_s)  G( Q_B\vp_r, A_{j_1} \ldots  ,A_{j_k}) \bigg]
\langle \vp_s^c | c_0^- b_0^{+} (L_0^+)^{-1} \GG | \vp_r^c\rangle
\nonumber \\
&& -   {g_s^2\over 2} \bigg[-G(A_1, \ldots  ,A_N ,Q_B\vp_s, \vp_r) 
- (-1)^{\gamma_s} G(A_1, \ldots  ,A_N ,\vp_s, Q_B\vp_r)\bigg]\, \langle \vp_s^c | c_0^- 
b_0^{+} (L_0^+)^{-1}  \GG 
| \vp_r^c\rangle
\, . \nonumber \\ 
\een 
The first two terms on the right hand side 
represent the contribution from the right hand side of \refb{evertex} when we
use \refb{evertex} to simplify the contribution from  
individual vertices of the Feynman diagram.
The other two terms on the right hand side come
from the fact that while using \refb{evertex} for a given vertex, 
we have to subtract the terms where $Q_B$ acts on the
legs of the vertex connected to internal propagators
since on the left
hand side of \refb{egid} $Q_B$ only acts on the external states. 
The third term represents the
contribution from a graph in which a propagator connects two 
otherwise disjoint Feynman diagrams and there is a $Q_B$ insertion on one of the
ends of the propagator. 
The last term represents the contribution from a graph in which a propagator
connects two external lines of a connected Feynman diagram, and there is a 
$Q_B$ insertion at one of the ends of the propagator. The overall minus signs
in front of the third and the fourth terms come from having to move these from the left
hand side of the equation, where they appear naturally, to the right hand side.
The minus signs inside the square brackets
come from the ones on the right hand sides of \refb{efa1} and \refb{efa2}. The
$(-1)^{\gamma_s}$ factors arise from having to move $Q_B$ through $\vp_s$.
In
the third term, we have included a factor of 1/2 to 
compensate for the double counting
associated with the $\{i_a\}\leftrightarrow \{j_b\}$ exchange. 
The 1/2 in the last factor arises from the right hand side of \refb{efa2}.

Using the completeness relation \refb{ecom}   and \refb{egrel}
we can now move $Q_B$ inside the
matrix element $\langle \vp_s^c | c_0^- 
b_0^{+} (L_0^+)^{-1}  \GG 
| \vp_r^c\rangle$ in the third and the fourth terms, {\it e.g.} we have 
\be \label{eman1}
Q_B |\vp_s\rangle \langle \vp_s^c | c_0^- 
b_0^{+} (L_0^+)^{-1}  \GG 
| \vp_r^c\rangle = Q_B b_0^{+} (L_0^+)^{-1}  \GG 
| \vp_r^c\rangle = |\vp_s\rangle \langle \vp_s^c | c_0^- 
Q_B b_0^{+} (L_0^+)^{-1}  \GG 
| \vp_r^c\rangle \, , 
\ee
and 
\ben \label{eman2}
&& (-1)^{\gamma_s} Q_B |\vp_r\rangle \langle \vp_s^c | c_0^- 
b_0^{+} (L_0^+)^{-1}  \GG 
| \vp_r^c\rangle =  Q_B |\vp_r\rangle \langle \vp_r^c 
| c_0^-  b_0^{+} (L_0^+)^{-1}  \GG 
| \vp_s^c\rangle 
=  Q_B 
b_0^{+} (L_0^+)^{-1}  \GG 
| \vp_s^c\rangle \nonumber \\ &=&
 |\vp_r\rangle \langle \vp_r^c | c_0^- Q_B 
b_0^{+} (L_0^+)^{-1}  \GG 
| \vp_s^c\rangle  = |\vp_r\rangle \langle \vp_s^c | c_0^-  
b_0^{+} Q_B (L_0^+)^{-1}  \GG 
| \vp_r^c\rangle
\, .
\een
 This allows us to express \refb{egidpre} as
\ben \label{egid} 
&&\hskip .2in \sum_{i=1}^N  G(A_1,\ldots  ,A_{i-1}, Q_B A_i,
A_{i+1} \ldots  ,A_N)  \nonumber \\
&& = -  
{1\over 2} \sum_{\ell,k\ge 0\atop \ell+k=N} \sum_{\{ i_a;a=1,
\ldots \ell\} , \{ j_b;b=1,\ldots k\} \atop
\{ i_a\} \cup \{ j_b\}  = \{ 1,\ldots N\}
}G( A_{i_1}, \ldots  ,A_{i_\ell} ,\vp_s)  G( \vp_r, A_{j_1} \ldots  ,A_{j_k})
\langle \vp_s^c | c_0^- \GG | \vp_r^c\rangle \nonumber \\
&& -   {1\over 2} \, g_s^2\, G(A_1, \ldots  ,A_N ,\vp_s, \vp_r) \, \langle \vp_s^c | c_0^- \GG 
| \vp_r^c\rangle \nonumber \\
&& + 
{1\over 2} \sum_{\ell,k\ge 0\atop \ell+k=N} \sum_{\{ i_a;a=1,
\ldots \ell\} , \{ j_b;b=1,\ldots k\} \atop
\{ i_a\} \cup \{ j_b\}  = \{ 1,\ldots N\}
}G( A_{i_1}, \ldots  ,A_{i_\ell} ,\vp_s)  G( \vp_r, A_{j_1} \ldots  ,A_{j_k})
\langle \vp_s^c | c_0^- \{Q_B, b_0^+\} (L_0^+)^{-1}\GG | \vp_r^c\rangle \nonumber \\
&& +  {1\over 2} \, g_s^2\, G(A_1, \ldots  ,A_N ,\vp_s, \vp_r) \, 
\langle \vp_s^c | c_0^- \{Q_B, b_0^+\} (L_0^+)^{-1}\GG | \vp_r^c\rangle \, . 
\een

Using the relations
$Q_Bb_0^++b_0^+ Q_B=L_0^+$ one can now
show that on the right hand side of  \refb{egid} the third term
cancels the first term and the fourth term cancels the second
term. This gives us the Ward identity for the 
off-shell Green's function
\be \label{efullG}
\sum_{i=1}^N  G(A_1,\ldots  ,A_{i-1}, Q_B A_i,
A_{i+1} \ldots  ,A_N) =0\, .
\ee
We remind the reader again that this identity is formal if there are massless tadpoles
present in the theory.

\subsection{Ward identity for 1PI amplitudes and 1PI action} \label{s1PI}

Historically, 1PI effective action was constructed before 
the introduction of the BV master
action\cite{1411.7478,1501.00988}, 
and the properties of the 1PI effective action were studied directly using the world-sheet
description of the interaction vertices. Here we shall follow a slightly different approach
in which we regard the 1PI action as the one derived from the BV master action 
following the procedure described in the last paragraph of \S\ref{sbvrev}, and derive
its properties from the properties of the BV master action. This makes it 
manifest that the Green's functions computed using tree graphs of the 1PI
action are identical to the ones computed using the full set of Feynman diagrams
of the master action, so that we can use either description for studying their properties.

The 1PI effective action described at the end of \S\ref{sbvrev} 
can be constructed using the 1PI amplitudes in the Siegel gauge, but
we take the external states to be
general elements of $\wh\HH_T$ without satisfying any
gauge condition. This implements 
the general gauge choice $\Xi_r^*=\bar\Xi_r^*$
described in \S\ref{sbvrev}. Let $\{A_1\ldots A_n\}$ denote the
1PI amplitude
of the external states $A_1,\ldots  ,A_n$ obtained by summing over all
the 1PI graphs. 
This is well defined even if the theory has massless tadpoles, since the 
sum of 1PI diagrams does not include the tadpole diagrams.
$\{A_1\ldots A_N\}$ will
satisfy an identity similar to \refb{egid} with $G(A_1,\ldots  ,A_n)$ replaced by
$\{A_1\ldots A_n\}$, and without the third term on the right hand side of 
\refb{egid}. This is due to the fact that by definition, 1PI amplitudes do not 
include sum over Feynman diagrams in which a single propagator connects two
other Feynman diagrams. Therefore the first term on the right hand side remains
uncanceled and we arrive at
the identity:
\ben\label{e1pi}
&&\sum_{i=1}^N \{A_1\ldots A_{i-1} (Q_B A_i)
A_{i+1} \ldots A_N\} \nonumber \\
&=& -  
{1\over 2} \sum_{\ell,k\ge 0\atop \ell+k=N} \sum_{\{i_a;a=1,\ldots \ell\}, \{j_b;b=1,\ldots k\}\atop
\{i_a\}\cup \{j_b\} = \{1,\ldots N\}
}\{A_{i_1} \ldots A_{i_\ell} \vp_s\} \{\vp_r A_{j_1} \ldots A_{j_k}\}
\langle \vp_s^c | c_0^- \GG | \vp_r^c\rangle\, .
\een
We can now construct the 1PI action by replacing
$\cL\cdots\cR$  by $\{\cdots\}$ in 
\refb{eactorg}: 
\be\label{eact1PI}
S_{1PI}={1\over g_s^2} \left[ -{1\over 2} \langle \wt\Psi | c_0^- Q_B \GG |\wt\Psi\rangle
+ \langle \wt\Psi | c_0^- Q_B |\Psi\rangle + \sum_{n=1}^\infty {1\over n!}
\{ \Psi^n\} \right]
\, .
\ee
The variables $\bar\Xi_r$, $\bar\Xi_r^*$ described at the end of
\S\ref{sbvrev} are identified as the components $\psi^r$, $\wt\psi^r$, $\psi_r^*$ and
$\wt\psi_r^*$  
of $\Psi$ and $\wt\Psi$ described in \refb{ephiexpan}. The anti-bracket is defined as
in \refb{eantib} and 
one can verify using \refb{e1pi} that the  action 
\refb{eact1PI} satisfies the
classical master equation
\be
\{S_{1PI}, S_{1PI}\} = 0\, .
\ee
This is in accordance with the general result in the BV formalism described at the
end of \S\ref{sbvrev}.

Using \refb{e1pi} one can also show that as expected from \refb{ebvgauge},
the action \refb{eact1PI} is
invariant under the gauge transformation 
\be  \label{egauge}
|\delta\Psi\rangle = Q_B|\Lambda\rangle + \sum_{n=0}^\infty {1\over n!} 
\GG[\Psi^n \Lambda]\, , \quad  
|\delta\wt\Psi\rangle = Q_B|\wt\Lambda\rangle +
\sum_{n=0}^\infty {1\over n!} 
[\Psi^n \Lambda]\, , 
\ee
where $|\Lambda\rangle$ is an arbitrary Grassmann odd
state in $\wh\HH_T$,  $|\wt\Lambda\rangle$ is an arbitrary Grassmann odd
state in $\wt\HH_T$, and
given a set of states $A_1,\ldots  ,A_N\in\wh\HH_T$, we define a state
$[A_1\ldots A_N]\in \wt\HH_T$ via the relation
\be \label{edefsq}
\langle A_0|c_0^- |[A_1\ldots A_N]\rangle = \{A_0A_1\ldots A_N\}\, ,
\ee
for any state $A_0\in \wh\HH_T$. 

The semi-amputated Green's functions $G$ introduced in \S\ref{sgintro} can be 
computed by summing over {\it tree level} Feynman diagrams of the 1PI 
action. This however can suffer from divergences associated with massless
tadpoles. We shall address this in \S\ref{svac}. 

The 1PI amplitude is given by an expression similar to \refb{edefcurl}
\be\label{edefcurl1PI}
\{ K_1\ldots K_m L_1\ldots L_n\} =\sum_{g=0}^\infty
g_s^{2g} \, \int_{\RR_{g,m,n}}
\Omega^{(g,m,n)}_{6g-6+2m+2n}(K_1,\ldots  ,K_m,L_1,\ldots  ,L_n)\, , 
\ee
where $\RR_{g,m,n}$ now denotes a subspace of  $\wt\PP_{g,m,n}$ 
given by the union of the section segments of
all 1PI Feynman diagrams of genus $g$, and $m$ NS and $n$ R punctures.
The relation 
\refb{e1pi} can also be derived using an identity similar to the one given in
\refb{eboundaryq}, with $\oR$ replaced by $\RR$ and the $\Delta$ terms 
being absent:
\ben  \label{eboundaryqR}
\p \RR_{g,m,n} &=& -{1\over 2} \sum_{g_1,g_2\atop g_1+g_2=g} 
\sum_{m_1,m_2\atop m_1+m_2 = m+2}
\sum_{n_1,n_2\atop n_1+n_2 = n}
{\bf S}[\{\RR_{g_1,m_1,n_1} , \RR_{g_2,m_2,n_2}\}] \nonumber \\ &&
-{1\over 2} \sum_{g_1,g_2\atop g_1+g_2=g} 
\sum_{m_1,m_2\atop m_1+m_2 = m}
\sum_{n_1,n_2\atop n_1+n_2 = n+2}
{\bf S}[\{\RR_{g_1,m_1,n_1} ; \RR_{g_2,m_2,n_2}\}] 
\, .
\een
This again follows from the requirement that the regions $\RR_{g,m,n}$ and
their plumbing fixture \refb{eplumb}, {\it with the two punctures that are
sewed now always lying
on different Riemann surfaces}, give the full section 
$\SSS_{g,m,n}$\cite{1411.7478,1501.00988}. The requirement of the punctures 
lying on different Riemann surfaces is a reflection of the fact that the {\it tree level}
graphs computed with 1PI vertices reproduce the full amplitude.

At the level of the 1PI action the theory admits a consistent truncation in which
we set all the R sector fields to zero. Furthermore, since now $\GG$ is the
identity operator, the $\wt\Psi$ equation
of motion takes the form $Q_B(\wt\Psi - \Psi)=0$, and can be satisfied by setting
$\wt\Psi=\Psi$. This gives an action of the form
\be
{1\over g_s^2} \left[ {1\over 2} \langle \Psi | c_0^- Q_B |\Psi\rangle
+ \sum_{n=1}^\infty {1\over n!}
\{ \Psi^n\} \right], \quad |\Psi\rangle\in \HH_{-1}\, 
\, ,
\ee
which can be used to compute all amplitudes involving only NS sector external
states. There is a similar truncation for the classical action and
also for the NSNS sector fields of type II string theories.

\subsection{Effective superstring field theory} \label{seff}

Let us suppose that we have a projection operator $P$ on a subset of string fields
satisfying the conditions
\be \label{epprop}
[P,b_0^\pm]=0, \quad [P,c_0^\pm]=0,
\quad [P,L_0^\pm]=0, \quad [P, \GG]=0, \quad [P,Q_B]=0\, .
\ee
An example of $P$ would be the projection operator on the mass$^2$ level zero fields
-- fields whose tree level
propagator in the Siegel gauge takes the form of that of a
massless field.  $P$ could also
be a projection operator into fields of any other fixed
mass$^2$ level or a set of mass$^2$ levels {\it e.g.} all fields below a 
certain mass$^2$ level.
Another example of $P$ in toroidally compactified string theory, 
relevant for possible construction of double field theory\cite{0904.4664},
is the projection on fields whose contribution to the mass comes only from
the momentum and winding modes but not from oscillator modes. 
In what follows we shall not assume any
property of $P$ other than the one given in \refb{epprop}.

Consider a set of $P$ invariant off-shell states $a_1,\ldots  ,a_N$. 
We denote by $\cL a_1\ldots a_N\cR_e$ the total contribution to
the amplitude with external states $a_1,\ldots  ,a_N$ 
from {\it all} the Feynman diagrams of superstring field theory, but with the
propagator factors appearing in \refb{efa1}, \refb{efa2} replaced by  
$\langle \vp_s^c | c_0^- b_0^{+} (L_0^+)^{-1} \GG (1-P) | \vp_r^c\rangle$.
This removes the contributions of $P$ invariant fields from the
propagator. Therefore $\cL a_1\ldots a_N\cR_e$ can be regarded as the contribution 
to the off-shell amplitude due to the elementary
$N$-point vertex of the effective theory, obtained by 
integrating out the $P$ non-invariant fields.
Even in the presence of tadpoles of 
mass$^2$ level zero fields, these
amplitudes do not suffer from tadpole
divergences of the kind mentioned at the
beginning of \S\ref{sgintro}
as long as $P$ invariant subspace includes the mass$^2$ level zero fields.
We can now repeat the argument leading to \refb{egid} with 
$G(\cdots)$ replaced by $\cL \cdots\cR_e$.
On the left hand side of \refb{egid} and the first two terms on the right hand side
of \refb{egid} we simply replace $G(\cdots)$ by $\cL\cdots\cR_e$, 
but in the last two terms 
of \refb{egid} the propagator factors will now have additional insertions of $(1-P)$ 
since this is the propagator used in the definition of $\cL\cdots\cR_e$. 
This gives 
\ben \label{ewid}
&&\sum_{i=1}^N  \cL a_1\ldots a_{i-1} (Q_B a_i)
a_{i+1} \ldots a_N\cR_e  \nonumber \\
&=& -  
{1\over 2} \sum_{\ell,k\ge 0\atop \ell+k=N} \sum_{\{ i_a;a=1,
\ldots \ell\} , \{ j_b;b=1,\ldots k\} \atop
\{ i_a\} \cup \{ j_b\}  = \{ 1,\ldots N\}
}\cL a_{i_1} \ldots a_{i_\ell} \vp_s\cR_e  \cL \vp_r a_{j_1} \ldots a_{j_k}\cR_e
\langle \vp_s^c | c_0^- \GG | \vp_r^c\rangle \nonumber \\
&& -  {1\over 2} \, g_s^2 \cL a_1 \ldots a_N \vp_s \vp_r\cR_e \, \langle \vp_s^c | c_0^- \GG 
| \vp_r^c\rangle \nonumber \\
&& +
{1\over 2} \sum_{\ell,k\ge 0\atop \ell+k=N} \sum_{\{ i_a;a=1,
\ldots \ell\} , \{ j_b;b=1,\ldots k\} \atop
\{ i_a\} \cup \{ j_b\}  = \{ 1,\ldots N\}
} \cL a_{i_1} \ldots a_{i_\ell}  \vp_s\cR_e  \cL \vp_r a_{j_1} \ldots a_{j_k}\cR_e 
\nonumber \\ &&  \hskip 3in \times
\langle \vp_s^c | c_0^- \{Q_B, b_0^{+}\} (L_0^+)^{-1} \GG (1-P)| \vp_r^c\rangle 
\nonumber \\
&& + {1\over 2} \, g_s^2 \cL a_1 \ldots a_N \vp_s \vp_r\cR_e
 \langle \vp_s^c | c_0^- 
\{Q_B, b_0^{+}\} (L_0^+)^{-1}  \GG (1-P)
| \vp_r^c\rangle
\, .  
\een  
\noindent 
Now the third and the fourth terms on the right hand side cancel the first and
the second terms only partially, leaving behind terms proportional to
$\langle \vp_s^c | c_0^- \GG 
\, P | \vp_r^c\rangle$:
\ben \label{ewidmid}
&&\sum_{i=1}^N  \cL a_1\ldots a_{i-1} (Q_B a_i)
a_{i+1} \ldots a_N\cR_e  \nonumber \\
&=& -  
{1\over 2} \sum_{\ell,k\ge 0\atop \ell+k=N} \sum_{\{ i_a;a=1,
\ldots \ell\} , \{ j_b;b=1,\ldots k\} \atop
\{ i_a\} \cup \{ j_b\}  = \{ 1,\ldots N\}
}\cL a_{i_1} \ldots a_{i_\ell} \vp_s\cR_e  \cL \vp_r a_{j_1} \ldots a_{j_k}\cR_e
\langle \vp_s^c | c_0^- \GG \, P| \vp_r^c\rangle \nonumber \\
&& -  {1\over 2} \, g_s^2 \cL a_1 \ldots a_N \vp_s \vp_r\cR_e \, \langle \vp_s^c | c_0^- 
\GG\, P 
| \vp_r^c\rangle\, .
\een
If we denote by $\{|\chi_\alpha\rangle\}$ and 
$\{|\chi_\alpha^c\rangle\}$
the basis states  in $P\wh\HH_T$
and $P\wt\HH_T$ respectively, satisfying
\be \label{echic}
\langle\chi_\alpha|c_0^-|\chi^c_\beta\rangle = \delta_{\alpha\beta}, \quad
\langle\chi^c_\beta|c_0^-|\chi_\alpha\rangle = \delta_{\alpha\beta},
\ee 
then we can express \refb{ewidmid} as 
\ben \label{ewidfin}
&&\sum_{i=1}^N  \cL a_1\ldots a_{i-1} (Q_B a_i)
a_{i+1} \ldots a_N\cR_e  \nonumber \\
&=& -  
{1\over 2} \sum_{\ell,k\ge 0\atop \ell+k=N} \sum_{\{ i_a;a=1,
\ldots \ell\} , \{ j_b;b=1,\ldots k\} \atop
\{ i_a\} \cup \{ j_b\}  = \{ 1,\ldots N\}
}\cL a_{i_1} \ldots a_{i_\ell} \chi_\alpha\cR_e  \cL \chi_\beta a_{j_1} \ldots a_{j_k}\cR_e
\langle \chi_\alpha^c | c_0^- \GG | \chi_\beta^c\rangle \nonumber \\
&& -  {1\over 2} \, g_s^2\, \cL a_1 \ldots a_N \chi_\alpha\chi_\beta\cR_e \, \langle 
\chi_\alpha^c | c_0^- \GG 
| \chi_\beta^c\rangle\, .
\een

Given the identity \refb{ewidfin}
one can now construct the effective string field theory action of $P$
invariant fields  $\Chi\in P\wh\HH_T$, $\wt\Chi\in P\wt\HH_T$ satisfying BV
master equation:
\be\label{emaster}
S_e={1\over g_s^2} \left[ -{1\over 2} \langle \wt\Chi | c_0^- Q_B \GG |\wt\Chi\rangle
+ \langle \wt\Chi | c_0^- Q_B |\Chi\rangle + \sum_{n=1}^\infty {1\over n!}
\cL \Chi^n\cR_e\right]
\, .
\ee
The proof that it satisfies the master equation follows from \refb{ewidfin} in a manner
identical to that described in \S\ref{smproof}. This action contains the full information 
about the amplitudes involving external $P$ invariant states.  Even 
though we shall carry out
our subsequent analysis with the full string field theory action, all the analysis can
be repeated with the effective action described here.

The utility of the 
effective action constructed above lies in the fact that if $P$ projects to finite 
dimensional subspaces of $\wh\HH_T$, $\wt\HH_T$ for a given momentum, 
then there are only a finite
number of fields and we do not have to deal with sum over infinite number of 
intermediate states in Feynman diagrams.  In
particular construction of the propagator in the shifted background, to be described in
\refb{edefDeltaI}, will require inverting a finite dimensional matrix. However for this
to be useful, we need to ensure that  that we do not integrate out any field that can
appear as initial or final state in the scattering amplitude. For a given amount of
center of mass energy $E_{cm}$,
this can be achieved if we integrate out all fields whose masses are
larger than  $E_{cm}$ but keep all fields whose masses are less than $E_{cm}$.  
 
Following a procedure similar to that for the original action, we can also construct
a 1PI effective action for the restricted string fields, with the 1PI action taking the
from
\be\label{e1PIeff}
S_{e,1PI}
={1\over g_s^2} \left[ -{1\over 2} \langle \wt\Chi | c_0^- Q_B \GG |\wt\Chi\rangle
+ \langle \wt\Chi | c_0^- Q_B |\Chi\rangle + \sum_{n=1}^\infty {1\over n!}
\{ \Chi^n\}_e\right]
\, ,
\ee
where the 1PI vertex $\{a_1\ldots a_N\}_e$ is given by the sum of all 1PI Feynman
diagrams derived from the action \refb{emaster} with external states $a_1,\ldots  ,a_N$.
It satisfies the identity
\ben \label{ewidfin1PI}
&&\sum_{i=1}^N  \{ a_1\ldots a_{i-1} (Q_B a_i)
a_{i+1} \ldots a_N\}_e  \nonumber \\
&=& -  
{1\over 2} \sum_{\ell,k\ge 0\atop \ell+k=N} \sum_{\{ i_a;a=1,
\ldots \ell\} , \{ j_b;b=1,\ldots k\} \atop
\{ i_a\} \cup \{ j_b\}  = \{ 1,\ldots N\}
}\{ a_{i_1} \ldots a_{i_\ell} \chi_\alpha\}_e  \{ \chi_\beta a_{j_1} \ldots a_{j_k}\}_e
\langle \chi_\alpha^c | c_0^- \GG | \chi_\beta^c\rangle 
\, .
\een

Finally note that although the construction of the vertices $\cL\cdots\cR_e$ 
and $\{\cdots\}_e$ described
above seems to require summing over infinite number of $P$ non-invariant
intermediate states in string field theory amplitudes, we could proceed differently,
namely take the off-shell amplitude for $P$ invariant states and subtract from this
the contribution from intermediate $P$ invariant states. 
As a simple example we can consider the tree level contribution to
$\cL a_1 \ldots a_4\cR_e$. This is given by
\ben
\cL a_1 \ldots a_4\cR_e &=& G_{tree}(a_1,\ldots  ,a_4) 
 \nonumber \\
&+& G_{tree}(a_1, a_2, \chi_\alpha) 
\, G_{tree}(\chi_\beta, a_3, a_4)\,  \langle \chi_\alpha^c | c_0^- 
b_0^{+} (L_0^+)^{-1}  \GG
| \chi_\beta^c\rangle  \nonumber \\
&+&  G_{tree}(a_1, a_3, \chi_\alpha) 
\, G_{tree}(\chi_\beta, a_2, a_4)\,  \langle \chi_\alpha^c | c_0^- 
b_0^{+} (L_0^+)^{-1}  \GG
| \chi_\beta^c\rangle \nonumber \\
&+&
G_{tree}(a_1, a_4, \chi_\alpha) 
\, G_{tree}(\chi_\beta, a_2, a_3)\,  \langle \chi_\alpha^c | c_0^- 
b_0^{+} (L_0^+)^{-1}  \GG
| \chi_\beta^c\rangle
\een
where $G_{tree}$ denotes the full tree level amplitude. This expresses 
$\cL a_1 \ldots a_4\cR_e$ in terms of amplitudes involving $P$ invariant states
only.
The detailed procedure
for doing this in the general case
has been described in \cite{1609.00459}. In this approach neither 
the construction of
the interaction vertex of the effective field theory nor further manipulations involving
it require having to explicitly deal with $P$ non-invariant states. Therefore the full
analysis may be carried out only with finite number of states.

\subsection{Field redefinition} \label{scomredef}

As discussed in \S\ref{scomstub}, there is  a lot of freedom in the choice of
the section segments $\oR_{g,m,n}$. 
This
includes in particular the freedom of adding stubs to
the vertices as described in \S\ref{scomstub}. These different choices
will lead to different superstring field theory action. It was shown in 
\cite{9301097} (in the context of bosonic string theory) that these
different actions are related to each other by a symplectic transformation of the
fields.  For 1PI effective action described in \S\ref{s1PI}, 
these different choices
correspond to ordinary field redefinitions\cite{1411.7478,1501.00988}, 
and therefore leave the 
physical quantities like the renormalized masses and S-matrix invariant. 
 This has been reviewed briefly in appendix \ref{safield}.

\sectiono{Vacuum shift, mass renormalization, unbroken (super)symmetry} \label{svac}

So far we have described the construction of string field theory / effective field theory
in the original background described by world-sheet superconformal field theory,
that solves the classical equations of motion of string field theory. 
In this section we shall describe, following \cite{1411.7478,1501.00988,1508.02481}, 
how to
systematically take into account the effect
of quantum corrections on the vacuum and mass spectrum, and also analyze the fate
of global symmetries under quantum corrections. Although we shall present our analysis
using the full 1PI effective action of superstring field theory, it holds also for the 1PI
effective action given in \refb{e1PIeff}
in which a subset of string fields have been integrated out.

\subsection{Equations of motion} \label{seom}

The equations of motion of the 1PI
effective string field theory, obtained by varying \refb{eact1PI} with respect to
the string field components, takes
the form
\ben \label{eeompre}
&& Q_B (|\Psi\rangle-\GG|\wt\Psi\rangle) = 0, \nonumber \\
&& Q_B|\wt\Psi\rangle + \sum_{n=1}^\infty {1\over (n-1)!} [\Psi^{n-1}] = 0\, ,
\een
with $[A_1\ldots A_N]$ defined as in \refb{edefsq}.
Multiplying the second equation by $\GG$ from the left and adding it to the
first equation we get
\be \label{eeom}
Q_B|\Psi\rangle + \sum_{n=1}^\infty {1\over (n-1)!} \GG[\Psi^{n-1}]=0\, . 
\ee
This is the interacting equation of motion for the $|\Psi\rangle$ field. Given a solution
to \refb{eeom}, the second equation of \refb{eeompre} determines $\wt\Psi$ up to
addition of free field equations of motion $Q_B |\delta\wt\Psi\rangle=0$. This shows that the
degrees of freedom contained in $\wt\Psi$ are free fields.

\subsection{Vacuum solution} \label{seomvac}

Our first task will be to look for solution(s) to \refb{eeom} that describes the 
quantum corrected vacuum state.\footnote{In most cases the
procedure described here yields results in agreement with the ad hoc procedure
described in \cite{1404.6254}.
}
Now it follows from the ghost number conservation law \refb{eghcon}, and the definition
of $[\cdots]$ given in \refb{edefsq}, that $[A_1\cdots A_N]$ has total ghost number
$3+\sum_{i=1}^N (n_i-2)$. Therefore we can look for solutions to \refb{eeom} with
string field carrying ghost number 2 only, i.e.\ only the matter fields setting all other
fields to zero, since in this case
both terms in eq.~\refb{eeom} will have ghost number 3. While looking for vacuum solutions
we shall focus on this sector. In the same spirit we shall set the R-sector fields to zero
and restrict to string field configurations carrying zero momentum
while looking for vacuum solution. In type II string theory we shall set
NSR, RNS and RR sector fields to zero, although in principle we could also look for
vacuum solutions with non-zero RR background. In this case once a solution to
\refb{eeom} has been found we can find solution to \refb{eeompre} by setting 
$\wt\Psi=\Psi$ since in the NS sector of the heterotic theory and NSNS sector 
of
type II theory, $\GG$ is the identity operator. So we focus on \refb{eeom}.

Since $[~]$ -- the $n=1$ term on the right hand side of \refb{eeom} --
gets non-zero contribution from Riemann surfaces of genus $\ge 1$ due to non-vanishing
one point function $\{A\}$, 
$|\Psi\rangle=0$ is not a solution to the equations of motion \refb{eeom}.
We shall now describe 
a systematic procedure for finding the vacuum solution 
$|\Psi_{\rm vac}\rangle$ -- a solution to \refb{eeom} in the NS sector
carrying zero momentum\cite{1411.7478}. This solution
is constructed iteratively as a power series in the string coupling $g_s$ 
starting at order $g_s$.\footnote{We 
are assuming here
that the vacuum solution admits an expansion in powers of
$g_s$. This includes the case of perturbative vacuum where the solution
will have expansion in powers of $g_s^2$ -- we simply will get
$|\Psi_{2k+1}\rangle=|\Psi_{2k}\rangle$ for all integer $k$.
However an interesting situation arises in SO(32) heterotic string theory compactified
on a Calabi-Yau manifold where the vacuum solution has leading contribution of
order $g_s$. Our analysis includes this case as well.
 There may
also be cases where the vacuum solution has an expansion in powers of
$g_s^\alpha$ for some $\alpha$ in the range $0<\alpha<1$. Our analysis 
can be extended to this case as well by replacing $g_s$ by $g_s^\alpha$
everywhere in this subsection. \label{fgs}} 
If $|\Psi_k\rangle$ denotes the solution to order $g_s^k$
then the solution to order $g_s^{k+1}$ is given by
\be \label{esoln}
|\Psi_{k+1}\rangle = -{b_0^+\over L_0^+} \sum_{n=1}^\infty {1\over (n-1)!} (1-{\bf P})
\GG [\Psi_k^{n-1}]
+ | \psi_{k+1}\rangle\, ,
\ee
where ${\bf P}$ the projection operator into zero momentum
$L_0^+=0$ states and
$| \psi_{k+1}\rangle$ satisfies\footnote{Since we are 
dealing with NS sector states, there is no distinction between 
$\wt\HH_T$ and $\wh\HH_T$. Therefore $\GG$ in \refb{esoln}-\refb{econdaa}
can be replaced by
identity operators.}
\be \label{esol1}
{\bf P}| \psi_{k+1}\rangle = | \psi_{k+1}\rangle, \qquad Q_B| \psi_{k+1}\rangle
= - \sum_{n=1}^\infty {1\over (n-1)!} {\bf P}
\GG [\Psi_k^{n-1}]+\OO(g_s^{k+2})\, .
\ee
Possible obstruction to solving these equations arises from the failure to find solutions
to \refb{esol1}. It can be shown that\cite{1411.7478} 
the solution to \refb{esol1} exists iff
\be \label{econdaa}
\EE_{k+1}(\phi)\equiv 
\sum_{n=1}^\infty {1\over (n-1)!}  \langle \phi| c_0^- \GG|[\Psi_{k}{}^{n-1}]\rangle 
=\OO(g_s^{k+2})
\, ,
\ee
for any BRST invariant zero momentum
state $|\phi\rangle\in\wt\HH_T$  of ghost number two and
$L_0^+=0$.
Therefore $\EE_{k+1}(\phi)$ represents an obstruction to extending
the vacuum solution beyond order $g_s^k$.
It was also shown in \cite{1411.7478} 
that as a consequence of $|\Psi_k\rangle$ satisfying the equations of motion
to order $g_s^k$,
the condition \refb{econdaa}
is trivially satisfied if $|\phi\rangle$ is BRST
exact. Hence the non-trivial constraints come from zero momentum
non-trivial elements of the BRST cohomology -- the zero momentum
mass$^2$ level zero physical
bosonic states. These obstructions correspond to the existence of massless tadpoles
in the theory. Therefore the absence of massless tadpoles to order $g_s^{k+1}$ will
correspond to \refb{econdaa}.

While finding solutions to \refb{esol1} we have the
freedom of adding to $|\psi_{k+1}\rangle$ any state of the form
\be \label{emarg}
\sum_\alpha a_\alpha |\vp_\alpha\rangle
\ee
where $\{|\vp_\alpha\rangle\}$ is a basis of zero momentum, NS sector
BRST invariant states in $\wh \HH_T$
and $a_\alpha$\plu\ are arbitrary coefficients.
Some of these $a_\alpha$\plu\ could get fixed while trying to ensure
\refb{econdaa} at higher order. Those that do not get fixed represent moduli
and can be given arbitrary values.

\subsection{Expansion around the shifted vacuum}

In this section we shall expand the action around the vacuum solutions and study
its properties. However first we need to define some new quantities that will make our task
easier.

Given a string field configuration $|\Psi_{\rm vac}\rangle$  satisfying
\refb{eeom}, we define\footnote{The bracket $\{ A_1 A_2 \}''$ and $[ A_1 ]''$ are defined to be zero in order to isolate the quadratic terms in \refb{eact}.}
\ben \label{eredefined}
&& \{ A_1\ldots A_k\}'' \equiv \sum_{n=0}^\infty {1\over n!} \, \{\Psi _{\rm vac}^n A_1\ldots A_k\}\, ,
\qquad \hbox{for $k\ge 3$}\, ,
\nonumber \\
&& [A_1\ldots A_k]'' \equiv \sum_{n=0}^\infty {1\over n!} \, [\Psi _{\rm vac}^n A_1\ldots A_k]\, ,
\qquad \hbox{for $k\ge 2$}\, , \nonumber \\
&& \{A_1\}'' \equiv 0, \qquad [~]''\equiv 
0,
\qquad \{A_1A_2\}''\equiv 0, \quad  [A_1]''\equiv 0\, , \nonumber \\
&& \wh Q_B|A\rangle \equiv Q_B|A\rangle + \sum_{n=0}^\infty {1\over n!}
\GG [\Psi_{\rm vac}^n A] \quad \hbox{for $|A\rangle\in\wh\HH_T$} \, .
\een
$[A_1\ldots A_N]''\in\wt\HH_T$ satisfies the relation
\be\label{edefsqpp}
\langle A_0| c_0^- | [A_1\ldots A_N]''\rangle = \{ A_0 A_1\ldots A_N\}'' \quad \forall\, A_0\in \wh\HH_T\, .
\ee
$\wh Q_B$ defined in \refb{eredefined} can be expressed
as
\be \label{edefK}
\wh Q_B = Q_B + \GG\, K\,  ,
\qquad
K\, |A\rangle \equiv \sum_{n=0}^\infty {1\over n!}
[\Psi_{\rm vac}^n A]\, .
\ee
$\wh Q_B$ and $K$
act naturally on states in $\wh\HH_T$.
We also define 
\be \label{edefqbt}
\wt Q_B = Q_B + K\, \GG \, .
\ee
$\wt Q_B$ acts naturally on states in $\wt\HH_T$.

Using the definition of $K$ given in \refb{edefK}, 
the equations of motion \refb{eeom} satisfied by $|\Psi_{\rm vac}\rangle$,
and the identities  \refb{egrel}, \refb{e1pi} one can prove the
following useful identities:
\be \label{eqbk}
Q_B K + K Q_B + K\GG K =0\, ,
\ee
\be \label{eqbtr}
\wh Q_B \GG =  \GG \wt Q_B\, .
\ee
\be \label{enil}
\wh Q_B^2=0, \quad \wt Q_B^2=0\, ,
\ee
and
\be \label{eabqb}
\langle A| c_0^- \wh Q_B|B\rangle = 
\langle \wt Q_B A| c_0^- |B\rangle\, , \qquad 
\langle B| c_0^- \wt Q_B|A\rangle =
\langle \wh Q_B B| c_0^- |A\rangle \, ,
\ee
where 
$\langle \wt Q_B A|$ and $\langle \wh Q_B B|$ are  respectively
the BPZ conjugates of
$\wt Q_B|A\rangle$ and $\wh Q_B|B\rangle$. Finally one can show
using \refb{e1pi}, \refb{eeom} that
\ben \label{eimpidpp}
&&  \sum_{i=1}^N \{ A_1\ldots  A_{i-1} (\wh Q_B  A_i)
 A_{i+1} \ldots  A_N\}'' \nonumber \\
&=& -  
{1\over 2} \sum_{\ell,k\ge 2 \atop \ell+k=N} \sum_{\{i_a;a=1,\ldots \ell\}, \{j_b;b=1,\ldots k\}\atop
\{i_a\}\cup \{j_b\} = \{1,\ldots N\}
}
\{ A_{i_1} \ldots  A_{i_\ell} \vp_s\}'' \{\vp_r A_{j_1} \ldots  A_{j_k}\}'' 
\langle \vp_s^c | c_0^- \GG | \vp_r^c\rangle
\,  .
\een
This generalizes similar results in tree level open and closed bosonic string field
theories\cite{aseq,wittensft}.

We are now ready to expand the action around the vacuum solution. Defining shifted
fields
\be \label{eshiftedphi}
|\Phi\rangle = |\Psi\rangle - |\Psi_{\rm vac}\rangle, \qquad
|\wt\Phi\rangle = |\wt\Psi\rangle - |\wt\Psi_{\rm vac}\rangle = 
|\wt\Psi\rangle - |\Psi_{\rm vac}\rangle
\ee
the 1PI action \refb{eact1PI} and the gauge transformation laws 
\refb{egauge} can be written as
\be \label{eact}
S_{1PI} = g_s^{-2}\left[
-{1\over 2} \langle\wt\Phi |c_0^-  Q_B \GG |\wt\Phi\rangle 
+ \langle\wt\Phi |c_0^- Q_B |\Phi\rangle +  {1\over 2} \langle \Phi| c_0^- K|\Phi\rangle
+ 
\sum_{n=3}^\infty {1\over n!} \{ \Phi^n\}''
\right] + S_{\rm vac}\, ,
\ee
\be \label{edelphi}
|\delta\Phi\rangle = \wh Q_B|\Lambda\rangle + \sum_{n=1}^\infty {1\over n!} 
\GG [\Phi^n \Lambda]''\, , \quad
|\delta\wt\Phi\rangle = Q_B |\wt\Lambda\rangle  + K|\Lambda\rangle 
+ \sum_{n=1}^\infty {1\over n!} 
[\Phi^n \Lambda]''\, ,
\ee
where $S_{\rm vac}$ is the value of the 1PI action \refb{eact1PI}
for the vacuum solution.
The equations of motion derived from \refb{eact} are
\be \label{e01}
Q_B (|\Phi\rangle - \GG|\wt\Phi\rangle) = 0\, ,
\ee
\be \label{e02}
Q_B |\wt\Phi\rangle + K|\Phi\rangle 
+ \sum_{n=3}^\infty {1\over (n-1)!} [\Phi^{n-1}]'' = 0\, .
\ee
Applying $\GG$ on \refb{e02} and using \refb{edefK}, \refb{e01} we get
\be \label{efull}
\wh Q_B |\Phi\rangle + \sum_{n=3}^\infty {1\over (n-1)!} \GG [\Phi^{n-1}]''  = 0\, .
\ee 
Therefore the linearized equations of motion for $|\Phi\rangle$ are
\be \label{eqexp}
\wh Q_B |\Phi_{\rm linear}\rangle=0\, .
\ee
There are 
families of solutions 
to \refb{eqexp} which exist for all momenta, -- these are pure gauge solutions
of the form $\wh Q_B|\Lambda\rangle$ for some $|\Lambda\rangle$. 
There are additional solutions which appear for definite values of $k^2$ -- these represent the
physical states and the values of $-k^2$ at which these solutions appear give the
physical mass$^2$ of the states. Of course in the Euclidean formalism, in
which we have been working so far, these
equations will have solutions for complex  
values of $k^0$, but in the Lorentzian space these solutions will
have real momenta when the corresponding particle is stable.

We shall now describe a systematic procedure for finding the solutions to \refb{eqexp}
in a power series expansion in $g_s$\cite{1411.7478,1501.00988}. 
It is clear that for perturbative solutions, the value of $-k^2$ should differ from the 
tree level values of the mass$^2$ -- that we have called mass$^2$ level -- by a term of
order $g_s$. Let us suppose that we want to find the solution to \refb{eqexp} for $-k^2$
close to some particular mass$^2$ level. We denote by $P_0$ the projection operator to
states in $\wh\HH_T$ carrying this particular mass$^2$ level. 
If $|\Phi_n\rangle$ denotes a solution to \refb{eqexp} 
to order $g_s^n$ then we determine
$|\Phi_n\rangle$ in the `Siegel gauge'\footnote{Siegel gauge here refers
to the gauge in which all states other than those projected by $P_0$
are annihilated by $b_0^+$. We shall shortly discuss the fate of $P_0|\Phi_n\rangle$. }
using the recursion relation:
\be \label{esa3}
|\Phi_0\rangle = |\phi_n\rangle, \qquad
|\Phi_{\ell+1}\rangle = -{b_0^+\over L_0^+} (1-P_0) \GG K |\Phi_\ell
\rangle + |\phi_{n}\rangle \, ,  \quad \hbox{for} \quad 0\le \ell\le n-1\, ,
\ee
where $|\phi_{n}\rangle$ satisfies
\be \label{esa4}
P_0|\phi_{n}\rangle = |\phi_{n}\rangle\, , 
\ee
\be \label{esa5}
Q_B|\phi_{n}\rangle = - P_0 
\GG K |\Phi_{n-1}\rangle +\OO(g_s^{n+1}) \, .
\ee
The projection operator 
$(1-P_0)$ in \refb{esa3} ensures that $L_0^+$ eigenvalue of the state is always of
order unity or larger in magnitude, and
the $(L_0^+)^{-1}$ operator in \refb{esa3} 
never gives any  inverse power of $g_s$. As a result \refb{esa3} leads 
to a well defined expansion of $|\Phi_n\rangle$ in powers of $g_s$, expressing 
it as a linear function of $|\phi_n\rangle$. 
After solving for $|\Phi_n\rangle$ this way we solve 
\refb{esa4}, \refb{esa5} to determine $|\phi_n\rangle$.
Since for given momentum $P_0$ projects onto a finite dimensional
subspace of $\HH_T$, \refb{esa5} gives a finite set of linear equations.
It will have a set of solutions which exist for all momenta. These are 
of the form $P_0\wh Q_B|\Lambda\rangle$ 
for some ghost number 1 state $|\Lambda\rangle$
carrying momentum $k$, and are associated with
pure gauge states. There is also another class of solutions which
exist for specific values of $-k^2$. These describe 
physical states, with the value of $-k^2$ at which the solution exists
giving the physical mass$^2$.

It may 
seem somewhat strange that we first
determine $|\Phi_{\ell+1}\rangle$ for all $\ell$ 
between 0 and $n-1$ iteratively in terms of
$|\phi_{n}\rangle$ and determine $|\phi_n\rangle$ at the end in 
one step by solving
a linear equation in the subspace projected by $P_0$. 
The reason for this is that for the
physical states the allowed value of $k^2$ changes 
at each order. Since a small change
in $k$ is not described by a small change in the vertex operator, it is better 
not to 
compute $|\phi_n\rangle$ iteratively but rather to 
compute it in one step at the very end.

An interesting question is whether $|\Phi_{n}\rangle$ can be chosen to satisfy the Siegel
gauge condition $b_0^+|\Phi_n\rangle =0$.
Eq.~\refb{esa3} ensures that $(1-P_0)|\Phi_n\rangle$
satisfies the Siegel gauge condition; so the question is whether by exploiting the 
gauge freedom of
choosing the solution $|\phi_n\rangle$ to \refb{esa5}, 
$P_0|\Phi_n\rangle=|\phi_n\rangle$  can also be made to satisfy the Siegel gauge condition.
This question was answered in the affirmative in \cite{1411.7478}, but we shall skip the details
of this analysis. 
In \S\ref{spropsieg} we shall describe how to identify renormalized 
masses of physical states in the Siegel  gauge fixed version of the theory.

\subsection{Global symmetries}

The gauge symmetries which 
preserve the vacuum solution 
$|\Psi_{\rm vac}\rangle$ correspond to global symmetries. Therefore they 
must have $|\delta\Phi\rangle=\OO(\Phi)$. 
Using the first equation in \refb{edelphi} this gives
\be \label{es.1}
\wh Q_B|\Lambda_{\rm global}\rangle\equiv 
Q_B|\Lambda_{\rm global}\rangle + 
\GG \, K\,  |\Lambda_{\rm global}\rangle=0\, .
\ee 
This does not guarantee that $|\delta\wt\Phi\rangle$ given by the second equation
in \refb{edelphi} vanishes at the vacuum, 
and hence one may
wonder whether \refb{es.1} itself is sufficient to declare $|\Lambda_{\rm global}\rangle$
to be a global symmetry. 
To this end note that at $\Phi=0$ we have
\be \label{e6.26}
|\delta\wt\Phi\rangle = Q_B |\wt\Lambda\rangle + K |\Lambda_{\rm global}\rangle
\,.
\ee
Therefore, using \refb{eqbk}, we have
\be 
Q_B |\delta\wt\Phi\rangle = Q_B \, K\, |\Lambda_{\rm global}\rangle
= - (K \, Q_B  + K\, \GG\, K) |\Lambda_{\rm global}\rangle
= - K\, \wh Q_B \, |\Lambda_{\rm global}\rangle = 0\, .
\ee
This shows that the transformation generated by 
$|\Lambda_{\rm global}\rangle$ adds a BRST invariant state to $|\wt\Phi\rangle$. 
As can be seen from \refb{e02}, for given
$\Phi$,
addition of BRST invariant states to  $|\wt\Phi\rangle$
generates new solutions to the
equations of motion, but this has no effect on the equations of motion 
\refb{efull} of $|\Phi\rangle$ describing the interacting part of the theory.
Therefore as far as the interacting part of the theory is concerned, 
$|\Lambda_{\rm global}\rangle$ acts as a generator of global 
symmetry.\footnote{In special cases, $|\Lambda_{\rm global}\rangle$ may have the
form $\GG\, |s\rangle$ with $|s\rangle\in\wt\HH$ satisfying $\wt Q_B|s\rangle=0$.
In that case if we choose $|\wt\Lambda\rangle = |s\rangle$ then the right hand side
of \refb{e6.26} will be given by $Q_B|s\rangle + K\GG |s\rangle =\wt Q_B |s\rangle=0$, and
hence the corresponding transformation will act as a global symmetry both in the
interacting sector and in the free sector.}

Such global symmetries
arising in the R-sector of heterotic string theory and RNS and NSR sectors
of type II string theories, carrying zero momentum,
correspond to global supersymmetries. 
Solutions to \refb{es.1} may be constructed more or less in the same way as the solutions
to \refb{eqexp}.  If $|\Lambda_k\rangle$ denotes the solution to \refb{es.1} to order $g_s^k$
then we can take
\be \label{esolam}
|\Lambda_k\rangle =   - {b_0^+\over L_0^+} (1 - {\bf P})
\GG \, K\,  |\Lambda_{k-1}\rangle + |\lambda_k\rangle\, ,
\ee
where ${\bf P}$ denotes the projection operator into $L_0^+=0$ states
and $|\lambda_k\rangle$ is an $L_0^+=0$  state satisfying
\be \label{etauk}
Q_B |\lambda_k\rangle =  -  {\bf P}
\GG \, K |\Lambda_{k-1}\rangle +\OO(g_s^{k+1})\, .
\ee
The possible obstruction to solving \refb{es.1} arises from \refb{etauk}. The latter equation
can be solved if and only if 
\be \label{esusy2}
\LL_k(\hat\phi) \equiv 
\langle \hat\phi |  c_0^-  \GG \, K  |\Lambda_{k-1}\rangle
=\OO(g_s^{k+1})\, ,
\ee
for any BRST invariant 
state $|\hat\phi\rangle\in\wt\HH_T$ of ghost number 3 and $L_0^+=0$. 
Therefore $\LL_k(\hat\phi)$ represents an obstruction to finding global
(super-)symmetry transformation parameter beyond order $g_s^{k-1}$.
A non-vanishing  $\LL_k(\hat\phi)$
signals spontaneous breakdown of the global (super-)symmetry
at order $g_s^{k}$, and the state, that is paired with
$|\hat\phi\rangle$ under the inner product \refb{ebas},
represents the candidate goldstone/goldstino state\cite{1209.5461}.

\subsection{Siegel gauge propagator}

Off-shell Green's functions of the fluctuating fields $\Phi$, $\wt\Phi$ are given by
tree graphs computed from the 1PI action \refb{eact}. The vertices of these graphs
are given by
$\{A_1\ldots A_N\}''$. For computing the propagator we shall use the Siegel gauge
$b_0^+|\Phi\rangle=0$,
$b_0^+|\wt \Phi\rangle=0$. Our goal in this subsection will be to compute the 
propagator in this gauge.

In the Siegel gauge $Q_B=c_0^+ L_0^+$ and after expanding $|\Phi\rangle$,
$|\wt\Phi\rangle$ as
\be
|\Phi\rangle = \sum_r \phi_r |\vp_r\rangle, \quad |\wt\Phi\rangle = 
\sum_r \wt\phi_r |\vp_r^c\rangle,
\ee
the kinetic operator of the action \refb{eact}  takes the form
(see \refb{e320} for notations) 
\be
\pmatrix{\langle \vp_s^c| & \langle \vp_s|}
\left[c_0^- c_0^+ L_0^+ \pmatrix{-\GG & 1\cr 1 & 0} + c_0^-
\pmatrix{0 & 0 \cr
0 & K}\right]\pmatrix{|\vp_r^c\rangle\cr  |\vp_r\rangle}
\, .
\ee
Inverting this and multiplying by $-1$ we get the propagator
\be \label{esignprop}
- \pmatrix{\langle \vp_s| c_0^- & \langle \vp_s^c| c_0^-}
\pmatrix{\check\Delta_F & \bar\Delta_F \cr \wt\Delta_F & \Delta_F}
\pmatrix{c_0^- |\vp_r\rangle\cr c_0^- |\vp_r^c\rangle}
\ee
where
\be \label{esiegel-}
\check\Delta_F = \left[ - {b_0^+\over L_0^+}  K {b_0^+\over L_0^+} 
+ {b_0^+\over L_0^+}  K {b_0^+\over L_0^+} \GG K {b_0^+\over L_0^+} 
+\cdots\right]
b_0^-\, ,
\ee
\be\label{esiegela}
\bar\Delta_F = \left[{b_0^+\over L_0^+}  - {b_0^+\over L_0^+}  K {b_0^+\over L_0^+} \GG 
+ {b_0^+\over L_0^+}  K {b_0^+\over L_0^+} \GG K {b_0^+\over L_0^+} \GG+\cdots\right]
b_0^- \, ,
\ee
\be \label{esiegelb}
\wt\Delta_F = \left[{b_0^+\over L_0^+} - {b_0^+\over L_0^+} \GG K {b_0^+\over L_0^+} 
+ {b_0^+\over L_0^+} \GG K {b_0^+\over L_0^+} \GG K {b_0^+\over L_0^+} +\cdots\right]
b_0^- \, , 
\ee
\be \label{esiegel}
\Delta_F = \left[{b_0^+\over L_0^+} \GG - {b_0^+\over L_0^+} \GG K {b_0^+\over L_0^+} \GG 
+ {b_0^+\over L_0^+} \GG K {b_0^+\over L_0^+} \GG K {b_0^+\over L_0^+} \GG+\cdots\right]
b_0^- \, .
\ee
The minus sign in \refb{esignprop} is a reflection of the fact that we use
$e^S$ as the weight factor in the path integral rather than $e^{-S}$.
$\check\Delta_F$, $\bar\Delta_F$, $\wt\Delta_F$ and $\Delta_F$ act naturally on states 
in $c_0^-\wh\HH_T$, $c_0^-\wt\HH_T$, $c_0^-\wh\HH_T$ and $c_0^-\wt\HH_T$
to produce states
in $\wt\HH_T$, $\wt\HH_T$, $\wh\HH_T$ and $\wh\HH_T$
 respectively.

Since the interaction term in the action \refb{eact}
depends only on $\Phi$, only the $\Phi-\Phi$ propagator
$\Delta_F$ will be relevant for our calculation.
It can be expressed in the compact form
\be \label{edefDeltaI}
\Delta_F = \GG (L_0^+ +  b_0^+ K\GG)^{-1} b_0^+ b_0^- 
= \GG \, b_0^+   (L_0^+ +  K\GG \, b_0^+ )^{-1}  b_0^-  
\quad \hbox{
acting on states in $c_0^-\wt\HH_T$}\, .
\ee
$\Delta_F$ satisfies
\be \label{eb0pm}
b_0^+ \Delta_F = 0, \quad \Delta_F \, b_0^+ = 0\, ,
\ee
\be \label{eboth}
\wh Q_B \Delta_F c_0^-+ \Delta_F c_0^- \wt Q_B =  \GG \quad \hbox{
acting on states in $\wt\HH_T$}\, .
\ee
This can be derived using \refb{edefK}, \refb{edefqbt},
\refb{eqbk} and
other well-known (anti-)commutators involving $Q_B$.

Naively, use of \refb{edefDeltaI} requires us to invert an infinite dimensional matrix.
However since in any given scattering process we can
use the procedure described in \S\ref{seff} to
integrate out the fields whose masses are sufficiently high so that they are not
produced in the scattering, we never have to deal with infinite dimensional matrices.
If we 
want to look for poles in $\Delta_F$ to compute renormalized masses,
we
can further simplify the analysis by taking 
one mass$^2$ level at a time and integrating 
out all fields other than those at the
chosen mass$^2$ level.

\sectiono{Ward identities in the shifted background} \label{swardshift}

In this section we shall describe the Ward identities satisfied by various quantities
in the shifted background. Our discussion will follow \cite{1508.02481}.

\subsection{Bose-Fermi degeneracy for global supersymmetry} \label{sbosefermi}

Let us suppose that we have a global supersymmetry
transformation parameter $|\Lambda_{\rm global}\rangle$ that 
preserves the vacuum solution $|\Psi_{\rm vac}\rangle$  
satisfying \refb{eeom}.
Therefore $|\Lambda_{\rm global}\rangle$ satisfies \refb{es.1}.
Let $|\Phi_{\rm linear}\rangle$ 
be a solution to the linearized equations of motion \refb{eqexp} around the background.
Then it follows from \refb{edefsqpp}, \refb{edefK}, \refb{eabqb}, \refb{eimpidpp}, 
\refb{eqexp} and \refb{es.1} that
\be \label{es.3}
\wh Q_B \GG\,  [\Lambda_{\rm global} \Phi_{\rm linear}]'' = 0\, .
\ee
Therefore $\GG[\Lambda_{\rm global}\Phi_{\rm linear}]''$ also 
satisfies the linearized equations of motion. Since 
$|\Lambda_{\rm global}\rangle$ is fermionic, this 
provides a map between the bosonic and 
fermionic solutions to the linearized equations of motion. Since 
$|\Lambda_{\rm global}\rangle$ carries
zero momentum, these solutions occur at
the same values of momentum.
Furthermore 
if the solution $|\Phi_{\rm linear}\rangle$ exists for all values of 
momenta so does the solution
$\GG[\Lambda_{\rm global}\Phi_{\rm linear}]''$ and if the solution 
$|\Phi_{\rm linear}\rangle$ exists for special values of $k^2$,
the solution $\GG[\Lambda_{\rm global} \Phi_{\rm linear}]''$ also exists 
for the same special values of $k^2$. 
Therefore this procedure pairs pure gauge solutions in the bosonic and fermionic sector
and also physical solutions in the two sectors. Furthermore, since the physical
solutions occur at the same values of $k^2$, it establishes the equality of
the masses of bosons and fermions (even though each of them
may get
renormalized by perturbative 
corrections of string theory).\footnote{The only exception
to this is the situation where 
$\GG[\Lambda_{\rm global}\Phi_{\rm linear}]''$ vanishes. 
However typically in such
situations one can identify another component of the supersymmetry transformation 
parameter which does the pairing.}

Note that the above analysis not only implies equality of the masses of the superpartners,
but also implies equality of the decay widths of the superpartners if they are unstable.
In this case the solution to $\wh Q_B|\Phi_{\rm linear}\rangle=0$
occurs at complex values of the momentum. It follows from the arguments given above 
that at the same complex value of the momentum we have a solution to the linearized
equation of motion carrying opposite Grassmann parity of $|\Phi_{\rm linear}\rangle$. 
Therefore they have the same imaginary part of the mass and hence the same decay width.

\subsection{Ward identities for local (super-)symmetry} \label{swardlocal}

In this subsection we shall
derive the Ward identities for S-matrix elements. 
Let $\Gamma$ denote the amputated Green's function with external
propagators removed.
$\Gamma$ differs from $G$ introduced in
\S\ref{sgintro} in that we are removing the full propagators from the external legs,
whereas in defining $G$ we only removed the tree level propagators. Also in computing
$\Gamma$ we take into account the effect of vacuum shift. The S-matrix 
elements can be computed from the amputated Green's functions 
$\Gamma$ by
setting the external states
on-shell and multiplying the result by appropriate wave-function renormalization 
factors for each external leg. 
We shall first show that the $\Gamma$\plu\ satisfy the identities:
\be \label{egrid}
\sum_{i=1}^N \Gamma(
| A_1\rangle, \ldots  ,| A_{i-1}\rangle,
\wh Q_B| A_i\rangle, | A_{i+1}\rangle, \ldots  ,| A_N\rangle) =0\, .
\ee

The proof of \refb{egrid} 
proceeds in a manner similar to the one used in \S\ref{sgintro}. 
We could take two different approaches -- either 
expand the original master action \refb{eactorg} around the vacuum
solution $|\Psi_{\rm vac}\rangle$ and use the Feynman rules derived from this action, 
or 
use the kinetic operator, interaction terms and propagators computed from
the 1PI action expanded around $|\Psi_{\rm vac}\rangle$, and sum over tree amplitudes
computed from these Feynman rules. Both give same results. We shall use the
second approach.
The analogue of \refb{egidpre} now takes the form:
\ben \label{egid1PIfull} 
&&\hskip .2in \sum_{i=1}^N  \Gamma(A_1,\ldots  ,A_{i-1}, \wh Q_B A_i,
A_{i+1} \ldots  ,A_N)  \nonumber \\
&& = -  
{1\over 2} \sum_{\ell,k\ge 2 \atop \ell+k=N} \sum_{\{ i_a;a=1,
\ldots , \ell\} , \{ j_b;b=1,\ldots  ,k\} \atop
\{ i_a\} \cup \{ j_b\}  = \{ 1,\ldots , N\}
}\Gamma( A_{i_1}, \ldots  ,A_{i_\ell} ,\vp_s)  \Gamma( \vp_r, A_{j_1} , \ldots  ,A_{j_k})
\langle \vp_s^c | c_0^- \GG | \vp_r^c\rangle \nonumber \\
&& -  
{1\over 2} \sum_{\ell,k\ge 2 \atop \ell+k=N} \sum_{\{ i_a;a=1,
\ldots , \ell\} , \{ j_b;b=1,\ldots , k\} \atop
\{ i_a\} \cup \{ j_b\}  = \{ 1,\ldots , N\}
}\bigg[ - \Gamma( A_{i_1}, \ldots  ,A_{i_\ell} ,\wh Q_B\vp_s)  
\Gamma( \vp_r, A_{j_1} , \ldots  ,A_{j_k})
\nonumber \\
&& \hskip 1in - (-1)^{\gamma_s}
\Gamma( A_{i_1}, \ldots  ,A_{i_\ell} ,\vp_s)  
\Gamma( \wh Q_B\vp_r, A_{j_1} , \ldots  ,A_{j_k}) \bigg]
\langle \vp_s^c | c_0^- \Delta_F c_0^- | \vp_r^c\rangle
\, . \nonumber \\ 
\een 
The analogue of the 
second term on the right hand side of \refb{egidpre} is absent due to the absence
of a similar term in \refb{eimpidpp} compared to \refb{evertex}, 
whereas the analogue of the last term on the right
hand side of \refb{egidpre} is absent since we need to compute only tree amplitudes 
using the 1PI vertices. The restriction $\ell, k\ge 2$ in the first term on the right hand side
has its origin in the corresponding restriction in \refb{eimpidpp}. On the other hand, the restriction
$\ell, k\ge 2$ in the second term has its origin in the absence of tadpoles. This is due to expanding the
action around the vacuum solution and the absence of self-energy insertions -- on internal lines
because we are using the full propagator $-\Delta_F$ and on external lines because we are working with
amputated Green's function. We can now use the analogue of \refb{eman1}, \refb{eman2} with $Q_B$ replaced
by $\wh Q_B$ and $b_0^+ (L_0^+)^{-1}\GG$ replaced by $\Delta_F c_0^-$.
For example the analogue of \refb{eman1} will be
\be \label{eman1new}
\wh Q_B |\vp_s\rangle \langle \vp_s^c | c_0^- 
\Delta_F c_0^- 
| \vp_r^c\rangle = \wh Q_B \Delta_F c_0^- 
| \vp_r^c\rangle = |\vp_s\rangle \langle \vp_s^c | c_0^- 
\wh Q_B \Delta_F c_0^- 
| \vp_r^c\rangle \, .
\ee
The analogue of \refb{eman2} can be derived using slightly different trick. We first use
\refb{ecom} to write
\be 
\langle \vp_s^c | c_0^- \Delta_F c_0^- | \vp_r^c\rangle \langle \vp_r|c_0^- \wt Q_B
= \langle \vp_s^c | c_0^- \Delta_F c_0^- \wt Q_B 
= \langle \vp_s^c | c_0^- \Delta_F c_0^- \wt Q_B  | \vp_r^c\rangle \langle \vp_r|c_0^-\, .
\ee
Now taking BPZ conjugate of both sides, and using $(-1)^{\gamma_r} = (-1)^{\gamma_s}$
and \refb{eabqb} we get
\be \label{eman2new}
(-1)^{\gamma_s} \wh Q_B |\vp_r\rangle \langle \vp_s^c | c_0^- 
\Delta_F c_0^- 
| \vp_r^c\rangle =   |\vp_r\rangle \langle \vp_s^c | c_0^-  
\Delta_F c_0^-  \wt Q_B
| \vp_r^c\rangle
\, , \nonumber \\
\ee
where we have used \refb{eabqb}, keeping in mind that the basis
state $|\vp_r^c\rangle$ is not necessarily Grassmann even and so in applying \refb{eabqb}
we have to account for the extra sign that comes from exchanging the relative positions
of $\wt Q_B$ and $\vp_r$. Using \refb{eman1new}, \refb{eman2new} 
to transfer $\wh Q_B$ inside the matrix element in the terms in the last
two lines of
\refb{egid1PIfull}, and using
\refb{eboth} one can show that the terms on the right hand side of 
\refb{egid1PIfull} cancel.
This leads to \refb{egrid}.

Let us now suppose that we have a set of physical 
external states $|\AAA_1\rangle,\ldots 
 ,|\AAA_N\rangle$ satisfying
\be \label{erenor}
\wh Q_B |\AAA_i\rangle = 0\, , \qquad \hbox{for} \qquad 1\le i\le N\, .
\ee
Let us also suppose that we have a local gauge  transformation parameter
$|\Lambda\rangle$ belonging either to the fermionic sector or to the bosonic sector.
Then $\wh Q_B|\Lambda\rangle$ represents a pure gauge state. It now follows
from \refb{egrid} with $N$ replaced by $N+1$ and the states $| A_1\rangle,\ldots
 ,| A_{N+1}\rangle$ replaced by $|\Lambda\rangle, |\AAA_1\rangle,\ldots  ,|\AAA_N\rangle$ 
that
\be \label{efinlocal}
\Gamma(\wh Q_B|\Lambda\rangle, |\AAA_1\rangle,\ldots  ,|\AAA_N\rangle)
= 0\, .
\ee
Since S-matrix elements with external states $\wh Q_B|\Lambda\rangle, |\AAA_1\rangle,
\ldots  ,|\AAA_N\rangle$ are given by multiplying 
$\Gamma(\wh Q_B|\Lambda\rangle, |\AAA_1\rangle,\ldots  ,|\AAA_N\rangle)$
by wave-function renormalization factors, vanishing of \refb{efinlocal} 
will also imply the
vanishing of this S-matrix element. This shows that pure gauge states of the form
$\wh Q_B|\Lambda\rangle$ decouple from the
S-matrix of physical states. Note that since we have taken $|\AAA_i\rangle$\plu\ to satisfy
\refb{erenor} which takes into account the effect of string loop corrections in the definition
of $\wh Q_B$, the decoupling of pure gauge states occurs even in the presence
of external states that
suffer mass renormalization.

\subsection{Ward identities for global (super-)symmetry}

\begin{figure}
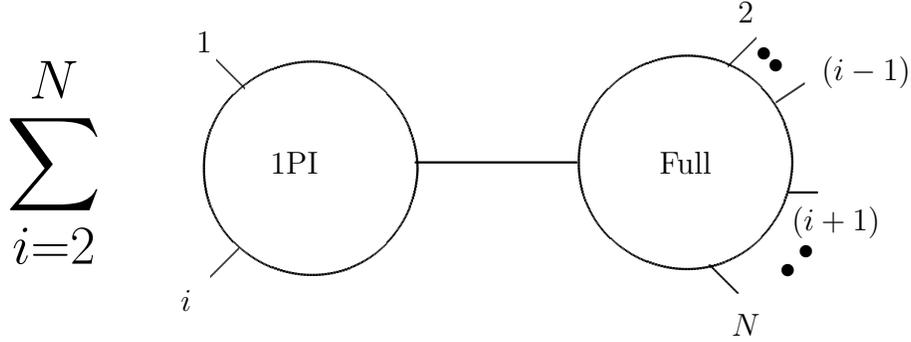


\begin{center}

\figwtgam

\end{center}

\vskip -.9in 

\caption{The contributions to be excluded from the definition of $\wt\Gamma$. 
Here the blob marked 1PI represents the 1PI vertex $\{\cdots\}''$, the blob marked Full represent
the full  amputated Green's function, 
the horizontal line connecting the two blobs
represent the full propagator $-\Delta_F$ 
and the short lines represent external states.
\label{figwtgam}}
\end{figure}

We shall now explore the consequences of global (super-)symmetry
on the S-matrix. As described in \refb{es.1}, the existence of
such a symmetry is signaled by a gauge transformation parameter 
$|\Lambda_{\rm global}\rangle$
satisfying
\be \label{eglobala}
\wh Q_B|\Lambda_{\rm global}\rangle=0\, .
\ee
Typically $|\Lambda_{\rm global}\rangle$ carries zero momentum. 
Now if we use \refb{efinlocal}
with $|\Lambda\rangle$ replaced by 
$|\Lambda_{\rm global}\rangle$ then the resulting 
identity is trivial. To get something
non-trivial we proceed somewhat differently. We first define a new object
$\wt \Gamma(| A_1\rangle,\ldots  ,| A_N\rangle)$ where the first argument
$| A_1\rangle$ plays a somewhat different role compared to the other arguments.
For this we begin with the expression for the amputated Green's 
function $\Gamma$ as sum of tree level
Feynman diagrams built from 1PI vertices and propagators, 
and
delete from this all terms where by removing a single propagator we can 
separate the external state $|A_1\rangle$ and one more $| A_i\rangle$ from
the rest of the $| A_i\rangle$\plu. 
This has been shown in Fig.~\ref{figwtgam}.
If we take the $| A_i\rangle$\plu\ to be states 
carrying fixed momenta $k_i$ then this means that we remove all terms where 
momentum conservation forces one
of the internal propagators to carry momentum $k_1+k_i$ for any $i$ between 2 and $N$.
We can now derive an identity analogous to \refb{egrid} for $\wt\Gamma$ using 
similar method, but now due to the special role played by $A_1$,
the identity \refb{egid1PIfull} will be modified to
\ben \label{egidnewer} 
&&\hskip .2in \sum_{i=1}^N \wt\Gamma(A_1,\ldots  ,A_{i-1}, \wh Q_B A_i,
A_{i+1} \ldots  ,A_N)  \nonumber \\
&& = -  
 \sum_{\ell\ge 1, k\ge 2 \atop \ell+k=N-1} \sum_{\{ i_a;a=1,
\ldots  ,\ell\} , \{ j_b;b=1,\ldots  ,k\} \atop
\{ i_a\} \cup \{ j_b\}  = \{ 2,\ldots  ,N\}
}\wt\Gamma( A_1,A_{i_1}, \ldots  ,A_{i_\ell} ,\vp_s)  \Gamma( \vp_r, A_{j_1} , \ldots  ,A_{j_k})
\langle \vp_s^c | c_0^- \GG | \vp_r^c\rangle \nonumber \\
&& -  
\sum_{\ell\ge 2, k\ge 2 \atop \ell+k=N-1} \sum_{\{ i_a;a=1,
\ldots  ,\ell\} , \{ j_b;b=1,\ldots  ,k\} \atop
\{ i_a\} \cup \{ j_b\}  = \{ 2,\ldots  ,N\}
}\bigg[ - \wt\Gamma( A_1, A_{i_1}, \ldots  ,A_{i_\ell} ,\wh Q_B\vp_s)  
\Gamma( \vp_r, A_{j_1} , \ldots  ,A_{j_k})
\nonumber \\
&& \hskip 1in - (-1)^{\gamma_s}
\wt \Gamma(A_1, A_{i_1}, \ldots  ,A_{i_\ell} ,\vp_s)  
\Gamma( \wh Q_B\vp_r, A_{j_1} , \ldots  ,A_{j_k}) \bigg]
\langle \vp_s^c | c_0^- \Delta_F c_0^- | \vp_r^c\rangle
\, . \nonumber \\ 
\een 
Note that the symmetry between the two sets $\{i_1,\ldots  ,i_\ell\}$ and $\{j_1,\ldots  ,j_k\}$
has been broken since the first set is always accompanied by 1. Consequently the factors
of 1/2 have disappeared. Furthermore, in the second term
on the right hand side the sum over $\ell$ has been
restricted to $\ell\ge 2$ since $\wt\Gamma$ excludes terms in which $A_1$ and
$A_i$ for any $i\ge 2$ can be separated from the rest by cutting a single propagator.
As a result the cancellation is incomplete, 
and we get
\ben \label{etildegam}
&& \sum_{i=1}^N  \wt\Gamma(
| A_1\rangle, \ldots  ,| A_{i-1}\rangle,
\wh Q_B| A_i\rangle, | A_{i+1}\rangle, \ldots  ,| A_N\rangle) 
\qquad \qquad \qquad \quad \nonumber \\ 
& =&  -  \sum_{i=2}^N
\wt\Gamma( A_{1},A_i, \vp_s)  \Gamma( \vp_r, A_{2}, \ldots   ,A_{i-1}, A_{i+1}, \ldots  ,A_{N})
\langle \vp_s^c | c_0^- \GG | \vp_r^c\rangle \nonumber \\
&=& - \sum_{i=2}^N \Gamma(
| A_2\rangle, \ldots  ,| A_{i-1}\rangle,
\GG[ A_1 A_i]'' , | A_{i+1}\rangle, \ldots  ,| A_N\rangle) 
\, .
\een
In arriving at the last step we have used the fact that for three arguments 
$\wt\Gamma(A,B,C)=\Gamma(A,B,C) =\{ABC\}''$.

Let us now suppose that we have a set of physical 
external states $|\AAA_1\rangle,\ldots  ,
|\AAA_N\rangle$ 
satisfying \refb{erenor} and a global (super-)symmetry transformation
parameter $|\Lambda_{\rm global}\rangle$  satisfying \refb{eglobala}. 
Then a direct application of \refb{etildegam} with $N$ replaced by $N+1$, and the states
$| A_1\rangle,\ldots  ,| A_{N+1}\rangle$ taken as 
$|\Lambda_{\rm global}\rangle, |\AAA_1\rangle,
\ldots  ,|\AAA_N\rangle$ gives
\be \label{eglobalid}
\sum_{i=1}^N 
\Gamma (|\AAA_1\rangle, \ldots  ,|\AAA_{i-1}\rangle, 
\GG[\Lambda_{\rm global} \AAA_i]'', |
\AAA_{i+1}\rangle, \ldots  ,|\AAA_N\rangle)=0\, .
\ee
Now, according to the analysis of \S\ref{sbosefermi},
$\GG[\Lambda_{\rm global} \AAA_i]''$ 
represents the on-shell state which is the transform of
$|\AAA_i\rangle$ under the infinitesimal global (super-)symmetry generated by 
$|\Lambda_{\rm global}\rangle$. Therefore we recognize \refb{eglobalid} 
as the Ward identity
associated with the global (super-)symmetry generated by 
$|\Lambda_{\rm global}\rangle$.

We again repeat that all the analysis in this and other sections could be
performed with the effective action with appropriately chosen projection
operator so that we have to deal with minimal number of fields.

\subsection{Changing the propagator} \label{echprop} 

\begin{figure}
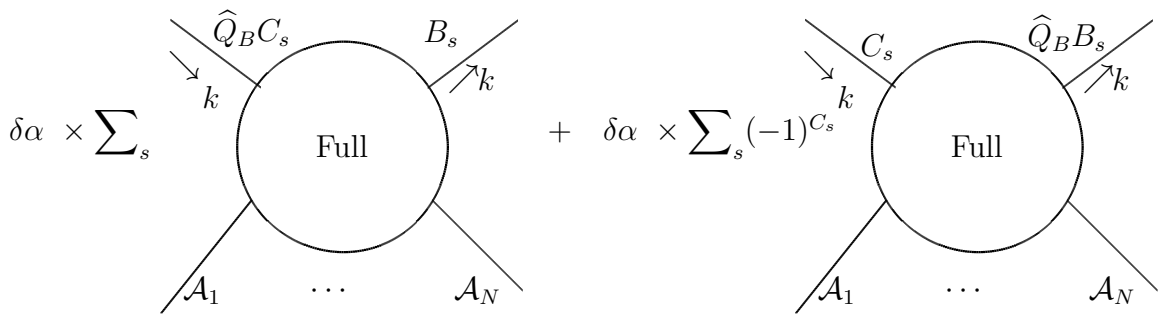


\begin{center}

\hbox{\figscsc \qquad  \figscscb}

\end{center}

\vskip -10pt

\caption{ Diagrams representing the change in the amplitude under an infinitesimal
change in the propagator $-\Delta_F^\alpha$ to 
$-\Delta_F^{\alpha+\delta\alpha}$.
\label{figsch}}
\end{figure}

In this section we shall consider a modified propagator $-\Delta_F^\alpha$ 
where
\be \label{edelpr}
\Delta_F^\alpha = \Delta_F - \alpha \sum_s \left( \wh Q_B|C_s\rangle \langle
B_s| +  (-1)^{C_s}  |C_s\rangle \langle
\wh Q_B B_s| \right)\, ,
\ee
for any positive 
constant $\alpha$ and any set of states $|C_s\rangle, |B_s\rangle\in \wh \HH_T$
with the ghost numbers of $C_s$ and $B_s$ adding up to 3 for each $s$. 
In this case, after taking into account the fact that the BPZ inner
product pairs states carrying total ghost number 6, 
we see that the additional terms act as operators carrying ghost number $-2$ like 
$\Delta_F$. $(-1)^{C_s}$ denotes 
the Grassmann parity of $C_s$. 
We could of course absorb $\alpha$ into the definition of $C_s$ or $B_s$
but we have kept it to facilitate our analysis below.
Using \refb{eabqb} -- keeping in mind that since
$C_s$ and $B_s$ are not in general Grassmann even there are extra signs
in moving the $\wh Q_B$ or $\wt Q_B$ past them --  one can see that 
the propagator modified this way continues to
satisfy \refb{eboth}. However it may not satisfy the Siegel gauge 
condition \refb{eb0pm}.  Our goal will be to show that amplitudes
computed with this modified propagator have the following properties:
\begin{enumerate}
\item The identity \refb{egrid} continues to hold.
\item  The 
amplitudes computed from the propagator $-\Delta_F$ are identical to those
computed with the modified propagator $-\Delta_F^\alpha$ if the external states
are all $\wh Q_B$ invariant.
\end{enumerate}
In the proof, we shall assume that the results are valid for certain value of 
$\alpha$ and  then show that it holds for $\alpha+\delta\alpha$ to first order
in $\delta\alpha$.  Since the results obviously hold for $\alpha=0$, this then
implies that they continue to hold for all $\alpha$.

Now the amplitude computed with the propagator 
$\Delta_F^{\alpha+\delta\alpha}$ differs from the one computed with the propagator
$\Delta_F^\alpha$ by the sum of Feynman diagrams shown schematically in
Fig.~\ref{figsch}. In this the blobs marked `Full' represent sum of all
Feynman diagrams -- with external propagators removed --
computed with the propagator $(- \Delta_F^\alpha)$, 
and may 
contain both connected diagrams and disconnected diagrams in which 
the lines $\wh Q_BC_s$ and $B_s$ (or  $C_s$ and $\wh Q_BB_s$)
are connected to different connected
components. Since by assumption
the amplitudes computed with the propagator 
$(-\Delta_F^\alpha)$ satisfy \refb{egrid}, we see first of all that
as long as the external states $|\AAA_i\rangle$ are all $\wh Q_B$ invariant,
the contribution from Fig.~\ref{figsch} vanishes by \refb{egrid}.  This shows 
that to first order in $\delta\alpha$, the on-shell amplitudes computed with the propagator 
$(-\Delta_F^{\alpha+\delta\alpha})$ are equal to those computed with the propagator
$(-\Delta_F^\alpha)$, which in turn are equal to those computed with the propagator
$(-\Delta_F)$ by our initial assumption. On the other hand if the external states
are not $\wh Q_B$ invariant, but our goal is to check \refb{egrid} for
$\alpha+\delta\alpha$, we see that the
change in the left hand side of \refb{egrid} induced by Fig~\ref{figsch} is given by 
\be \label{erew1}
\delta\alpha \,  \sum_s \sum_{i=1}^N \bigg[\Gamma^\alpha(\wh Q_B C_s, 
B_s, \{\AAA_\ell, \, \, \ell\ne i\}, 
\wh Q_B \AAA_i)
+ (-1)^{C_s} \Gamma^\alpha(C_s, \wh Q_B B_s,
\{\AAA_\ell, \, \, \ell\ne i\},  \wh Q_B \AAA_i)
\bigg]\, ,
\ee
where $\Gamma^\alpha$ denotes the  amplitude computed with
the propagator $(-\Delta_F^\alpha)$. In \refb{erew1}
we have used the evenness of all the $\AAA_i$'s to push the 
$\wh Q_B\AAA_i$ to the end of the argument of $\Gamma^\alpha$.
Since the amplitudes computed with the propagator $(-\Delta_F^\alpha)$ satisfy
\refb{egrid} by assumption, we can now use \refb{egrid} to rewrite 
\refb{erew1} as
\be \label{erew2}
\delta\alpha \,  \sum_s \sum_{i=1}^N \sum_{j=1\atop j\ne i}^N
\Gamma^\alpha (C_s, 
B_s, \{\AAA_\ell, \, \ell\ne i,j\},
 \wh Q_B  \AAA_j, \wh Q_B \AAA_i)\, .
\ee
In arriving at \refb{erew2} we have taken into account the fact that $C_s$ and
$B_s$ are not Grassmann even and therefore there are extra signs $(-1)^{C_s}$ and
$(-1)^{B_s}$ as we pass $\wh Q_B$ through them. We have also used the fact the
$(-1)^{C_s} (-1)^{Bs}=-1$ since the ghost number of $C_s$ and $B_s$ add up to three.
We now note that the summand in \refb{erew2} is odd under the exchange of $i$ and $j$.
Therefore the result vanishes after summing over $i$ and $j$, establishing that the
identity \refb{egrid} continues to hold with the propagator $(-\Delta_F^{\alpha+\delta\alpha})$
to first order in $\delta\alpha$.
This in turn establishes the two assertions made at the beginning of this subsection.

This result has the following important application. 
While computing the renormalized propagator we may 
sometime encounter situations in which the
propagator is found to have spurious double poles with coefficients
proportional to pure gauge states. 
For example in a general Lorentz
covariant gauge the propagator of  a gauge field has the form
$(\eta^{\mu\nu} - \alpha \, k^\mu k^\nu /k^2) / k^2$ for some constant
$\alpha$. In such cases we can define a modified propagator
of the form given in \refb{edelpr} in which we subtract the contribution
from the double pole terms. 
From now on it will be understood that we always work with a
propagator
that does not have these double poles.

\subsection{Supersymmetry and massless tadpoles} \label{stad}

So far in our analysis we have assumed that we have a vacuum
solution $|\Psi_{\rm vac}\rangle$ to the equations of motion of 1PI effective
superstring field theory to all orders in $g_s$, and then derived the Ward
identities for the field theory expanded around this background. But we could ask a
slightly different question: if we assume that the vacuum solution exists to certain order
in $g_s$ (say to order $g_s^k$) 
and that to this order there exists a global supersymmetry transformation
parameter, can we determine if the solution can be extended to the next order?
Since the obstruction to extending the solution to order $g_s^{k+1}$ arises from
possible failure of \refb{econdaa}, the relevant question is: can we use existence
of supersymmetry to order $g_s^k$, encoded in \refb{esusy2}, to prove
\refb{econdaa}? 
This was addressed in the context of perturbative vacuum using the world-sheet
approach in \cite{1209.5461}; here we want to ask whether string field theory can extend this
also to non-trivial 
vacua where the string field expectation value is of order $g_s$ (or given by
some other power of $g_s$ less than 2).
It turns out that this is indeed possible. 
The details can be found in \cite{1508.02481}.

\subsection{Application to SO(32) heterotic string theory on Calabi-Yau manifolds}

So far we have discussed superstring field theory in an abstract formalism. A concrete
class of examples where the full power of this formalism can be displayed is in
SO(32) heterotic string theory compactified on a Calabi-Yau manifold with spin connection
identified to gauge connection. The low energy effective field theory for this class of
compactifications is described by $N=1$ supergravity coupled to matter
fields. An important feature of these theories is 
the existence of
a U(1) gauge field for which a Fayet-Iliopoulos D-term is generated at
one loop\cite{DSW,ADS,DIS,greenseiberg,AtickS,1304.2832,1404.5346,1408.0571}. 
As a result a scalar field $\phi$ charged under this gauge field acquires a 
potential of the form
\be\label{epotV}
V = {1\over 2} (\phi^* \phi - c \, g_s^2)^2\, ,
\ee
where $c$ is a constant that can be computed and shown to be positive\cite{ADS,DIS}.
This has several interesting consequences:
\begin{enumerate}
\item The original perturbative vacuum $\phi=0$ breaks supersymmetry at one loop.
\item At this vacuum the scalar field $\phi$ acquires a negative mass$^2$ 
given by $- c g_s^2/2$.
\item At two loop order the perturbative vacuum $\phi=0$ acquires  a  non-zero cosmological
constant $c^2 g_s^4 /2$. This also leads to a dilaton tadpole, but it is not
visible in \refb{epotV} since we have not displayed the coupling to the dilaton field.
\item There is a shifted vacuum at $|\phi|=g_s\sqrt c$ where supersymmetry is restored.
\item The two loop cosmological constant and the dilaton tadpole
vanishes at the shifted vacuum. 
\end{enumerate}

One can formulate superstring field theory in the background of SO(32) string
theory on Calabi-Yau manifolds and try to verify these predictions of effective field theory.
One indeed finds perfect agreement\cite{1508.02481}.  We refer the reader to the original
reference for details, but summarize here the main results:
\begin{enumerate}
\item The scalar field acquires a negative mass$^2$ at one 
loop\cite{DIS,ADS,DIS,1304.2832,1408.0571}. This computation does not require
use of string field theory, but requires carefully ensuring that near the boundary of
the moduli space, where two of the punctures on the torus come together, the
PCO locations are arranged correctly in accordance 
with the factorization rules
described around \refb{epcoarrange}.
\item Eq.~\refb{es.1} for supersymmetry transformation fails to have a solution
at order $g_s^2$ due to the failure of \refb{esusy2} for $k=2$.
\item At the original vacuum, one generates a dilaton tadpole and cosmological constant
at two loop order, with values that are precisely in agreement with the predictions of
effective field theory. These show up as the failure of \refb{econdaa} for $k=3$.
\item One can find a non-trivial vacuum solution 
of the string field theory equation
\refb{eeom}, whose expansion begins at order $g_s$. For this 
solution \refb{es.1} has a solution for global supersymmetry transformation parameter
at order $g_s^2$.
\item At this vacuum the mass$^2$ of the scalar $\phi$ and its superpartner fermion are 
equal to order $g_s^2$
and are in agreement with the predictions of the effective field theory. The scalar
mass$^2$ is of order $g_s^2$ and the fermion mass is of order $g_s$. The latter arises
from genus zero cubic interaction term after taking into 
account the order $g_s$ shift of the background. 
\item At the shifted vacuum the two loop cosmological constant and the dilaton tadpole
vanishes to order $g_s^4$.
\end{enumerate}

It is worth emphasizing that
even though one can derive the above results using superstring field theory, this
is not the most efficient way of arriving at these results -- effective field theory
based on the potential \refb{epotV} is clearly more efficient. What string field theory
achieves however is that
once superstring field theory around such a vacuum has been formulated, one can
use it to carry out computations beyond what is possible for effective field theory. For
example one can in principle 
compute the masses of the massive string states or the
analogue of the Virasoro-Shapiro amplitude in the
shifted background, and use the general result of \S\ref{sunitary} to prove unitarity
of the theory in the shifted background. These are not possible within effective field
theory.

\sectiono{String field theory in the momentum space} \label{smom}

So far we have described the amplitudes in string field theory in the Schwinger
parameter representation since this was useful in making contact with the standard
description as integrals over the moduli space of Riemann surfaces. In this section
we shall describe them as integrals over loop momenta that is more 
conventional in a quantum field theory. We shall do this analysis in the 
Lorentzian formalism that requires using a weight factor of $e^{iS}$ in the
path integral. As mentioned in \S\ref{slorentz}, this will require 
multiplying the vertices by $i$ and propagators by
$-i$ relative to the Feynman rules described in \S\ref{ssft}. Our discussion will mainly
follow \cite{1604.01783,1607.06500}.

\subsection{Loop energy integration contour} \label{sloopcon}

Consider an off-shell $n$-point interaction vertex $\cL A_1\ldots A_n\cR$ of 
string field theory with external legs of
mass $m_1,\ldots  ,m_n$ and momenta $k_1,\ldots  ,k_n$. 
Our focus will be on the momentum dependence of the interaction vertex. Inside the
correlation function in \refb{edefomega}, which enters the definition of the 
interaction vertex \refb{edefcurl}, the momentum dependence comes 
from the $e^{ik_i\cdot X}$
factors in the vertex operators $A_i$ and possibly explicit
powers of momenta
coming from the vertex operators. If $\{y_a\}$ denotes collectively the
parameters labelling points on the relevant $\oR_{g,m,n}$
-- which can be chosen to be the coordinates of the projection of $\oR_{g,m,n}$
on the base $\MM_{g,m,n}$ -- then the momentum dependence of the integrand in 
\refb{edefcurl} has the form of the exponential of a quadratic expression in 
momenta,\footnote{The correlation function 
$\left\langle \prod_i e^{i k_i\cdot X(y_i)}\right\rangle$ is given by 
$\exp[-k_i\cdot k_j \, G(y_i, y_j]$ where $G(y_i,y_j)$ denotes the Green's function
$\langle X(y_i) X(y_j)\rangle$. Additional factors of derivatives of $X$ in the vertex
operators will generate multiplicative factors of momenta in the correlation function.}
with coefficients depending on $y$, multiplied by a polynomial in $k_i$. Therefore
the general form of the interaction vertex is given by 
\be \label{ever1}
\int [dy] \, \exp\left[- \sum_{i,j} g_{ij}(y) k_i\cdot k_j\right]
P(y, \{k_i\})\, .
\ee
Here $g_{ij}(y)$ is some function of $\{y_a\}$ and $P(y, \{k_i\})$ is a
 polynomial in the 
$\{k_i^\mu\}$, with $\{y_a\}$ dependent coefficients. Now the effect of adding stubs to the vertex 
-- as discussed in 
\S\ref{scomstub} -- has the effect of multiplying the integrand in \refb{ever1} by 
a factor of $\exp[-\sum_i \Lambda_i(y) (k_i^2+m_i^2)]$ for some positive constants
$\Lambda_i(y)$. By absorbing the $\exp[-\sum_i \Lambda_i(y) k_i^2]$ term into
the definition of $g_{ij}(y)$ 
\be \label{evertexexp}
g_{ij}(y)\to g_{ij}(y) + \Lambda_i(y) \delta_{ij}\, ,
\ee
we can ensure that $g_{ij}(y)$ is a positive definite
matrix.\footnote{Since the effect of adding stubs also requires rearranging
the section segments $\oR_{g,m,n}$, one might wonder whether we can consistently 
make all the $g_{ij}$\plu\ positive definite by adding stubs. To this end note that when we
add stubs to a given interaction vertex, it forces us to modify the section segments of
{\it higher order interaction vertices}, containing more 
punctures or higher genus surfaces
or both. Therefore to any given order in perturbation theory, 
we can systematically add stubs to all the relevant interaction vertices and make
the corresponding $g_{ij}$\plu\ positive definite.
}
As we shall discuss shortly, this makes the momentum integrals converge.
On the other hand  the $\exp[-\sum_i \Lambda_i(y) m_i^2]$ factor makes
the sum over infinite number of intermediate states, whose number grows as
$\exp(c\, m)$ for some positive constant $c$, converge.
We shall absorb this factor into the expression for $P(y, \{k_i\})$.

We can now compute contributions from Feynman diagrams using these 
interaction vertices.
The propagator has the standard form $(k_i^2+m_i^2)^{-1}$,
possibly multiplied by some polynomial
in $k_i$. If we denote by $\{\ell_s\}$ the independent loop momenta, by $\{p_\alpha\}$
the external momenta and by $\{k_i\}$ the momenta carried by individual internal
propagators, given by linear combinations of $\{\ell_s\}$ and $\{p_\alpha\}$, then
the contribution to the Feynman diagram takes the general form
\ben  \label{e1}
&& \int [dY] \int \prod_{s}d^D \ell_s \, \exp\left[-G_{rs}(Y) \ell_r \cdot 
\ell_s - 2 H_{s\alpha}(Y) \ell_s \cdot p_\alpha
- K_{\alpha\beta}(Y) p_\alpha \cdot p_\beta\right]\nonumber \\
&& \hskip 1in \times  \prod_i (k_i^2+m_i^2)^{-1} Q(Y, \ell, p) \, ,
\een
where $Y$ denotes collectively all the integration parameters $y$ from 
all the vertices, and $G_{rs}$,
$H_{s\alpha}$ and $K_{\alpha\beta}$ are matrices that arise by combining the
exponential factors \refb{ever1} from all the interaction vertices 
after expressing the momenta
$k_i$ carried by various propagators in terms of loop momenta and external momenta.
$Q(Y, \ell, p)$ is a function of the moduli $Y$ and a polynomial in the $\ell_i$\plu\ and $p_\alpha$\plu,
arising from the products of the factors of $P$ from each 
interaction vertex and the numerator
factors in various propagators.
Positive definiteness of $g_{ij}(y)$ in \refb{ever1} ensures that the matrix
$\pmatrix{G & H\cr H^T & K}$ is positive definite and hence $G$ and $K$
themselves are
positive definite.

Positive definiteness of $G_{rs}(Y)$ guarantees that the integration over the spatial
components of the loop momenta are free from UV divergence. 
However
if we regard the integration over the loop energies to be running along the real axis then
the  $\ell_i^0$ dependent
quadratic term in the exponent is given by $\exp[G_{rs}(Y) \ell_s^0 \ell_r^0]$, and
since
$G_{rs}$ is positive definite, the
integral diverges. The remedy suggested  in \cite{1604.01783} is to 
define the amplitude as analytic continuation of Euclidean Green's function. A
systematic procedure for doing this was described in \cite{1604.01783} and goes
as follows.
\begin{enumerate}
\item First we 
multiply all the  external
energies by a common complex number $u$ lying in the first quadrant of the complex 
plane.
\item For $u$ lying on the imaginary axis, we take all loop energy integration contours to be
along the imaginary axis -- starting at $-i\infty$ and ending at $i\infty$. In 
this case 
the energies carried by all the internal propagators are imaginary and therefore the 
$(k_i^2+m_i^2)^{-1}$ factors in \refb{e1} do not have any poles on the integration contours.
Furthermore the integrand is exponentially suppressed as the loop energies approach 
$\pm i\infty$ due to the exponential suppression factor from the vertices. Therefore the
integral is well defined.
\item We now deform $u$ towards 1 along the first quadrant. During this deformation some 
of the poles of the propagators may approach the loop energy integration contours. If we let
the poles cross the integration contour then the integral jumps discontinuously and
the result can no longer be regarded as the analytic continuation of the result from the
imaginary $u$-axis. Therefore we must deform the integration contours away from the
poles so that the poles never touch the integration contour. However
during this deformation we
must keep the ends of each loop energy integration contours at $\pm i\infty$ so that the
integral converges. The spatial components of loop momenta are always integrated along
the real axis.
\item The final result is taken to be the $u\to 1$ limit of the above result from the first
quadrant. In this limit
the integration 
contour over each loop energy
begins at $-i\infty$ and ends at $i\infty$ but has complicated shape in the interior. 
\item Since the poles do not cross the loop energy integration contours during the deformation
of $u$, given any contour we can determine on which side of it a 
given pole lies in the $u\to 1$ limit by knowing  the 
corresponding data for imaginary $u$. This leads to the following simple 
prescription\cite{1604.01783}
-- replace $k^2+m^2$ factors in the denominator by $k^2+m^2-i\eps$ and 
{\it pretend} that
near this pole the loop energy contours lie along the real axis
from $-\infty$ to $\infty$. Then the side of the contour
to which the pole of $(k^2+m^2-i\eps)^{-1}$ lies correctly determines the required 
information.
\end{enumerate}
Appendix \ref{sacutkosky}  illustrates  this procedure for choosing the loop energy
integration contour  for a simple
Feynman diagram. This procedure was used in \cite{1607.06500} to
compute the
real and imaginary parts of the renormalized mass$^2$ of a massive
particle in superstring theory at one loop order.

One worry one may have is whether this procedure is well defined.  As we are deforming 
$u$ from the imaginary axis to 1, it may happen that two poles of the 
integrand approach
each other from opposite sides of a loop energy contour, and prevent further
deformation without crossing one of the poles. It was shown in \cite{1604.01783} that this does
not happen; for any path in the complex $u$-plane from the imaginary axis to 1,
it is always possible to deform the loop energy integration contours while keeping it
away from
the poles.
This means that the result of integration is
an analytic function in the first quadrant of the 
$u$-plane.

There is an alternate prescription\cite{berera,1307.5124} 
in which we replace each of the propagator factors
$(k_i^2+m_i^2)^{-1}$  in \refb{e1} by its Schwinger parameter representation, 
but
with the Schwinger parameter integration running along the imaginary axis. More precisely,
we make the replacement
\be
(k_i^2+m_i^2)^{-1} \Rightarrow \int_0^{i\infty} dt_i 
e^{-t_i (k_i^2+m_i^2-i\eps)}
\ee
for some small positive constant $\eps$. We then carry out the 
loop momentum integrals in \refb{e1} using
the rules of gaussian integration pretending that they converge and express the
result as
\be 
\prod_i \int_0^{i\infty} dt_i \, e^{i\epsilon t_i}\, \, \int [dY] \, F(Y, \{t_i\}, \{p_\alpha\})
\ee
for some function $F$. This integral can be shown to give finite result in the $\eps\to 0^+$
limit. It was shown in \cite{1610.00443} that this prescription 
gives the same result as the one described
above involving non-trivial choice of loop energy integration contours.

\subsection{Wilsonian effective action} \label{swilson}

Before moving on, we shall comment on an interesting consequence of the prescription
involving non-trivial choice of loop energy integration contours.
From \refb{evertexexp} it would seem that by increasing the stub length we can increase
$\Lambda_i(y)$ arbitrarily, and this will bring down the effective UV cut-off of the
theory. At first sight this may seem surprising since one expects that in string theory
the UV cut-off should be given by 
the string scale. Now since for real external
energies the loop energy integration 
contours cannot be taken
fully along the imaginary axis, and since the exponential factor in \refb{ever1}
grows for real $k_i^0$, we cannot really bring down the UV cut-off to arbitrarily low
values -- the minimum is set by the spread of the loop energy integration contour 
along the real axis. In any scattering process it follows from simple scaling argument
that generically the spread of the loop energy integration contours along the real axis
will be of the order of the center of mass energy of the incoming particles and therefore
the cut-off cannot be reduced below this value. For scattering of massive particles
this is of the order of the string scale. However for mass$^2$ level zero particles the total center
of mass energy can be much lower than the string scale and for this case the UV cut-off
can indeed be made much lower than the string scale by taking
$\Lambda_i(y)$ to be sufficiently large. A physical explanation of this
was given in \cite{1609.00459} based on the identification of the effective master action 
of the mass$^2$ level zero
fields, obtained by integrating out the massive fields, as a Wilsonian effective action. 
The main idea is that as we increase the stub length, we are transferring some
contributions that were earlier in the Feynman diagrams with propagators into
the elementary vertex. As already remarked in \S\ref{scomstub},
in the limit of very large stub length, most of the contribution
to an amplitude comes from just the elementary vertex, and only contributions
very close to the boundary of the moduli space are captured by the Feynman diagrams
with propagators. Therefore the effective master action of
mass$^2$ level zero field described in \S\ref{seff} represents a Wilsonian effective
action\cite{wilson1,wilson2,polchinski} 
in which all the massive fields as well as modes of the mass$^2$ level zero fields above
a certain energy scale have been integrated out\cite{br1,br2}. 
From this point of view it is not surprising that the UV cut-off is
also controlled by the stub length and not by the string scale.

\sectiono{Unitarity of superstring theory} \label{sunitary}

Let $S={\bf 1}-iT$ denote the S-matrix of string theory. The unitarity constraint
$S^\dagger S={\bf 1}$ gives us 
\be \label{esumn}
i(T - T^\dagger) = T^\dagger T =  T^\dagger|n\rangle \langle n| T\, ,
\ee
where the sum over $n$ represents sum over complete set of asymptotic states
in the theory. 
We shall now discuss how superstring field theory establishes this relation
following the analysis of \cite{1604.01783,1606.03455,1607.08244}.
Alternative approach based on light-cone string
field theory has been pursued in \cite{berera}, but this suffers from the contact
term ambiguities\cite{gr1,gr2,gr3,greenseiberg,dhoker,1605.04666}. 
Ref.~\cite{dhoker} attempted to resolve this by showing the equivalence
of the covariant and light-cone string theories, but these arguments have not
been revisited in the light of recent understanding of the supermoduli 
space\cite{1304.7798,1404.6257}.

Throughout this section we shall work in Lorentzian space-time and with on-shell
external states carrying real energy and momentum. As mentioned in \S\ref{slorentz},
this requires multiplying the propagators by $-i$ and the
vertices by $i$. The Feynman diagrams computed
with these rules give matrix elements of $-i\, T$.

\subsection{Cutkosky rules} \label{scut}

Based on the prescription for integration over loop energies given in \S\ref{sloopcon},
\cite{1604.01783} proved Cutkosky rules for the amplitudes of superstring field theory,
namely,
the contribution to $i(T -T^\dagger)$ is given by the sum over cut 
diagrams\cite{Cutkosky,fowler,1512.01705}.  
We shall
begin by explaining these rules.
Let us represent a Feynman diagram with the incoming states to the left and the
outgoing states to the right.  Any
of its cut diagrams is represented by a line -- known as the cut -- passing
through the original diagram that separates
the incoming states from the outgoing states, and crosses one or more propagators.
The rules for computing the contribution from such a cut diagram are as follows:
\begin{enumerate}
\item The $-i(k^2+m^2)^{-1}$ factor of a cut internal propagator is replaced by 
$2\pi \, \delta (k^2+m^2) \, \Theta(k^0)$, where $k$ denotes the momentum 
flowing from the
left to the right of the cut and $\Theta$ is the step function. 
Cuts of external lines have no effect on the diagram.
\item Part of the diagram to the left of the cut is evaluated using the usual Feynman
rules. This gives the matrix element of $-i \langle n|T|b\rangle$ with $\langle n|$
representing the states associated with the cut propagator and $|b\rangle$
representing the incoming states.
\item Part of the diagram to the right of the cut is evaluated using the usual Feynman
rules with the following difference. First of all the parameter $u$ introduced in 
\S\ref{sloopcon}
is to be complex conjugated, i.e.\ we take the $u\to 1$ limit from the fourth quadrant.
Furthermore, all the loop energy integration contours are also complex conjugated.\footnote{For 
multi-component complex fields, one also needs to complex conjugate the indices carried by the fields and for fermions, one also needs to account for some additional signs. See \cite{1604.01783} for details.}
It was shown in \cite{1604.01783} that this is equivalent to evaluating the matrix element
$i \langle a|T^\dagger|n\rangle$ with $\langle a|$ representing the outgoing states and 
$|n\rangle$ representing the states associated with the cut propagator.
\end{enumerate}
The reality of the action, discussed in \S\ref{sreal},
is essential for the proof. 
We shall not give the details of the proof; the interested
reader may consult the original paper \cite{1604.01783}. 
A simple illustration of how it works can be found
in appendix \ref{sacutkosky}.

\begin{figure}
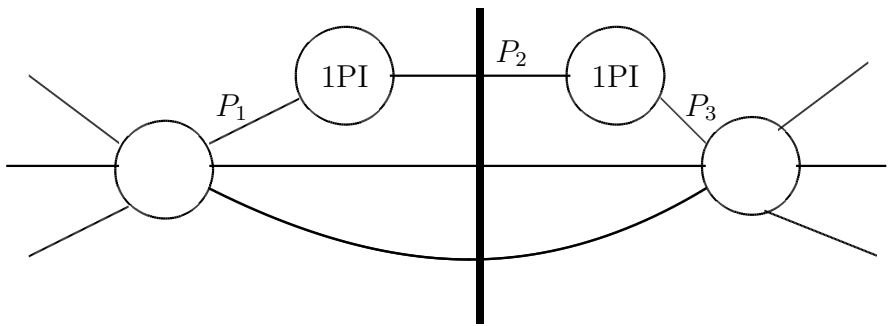


\begin{center}

\figblind

\end{center}

\vskip -1.5in
\caption{A problematic cut diagram, with  the vertical thick line representing the cut.
\label{fblind}
}
\end{figure}

 Naively, this establishes the unitarity relation \refb{esumn}
in component form:
\be 
i\langle a|(T - T^\dagger)|b\rangle =  \langle a|T^\dagger|n\rangle \langle n| T|b\rangle\, .
\ee
 However there are some
subtle points that need to be addressed. First of all, 
a blind application of these rules can lead to ambiguous results as can be seen from the
example shown in Fig.~\ref{fblind}. In this diagram the propagators $P_1$, $P_2$ and
$P_3$ carry the same momentum $k$. Since $P_2$ is cut, we have a factor of
$2\pi \, \delta(k^2+m^2)\Theta(k^0)$, but $P_1$ and $P_3$, being ordinary propagators,
give factors of $\mp i\, (k^2+m^2\mp i\eps)^{-1}$.
Therefore the product of their contributions is
ill defined in the $\eps\to 0$ limit, 
making the contribution of the cut diagram ill defined. The 
remedy\cite{veltman,diagrammar} is to sum over all cut diagrams that differ from each other in
where the cut intersects the top segment. This includes the cases where 
the cut passes through
the propagators $P_1$, $P_2$ or $P_3$, or one of the 1PI blobs. 
Using the Cutkosky rules, we can express the result
as the sum of  the full propagator 
and its hermitian conjugate, which we shall call the cut full
propagator.\footnote{During the analysis of 
\cite{1604.01783}
one actually first arrives at the result expressed in terms of
hermitian part of the full propagator and then
carries out further manipulation to express the result as the sum of all individual
cut diagrams. So all we need to do is to halt the analysis of \cite{1604.01783} after
one gets the result in terms of the hermitian part of the full propagator.}
This is the general procedure we shall follow for all cut diagrams, 
i.e. instead of allowing self
energy insertions on a cut propagator, we shall regard all cut propagators as
cut full propagators.
If the full propagator (after resummation of arbitrary number of insertions of
1PI blobs)  is given by $-i (k^2+m^2 - \Sigma(k)-i\eps)^{-1}$ 
where $i\Sigma(k) $ represents the contribution from the 1PI blobs, then in the cut full
propagator this factor is replaced by
\be \label{edisc}
-i \, \{ k^2 + m^2 -i\eps -\Sigma(k)\}^{-1} + i\, \{k^2 + m^2 +i\eps -\Sigma(k)^*\}^{-1} \, .
\ee
If $\Sigma(k)$ is real then the above expression is non-zero only when
$(k^2 + m^2  -\Sigma(k))$ vanishes. Let us suppose that this happens at $k^2=-M^2$ and
that near $k^2=-M^2$, we have
\be
(k^2 + m^2  -\Sigma(k))^{-1} = Z \, (k^2+M^2)^{-1} +\hbox{non-singular}\, .
\ee
In this case we have
\be 
-i\,  (k^2 + m^2 -i\eps -\Sigma(k))^{-1} +i\,  (k^2 + m^2 +i\eps -\Sigma(k)^*)^{-1}
= 2\pi \,  Z\, \delta(k^2+M^2) \, .
\ee
This is analogous to the rules for a cut propagator, except that this now applies to the
full propagator near its pole on the real $k^2$ axis.
On the other hand if $\Sigma(k)$ is complex, then the $i\eps$ terms in 
\refb{edisc} are irrelevant, and we may rewrite \refb{edisc} by
\be \label{ecutdiff}
-i \{k^2 + m^2 -\Sigma(k)\}^{-1} \, (-i)  \, \{\Sigma(k)  - \Sigma(k)^*\} \, i\, \{k^2 + m^2 -
\Sigma(k)^* \}^{-1}\, .
\ee
Pictorially this may be represented by Fig.~\ref{fcutPI}.

\begin{figure}

\begin{center}

\figcutPI

\end{center}

\vskip -1in

\caption{A pictorial representation of \refb{ecutdiff}. \label{fcutPI}}

\end{figure}

\begin{figure}
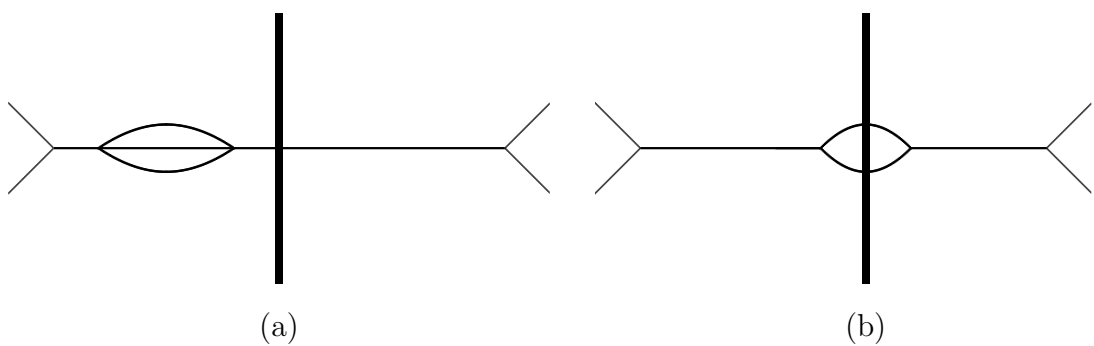


\begin{center}

\figoneuni

\end{center}

\vskip -.2in

\caption{Fig.~(a) shows the example of a disallowed cut diagram and Fig.~(b) shows the
example of an allowed cut diagram. In both examples the thick vertical line denotes the 
cut.
\label{f1C}}
\end{figure}

Therefore the procedure for summing over cut diagrams can be stated as follows.
\begin{enumerate}
\item In the internal uncut lines of a Feynman diagram we use the 
full propagator
$-i\, \{k^2 + m^2 -i\eps-\Sigma(k)\}^{-1}$, with $\Sigma(k)$ computed to
the desired order in perturbation theory.
\item If the full
propagator has a pole at $k^2=-M^2$ 
on the real $k^2$ axis with residue $-i\, Z$, then the cut full
propagator has a contribution $2\, \pi \,  Z\, \delta(k^2+M^2)$. Therefore the internal
states $|n\rangle$ over which we sum have renormalized mass. This is clearly
a desired result since asymptotic states carry renormalized mass.
\item If the full
propagator has a pole in the complex $k^2$ plane off the real axis, then
we do not need to include any additional contribution in the expression for the
cut propagator. This corresponds to the case of complex $\Sigma(k)$ and is included
in diagrams of the form shown in Fig.~\ref{fcutPI}. This is in accordance with the fact
that complex poles in the $k^2$ plane represent unstable particles, and they are not
genuine asymptotic states.
\item Since it is understood that each (cut) propagator is the
(cut) full propagator, we do not
include separately diagrams with self-energy insertions on a cut
propagator like the one shown in 
Fig.~\ref{f1C}(a). However a cut could pass through the 
1PI blob of a self-energy insertion diagram, {\it e.g.} a cut diagram of the form shown 
in Fig.~\ref{f1C}(b) is allowed, and represents part of the contribution to
Fig.~\ref{fcutPI}. It is again understood that the internal uncut propagators are
full propagators.
\end{enumerate}

These results are well suited for being adapted to superstring field theory, since the
propagator $i\Delta_F$ with $\Delta_F$ given in \refb{edefDeltaI} is the full propagator after inclusion
of self-energy corrections. Let us suppose that $\Delta_F$ given in \refb{edefDeltaI}
has a pole at $k^2+M^2=0$ with real $M^2$. 
Then near $k^2=-M^2$ we have
\be  \label{edfres}
i\, \Delta_F = -i\, (k^2+M^2-i\eps)^{-1} \Xi_0 + \hbox{non-singular}\, .
\ee
Even though
$\Delta_F$ is an infinite dimensional matrix,\footnote{Since
in any scattering process with given set of incoming particles, there is an upper bound
on the maximum mass$^2$ level particle that may be produced, one can always work with
the effective action of \S\ref{seff} obtained by integrating out fields above that mass$^2$ level.
This way one never has to deal with infinite dimensional matrices.} 
$\Xi_0$ is a matrix of finite rank
since for given momentum we expect only a finite number of states 
for which the propagator develops a pole at $k^2=-M^2$.
Then the cut propagator is given by
\be\label{epole}
2\pi \, \delta(k^2+M^2) \, \Theta(k^0) \, \Xi_0\, .
\ee

We now turn to the second subtlety in going from Cutkosky rules 
to the proof of unitarity of the S-matrix.
If all poles of $\Delta_F$ had represented physical states then the result quoted
above would imply \refb{esumn}, with the integration over the momenta of the cut propagator
representing sum over intermediate states $|n\rangle$, and $\Xi_0$
representing the effect of wave-function renormalization. However not every pole of
the Siegel gauge propagator represent physical states. Therefore we need to show
that the contribution from the additional states cancel among themselves. This is the
task to which we now turn.

\subsection{Properties of the propagator} \label{spropsieg}

It should be clear from  \refb{epole} that for analyzing the 
contribution from cut diagrams
we need to focus on the properties of $\Xi_0$ associated
with the poles that occur at real momenta. 
Multiplying both sides of \refb{eboth} by 
$k^2+M^2$, using \refb{edfres} and taking the limit $k^2\to -M^2$, we get
\be \label{eqdid}
\wh Q_B \Xi_0 c_0^-+ \Xi_0 c_0^- \wt Q_B = 0\, .
\ee
Now, from the property of $\Delta_F$ mentioned below \refb{esiegel} and the definition
\refb{edfres} of $\Xi_0$ it follows that 
$\Xi_0$ acts on states in 
$c_0^-\wt\HH_T$ to produce states in $\wh\HH_T$.
This, together with the fact that BPZ inner product pairs states
in $\wh\HH_T$ with states in $c_0^-\wt\HH_T$, allows us to express
$\Xi_0$ as
\be \label{e9.12}
\Xi_0=\sum_{m=1}^R |\Phi_m\rangle \langle \Psi_m|,
\qquad |\Phi_m\rangle\in\wh\HH_T, \quad |\Psi_m\rangle\in\wh\HH_T\, ,
\ee
where $R$ is the rank of $\Xi_0$ and $\{|\Phi_m\rangle\}$ and $\{\langle\Psi_m|\}$
are a set of linearly independent states. 
While the forms of individual $|\Phi_m\rangle$'s (and $\langle\Psi_m|$'s) 
are ambiguous since we can take linear combinations of these states and declare
them as our new $|\Phi_m\rangle$'s (and $\langle\Psi_m|$'s), the linear span
of the subspace of $\wh\HH_T$ spanned by the $|\Phi_m\rangle$'s 
(and independently the $\langle\Psi_m|$'s) is unambiguous. We now divide the
states $|\Phi_m\rangle$ into the following categories\footnote{This classification
agrees with the prescription given in \cite{1311.1257,1401.7014}.}
\begin{enumerate}
\item Unphysical states: These are linearly independent 
states $|U_r\rangle$ satisfying 
\be \label{eun}
\wh Q_B \sum_r a_r |U_r\rangle \ne 0
\quad \hbox{for any choice of $\{a_r\}$ other than $a_r=0$ for every $r$.}
\ee
\item Physical states: These are states satisfying 
\ben \label{eph}
&& \wh Q_B |P_a\rangle =0,
 \\
&& \sum_a c_a |P_a\rangle \ne \wh Q_B|\Lambda\rangle, 
\quad \hbox{for any choice of $|\Lambda\rangle$ or $\{c_a\}$ other than
$c_a=0$ for every $a$.} \nonumber 
\een
These represent states that satisfy
the linearized equations of motion \refb{eqexp}, but are not pure gauge in the sense
described below \refb{eqexp}.
\item Pure gauge states: These are linearly independent states of the form 
$\wh Q_B |\Lambda_\alpha\rangle$, with $\alpha$ running over a certain range
of values.
\end{enumerate}
The  candidates for $\langle\Psi_m|$ can be similarly classified, although we shall
not directly make use of this below. Each of the states $|\Phi_m\rangle$ and 
$\langle\Psi_m|$ are also annihilated by $b_0^+$ due to the Siegel gauge condition
\refb{eb0pm}, but in the light of the discussion in \S\ref{echprop}, 
we shall proceed without making this assumption so that our results are valid
also for the modified propagator $i\, \Delta_F^\alpha$ defined there. 

Let us
suppose that at some given momentum at which $\Delta_F$ has a pole,
there are a certain number of
linearly independent 
physical states $\{|P_a\rangle\}$, unphysical states $\{|U_r\rangle\}$ 
and pure gauge states $\{\wh Q_B |\Lambda_\alpha\rangle\}$. Then $\Xi_0$
can be expressed as 
\be \label{eDel}
\Xi_0 = \sum_a |P_a\rangle \langle B_a| 
+ \sum_r |U_r\rangle \langle C_r| + \sum_\alpha \wh Q_B |\Lambda_\alpha
\rangle \langle \Sigma_\alpha|\, ,
\ee
for some linearly independent 
states $|B_a\rangle,|C_r\rangle,|\Sigma_\alpha\rangle\in\wh\HH_T$. Our
goal will be to determine the general form of these states. 
Substituting \refb{eDel} into
\refb{eqdid} and using  \refb{eph} we get
\be \label{e9.18}
\sum_r \wh Q_B |U_r\rangle \langle C_r| c_0^-+
 \sum_a |P_a\rangle \langle B_a| c_0^-  \wt Q_B + 
 \sum_r |U_r\rangle \langle C_r|  c_0^-  \wt Q_B+ \sum_\alpha \wh Q_B |
 \Lambda_\alpha\rangle \langle \Sigma_\alpha| 
  c_0^-  \wt Q_B= 0\, .
\ee
Applying $\wh Q_B$ from the left and using \refb{eun} 
we see that the coefficient of $|U_r\rangle$ in
the third term in \refb{e9.18} must
vanish by itself. Using \refb{eph} we now see that the coefficient of $|P_a\rangle$
in the second term must also vanish.  This gives, using \refb{eabqb},
\ben \label{erelnew}
\langle B_a| c_0^-  \wt Q_B =0 \quad &\Rightarrow&
\quad \langle\wh Q_B B_a|=0, 
\nonumber \\  
\langle C_r| c_0^- \wt Q_B=0  \quad &\Rightarrow&
\quad \langle \wh Q_B C_r|=0\, .
\een
Therefore we are left with the first and
the last term in \refb{e9.18}.  Now let us suppose that the linear span of the states 
$\{\wh Q_B|\Lambda_\alpha\rangle\}$ contains linearly independent states 
$\{\wh Q_B|\lambda_k\rangle\}$ that are outside the linear span of the 
states $\{ \wh Q_B|U_r\rangle\}$. Then we can write
\be  \label{erelze}
\wh Q_B|\Lambda_\alpha\rangle = \sum_r A_{\alpha r} \wh Q_B |U_r\rangle
+ \sum_k  S_{\alpha k} \wh Q_B |\lambda_k\rangle\, ,
\ee
for some coefficients $A_{\alpha r}$ and $S_{\alpha k}$. Substituting this into
\refb{e9.18} and using \refb{erelnew} and
the fact that $\{\wh Q_B|\lambda_k\rangle\}$  and 
$\{ \wh Q_B|U_r\rangle\}$ are linearly independent states, we get
\be \label{erelfi}
\langle \tau_k |  c_0^- \wt Q_B = 0, 
\quad \langle \tau_k| \equiv 
\sum_\alpha S_{\alpha k} \langle \Sigma_\alpha|\, , \qquad \Rightarrow \qquad
\langle \wh Q_B\tau_k|=0\, ,
\ee
and
\be \label{erelmid}
\langle C_r| c_0^- = - \langle D_r|  c_0^-  \wt Q_B, \quad 
\langle D_r| \equiv \sum_\alpha A_{\alpha r} \langle \Sigma_\alpha|,
\qquad \Rightarrow
\qquad \langle C_r| =
- (-1)^{{D_r}} \langle \wh Q_B D_r| \, ,
\ee
where $(-1)^{D_r}$ takes value 1 if $D_r$ is 
Grassmann even  and $-1$ if $D_r$ is
Grassmann odd. Using \refb{erelze}, \refb{erelfi} and \refb{erelmid} we can 
express \refb{eDel} as
\be \label{edel0}
\Xi_0 = \sum_a |P_a\rangle \langle B_a| 
- \sum_r (-1)^{  {D_r}}  |U_r\rangle \langle \wh Q_B D_r|
+ \sum_r \wh Q_B |U_r\rangle \langle D_r| + \sum_k \wh Q_B|\lambda_k\rangle 
\langle \tau_k|\, .
\ee
Since $\Delta_F$ given in \refb{edefDeltaI}  carries total ghost number $-2$ and
since the BPZ inner product pairs
states carrying total ghost number 6,  we have
\be \label{enghost}
n_{P_a}+n_{B_a}=4, \quad n_{U_r}+n_{D_r}=3, \quad n_{\lambda_k} + n_{\tau_k}
=3\, ,
\ee
where for any state $|A\rangle$, $n_A$ denotes its ghost number. In the following we
shall for simplicity of notation absorb the last term $\sum_k \wh Q_B|\lambda_k\rangle 
\langle \tau_k|$ into the sum $ \sum_a |P_a\rangle \langle B_a|$ since, like
$P_a$ and $B_a$, $\wh Q_B \lambda_k$ and $\tau_k$ are annihilated by 
$\wh Q_B$. Later we shall argue that the contribution from the 
$\sum_k \wh Q_B|\lambda_k\rangle 
\langle \tau_k|$ term to the cut
diagram actually vanishes.

\subsection{Unitarity} \label{sproofA}

In the analysis of this section the main players will be the 
 amputated Green's function $ \Gamma(
| A_1\rangle, \ldots | A_N\rangle)$
introduced in \S\ref{swardlocal} and the residue $\Xi_0$ of $\Delta_F$ introduced 
in \refb{edfres}.
$\Gamma$ satisfies the identity \refb{egrid}, and gives the matrix elements of 
$-iT$ up to
wave-function renormalization constants when the external states are physical states, annihilated by
$\wh Q_B$.

\begin{figure}
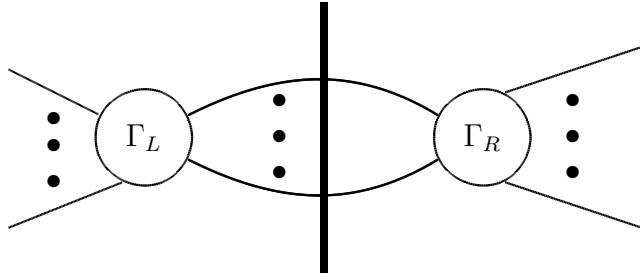

\begin{center}

\figtwouni

\end{center}

\vskip -.8in

\caption{A cut diagram in superstring field theory. \label{f2C}}

\end{figure}

Now, in a cut diagram like the one shown in Fig.~\ref{f2C},
each cut propagator is replaced by the factor $2\,\pi\,\delta(k^2+M^2)\, \Theta(k^0)\Xi_0$, where $\Xi_0$ is given by the right hand side of equation \refb{edel0}.
Let us suppose that we have a cut diagram with $N$ cut propagators.
Using the
superscript $(i)$ to label the states associated with the $i$-th cut propagator
and the operators acting on these states, we have a net factor of
\be \label{ebbi}
\prod_{i=1}^N (\Xi_0)^{(i)} =
\prod_{i=1}^N  \left[ \sum_a |P_a^{(i)}\rangle \langle B^{(i)}_a| 
- \sum_r (-1)^{  {D_r^{(i)}}} |U_r^{(i)}\rangle 
\langle \wh Q_B^{(i)} D^{(i)}_r|  \,
+ \sum_r \wh Q_B^{(i)} |U_r^{(i)}\rangle 
\langle D^{(i)}_r|  
 \right]\, ,
\ee
associated with all the cut propagators.\footnote{The range of $a$ and $r$ in \refb{ebbi} 
are in general different for different $i$.} It will be understood that the sum over
$a$ includes the $\sum_k \wh Q_B^{(i)}|\lambda^{(i)}_k\rangle 
\langle \tau^{(i)}_k|$ term as well. Since the 
incoming states are drawn to the left and the outgoing
state are drawn to the right, the natural convention is that
the
ket states of \refb{ebbi} are inserted into the amplitude $\Gamma_R$ 
on the right side of the cut and the bra states are inserted into
the amplitude  $\Gamma_L$ on the left side of the cut.
Besides these $\Gamma_L$ and $\Gamma_R$ have insertions of
external incoming and outgoing states respectively, which are all annihilated by
$\wh Q_B$.

We now expand \refb{ebbi} as a sum of $3^N$ terms. 
There is one term given by 
\be \label{edesd}
\prod_{i=1}^N \left\{ \sum_a |P_a^{(i)}\rangle \langle B^{(i)}_a| \right\} \, .
\ee
We shall now show that the contribution to the cut diagram from all 
other terms in the right hand side of equation \eqref{ebbi} cancel among themselves.
A quick way to prove this would be to note that by modifying the propagator
in the way described in \S\ref{echprop} we could remove the second and the
third term inside the square bracket on the right hand side of \refb{ebbi}.  
However we
shall provide a direct combinatorics proof below.
On the other hand \refb{edesd} gives the term required for proving unitarity of the
S-matrix.

 Before considering the general case, let us illustrate how this works using 
some simple examples. First consider the term in \refb{ebbi} where $(N-1)$ of
the terms are of the form $\sum_a |P_a^{(i)}\rangle \langle B^{(i)}_a|$:
\be \label{eesq}
\sum_{j=1}^N 
\prod_{i=1\atop i\ne j}^N \left\{ \sum_a |P_a^{(i)}\rangle \langle B^{(i)}_a| \right\} 
\left[
- \sum_r (-1)^{  {D_r^{(j)}}} |U_r^{(j)}\rangle 
\langle \wh Q_B^{(j)} D^{(j)}_r|  \,
+ \sum_r \wh Q_B^{(j)} |U_r^{(j)}\rangle 
\langle D^{(j)}_r|  
\right]\, .
\ee
If we pick the first term inside the square bracket then $\Gamma_L$ will have an
external state $\wh Q_B^{(j)} D^{(j)}_r$. All other external states of $\Gamma_L$ 
are annihilated by $\wh Q_B$. The Ward identity \refb{egrid} now tells us that this
amplitude vanishes. Similarly for the second term inside the square bracket in
\refb{eesq}, the $\Gamma_R$ will have one insertion of $\wh Q_B^{(j)} U_r^{(j)}$ and
other insertions of $\wh Q_B$ invariant states. This again vanishes by \refb{egrid}.
Therefore the term given in \refb{eesq} does not contribute to \refb{ebbi}.

 The next complicated case is when $(N-2)$ terms are of the form 
$\sum_a |P_a^{(i)}\rangle \langle B^{(i)}_a|$:
\ben \label{eesq2}
&& \sum_{j,k=1\atop j<k}^N 
\prod_{i=1\atop i\ne j,k}^N \left\{ \sum_a |P_a^{(i)}\rangle \langle B^{(i)}_a| \right\} 
\left[
- \sum_r (-1)^{  {D_r^{(j)}}} |U_r^{(j)}\rangle 
\langle \wh Q_B^{(j)} D^{(j)}_r|  \,
+ \sum_r \wh Q_B^{(j)} |U_r^{(j)}\rangle 
\langle D^{(j)}_r|  
\right] \nonumber \\ && \hskip 1in
\left[
- \sum_s (-1)^{  {D_s^{(k)}}} |U_s^{(k)}\rangle 
\langle \wh Q_B^{(k)} D^{(k)}_s|  \,
+ \sum_s \wh Q_B^{(k)} |U_s^{(k)}\rangle 
\langle D^{(k)}_s|  
\right]\, .
\een
If we pick the first term from inside each square bracket then $\Gamma_L$ will have
two insertions of $\wh Q_B$ exact states and other insertions of $\wh Q_B$
invariant states. This vanishes by \refb{egrid}. If we pick the second term from inside
each square bracket then $\Gamma_R$ will vanish due to similar reasons. 
Therefore the only combination of terms in the product of the two square 
brackets that could give non-zero contribution is:
\ben \label{eesq3}
&& - \sum_r (-1)^{  {D_r^{(j)}}} |U_r^{(j)}\rangle 
\langle \wh Q_B^{(j)} D^{(j)}_r|  \, \sum_s \wh Q_B^{(k)} |U_s^{(k)}\rangle 
\langle D^{(k)}_s|  \nonumber \\ &&
- \sum_r \wh Q_B^{(j)} |U_r^{(j)}\rangle 
\langle D^{(j)}_r|  \sum_s (-1)^{  {D_s^{(k)}}} |U_s^{(k)}\rangle 
\langle \wh Q_B^{(k)} D^{(k)}_s| \, .
\een
Let us examine the contribution from the first term. For this term $\Gamma_L$
has insertions of $\wh Q_B^{(j)} D^{(j)}_r$, $ D^{(k)}_s$ and other $\wh Q_B$
invariant states. Using \refb{egrid} we can move the $\wh Q_B$ operator from
$D^{(j)}_r$ to $D^{(k)}_s$ at the cost of picking up a sign of
$(-1)^{D^{(j)}_r+U_s^{(k)}}$. Similarly for the second term, on
$\Gamma_R$ we have insertions of $\wh Q_B^{(j)} U_r^{(j)}$, $U_s^{(k)}$ and
other $\wh Q_B$ invariant states, and using \refb{egrid}
we can move $\wh Q_B$
from $U_r^{(j)}$ to $U_s^{(k)}$ at the cost of picking up a sign of
$(-1)^{U^{(j)}_r+D_r^{(j)}+1}$. This makes the contribution from the 
first term in \refb{eesq3} identical to the second term up to a sign. 
The relative sign can easily be seen to be $-1$ once we use the fact that
$U_r$ and $D_r$ have opposite Grassmann parities as a consequence of
\refb{enghost}. Therefore the two 
contributions cancel, showing that \refb{eesq2} does not
contribute to \refb{ebbi}.

Let us now turn to the general case.
 In the following we use the convention that
for any set $S_1$ and its subset $S_2$, $S_1-S_2$ denotes
the complement of $S_2$ in $S_1$.
We now 
group together all terms  in the expansion of \refb{ebbi} with
the same factors of $ \sum_a |P_a^{(i)}\rangle \langle B^{(i)}_a|$,
and in any given
group denote by $S$ the set
of labels $i$ carried by the rest of the factors.
Let $\alpha$ be a particular label of $S$ -- for
definiteness we can take this to be the lowest element of $S$.
For any $A\subseteq S -\{\alpha\}$  we introduce two amplitudes 
$F_S(\alpha; A)$ and $G_S(\alpha;A)$ as follows.
Both in $F_S(\alpha;A)$ and $G_S(\alpha;A)$ 
the labels $i$ in $A$ are carried by $-\sum_r 
(-1)^{D^{(i)}_r}|U_r^{(i)}\rangle \langle  \wh Q_B^{(i)}D^{(i)}_r| $ and
the labels $i$ 
in $S-\{\alpha\}-A$ are carried by 
$\sum_r \wh Q_B^{(i)}|U_r^{(i)}\rangle \langle D^{(i)}_r|$.
In $F_S(\alpha;A)$  the label $\alpha$ is carried by the factor 
$ - \sum_r (-1)^{  {D_r^{(\alpha)}}}  |U^{(\alpha)}_r\rangle 
\langle\wh Q_B^{(\alpha)}D^{(\alpha)}_r|  $, whereas in 
$G_S(\alpha; A)$ the  label $\alpha$ is carried by $\sum_r
\wh Q_B^{(\alpha)}|U^{(\alpha)}_r\rangle 
\langle D^{(\alpha)}_r|  $.  
Then the sum of all terms with a fixed set of labels $i\in 
\{1,\ldots N\}-S$ carrying $ \sum_a |P_a^{(i)}\rangle \langle B^{(i)}_a|  $ 
factors,
is given by
\be \label{epair}
\sum_{A\subseteq S-\{\alpha\}} \left[ F_S(\alpha; A) + G_S(\alpha; A)\right]  \, .
\ee
The sum is clearly independent of the choice of $\alpha$. 

Now in $F_S(\alpha; A)$
the external states of 
the amplitude $\Gamma_L$
on the left of the cut are
$\langle B^{(i)}_a|  $ for $i\not\in S$, 
$-(-1)^{D^{(\alpha)}_r}\langle \wh Q_B^{(\alpha)}D^{(\alpha)}_r| $, 
$-(-1)^{D^{(i)}_r} \langle  \wh Q_B^{(i)} D^{(i)}_r|$ with $i\in A$, 
$\langle D^{(i)}_r|$ for $i\in S-\{\alpha\}-A$ and 
the incoming physical states.  Using 
\refb{egrid} we can express this amplitude as a
sum of terms in which $\wh Q_B^{(\alpha)}D^{(\alpha)}_r$ is replaced 
by $D^{(\alpha)}_r$, but $\wh Q_B$ acts in turn on the other states.
Since the incoming states as well as $B_a^{(i)}$ and 
$\wh Q_B^{(i)} D^{(i)}_r$ are all annihilated by $\wh Q_B$, the only
non-vanishing contribution comes from the terms where 
$\wh Q_B$ acts on one of the states 
$\langle D^{(j)}_r|  $ for $j\in S-\{\alpha\}-A$. This gives,
\be \label{efid}
F_S(\alpha; A) = \sum_{j\in S-\{\alpha\}-A} \, s(\alpha; j; A) \, H_S(\alpha; j; A)
\ee
where $s(\alpha; j; A)$ takes value $\pm 1$ and
$H_S(\alpha; j; A)$ denotes the contribution from a cut diagram where  
the label $j$ is carried by
$-\sum_r (-1)^{D^{(j)}_r}
\wh Q_B^{(j)}|U^{(j)}_r\rangle \langle \wh Q_B^{(j)}D^{(j)}_r|$, the 
label $\alpha$ is carried by
$\sum_r |U^{(\alpha)}_r\rangle 
\langle D^{(\alpha)}_r|$, the labels $i$ in $A$ are
carried by
$-\sum_r (-1)^{D^{(i)}_r} |U_r^{(i)}\rangle \langle \wh Q_B^{(i)}D^{(i)}_r|$
and the labels $i$ in $S-\{\alpha\}-A-\{j\}$ are carried by 
$\sum_r \wh Q_B^{(i)}|U_r^{(i)}\rangle \langle D^{(i)}_r|$. 
The sign $s(\alpha;j;A)$ can be computed by 
keeping track of the movement
of $\wh Q_B^{(\alpha)}$ inside the expansion of
\refb{ebbi} and the extra minus sign that comes from having to move part of the
contribution from the left hand side to the right hand side of \refb{egrid} in applying
the Ward identity.
This gives
\ben \label{eone}
\sum_{A\subseteq S-\{\alpha\}} F_S(\alpha; A) 
&=& \sum_{A\subseteq S-\{\alpha\}} \sum_{j\in S -A-\{\alpha\}} 
s(\alpha; j; A) \, H_S(\alpha; j; A)
\nonumber \\
&=& \sum_{j\in S-\{\alpha\}} \sum_{A\subseteq S-\{\alpha, j\}}
s(\alpha; j; A) \, H_S(\alpha; j; A)\, .
\een
Carrying out a similar manipulation of the amplitude 
$\Gamma_R$ on the right of the cut, moving
$\wh Q_B$ from $U^{(\alpha)}_r$ to one of the states $U^{(j)}_r$ for $j\in A$, we get
\be \label{egidA}
G_S(\alpha; A) = \sum_{j\in A} s' (\alpha; j; A-\{j\}) H_S(\alpha; j; A-\{j\})\, ,
\ee
where $s' (\alpha; j; A-\{j\})$ takes value $\pm 1$. Therefore
\ben \label{etwo}
\sum_{A\subseteq S-\{\alpha\}} G_S(\alpha; A) 
&=& \sum_{A\subseteq S-\{\alpha\}} \sum_{j\in A} s'(\alpha; j; A-\{j\})\,  H_S(\alpha; j; A-\{j\})
\nonumber \\
&=& \sum_{j\in S-\{\alpha\}} \sum_{A\subseteq S-\{\alpha, j\}} s'(\alpha; j; A)\, 
H_S(\alpha; j; A)\, ,
\een
where in the last step we have relabelled $A-\{j\}$ as $A$.
The right hand sides of
 \refb{eone} and \refb{etwo} are the same up to signs. It was shown in
 \cite{1607.08244} that we always have
 \be
 s'(\alpha; j; A) = - s(\alpha; j; A)\, .
 \ee
This in turn shows that the right hand sides of \refb{eone} and
\refb{etwo} cancel, 
making \refb{epair} vanish. Therefore 
the only term that contributes to the cut diagram is the one 
where \refb{ebbi} is replaced by \refb{edesd}.

Now the
first equation in \refb{erelnew} shows 
that $\langle B_a|$ is annihilated by $\wh Q_B$.
This allows 
$\langle B_a|$ to be either a physical state or
a pure gauge state of the form $\langle \wh 
Q_B E_a|$ for some $\langle E_a|$.
However since all other states entering in the argument of
$\Gamma_L$ are annihilated by $\wh Q_B$, the 
amplitude with one or more $\langle B_a|$
having the form $\langle\wh Q_B E_a|$ will
vanish due to \refb{egrid}. This shows that $\langle B_a|$ must be a physical state.
Similarly the term $\sum_k \wh Q_B|\lambda_k\rangle \langle \tau_k|$, which
was included in the sum  $\sum_a |P_a\rangle\langle B_a|$, 
will also give vanishing contribution. It now follows from 
\refb{edesd} that only physical states contribute to the cut propagators.
The states $|P_a\rangle$ and $\langle B_a|$ are not 
normalized, but the residue
$\sum_a |P_a\rangle \langle B_a|$ may be expressed as 
$\sum_{i,j} |N_i\rangle Z_{ij} \langle N_j|$  in terms of normalized physical
states $|N_i\rangle$,
and the 
normalization matrix $Z_{ij}$ can be absorbed into the
definition of the T-matrix elements on the two sides of the cut by the
standard LSZ rules. Finally,
the $2\pi \delta(k^2+M^2)\theta(k^0)$ factor in the cut propagator produces the
correct phase space integral over the momenta carried by the intermediate states.

This establishes \refb{esumn}
and hence the unitarity of the amplitude.

\sectiono{Other approaches} \label{sother}

The formulation of superstring field theory 
described in this review makes manifest the infrared 
divergences in superstring theory,
and allows us to use standard techniques of quantum field theory to address them. This 
is useful for proving general results on perturbative superstring theory, {\it e.g.} the unitarity of the
theory as discussed in \S\ref{sunitary}. 
We expect that this formulation can be used to address various other conceptual
issues in superstring theory -- {\it e.g.} infrared divergences in four dimensions,
analytic properties of S-matrix etc.\  using standard tools of quantum field 
theory.  It may also be useful in addressing the question of 
background independence of superstring theory.
However as it stands, this formalism 
does not provide us with an 
efficient way of computing amplitudes of superstring theory. Before embarking on any computation,
one has to choose local coordinates at the punctures and the PCO locations on the Riemann 
surfaces consistent with the procedure described in \S\ref{ssft}, since the intermediate steps in
the analysis depend on this choice. While it is in principle possible to carry on this program with the
help of numerical codes, any analytic computation will require a 
choice of the data 
mentioned above that can be specified in closed form. At present no natural 
choice is known. For part of the data -- the
choice of local coordinate system -- one can use the  
prescription given in \cite{9206084} 
for bosonic string field theory in which we use the minimal area 
metric to fix the local coordinates around the punctures. Even 
for this prescription explicit form of the minimal area metric is not known and
therefore explicit analytic computation of off-shell amplitudes has not
been possible. Another possible approach, in which we choose local
coordinates around the punctures by making use of the
constant  negative curvature metric on the Riemann surface, 
has been explored recently in
\cite{roji,1706.07366}. It will clearly be desirable to have explicit closed form solutions
to the constraints given in \S\ref{ssft}, giving specific choice of local coordinates
around the punctures and specific choice of PCO insertions.

For {\it tree level} open and closed bosonic string field theories, 
canonical choices for local coordinates around the punctures 
are available. For
open strings the star product defined by the Witten interaction 
vertex\cite{wittensft} provides us with 
an explicit description of the three
string interaction vertex, and we do not need interaction vertices with four or more strings. For closed strings we need
interaction vertices with arbitrary number of external strings, but explicit form of these interaction vertices -- i.e. the region of the
moduli space they cover as well as the choice of local coordinate system on the corresponding
Riemann surfaces -- can be provided by the
Strebel quadratic differential\cite{saadi,kugo}. 

At tree level, 
there have been various recent (and not so recent) proposals for
open and closed superstring field theories. These approaches may be 
classified according to their off-shell field content. In the approach described
in this review the off-shell closed string field is an arbitrary GSO even state in
the small Hilbert space, annihilated by $L_0^-$ and $b_0^-$, and
carrying picture numbers $-1$, $-1/2$ and $-3/2$. Analogous
formulation of open superstring field theory will involve arbitrary GSO even states
in the small Hilbert space of picture numbers $-1$, $-1/2$ and $-3/2$.
However not all approaches to superstring field theory use the same set
of fields. 

One of the earliest and fully consistent and manifestly
Lorentz covariant formulation
of tree level open superstring field theory for the NS sector states was given in \cite{9503099}.
This approach takes the string field to be in the large
Hilbert space without any restriction on the picture number. 
In \cite{0406212,0409018}, this 
procedure was combined with the formulation of closed bosonic string field theory given in \cite{saadi,kugo} to give a
consistent and Lorentz covariant formulation of tree level heterotic string field theory for the
NS sector fields.  

A formulation of tree level 
open string field theory that uses GSO even states in the small Hilbert
space carrying picture numbers $-1$, $-1/2$ and $-3/2$ 
was given in \cite{1312.2948,1506.05774,1602.02582,1602.02583}.
Its generalization to tree level closed
string field theory of NSNS sector fields, using GSO even states in the small Hilbert space annihilated
by $b_0^-$, $L_0^-$ and carrying picture number $-1$
was given in \cite{1403.0940}. These theories are closely related
to the ones described in this review. However the former
give explicit prescriptions for inserting PCOs
satisfying the constraints described in \S\ref{ssft} (and analogous constraints for
open string field theories).  Also all PCOs are inserted via line integrals of
the type given in \refb{edefggr}. 
Of course this choice is not unique, but our general 
analysis reviewed in appendix \ref{safield} suggests that 
any other choice will be related to these by
appropriate field redefinition.
Furthermore 
refs.\cite{1505.01659,1505.02069,1510.00364} also describe
ways of relating 
the tree level NS sector open string field theory, formulated in \cite{1312.2948}, 
to that based on Berkovits
formalism\cite{9503099} after partially gauge fixing the latter.
It will be interesting to explore if these constructions
can be generalized to loop level. 

Another approach to the construction of tree level open superstring field theory
was given in \cite{1508.00366}.  
In this formulation the NS sector states are in the large Hilbert space without 
any restriction on the picture number as in \cite{9503099}, but the R
sector states span a GSO even  proper subspace of the  small
Hilbert space of picture number $-1/2$. In \cite{1602.02582,1602.02583} 
a modified version of this formalism was given, in which the NS sector
fields are in the small Hilbert space carrying picture number $-1/2$ and the
R sector states are in a  GSO even proper subspace 
of the small Hilbert space carrying
picture number $-1/2$. An advantage of this approach 
over the one described
in this review is that one does not need to introduce extra free fields in
the Ramond sector. 
It is conceivable that this type of action can be obtained as
a result of partially gauge fixing a theory in which R sector fields take value
in the full  GSO even subspace of  the small Hilbert space carrying
picture number $-1/2$ (and possibly $-3/2$), 
but this has not yet been proven. At present the
generalization of this formalism to closed strings or loop amplitudes is
not known, but it will be worth exploring the possibility

\bigskip

\noindent{\bf Acknowledgements:}  We wish to thank Ted Erler, Rajesh Gopakumar,
R.~Loganayagam, Faroogh Moosavian, Yuji Okawa,
Roji Pius, Arnab Rudra, Ivo Sachs and Barton Zwiebach for useful discussions.
H.E. acknowledges support from Cefipra
under project 5204-4.
The work of SPK, AS and MV 
was
supported in part by the 
DAE project 12-R\&D-HRI-5.02-0303. The work of AS was also supported
by the J. C. Bose fellowship of 
the Department of Science and Technology, India.
The work of MV was also supported by the SPM fellowship of CSIR.

\appendix

\newcommand{\g}{\bar L^G}

\newcommand{\pgmn}{\wt\PPP_{g,m,n}}

\sectiono{Summary of conventions} \label{saconv}

In this appendix we shall give a summary of some of the notations and
conventions we use. We begin by giving a summary of our notations for the world-sheet
superconformal field theory: 
\begin{itemize}
\item We use the acronym CFT to mean world-sheet 
	conformal field theory.
	\item We use the acronym SCFT  to mean world-sheet 
	superconformal field theory.
 
	\item We call the holomorphic fields in the world-sheet theory 
	right movers and the 
	anti-holomorphic fields left
       movers.

	\item Conformal dimension of an operator is denoted by $(\bar{h},h)$, $\bar{h}$ denoting the left 
       conformal dimension and $h$ denoting the right conformal dimension.
	
	\item We use the acronym OPE to mean operator product expansion.
	
	\item For a primary field $\phi(z)$ with conformal dimension $(0,h)$, we take the mode
       expansion 
\be
\phi(z)=\sum_{n=-\infty}^{\infty}\phi_n z^{-n-h}\, .
\ee
On the other hand, for a 
primary field $\bar\phi(\bar z)$ with conformal dimension $(\bar h,0)$, we take the mode
       expansion 
\be
\bar\phi(\bar z)=\sum_{n=-\infty}^{\infty} \bar \phi_n \bar z^{-n-\bar h}\, .
\ee
	\item An expression like $QV(w)$, where $Q=\oint j(z)dz$, denotes the following contour integral    
 \be \label{ea3} 
 QV(w)=\oint_w dz j(z)V(w)\, .
 \ee  
       Here the subscript $w$ implies that the contour surrounds $w$.
In order to evaluate \refb{ea3} we take 
the OPE of $j(z)V(w)$ and pick up only the coefficient of the single pole
       (i.e. use residue theorem of complex analysis).	

       \item In our convention, the contour integral measure $dz$ implicitly 
       includes a factor of $1/2\pi i$ so that we have, e.g.,
\be
\oint_w \frac{dz}{z-w}=1  
\ee 

 	\item We shall take $\alpha'=1$, so that string tension is $1/2\pi$.
\item The mass$^2$ level of a state carrying momentum $k$ is given by
its $2L_0^+-k^2$ eigenvalue. 

\item Given any operator $\phi(z,\bar z)$ in the world-sheet SCFT, we associate with it the
state 
\be
|\phi\rangle = \phi(0) |0\rangle\, ,
\ee
where $|0\rangle$ is the SL(2,C) invariant vacuum. 
We shall use the symbol $\phi$ to denote the operator $\phi$, as well as a short-hand
notation for the corresponding state $|\phi\rangle$.
\item Given a conformal map $f(z)$, we denote by $f\circ \phi$ the conformal transform of the
operator $\phi$. For example if $\phi$ is a primary operator of dimension $(\bar h,h)$,
we have $f\circ\phi(z) = (f'(z))^h \overline{f'(z)}^{\bar h} \phi(f(z), \overline{f(z)})$. With this
notation, we denote the BPZ conjugate of the state $|\phi\rangle$ by 
\be 
\langle\phi|=\langle 0| I\circ \phi(0)\, ,
\ee
where $I(z)$ denotes the conformal transformation
\be
I(z) = 1/z\, .
\ee
\item  $\HH_T$ 
denotes states in the Hilbert space of SCFT that are annihilated by
$(b_0-\bar b_0)$ and $(L_0-\bar L_0)$. In the heterotic string theory,
$\wt\HH_T$ is the subspace of
$\HH_T$ carrying picture number $-1$ and $-1/2$ and $\wt\HH_T$ is 
the subspace of
$\HH_T$ carrying picture number $-1$ and $-1/2$. In type II string theory
$\wt\HH_T$ is the subspace of
$\HH_T$ carrying picture number $(-1,-1)$, $(-1/2,-1)$, $(-1,-1/2)$ and
$(-1/2,-1/2)$ and $\wt\HH_T$ is 
the subspace of
$\HH_T$ carrying picture number $(-1,-1)$, $(-3/2,-1)$, $(-1,-3/2)$ and
$(-3/2,-3/2)$.  
\item A state $|s\rangle\in \HH_T$ is called BRST invariant if $Q_B|s\rangle=0$ and
BRST exact if $|s\rangle = Q_B|t\rangle$ for some $|t\rangle\in \HH_T$.
BRST cohomology is the space of BRST invariant states modulo addition of BRST
exact states, {\it e.g.} if two BRST invariant states $|s\rangle,|s'\rangle\in\HH_T$
differ by $Q_B|t\rangle$ for some $|t\rangle\in \HH_T$, they describe the same
element of BRST cohomology.  
\end{itemize}

Next we shall describe some notations and conventions that are used in the
construction of superstring field theory:
\begin{itemize}
	\item 1PI will stand for one particle irreducible. This will refer to 
	any Feynman diagram that cannot be split
       into two disconnected diagrams by cutting a single internal line.
\item $\wh\PP_{g,m,n}$ will denote a fiber bundle whose base is the
moduli space  $\MM_{g,m,n}$ of Riemann surfaces (including information about
spin structure) with $m$ NS and $n$ R punctures, and whose
fiber describes possible choices of local coordinate systems at the punctures.
\item $\wt\PP_{g,m,n}$ will denote a fiber bundle whose base is $\MM_{g,m,n}$,
and whose
fiber describes possible choices of local coordinate systems at the punctures
and PCO locations.  
\item Given $A_1,\ldots  ,A_N\in \wh\HH_T$, 
$\Omega^{(g,m,n)}_p(A_1,\ldots  ,A_N)$ is a 
$p$-form on $\wt\PP_{g,m,n}$ constructed from the correlation functions
of $A_1,\ldots  ,A_N$ and other universal operators on the Riemann surface. 
\item For any Feynman diagram of string field theory we assign a section of
$\wt\PP_{g,m,n}$
over a codimension zero subspace of $\MM_{g,m,n}$. This is called the
section segment of the corresponding Feynman diagram.  
\item  We denote by $\oR_{g,m,n}$ 
the section segment of the interaction vertex of
the BV master action at genus $g$, with $m$ external NS sector states and $n$
external R-sector states. Here section segment of an interaction vertex
refers to the section segment of a Feynman diagram containing a single interaction vertex
and no internal propagator. 

\item  We denote by $\RR_{g,m,n}$ 
the section segment of the sum of 1PI Feynman diagrams at 
genus $g$, with $m$ external NS sector states and $n$
external R-sector states.

\item  We denote by $\cL A_1\ldots A_N\cR$  
the contribution to the amplitude
of external states $A_1,\ldots  ,A_N$ from the $N$-point interaction vertex of the 
BV master action.

\item  We denote by $\{ A_1\ldots A_N\}$ 
the 1PI amplitude with
external states $A_1,\ldots  ,A_N$. 

\item  We denote by $\cL a_1\ldots a_N\cR_e$ 
the contribution to the amplitude of external 
states $a_1\ldots  ,a_N$ from the $N$-point interaction vertex of the 
BV master action obtained {\it after integrating out a subset of the fields}.

\item  We denote by $\{ a_1\ldots a_N\}_e$ 
the 1PI amplitude of external 
states $a_1\ldots  ,a_N$ computed from  the effective BV master
action obtained {\it after integrating out a subset of the fields}. 

\item We denote by $\{ A_1\ldots A_N\}''$ 
the interaction vertex of the 1PI action, expanded around the quantum
corrected vacuum solution,
for
external states $A_1,\ldots  ,A_N$.

\item We denote by $G(A_1,\ldots ,A_N)$ the Green's function of external states
$A_1, \ldots ,A_N$, computed around the perturbative vacuum, with the {\it tree
level} external propagators removed. 

\item We denote by $\Gamma(A_1,\ldots\, ,A_N)$ the Green's function of 
external states
$A_1,\ldots ,A_N$, computed around the quantum corrected vacuum, 
with the {\it full} external propagators removed.

\end{itemize}

\sectiono{Some examples  of off-shell amplitudes}  \label{saexamples}

\newcommand{\non}{\nonumber}

In this appendix we shall illustrate the procedure for defining off-shell amplitudes, as given in
\S\ref{soffsub}, using two examples: four punctured sphere and two punctured torus.

\begin{figure}
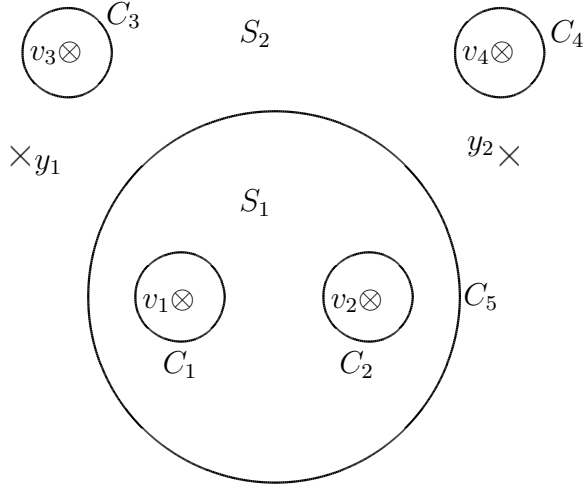


\begin{center}

\figfourpoint

\end{center}

\caption{Four punctured sphere} \label{four_pun_sphere_1}
\end{figure}

\subsection{ Four punctured sphere}

We begin with the example of the four point function of NS sector states on the sphere. This requires the insertion of
two PCOs. The relevant geometry has been shown in Fig.~\ref{four_pun_sphere_1}, with $v_1,v_2,v_3,v_4$
labelling the locations of the punctures and $y_1$ and $y_2$ labelling the PCO locations. 
We denote by $D_a$ the disk around the  $a^{th}$ puncture (not marked explicitly in the figure), by
$C_a$ the boundary of $D_a$ for $1\le a\le 4$, and by $C_5$ another circle that encloses $D_1$ and
$D_2$ but not $D_3$ and $D_4$. These curves divide the original sphere into four disks $D_1,\ldots ,D_4$ and
two spheres $S_1$ and $S_2$, each with three holes, as marked in the figure. Note that
$S_2$ acquires the topology of a sphere with three holes after identifying the points at infinity.
Let $z$ denote the global complex coordinate on the whole plane. Then we take the local
coordinates around the punctures to be
\be
w_a = z - v_a, \quad \hbox{for $1\le a\le 4$}\, .
\ee
Furthermore we choose the coordinate system on $S_1$ and $S_2$ to be
\be 
z_1=z, \quad z_2=z^{-1} \, ,
\ee
respectively. Then the transition functions across the various circles are as follows:
\ben \label{tran_fun_4_sphere}
 C_1  & :&\quad  w_1=z_1-v_1 \non\\
 C_2 & : & \quad w_2=z_1-v_2 \non\\
  C_3 & : &\quad  w_3=z_2^{-1} -v_3 												\\
   C_4 & : &\quad  w_4=z_2^{-1} -v_4 \non\\
   C_5 & : &\quad  z_1=z_2^{-1}  \non
\een
It is important to keep in mind that all that matters are the transition functions 
and not how we got them. For instance to get the above 
transition functions we made use of the global $z$ coordinate on a plane. Once we have stated the transition functions we can forget about the global $z$ coordinate. Finally, since both PCOs are situated on $S_2$,
their locations $y_1$ and $y_2$ are measured in the $z_2$ coordinate system.

Since the moduli space $\MM_{0,4,0}$ is two dimensional, the amplitude is given by integration over a two
dimensional section of $\wt\PP_{0,4,0}$. Let us denote by $t_1$ and $t_2$ the coordinates labelling this
section. Then $v_a$ for $1\le a\le 4$ and $y_\alpha$ for $\alpha=1,2$ are functions
of $t_1$ and $t_2$. It now follows from \refb{edefbbi}, \refb{emeasure},
\refb{eoffshellamp} that the off-shell amplitude for external states
$A_1,\ldots ,A_4\in \HH_{-1}$ is given by
\be \label{emeasureexample}
(-2\pi i)^{-1} \int \left\langle \BB_1 \, dt_1\wedge \BB_2 \, dt_2 \,  \XX(y_1) \XX(y_2)
\, A_1(v_1)\ldots A_4(v_4)\right\rangle_{\Sigma_{0,4,0}}\, ,
\ee
where $A_i$ is inserted using the coordinate system $w_i$ and
\be 
\BB_i = -\sum_{s=1}^4 \ointop_{C_s} {\p v_s\over \p t_i} dw_a b(w_a) -\sum_{s=1}^4 \ointop_{C_s} 
{\p \bar v_s\over 
\p t_i} d\bar w_a \bar b(\bar w_a) - \sum_{\alpha=1}^2 {1\over \XX(y_\alpha)}
{\p y_\alpha\over \p t_i} \p\xi(y_\alpha)\, .
\ee
According to the convention described in \S\ref{soffsub} the coordinate systems $w_a$ will be on the
left of $C_a$ for $1\le a \le 4$. Therefore, the contour $C_a$ runs anti-clockwise around $v_a$. Integration
over holomorphic coordinates is accompanied by a factor of $(2\pi i)^{-1}$ and integration over
anti-holomorphic coordinates is accompanied by a factor of $(-2\pi i)^{-1}$.

Dependence of the $v_s$ and $y_\alpha$ on  the parameters $t_1,t_2$ can be chosen arbitrarily. 
As an example we can consider the choice in which
$v_1,v_2,v_3$ are independent of $t_i$ and
$v_4=t_1$, $\bar v_4=t_2$. In this case the vertex operators $A_1$, $A_2$ and $A_3$ are
inserted at fixed locations $v_1, v_2, v_3$ and we integrate over $v_4$. We have
\be
\BB_1=-\ointop_{C_4} dw_4 \, b(w_4) - \sum_{\alpha=1}^2 {1\over \XX(y_\alpha)}
{\p y_\alpha\over \p v_4} \p\xi(y_\alpha)\, ,
\quad \BB_2=-\ointop_{C_4} d\bar w_4 \, \bar b(\bar w_4) - \sum_{\alpha=1}^2 {1\over \XX(y_\alpha)}
{\p y_\alpha\over \p \bar v_4} \p\xi(y_\alpha)
\ee
Therefore, the off-shell amplitude is given by
\ben
&&(-2\pi i)^{-1} \int dv_4\wedge d\bar v_4 \left\langle \left\{
\ointop_{C_4} dw_4 \, b(w_4) + \sum_{\alpha=1}^2 {1\over \XX(y_\alpha)}
{\p y_\alpha\over \p v_4} \p\xi(y_\alpha)\right\} \right. \nonumber \\ &&
\left. \left\{ \ointop_{C_4} d\bar w_4 \, \bar b(\bar w_4) + \sum_{\alpha=1}^2 {1\over \XX(y_\alpha)}
{\p y_\alpha\over \p \bar v_4} \p\xi(y_\alpha)
\right\} \XX(y_1) \XX(y_2) A_1(v_1)\ldots A_4(v_4)\right\rangle\, .
\een
Globally we cannot take the $y_\alpha$ to be independent 
of $v_4,\bar v_4$, since, for
example, as $v_4\to y_\alpha$ for $\alpha=1,2$  we have a spurious singularity that needs to
be avoided by moving the PCOs away from $v_4$. But if for some range of integration over $v_4,\bar v_4$
we take $y_\alpha$ to be constant then the integrand simplifies and we get
\be
\int dv_4\wedge d\bar v_4 \langle \XX(y_1) \XX(y_2) A_1(v_1) A_2 (v_2)A_3(v_3) (b_{-1} \bar b_{-1} A_4
(v_4))\rangle \, ,
\ee
where $b_{-1} \bar b_{-1} A_4$ denotes the vertex operator of the state $b_{-1} \bar b_{-1} |A_4\rangle$.

\begin{figure}
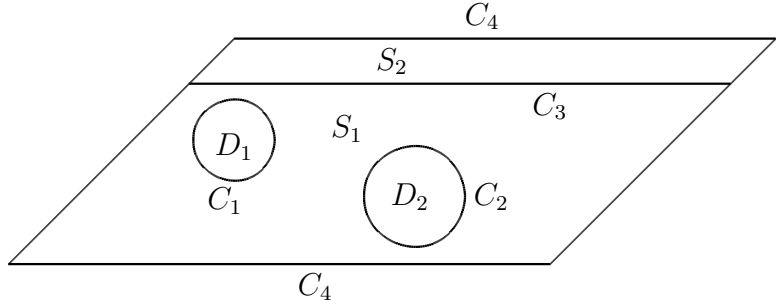


\begin{center}

\figtorusamp

\vskip -1in

\caption{Torus with two punctures represented by a parallelogram with 
diametrically opposite sides identified.
The punctures (not shown) are 
situated at the centers of the disks $D_1$ and $D_2$.
$S_1$ represents a sphere with four holes and $S_2$ represents a sphere with two holes.
$C_s$ are circles separating the disks and the spheres. \label{ffigtor}}
\end{center}

\end{figure}

\subsection{Two punctured torus}
We shall now illustrate how to compute the two 
point amplitude of two NS sector states 
on torus. We regard the
torus as a parallelogram with opposite sides identified.
This can be obtained from the complex plane by making the following identifications 
\be \label{edefglobalz}
z\simeq z+1\simeq z+\tau\, .
\ee

Next we partition the torus to view it as a collection of two disks, one around each puncture, and
 two spheres, one with four holes and the other with two holes. This is shown in Fig.~\ref{ffigtor}. This 
 is not quite the way we carried out our discussion in \S\ref{soffsub} where the 
components were  disks
around the punctures and a collection of spheres each with three holes. However, as was mentioned there,
this was not necessary, and the general formalism of \S\ref{soffsub} holds for the partitioning used here as
well. We denote by $v_1$ and $v_2$ the locations of the punctures in the $z$ coordinate.

We choose the coordinate system $z_1$
on $S_1$, $z_2$ on $S_2$, $w_1$ on $D_1$ and $w_2$ on $D_2$ in terms of
the coordinate $z$  in \refb{edefglobalz} as follows:
\be \label{edeftorcor}
z_1=z\;,\; z_2=z\;,\; w_1=f(\tau, \bar\tau)(z-v_1)\;,\; w_2=f(\tau, \bar\tau)(z-v_2)\, ,
\ee
where $f(\tau,\bar\tau)$ has modular transformation properties
\be
f(\tau+1,\bar\tau+1) = f(\tau,\bar\tau), \quad f(-\tau^{-1},-\bar\tau^{-1})
= \tau \, f(\tau,\bar\tau)\, .
\ee
The scaling by $f$ ensures that the local coordinates $w_1$, $w_2$ around the punctures
are modular invariant up to overall phases. Note 
that the coordinates $z_1$ and $z_2$ have identification under translation by 1. This is
okay as long as
the period is 
independent of the parameters over which we shall integrate. The 
 transition functions on the circles $C_1,C_2,C_3$ and $C_4$, separating the disks and spheres, can be
 determined from \refb{edeftorcor} and the $z\equiv z+\tau$ identification.
 They are as follows:
\ben\label{tran_fun_2_torus}
 C_1  & :&\quad  w_1=f(\tau, \bar\tau)(z_1-v_1) \non\\
 C_2 & : & \quad w_2=f(\tau, \bar\tau)(z_1-v_2) \non\\
  C_3 & : &\quad  z_1=z_2  \non \\
   C_4 & : &\quad  z_1=z_2-\tau \, .
\een

We also need two PCO insertions for this amplitude; let us label their
coordinates by $y_1$ and $y_2$. Recall that $y_\alpha$ need to be measured in the coordinate
system used on the component where they are situated, {\it e.g.} if they are located on $S_1$ then we must use
the $z_1$ coordinate system.

Now, for this amplitude the relevant moduli space $\mathcal{M}_{1,2,0}$
has 4 real dimensions. This requirement can be fulfilled by two complex parameters. We could
give the result for a 
general choice of parameters as in the last example, but let us be specific and choose them 
to be the coordinate $v_2$ of the second puncture and the modular parameter $\tau$ of the torus. We further assume that the coordinate $v_1$ is fixed at some position
independent of $\tau,\bar\tau,v_2,\bar v_2$.
Using \refb{edefbbi}, \refb{emeasure}  
and \refb{eoffshellamp} we now see that
the off-shell amplitude of two NS sector states $A_1$ and $A_2$ is given by
\be
(-2\pi i)^{-2}\int d\tau \wedge d\bar{\tau}\wedge dv_2\wedge d\bar{v}_2
\left\langle \BB_\tau \, \BB_{\bar\tau} \, \BB_{v_2} \, \BB_{\bar v_2} \, 
\XX(y_1) \XX(y_2) \, A_1(v_1) A_2(v_2)
\right\rangle_{\Sigma_{1,2,0} }
\ee
where
\ben \label{eb14}
\BB_\tau &=& -\oint_{C_4}dz b(z) 
+ {1\over f}{\p f\over \p\tau} \oint_{C_1} (z-v_1) b(z) dz
+ {1\over f}{\p f\over \p\tau} \oint_{C_2} (z-v_2) b(z) dz 
 \nonumber \\ &&
+ {1\over \bar f}{\p \bar f\over \p\tau} \oint_{C_1} (\bar z-\bar v_1) \bar b(\bar z) d\bar 
z
+ {1\over \bar f}{\p \bar f\over \p \tau} \oint_{C_2} (\bar z-\bar v_2) \bar b(\bar z) 
d\bar z - \sum_{\alpha=1}^2 {1\over \XX(y_\alpha)}
{\p y_\alpha\over \p \tau} \p\xi(y_\alpha)
\, , \nonumber \\
\BB_{\bar\tau} &=& -\oint_{C_4}d\bar z \bar b(\bar z)
+ {1\over f}{\p f\over \p\bar\tau} \oint_{C_1} (z-v_1) b(z) dz 
+ {1\over f}{\p f\over \p\bar\tau} \oint_{C_2} (z-v_2) b(z) dz 
 \nonumber \\ &&
+ {1\over \bar f}{\p \bar f\over \p\bar\tau} \oint_{C_1} (\bar z-\bar v_1) \bar b(\bar z) d\bar 
z
+ {1\over \bar f}{\p \bar f\over \p\bar \tau} \oint_{C_2} (\bar z-\bar v_2) \bar b(\bar z) 
d\bar z - \sum_{\alpha=1}^2 {1\over \XX(y_\alpha)}
{\p y_\alpha\over \p \bar\tau} \p\xi(y_\alpha)\, , \nonumber \\
\BB_{v_2} &=& -\oint_{C_2}dz b(z) - \sum_{\alpha=1}^2 {1\over \XX(y_\alpha)}
{\p y_\alpha\over \p {v_2}} \p\xi(y_\alpha)\, , \nonumber \\
\BB_{\bar{v}_2} &=& -\oint_{C_2}d\bar z \bar b(\bar z) - \sum_{\alpha=1}^2 {1\over \XX(y_\alpha)}
{\p y_\alpha\over \p \bar v_2} \p\xi(y_\alpha)\, .
\een
In writing \refb{eb14} we 
have converted all integrals over $w_1$, $w_2$, $z_1$, $z_2$
and their complex conjugates
to integrals over $z,\bar z$ using conformal transformation properties of $b$, $\bar b$.
The contour $C_4$ runs from left to right and the contour $C_2$
runs anti-clockwise around $v_2$. 
We cannot take $\{y_\alpha\}$ and $f$ to be independent of $v_2$,
$\bar v_2$, $\tau$, $\bar\tau$ globally, but if we assume that in a local patch
$\{y_\alpha\}$ and $f$ are independent of $v_2$,
$\bar v_2$, $\tau$, $\bar\tau$, then in this patch the integrand reduces to
\be
(-2\pi i)^{-2} d\tau\wedge d\bar\tau\wedge d v_2\wedge d\bar v_2 
\left\langle \, \ointop_{C_4} dz\, b(z) \ointop_{C_4} d\bar z \, \bar b(\bar z) \, 
\XX(y_1) \XX(y_2) A_1(v_1) \, b_{-1} \bar b_{-1} A_2(v_2)\right\rangle_{\Sigma_{1,2,0}}
\, .
\ee

\sectiono{Spurious poles  and vertical integration} \label{saspurious}

Any singularity of the $p$-form $\Omega_p^{(g,m,n)}$ in 
$\mathcal{\wt P}_{g,m,n}$, which
does not arise from the degeneration limit of Riemann surfaces, is called spurious singularity. 
Unlike the singularities associated with the degenerate Riemann surfaces, the 
spurious singularities can occur in the interior of the moduli space.  
In that case not all the divergences will have 
interpretation as the usual infrared divergences in superstring field theory arising in 
the limit
of large Schwinger parameters. For this reason we need to ensure that the 
(generalized) sections $\SSS_{g,m,n}$ used in defining off-shell amplitudes 
are free from spurious singularities.

Spurious poles can arise from different sources. First of all, they 
can arise from the collision of two PCOs 
since the OPE of two PCOs is singular.
They can also result from the collision of a PCO with a 
vertex operator. Finally, for
genus $g\ge 1$, they may arise at points in the moduli space where no
operators coincide. Since the last one 
is an unusual type of singularity we shall discuss its
origin in some detail.

Let us consider the 
correlators involving $\xi,\eta$ and $\phi$ fields in the 
large Hilbert space. 
On any Riemann surface this vanishes unless there is precisely one
extra $\xi$ insertion compared to the number of $\eta$ insertions. On the
torus the correlation function is given by\cite{verlinde,lechtenfeld,morozov}:
\ben \label{ed1}
&& \left\la \prod_{i=1}^{n+1}\xi(x_i)\prod_{i=1}^n\eta(y_i)\prod\limits_{k=1}^m e^{q_k\phi(z_k)}\right\ra_{\delta}\nonumber\\
&=&\frac{\prod\limits_{j=1}^{n}\vartheta_\delta\left(-y_j+\sum\limits_i x_i -\sum\limits_i y_i+\sum\limits_k q_kz_k\right)}{\prod\limits_{j=1}^{n+1}\vartheta_\delta\left(-x_j+\sum\limits_ix_i-\sum\limits_iy_i+\sum\limits_kq_kz_k\right)}\times 
\frac{\prod\limits_{i<i'}E(x_i,x_i')\prod\limits_{j<j'}E(y_j,y_j')}{\prod\limits_{i,j}E(x_i,y_j)\prod\limits_{k< l}E(z_k,z_l)^{q_kq_\ell}}
\label{ghost_correlator}
\een
where $\delta$ denotes the spin structure, $\vt_\delta$\plu\ are Jacobi theta functions with
$\vt_1$ denoting the unique odd theta function, 
and
\be 
E(x,y)=\frac{\vartheta_{1}(x-y)}{\vt_1'(0)} \quad ;  
\quad E(x,y)\sim x-y \quad \hbox{for} \; x\simeq y\, .
\ee
 This formula has simple generalization at higher genus.
 To satisfy the (anomalous) $\phi$-charge conservation law mentioned below 
 \refb{e2.4},
 we must have
 $
 \sum_k q_k=0$ on the torus.

The correlator in \eqref{ghost_correlator} is in large Hilbert space. One 
of the properties of this correlator, which is important for computations, 
is that the object
\be
I\equiv \prod\limits_{i=2}^{n+1} 
\left(\frac{\p}{\p x_i}\right)^{\ell_i}\left\langle\prod\limits_{i=1}^{n+1}\xi(x_i)\prod\limits_{i=1}^{n}\eta(y_i)\prod\limits_{k=1}^{m}e^{q_k\phi(z_k)}\right\rangle_\delta\qquad, \qquad (\ell_{i}\ge 1)
\label{imp_property}
\ee
is independent of $x_1$\cite{verlinde,lechtenfeld,morozov}. This 
is a reflection of the fact that the derivatives of $\xi$ do not contain the zero
mode of $\xi$. But in the large Hilbert space, we need to soak up the $\xi$ zero mode
by introducing an explicit factor of $\xi$ without derivative in the correlation function.
Therefore once the last $n$ 
$\xi$\plu\ are accompanied by derivative operators, the zero
mode of $\xi$ must come from the $\xi(x_1)$ factor, making the result independent
of $x_1$. While writing the correlator in the small Hilbert space we shall not
explicitly write the
$\xi(x_1)$ factor inside the correlator.

All the zeros and poles of the correlator on the left hand side of \refb{ed1}, 
which are expected from operator product expansion, 
are encoded in the functions $E(x,y)$.
We now note that 
the correlator in \eqref{ghost_correlator} develops additional 
singularities when the 
function $\vartheta_\delta$ in the denominator vanishes.  If the $\xi$ without 
any derivative
is inserted at $x_1$ then the location of the singularity is at
\be \label{epcloc}
\vt_\delta\left(\sum_{i=2}^{n+1} x_i-\sum_{i=1}^n y_i+\sum\limits_k q_kz_k
\right) = 0\, ,
\ee
since this is the only combination in the arguments of $\vt_\delta$\plu\
in the denominator that is independent of $x_1$. This 
singularity is not implied by any OPE and corresponds to the spurious singularities of the third type mentioned above. Using \eqref{epcloc}, we can deduce the following useful properties of these singularities:
\begin{enumerate}
\item \refb{epcloc} shows that if we have a vertex operator
containing $m$ factors of $\p\xi$ or its derivatives, $n$ factors of $\eta$ or its derivatives,
a factor of $e^{p\phi}$, and arbitrary number of derivatives of $\phi$, then the location
of the PCO depends on the location $z$ of the vertex operator through the combination
$(m-n+p)z$.
Since $m-n+p$ is the picture number of the vertex operator, this  shows that
the location of the spurious pole depends on the location of a vertex operator only through
its picture number. This has an important consequence that once we have chosen the
PCO locations to avoid spurious poles for one set of vertex operators, it will also avoid
spurious poles for any other set of vertex operators as long as the new vertex operators
carry the same picture number as the original vertex operators.  This feature continues to
hold for higher genus amplitudes. 
\item This also means that if the correlation function
contains insertions of
$\beta$ and $\gamma$ fields,
then the locations of the spurious poles do not depend on
the arguments of these fields\cite{lechtenfeld,morozov}. 
This is important since the BRST current depends on the
superconformal ghost system through
$\beta$ and $\gamma$ fields, and the above property implies that in a correlation
function with insertions of the 
BRST current, the spurious poles locations do not depend on the
argument of the BRST current. This is turn means that while deforming the integration contours
over the BRST current, we do not need to worry about possible residues from spurious poles. 
\end{enumerate}

A common feature of all three types of 
spurious poles is that they occur on a subspace of $\pgmn$
of complex co-dimension 1 (or real co-dimension 2) since they involve a complex condition
relating the locations of the PCOs and the moduli of the punctured Riemann surfaces. 
Typically this subspace depends non trivially on the locations of vertex operators, 
locations of PCOs as well as the other moduli parameters, but not on the choice of local
coordinates at the punctures.

\begin{figure}
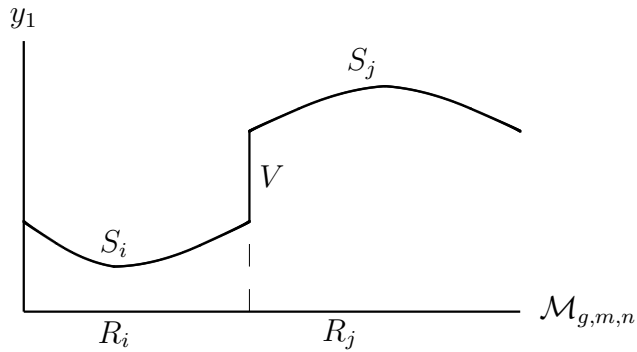

\begin{center}
\figvert 
\end{center}
\caption{The vertical segment $V$ {\it filling the gap} between two section segments
$S_i$ and $S_j$ in $\wt\PP_{g,m,n}$. 
\label{figvert}}
\end{figure}
 
Since spurious poles occur on real codimension 2 subspaces of $\pgmn$, 
a section $\SSS_{g,m,n}$ will typically intersect the loci of spurious poles on real
codimension 2 subspaces of $\SSS_{g,m,n}$. 
This will make the integral of $\Omega^{(g,m,n)}_{6g-6+2m+2n}$ over such sections
ill defined, not only for off-shell amplitudes, but also for on-shell amplitudes.
Our goal now will be to describe how to avoid this situation. Since the 
locations of the spurious poles depend on
the locations of the PCOs, at any point in $\MM_{g,m,n}$ we can avoid spurious poles
by appropriate choice of PCO locations. It follows from this 
that if we consider a sufficiently small
region of $\MM_{g,m,n}$, then we can choose a section segment on that region that avoids
spurious poles. Our strategy will be to divide $\MM_{g,m,n}$ into such sufficiently small
regions $\{R_i\}$, and on each such region, choose section segments avoiding spurious poles.
Furthermore we shall choose these section segments such that the local coordinates vary
continuously across the boundary of two such regions -- only the PCO locations can
have
possible discontinuities. 
Our goal will be to show that we can compensate for
the discontinuities of the section segments in $\wt\PP_{g,m,n}$ 
by adding appropriate correction terms so that for all practical 
purpose we can pretend {\it as if} the integration is performed over a continuous subspace
of $\pgmn$.

Let us first consider the situation where there is a single PCO. Let us denote its location
by $y_1$ and suppose that at some fixed point in $\MM_{g,m,n}$ at the boundary between
two regions $R_i$ and $R_j$, the PCO location jumps from $y^{(i)}_1$ to $y^{(j)}_1$ as
we move from $R_i$ to $R_j$. This has been shown in Fig.~\ref{figvert},
with $S_i$ and $S_j$ denoting the sections over $R_i$ and $R_j$. Now 
let us {\it fill the gap} between the section
segments $S_i$ and $S_j$ on $R_i$ 
and $R_j$ by drawing a {\it vertical segment} $V$ in
$\pgmn$ that connects $y^{(i)}_1$ to $y^{(j)}_1$ for each point on the boundary separating
$R_i$ and $R_j$, and integrate $\Omega^{(g,m,n)}_{6g-6+2m+2n}$ along 
this vertical
segment. The integral can be performed by first integrating $y_1$ from $y^{(i)}_1$ to
$y^{(j)}_1$ for fixed values of the other coordinates, and then integrating over the other
coordinates. 
Now since we are integrating along $y_1$, 
the integrand will involve contraction of
$\Omega^{(g,m,n)}_{6g-6+2m+2n}$ with $\p/\p y_1$. 
According to \refb{edefomega}, this inserts a factor of $-\p\xi(y_1)$
into the correlation function and at the same time removes the $\XX(y_1)$ factor. Since 
there is no $y_1$ dependence in the rest of the correlation function, the integration over
$y_1$ can be performed explicitly to give $\xi(y^{(i)}_1)-\xi(y^{(j)}_1)$. 
Eq.~\refb{ed1} and its higher genus generalization shows that 
this correlation function is manifestly
free from spurious poles as long as there are no spurious poles for the locations $y^{(i)}_1$
and $y^{(j)}_1$ of the PCO, even if the integration contour over $y_1$ 
passes through the spurious
pole. Once we add the integral over the vertical segment $V$ defined this
way to the integrals over the section segments $S_i$ and $S_j$ over $R_i$ and
$R_j$, the result behaves as if we have an integral over a continuous subspace
of $\wt\PP_{g,m,n}$, and obeys all the identities that are satisfied by the integrals
over continuous subspaces. 

For one PCO this is the end of the story. When there are more than one PCOs, we have to
be somewhat careful about how we erect the vertical segment. The general rule is that
we always move the PCOs one at a time, e.g. if we have $K$ PCOs $y_1,\ldots ,y_K$
then we may choose the convention that we first move $y_1$ from its value in $R_i$ to
its value in $R_j$ keeping all other $y_\alpha$\plu\ fixed at their value in $R_i$, then we
move $y_2$ from its value in $R_i$ to its value in $R_j$ and so on. Of course,
any other order is also acceptable.  But now when different boundaries meet, {\it e.g.}
on the codimension two subspace of $\MM_{g,m,n}$ describing the 
common intersection of $R_i$, $R_j$ and $R_k$, the vertical segments between 
$R_i$ and $R_j$, between $R_j$ and $R_k$ and between $R_k$ and $R_i$ may not fit
together to give a continuous subspace of $\wt\PP_{g,m,n}$.  We now have to `fill the gap'
by adding new two dimensional vertical segments on the codimension two subspace of 
$\MM_{g,m,n}$ describing the intersection of $R_i$, $R_j$ and $R_k$. A systematic
procedure for doing this was given in \cite{1504.00609}. 
Even though these vertical segments
pass through spurious poles and hence the integral over these segments is not 
strictly defined,
we can formally perform the integral and express the result as a correlation function of the
differences in $\xi$ fields evaluated at the corner points of the segment, representing PCOs
locations for the section segments on the $R_i$\plu. By construction, these are kept away
from spurious poles. 
Furthermore the final expression satisfies all the usual identities {\it as if} we had
integrated $\Omega^{(g,m,n)}_{6g-6+2m+2n}$ along a continuous subspace of 
$\pgmn$ without encountering any divergence.\footnote{A similar construction in
the context of topological string theory can be found in \cite{9510003}.}

The procedure described above assumes that we have the complete freedom 
of choosing the section $\SSS_{g,m,n}$.
For the construction of superstring field theory we only have the freedom of choosing the
section segments $\oR_{g,m,n}$ 
of elementary interaction vertices of the field theory, and for these
we use the procedure described above for avoiding spurious poles. It has been argued
in \S\ref{scomstub} that once this is done, the section segments of other Feynman
diagrams will be manifestly free from spurious poles as long as
the interaction vertices contain large stubs. We shall
see an example of this in appendix \ref{saspdeg}.

 \sectiono{Spurious poles near degeneration} \label{saspdeg}

\begin{figure}
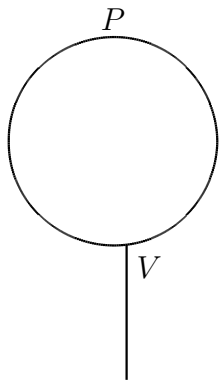


\begin{center}

\hbox{~\hskip 1.5in \figspdeg}

\end{center}

\caption{A one loop one point function, obtained by joining two external legs of 
a tree level 3-point vertex $V$ by a propagator $P$. \label{fspdeg}}

\end{figure}

We have argued in \S\ref{scomstub} that once we choose the section segments of
elementary vertices avoiding spurious poles and containing sufficiently long stubs, 
the section segments of general Feynman
diagrams will also be free from spurious poles. 
In this appendix we shall verify this in a simple example.

The example we consider is that of one loop tadpole graph of an NS sector state, 
obtained by starting with
a tree level three point vertex $V$ and gluing two of its external lines by a propagator $P$.
This has been shown in Fig.~\ref{fspdeg}. This represents a one point function on the
torus.  
By adding long stubs to the three point vertex
one can ensure that the torus associated with this graph is near 
degeneration i.e.\ its modulus $\tau$ has large imaginary part. Our goal
will be to show that the section segment associated with this diagram does not suffer
from any spurious singularity.
Since our goal
is limited, we shall not try to carry out a complete computation with fully symmetric
three string vertex, but instead take a convenient form of the vertex that simplifies the
various expressions. A complete analysis of this problem based on fully symmetric vertex has
been carried out recently in \cite{erler}.

For our analysis it will be useful to understand the relationship between the coordinate
system on the original three punctured sphere and the torus in the limit of large
stubs. Let us suppose that on the
original sphere, labelled by the coordinate $z$ on a complex plane, the punctures 
are at $v_1$, $v_2$, $v_3$ and the local coordinates around the first two punctures are
chosen respectively as 
\be
w_1 = e^\Lambda\,  {z-v_1\over z-v_2}, \qquad w_2 = 
e^\Lambda\,  {z-v_2\over z-v_1}\, ,
\ee
for some constant $\Lambda$. Large $\Lambda$ corresponds to
large stubs. We now consider the torus obtained by sewing the first and
the second punctures by the relation 
\be \label{ed.1}
w_1 = e^{- s -i\theta}/w_2\, ,
\ee
where $s,\theta$ are the sewing parameters.
If we now define 
\be \label{ed1.5}
\wt z = -{i\over 2\pi} \, \left[ \ln {z-v_1\over z-v_2} - \ln {v_3 - v_1 \over v_3 - v_2}\right]\, ,
\ee
then we have the identifications
\be \label{ed.2}
\wt z \equiv \wt z +1 \equiv \wt z +\tau, \quad \tau \equiv 
{i\over 2\pi} (2\Lambda + s
+ i\theta) \, ,
\ee 
following from single-valuedness of the $z$ coordinate and the identification
\refb{ed.1}. Therefore $\wt z$ describes the standard
coordinate system on the torus. It is now clear from \refb{ed.2} that
for large $\Lambda$, $\tau$ acquires a large imaginary part. 

Using \refb{ed1.5} we see that the external NS sector 
vertex operator at $z=v_3$ is
sitting at the origin $\wt z=0$ of the torus. In any case, this can always be
achieved by translational invariance on the torus.
Since this vertex operator carries picture number $-1$, we need
to insert a single PCO on the torus. Let $y_1$ denote the location
of this PCO. Then it follows from \refb{ed1} that the spurious pole is at the location
\be \label{evtdel}
\vt_\delta(y_1)=0\, ,
\ee
where $\delta$ denotes the spin structure. We shall now consider two cases separately:
\begin{enumerate}
\item First consider the case where $V$ is 
an interaction vertex of three NS sector states, and $P$ is
an NS propagator. Since the tree level
three point vertex of three NS sector states, each carrying picture number $-1$, 
requires a
PCO insertion, we have a single PCO inserted on the sphere. 
This corresponds to inserting the PCO at some point in the $z$ plane
away from $v_1$, $v_2$, $v_3$. By \refb{ed1.5} this translates to a finite point in the
$\wt z$ coordinate system.
The NS propagator 
on the other hand does not have any PCO insertion. Therefore in the convention 
we have adapted, the PCO location
$y_1$ remains within finite distance from 
zero 
as the modulus $\tau$ of the torus goes to $i\infty$.

On the other hand,  for NS sector propagator, the spin structure $\delta$ appearing in
\refb{evtdel}  corresponds to imposing anti-periodic boundary condition along the
$a$-cycle. Therefore $\delta$ takes value 3 or 4, and
the location of the spurious pole given in
\refb{evtdel}  is at $y_1=\tau/2$ or $(\tau+1)/2$. This is incompatible with $y_1$ remaining
close to zero in the limit of large ${\rm Im}(\tau)$, showing 
that the amplitude is free from spurious
poles.

\item Next consider the case where $V$ represents a vertex with one NS and two
R-sector states, and $P$ is a Ramond
propagator joining the two R-sector states of the vertex.
In this case the vertex does not have any PCO insertion since the total picture number of
the external states add up to $-1 -1/2 -1/2=-2$, 
but a zero mode of the PCO is
inserted on the propagator around $|w_1|=|q|^{1/2}$ (or equivalently 
$|w_2|=|q|^{1/2}$). In the $\wt z$ coordinate
system of the torus, this is mapped to a curve along the $a$-cycle around
$y_1=\tau/2$ (or equivalently $y_1=-\tau/2$). On the other hand
$\delta$ appearing in \refb{evtdel} now takes value 1 or 2 due to periodic boundary
condition along the $a$-cycle, and therefore the spurious pole,
obtained by solving \refb{evtdel}, lies at $y_1=0$ or $1/2$. This is again incompatible with
the actual location of the PCO around $\tau/2$, showing that the amplitude is free from
spurious poles.

\end{enumerate}

\sectiono{Field redefinition} \label{safield}

The superstring field theory action depends on the choice of the section segments
$\oR_{g,m,n}$ of the interaction vertices. Therefore it is natural to examine
whether the final results for the physical quantities computed from this action
depend on this choice. Instead of working with the section segments of elementary
vertices, we shall find it easier to work with the union of the section segments of 
all 1PI diagrams, denoted by $\RR_{g,m,n}$ in \refb{edefcurl1PI}.
Let us denote by $\RR_{g,m,n}$ and $\RR'_{g,m,n}$ two different choices of
these section segments. We shall consider
infinitesimal deformations so that $\RR_{g,m,n}$ and $\RR'_{g,m,n}$ are close in
$\wt\PP_{g,m,n}$. Then we can write
\be \label{edeltaactionpre}
\delta S_{1PI} = \sum_{g=0}^\infty  g_s{}^{2g-2}  \sum_{m,n} {1\over m! \, n!} \,
\left[\left(\int_{\RR'_{g,m,n}} - \int_{\RR_{g,m,n}}\right)
\, \Omega^{(g,m,n)}_{6g-6+2m+2n}(|\Psi_{NS}\rangle^{\otimes m}, |\Psi_{R}\rangle^{\otimes n})
\right]\, ,
\ee
where $|\Psi_{NS}\rangle$ and $|\Psi_R\rangle$ denote the NS and R components of 
$|\Psi\rangle$ and $|\Psi_{NS}\rangle^{\otimes m}$ and  $|\Psi_{R}\rangle^{\otimes n}$
denote that we have $m$ entries of $|\Psi_{NS}\rangle$ and $n$ entries 
of $|\Psi_R\rangle$. 
Let $\hat U_{g,m,n}$ be an infinitesimal vector field that takes a point in $\RR_{g,m,n}$ to a
neighboring point in $\RR'_{g,m,n}$. The definition of $\hat U_{g,m,n}$ is ambiguous up to 
addition of infinitesimal tangent vectors of $\RR_{g,m,n}$, but this will not affect the
final result. 
In this case \refb{edeltaactionpre} can be expressed as\cite{9301097}
\ben \label{edeltaaction}
\delta S_{1PI} &=& \sum_{g=0}^\infty  g_s{}^{2g-2}  \sum_{m,n} {1\over m!\, n!} \, 
\left[\int_{\RR_{g,m,n}}
\, d\Omega^{(g,m,n)}_{6g-6+2m+2n}[\hat U_{g,m,n}] (|\Psi_{NS}\rangle^{\otimes m}, 
|\Psi_{R}\rangle^{\otimes n}) \right.  \nonumber \\ &&
 \left. +
\int_{\p \RR_{g,m,n}} \Omega^{(g,m,n)}_{6g-6+2m+2n}[\hat U_{g,m,n}] 
(|\Psi_{NS}\rangle^{\otimes m}, |\Psi_{R}\rangle^{\otimes n}) 
\right]\, ,
\een
where for any $p$-form $\omega_p$, 
$\omega_p[\hat U]$ denotes the contraction of 
$\omega_p$ with the vector field $\hat U$:
\be
\omega_{i_1\ldots i_p}dy^{i_1} \wedge \cdots \wedge dy^{i_p}[\wh U] \equiv \wh U^{i_1} 
\omega_{i_1 i_2\ldots i_p}dy^{i_2} \wedge \cdots \wedge dy^{i_p}\, .
\ee
A pictorial representation of this
can be found in Fig.~\ref{fonly}. The first term on the right
hand side of \refb{edeltaaction}  represents 
the integral of $d\Omega^{(g,m,n)}_{6g-6+2m+2n}$ over a $6g-5+2m+2n$
dimensional region $\wt \RR_{g,m,n}$
bounded by $\RR_{g,m,n}$ and $\RR'_{g,m,n}$. This can be integrated
to give \refb{edeltaactionpre}, represented by integral of
$\Omega^{(g,m,n)}_{6g-6+2m+2n}$ along the horizontal boundaries
of $\wt\RR_{g,m,n}$ in Fig.~\ref{fonly}, and an integral of
$\Omega^{(g,m,n)}_{6g-6+2m+2n}$ along the
boundary of $\wt\RR_{g,m,n}$ that joins $\p\RR'_{g,m,n}$ to 
$\p\RR_{g,m,n}$,  shown by the vertical lines
in Fig.~\ref{fonly}. The second
term in \refb{edeltaaction} subtracts the latter contribution. 

\begin{figure}

\begin{center}

\figfieldredef

\end{center}

\vskip -1.3in

\caption{A pictorial representation of eqs.~\refb{edeltaactionpre} 
and \refb{edeltaaction}.
The right hand side of \refb{edeltaactionpre} is the contribution to the 
integral of
$\Omega^{(g,m,n)}_{6g-6+2m+2n}$ from the upper and lower horizontal
edges of the rectangle.
The first term on the right hand side of \refb{edeltaaction} is the volume integral
of $d\Omega^{(g,m,n)}_{6g-6+2m+2n}$ over the interior $\wt\RR_{g,m,n}$ of the rectangle. 
Since the height of the rectangle is infinitesimal we can replace the effect of
integration along the vertical direction by contraction with $\wh U_{g,m,n}$. 
Finally the last
term of \refb{edeltaaction} represents the opposite of the contribution to the boundary integral
of $\Omega^{(g,m,n)}_{6g-6+2m+2n}$ from the vertical edges of the rectangle. Thus 
\refb{edeltaaction} follows from \refb{edeltaactionpre} via Stokes' theorem. 
Although we have taken the height of the rectangle to be constant for the ease of drawing
the figure, this is certainly not necessary. Finally note that here we have drawn $\RR_{g,m,n}$
and $\RR'_{g,m,n}$ as one dimensional horizontal lines, but the general case corresponds to
them being multidimensional, with the whole figure stretching out of the plane of the 
paper / screen.
\label{fonly}}

\end{figure}

It was shown in \cite{1411.7478,1501.00988} that\footnote{Ref.~\cite{1501.00988} 
worked at the level of equations of motion and considered only the
field redefinition of $|\Psi\rangle$.
But following \cite{1411.7478} the analysis can 
easily be generalized to that for the 1PI action by choosing
$|\delta\wt\Psi\rangle$ such that
$\GG|\delta\wt\Psi\rangle=|\delta\Psi\rangle$. }
the change in action given in \refb{edeltaaction} 
can be regarded as the result of a redefinition of the fields $|\Psi\rangle$ and 
$|\wt\Psi\rangle$ to
$|\Psi\rangle+|\delta \Psi\rangle$ and $|\wt\Psi\rangle+|\delta \wt\Psi\rangle$ 
respectively, 
if we  take $|\delta\Psi\rangle$ and  $|\delta\wt\Psi\rangle$ to be of the form
\ben \label{edeltafield}
&& \langle \phi| c_0^- |\delta \Psi\rangle \nonumber \\
&=& - \sum_{g=0}^\infty g_s{}^{2g} \, \sum_{m,n=0}^\infty  {1\over m!n!} \, 
\int_{\RR_{g,m+1,n}} \, \Omega^{(g,m+1,n)}_{6g-5+2m+2n+2}[\wh U_{g,m+1,n}](\GG|\phi_{NS}
\rangle, 
|\Psi_{NS}\rangle^{\otimes m}, |\Psi_R\rangle^{\otimes n})\nonumber \\ &&
- \sum_{g=0}^\infty g_s{}^{2g} \, \sum_{m,n=0}^\infty  {1\over m!n!} \, 
\int_{\RR_{g,m,n+1}} \, \Omega^{(g,m,n+1)}_{6g-5+2m+2n+2}[\wh U_{g,m,n+1}](
|\Psi_{NS}\rangle^{\otimes m}, \GG|\phi_{R}
\rangle,  |\Psi_R\rangle^{\otimes n})\nonumber \\ \, ,
\een
and
\ben \label{edeltafieldb}
&& \langle \chi| c_0^- |\delta \wt\Psi\rangle \nonumber \\
&=& - \sum_{g=0}^\infty g_s{}^{2g} \, \sum_{m,n=0}^\infty  {1\over m!n!} \, 
\int_{\RR_{g,m+1,n}} \, \Omega^{(g,m+1,n)}_{6g-5+2m+2n+2}[\wh U_{g,m+1,n}](|\chi_{NS}
\rangle, 
|\Psi_{NS}\rangle^{\otimes m}, |\Psi_R\rangle^{\otimes n})\nonumber \\ &&
- \sum_{g=0}^\infty g_s{}^{2g} \, \sum_{m,n=0}^\infty  {1\over m!n!} \, 
\int_{\RR_{g,m,n+1}} \, \Omega^{(g,m,n+1)}_{6g-5+2m+2n+2}[\wh U_{g,m,n+1}](
|\Psi_{NS}\rangle^{\otimes m}, |\chi_{R}
\rangle,  |\Psi_R\rangle^{\otimes n})\nonumber \\ \, ,
\een
for any Grassmann odd\footnote{The result for Grassmann even state can be 
read out by multiplying both sides
of \refb{edeltafield} by a Grassmann odd number and moving it through various factors so that it 
multiplies $|\phi\rangle$. 
This gives extra minus signs in both terms on the right hand side of \refb{edeltafield}
since we have to move the Grassmann number through the $6g-5+2m+2n+2$ insertions of
$b$-ghost field associated with $  \Omega^{(g,m+1,n)}_{6g-5+2m+2n+2}$.}
state $|\phi\rangle=|\phi_{NS}\rangle + |\phi_R\rangle\in\wt\HH_T$ and 
$|\chi\rangle =|\chi_{NS}\rangle + |\chi_R\rangle\in\wh\HH_T$.
Since
field redefinition does not change the values of physical quantities, we conclude
from this that the different choices of $\RR_{g,m,n}$ lead to the same results for all
physical quantities.

This result can also be proved if we consider 
generalized section segments of the form 
$\RR_{g,m,n}=\sum_i w_i \RR^{(i)}_{g,m,n}$ where $\RR^{(i)}_{g,m,n}$ are
regular section segments and $w_i$ are weight factors. Now, instead of deforming
the regular section segments $\RR^{(i)}_{g,m,n}$, we deform the weight factors $w_i$ preserving the
$\sum_i w_i=1$ constraint (see \cite{1411.7478}, appendix A). 
This is important in the presence of vertical section
segments.
If we have two choices of $\RR_{g,m,n}$ which differ from each other in 
the order in which
we move the PCOs in a vertical segment, say in one
we move $y_1$ first and then $y_2$ while in the other we move $y_2$ first and
then $y_1$, then we cannot continuously deform these 
$\RR_{g,m,n}$ to each other. 
However we can take a generalized section segment
parametrized by weight 
factors such that by varying the weight factors we can continuously interpolate 
between these two $\RR_{g,m,n}$. The previous result can now be used
to show that the superstring field theories corresponding to the 
two different choices of $\RR_{g,m,n}$ are related by field redefinition.

\sectiono{Reality condition on the string fields} \label{sareal}

Reality of the superstring field theory action is necessary for proving unitarity of the 
theory. This was proved in \cite{1606.03455} 
for ten dimensional heterotic and type II string theories,
and also for compactified theories with NS
background  (NSNS background in type II string theory) where the compact part of the 
theory is described by a unitary superconformal field theory. In this appendix we shall
describe the reality condition on the string fields -- necessary for the reality of the action
-- for ten dimensional heterotic and type II string theories.

We shall first describe the results for heterotic string theory.  The ghost sector of
the world-sheet theory has been defined in \S\ref{s2.1}, but for describing the reality
condition we also need to fix the conventions in the matter sector. The matter fields
consist of 
10 scalars
$X^\mu(z,\bar z)$ and 10 right-moving Majorana-Weyl
fermions $\psi^\mu(z)$ for $0\le\mu\le 9$, 
and a CFT of left-movers of central charge 16, describing 
either $E_8\times E_8$ or SO(32) current algebra. 
We shall denote the last CFT by $CFT_G$.
 The operator product
expansions of these fields have the form:
\ben \label{exxope}
&& \p X^\mu(z) \p X^\nu(w) = -{\eta^{\mu\nu}\over 2 (z-w)^2}+\cdots,
\quad \bar\p X^\mu(\bar z) \bar\p X^\nu(\bar w) = -{\eta^{\mu\nu}
\over 2 (\bar z-\bar w)^2}+\cdots\, , \nonumber \\&& 
\psi^\mu(z) \psi^\nu(w) = -{1\over 2(z-w)}\, \eta^{\mu\nu}+\cdots \,  ,
\een
where $\cdots$ denotes non-singular terms.
For $CFT_G$
we shall not use any explicit representation, but 
denote by $|\bar V_K\rangle=\bar V_K(0)|0\rangle$ a 
basis of Virasoro primary states satisfying
\be \label{ecftg}
\langle \bar V_K | \bar V_L\rangle =\delta_{KL}, \quad \langle \bar V_K | 
\bar V_J(1) | \bar V_L\rangle = \hbox{real} \, .
\ee
The full set of states in this CFT are obtained by acting on these
primary states with the Virasoro generators $\g_{-n}$ of this CFT. 
We shall denote the anti-holomorphic stress tensor of $CFT_G$ by
$\bar T^G$.

Construction of the vertex operators in the Ramond sector also requires introduction
of spin fields. The spin fields are of two types: chiral fields $S_\alpha$ and anti-chiral
fields $S^\alpha$. The mutually local GSO even 
combinations of spin fields in the matter and
ghost sector are
\be \label{ee.3}
e^{-(4n+1)\phi/2} S_\alpha, \quad e^{-(4n-1)\phi/2} S^{\alpha}, 
\ee
and their derivatives and products with the NS sector GSO even operators. The spin fields
will be normalized so that they have the basic operator product expansions:
\ben \label{emainope1}
&& \psi^\mu(z) \, e^{-\phi/2} S_\alpha(w)={i\over 2} (z-w)^{-1/2}
(\gamma^\mu)_{\alpha\beta} e^{-\phi/2}S^\beta(w)+\cdots, \nonumber \\ &&
\psi^\mu(z) \, e^{-\phi/2} S^\alpha(w)={i\over 2} (z-w)^{-1/2}
\gamma^{\mu\alpha\beta} e^{-\phi/2}S_\beta(w)
+\cdots,  \nonumber \\ &&
e^{-\phi/2} S_\alpha(z) \, \,  e^{-3\phi/2}S^\beta(w) 
= \delta_{\alpha}^{~\beta} (z-w)^{-2} e^{-2\phi}(w)+\cdots \, , 
\een  
where
$\gamma^\mu$ are ten dimensional $\gamma$-matrices, normalized as
\be
\{\gamma^\mu, \gamma^\nu\} = 2 \eta^{\mu\nu} \, {\bf 1}\, ,
\ee
where $(\gamma^\mu \, \gamma^\nu)_\alpha^{~\beta} 
\equiv \gamma^\mu_{\alpha\delta} \gamma^{\nu\delta\beta}$ etc.
We shall use a representation in which all the $\gamma^\mu$ are 
purely imaginary and symmetric:
\be \label{sogamma}
(\gamma^\mu_{\alpha\beta})^*=-\gamma^\mu_{\alpha\beta},
\quad (\gamma^{\mu\alpha\beta})^* =- \gamma^{\mu\alpha\beta}, \quad
\gamma^\mu_{\alpha\beta}=
\gamma^\mu_{\beta\alpha}, \quad 
\gamma^{\mu\alpha\beta}=
\gamma^{\mu\beta\alpha}
\, .
\ee
In this representation, the right hand sides of \refb{emainope1} have real coefficients.

In order to facilitate the discussion on the reality conditions on various components
of the string field, it will be useful to fix some convention on the choice of basis states
in $\wh\HH_T$ and $\wt\HH_T$.
In the NS sector we construct
the basis of states $|\vp_r(k)\rangle$ of $\HH_{-1}$ such that the corresponding vertex
operators $\vp_r(k)$ can be built from linear combinations
of GSO even products of (derivatives of)
$\partial X^\mu$, $\bar\partial X^\mu$, $\psi^\mu$, $e^{ik\cdot X}$,
$b$, $c$, $\bar b$, $\bar c$,
$e^{q\phi}$, $\p\phi$, $\p\xi$, $\eta$, $\bar T^G$ and $\bar V_K$, without 
any explicit factor of 
$i$.
We shall choose the basis of states $|\wh\vp_s(k)\rangle$ of $\HH_{-1/2}$ and
$|\wt\vp_s(k)\rangle$ of
$\HH_{-3/2}$
such that
their vertex operators are constructed from products of (derivatives of) the
operators appearing in \refb{ee.3}, and 
other GSO even operators that were used to construct vertex
operators for the basis states in the NS sector,
without any explicit factor of $i$. 
In this case all the coefficients appearing in the operator
product expansion of operators representing GSO even basis states
in the NS and R sectors are manifestly
real except for the factor of $i$ multiplying each factor of $k^\mu$.

Let us now expand $|\Psi\rangle$ and $|\wt\Psi\rangle$ as
\be 
|\Psi\rangle = \sum_r  \int {d^{{10}} k \over (2\pi)^{10}}
\, \psi_r(k) |\vp_r(k)\rangle
+ \sum_s \int {d^{{10}} k \over (2\pi)^{10}}\, \wh\psi_s(k) |\wh\vp_s(k)
\rangle\, ,
\ee
and
\be
|\wt\Psi\rangle = \sum_r \int {d^{10} k \over (2\pi)^{10}}\, 
\xi_r(k) |\vp_r(k)\rangle
+ \sum_s  \int {d^{10} k \over (2\pi)^{10}}
\, \wh\xi_s(k) |\wt\vp_s(k)\rangle\, .
\ee
It was shown in \cite{1606.03455} that the action \refb{eactorg} is real if we
impose the following reality condition on the coefficient of expansion of  
$|\Psi\rangle$ and $|\wt\Psi\rangle$: 
\be \label{e2.16}
\psi_r(k)^* = (-1)^{n_r (n_r+1) /2 + 1}\psi_r(-k)\, ,
\quad
\wh\psi_s(k)^* = -i \, (-1)^{(\wh n_s+1) (\wh n_s+2) /2}\wh\psi_s(-k)\, ,
\ee
\be \label{e2.30}
\xi_r(k)^* = (-1)^{n_r (n_r+1) /2 + 1} \xi_r(-k)\, ,
\quad
\wh\xi_s (k)^* = -i \, (-1)^{(\wt n_s+1) (\wt n_s+2) /2 + 1}\wh\xi_s(-k)\, .
\ee
where $n_r$, $\wh n_s$ and $\wt n_s$ are ghost numbers
of $\vp_r$, $\wh \vp_s$ and $\wt\vp_s$ respectively.  

The reality condition on the fields of type II string theories is similar. We choose the basis of
states for type II world-sheet theory in a manner similar to that in the case of heterotic
string theory, so that the coefficients in the operator product expansion of the basis states
are real except for the factors of $i$ multiplying each factor of momentum. Then the reality
condition on the fields in the NSNS and RR sectors are the same as that for the NS sector
fields $\psi_r(k)$ and $\xi_r(k)$ of the heterotic string theory. On the other hand the reality
condition on the NSR and RNS sector fields in type II string theory are the same as that
on the R sector fields $\wh\psi_s(k)$ and $\wh\xi_s(k)$ of the heterotic string theory.

Finally we would like to mention that the reality condition on the fields is not completely fixed 
by demanding the reality of the action. For example since the action always contains 
an even
number of fermion fields, we could always include an extra factor of $-1$ in the reality condition
on each fermion field. Similarly, using ghost charge conservation one can show that we can
include in the reality condition of a field, that accompanies a ghost number $n$ state in the
world-sheet theory, a factor of $e^{i\phi(n-2)}$ where $\phi$ is some real number. 

After imposing the reality condition we need to examine if the action \refb{eactorg} has the correct
sign for the kinetic term. It turns out that for Euclidean path integral, using the weight factor $e^S$
in the path integral as we have been doing, the action has the correct sign in the heterotic string
theory, but has the wrong sign in type II string theory. This can be rectified by changing $g_s^2$
to $-g_s^2$ everywhere in the analysis of type II string theory.

\sectiono{Cutkosky rules} \label{sacutkosky}

A general proof of Cutkosky rules stated in \S\ref{scut} was given in \cite{1604.01783}. In this appendix
we shall illustrate this using a simple example. 

\begin{figure}
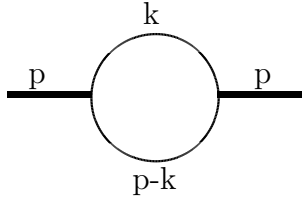


\begin{center}

\figqft

\end{center}

\vskip -.5in 
\caption{One loop mass renormalization of heavy particle of mass $M$ (denoted by thick line) due
to a loop of light particle of mass $m$ (denoted by thin line). All momenta flow from left to right.
\label{fqft}}
\end{figure}

We consider a quantum field theory in $D$ space-time dimensions
with two particles, one of mass $M$ and the other of mass $m$, with $M>2m$.
We assume further that there is a three point coupling between one heavy particle and
two light particles. In this theory we shall analyze the one loop mass
renormalization diagram shown in Fig.~\ref{fqft}. As in string field
theory, we shall assume that the vertex 
contains a factor of $\exp[- {1\over 2} A \{k^2+m^2\} - {1\over 2}
A \{(p-k)^2 + m^2\}]$ for some positive constant $A$ that makes the diagram
ultraviolet finite. 
Then the contribution of this diagram to the mass$^2$ of the
heavy particle can be expressed as
\be \label{e1aaa}
\delta M^2 = i\, B \, \int{d^D k \over (2\pi)^D} \, \exp[-A \{k^2+m^2\} -
A \{(p-k)^2 + m^2\}]\, \{k^2+m^2\}^{-1} \{(p-k)^2 + m^2\}^{-1}\, ,
\ee
where $B$ is another positive constant that includes multiplicative constant
contributions to the vertices, and $p$ is an on-shell external momentum
satisfying  $p^2=-M^2$. 

Using $k^2 = -(k^0)^2 + \vec k^2$ where $\vec k$ denotes $(D-1)$-dimensional
spatial momenta, we see that the exponential factor falls off exponentially as
$|\vec k|\to\infty$ but grows exponentially as $k^0\to \pm\infty$. This shows that
we cannot take the $k^0$ integral to run along the real axis. 
As discussed in \S\ref{sloopcon}, 
we resolve this problem by taking the ends of the $k^0$ integral to
be at $\pm i\infty$, but the
integration contour may take
complicated form in the interior of the complex
$k^0$ plane to avoid poles of the propagator. We shall now see how this is done
in this particular example.
The integrand of \refb{e1aaa} has poles in the $k^0$
plane at
\be  \label{e2}
Q_1 \equiv \sqrt {\vec k^2 + m^2}, \quad Q_2 \equiv -\sqrt{\vec  k^2 + m^2}, \quad
Q_3 \equiv p^0 + \sqrt{(\vec p - \vec  k)^2 + m^2} , \quad Q_4 \equiv
p^0 - \sqrt{(\vec p - \vec  k)^2 + m^2}\, .
\ee
For imaginary $p^0$, and $k^0$ contour running along the imaginary axis from
$-i\infty$ to $i\infty$, 
the poles $Q_1$ and $Q_3$ are to the right of the integration
contour whereas the poles $Q_2$ and $Q_4$ are to the left of the integration
contour. When $p^0$ is continued to the real axis along the first quadrant, the contour
needs to be deformed appropriately so that $Q_1$ and $Q_3$ continue to lie
on the right and $Q_2$ and $Q_4$ continue to lie on the left. There
are different possible configurations depending on the value of $\vec k$.

\begin{figure}
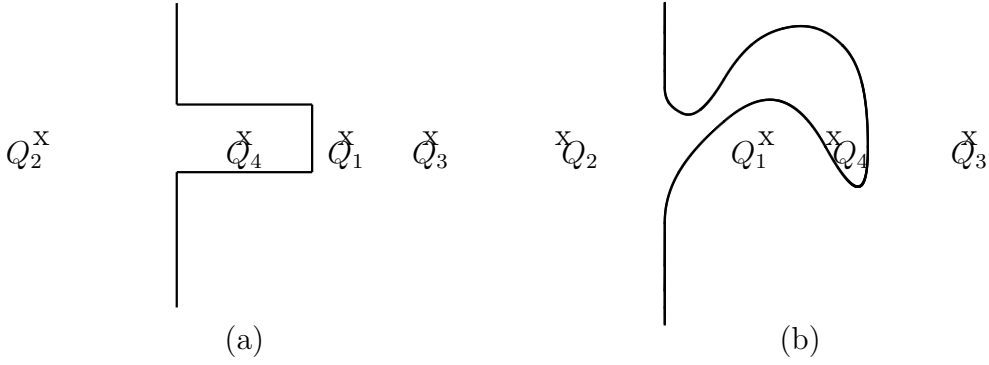


\begin{center}

\hbox{\figtwocut \hfill \figfourcut}

\vskip 25pt 

\caption{The integrations contours in the $k^0$ plane.
\label{f2cut}}

\end{center}

\end{figure}

For $p^0< \sqrt{\vec  k^2+m^2} + 
\sqrt{(\vec p-\vec  k)^2+m^2}$,  
$Q_4$ lies to the left of $Q_1$ and the contour can be taken 
as shown in Fig.~\ref{f2cut}(a). 
On the other hand for $p^0 > \sqrt{\vec  k^2+m^2} + 
\sqrt{(\vec p-\vec  k)^2+m^2}$, $Q_4$ is to the right of $Q_1$ and the 
deformed contour takes the form shown in Fig.~\ref{f2cut}(b). In drawing this
we have used the fact that when $p^0$ lies in the first quadrant,
$Q_4$ remains above $Q_1$ as it 
passes $Q_1$ and that during this process the contour needs to be
deformed continuously without passing through a pole. At the
boundary between these two regions $Q_4$ approaches
$Q_1$. In this case we have to use a limiting procedure to determine the contour, and the
correct procedure will be to take $p^0$ in the first quadrant, evaluate the integral
and then take the limit of real $p^0$. This in particular means that $Q_4$ approaches
$Q_1$ from above in this limit.

\begin{figure}
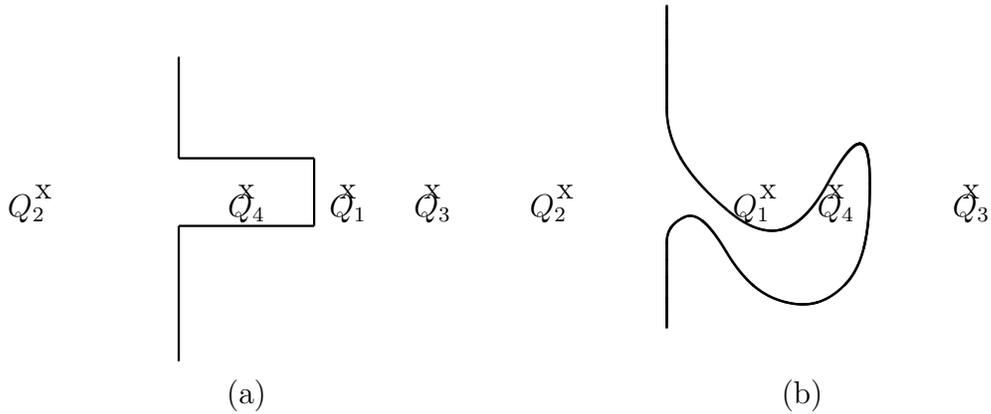


\begin{center}

\hbox{\figtwocut \hfill \figthreecut}

\vskip 25pt 

\caption{The complex conjugate integrations contours in the $k^0$ plane.
\label{f3cut}}

\end{center}

\end{figure}

Our goal will be to evaluate the imaginary part of \refb{e1aaa}.
In this case this can be done by explicit computation. But we shall use this
example to verify some of the steps in the analysis of \cite{1604.01783}.
\begin{enumerate}
\item {\bf The complex conjugate contribution to an amplitude is given by the same
expression as the original amplitude with  
all the external momenta complex conjugated
and the choice of contour given by the complex
conjugate of the original contour.} To prove this for the amplitude \refb{e1aaa}
note that under complex conjugation the explicit factor of
$i$ changes sign and the end points $\pm i\infty$ of the $k^0$ integration contour
get exchanged. These two minus signs cancel against each other. Therefore the
net effect of complex conjugation is to take the complex conjugate of the integrand.
This in particular replaces $k^0$ by $(k^0)^*$ and $p^0$ by $(p^0)^*$ in the integrand.
Once we relabel $(k^0)^*$ as $k^0$, it 
automatically sends the original integration contour to its complex conjugate,
proving the desired result. The complex conjugates of the integration contours
of Fig.~\ref{f2cut} are shown in Fig.~\ref{f3cut}.
\item {\bf The contribution to the imaginary part vanishes when the spatial components of
loop momenta
are such that the loop energy integration contour is away from the pinch singularity.} Here
pinch singularity refers to the situation where two poles approach each other from
opposite sides of the integration contour. In Fig.~\ref{f2cut} the pinch singularity corresponds
to the limit in which $Q_1$ and $Q_4$ approach each other.
Now away from the pinch
singularity we have either $Q_1<Q_4$ or $Q_1>Q_4$. For $Q_1<Q_4$ the 
integration contour shown in Fig~\ref{f2cut}(a) clearly matches its complex conjugate
contour shown in Fig.~\ref{f3cut}(a). Therefore for real $p^0$
the contribution to the integral has vanishing imaginary part.
For $Q_1>Q_4$ the contour shown in Fig.~\ref{f2cut}(b) is not deformable to
its complex conjugate contour shown in Fig.~\ref{f3cut}(b) without crossing a pole. But
the former can be deformed to the latter by making two segments 
of the integration
contour pass through the pole at $Q_1$. It is easy to verify that the residues picked up
at $Q_1$ from these two segments have opposite sign. Therefore they cancel each other
and again the results of integration over the contours in Fig.~\ref{f2cut}(b) and \ref{f3cut}(b)
are identical for real $p^0$, 
showing that the contribution to the imaginary part of the amplitude vanishes.
Therefore the only possible contribution to the imaginary part can come from the 
$Q_4\to Q_1$ limit, i.e. at the pinch singularity.
\item {\bf The contribution to the imaginary part of the amplitude from the pinch singularities
is given by Cutkosky rules.}  To verify this in this example we can deform the integration
contour through the pole at $Q_4$ to make it into a contour along the imaginary axis and 
a contour around $Q_4$ in
all cases shown in Figs.~\ref{f2cut} and \ref{f3cut}. The resulting contour along the
imaginary axis is clearly invariant
under complex conjugation and hence does not contribute to the imaginary part.
Therefore we only need to evaluate the residue at the pole $Q_4$.  For simplicity
let us consider the case where the spatial component of the external momentum
vanishes, i.e. $\vec p=0$. In this case the residue at $Q_4$ is easily evaluated and the
result for Fig.~\ref{f2cut} is given by
\ben \label{ei2-}
&& -B  \int{d^{D-1}k\over (2\pi)^{D-1}}
\exp\left[A \left(p^0 - \sqrt{\vec k^2+m^2}\right)^2 - A (\vec k^2 + m^2)
\right] \Theta \left({\rm Re}(p^0) - \sqrt{\vec k^2+m^2}\right)
\nonumber \\ && 
\left(2  \sqrt{\vec k^2 + m^2} \right)^{-1}
(p^0)^{-1}
 \left\{ 2\sqrt{\vec k^2 + m^2} - p^0\right\}^{-1}\, .
\een
The step function $\Theta$ reflects the fact that if 
$Q_4$ lies on the left of the origin then we do not
pick up any residue from the
pole at $Q_4$ while making the contour lie along the imaginary axis.  This contribution
looks real, but that is deceptive since the $\left\{ 2\sqrt{\vec k^2 + m^2} - p^0\right\}^{-1}$
can become singular and has to be defined by taking the $p^0\to M$ limit from
the first quadrant. This is achieved by replacing $p^0$ by $M+i\eps$ and defining the
amplitude by taking the $\eps\to 0^+$ limit. Since all other  quantities in the integrand are
non-singular in this limit, we may express \refb{ei2-} as
\ben \label{ei2+}
&& -B  \int{d^{D-1}k\over (2\pi)^{D-1}}
\exp\left[A \left(M- \sqrt{\vec k^2+m^2}\right)^2 - A (\vec k^2 + m^2)
\right] \Theta \left(M - \sqrt{\vec k^2+m^2}\right)
\nonumber \\ && 
\left(2  \sqrt{\vec k^2 + m^2} \right)^{-1}
(p^0)^{-1}
 \left\{ 2\sqrt{\vec k^2 + m^2} - M -i\eps\right\}^{-1}\, .
\een
The difference between \refb{ei2+} and its complex conjugate,
which is also the contribution to $M^2 -  (M^2)^*$, is given by
\ben \label{ei2++}
&& - i\, B  \int{d^{D-1}k\over (2\pi)^{D-2}}
\exp\left[A \left(M- \sqrt{\vec k^2+m^2}\right)^2 - A (\vec k^2 + m^2)
\right] \Theta \left(M - \sqrt{\vec k^2+m^2}\right)
\nonumber \\ && 
\left(2  \sqrt{\vec k^2 + m^2} \right)^{-1}
(p^0)^{-1}
\delta\left(2\sqrt{\vec k^2 + m^2} - M\right)\, .
\een
By rewriting this as
\ben
&& - i B  \int{d^{D}k\over (2\pi)^{D}}
\exp[-A \{k^2+m^2\} -
A \{(p-k)^2 + m^2\}]\, 2\pi \delta(k^2+m^2) \, \Theta(k^0) \nonumber \\
&& \hskip 2in 2\pi \delta((p-k)^2 + m^2) \, \Theta(p^0-k^0)
\, ,
\een
with all integrations performed along the real axis,
one can easily verify that this is in agreement with the Cutkosky rule for $T-T^\dagger$
stated in \S\ref{scut}.
\end{enumerate}

\end{document}